\documentclass[aps,rmp,superscriptaddress,
twocolumn,longbibliography]{revtex4-1} 
\usepackage[T1]{fontenc}
\usepackage{array}
\usepackage{float}
\usepackage{adjustbox}
\usepackage{booktabs}
\usepackage{multirow}
\usepackage{amsmath}
\usepackage{amssymb}
\usepackage{graphicx}
\usepackage{MnSymbol}
\usepackage{color}
\usepackage[colorlinks=true, urlcolor=blue, citecolor=blue,linkcolor=blue,citebordercolor={1 0 0},linkbordercolor={0 0 1}]{hyperref}
\usepackage{subeqnarray}
\usepackage{verbatim}
\usepackage{todonotes}
\usepackage{tabularx}
\usepackage{braket}
\usepackage{bbold}
\usepackage{comment}

\usepackage[toc,page]{appendix}

\newcolumntype{L}[1]{>{\hsize=#1\hsize\raggedright\arraybackslash}X}%
\newcolumntype{R}[1]{>{\hsize=#1\hsize\raggedleft\arraybackslash}X}%
\newcolumntype{C}[1]{>{\hsize=#1\hsize\centering\arraybackslash}X}%

\newtheorem{theorem}{Theorem}

\newtheorem{lem}[theorem]{Lemma}

\newtheorem{definition}[theorem]{Definition}
\newtheorem{defn}[theorem]{Definition}

\def\be{\begin{equation}}
\def\ee{\end{equation}}
\newcommand{\dbra}[1]{\llangle #1|}
\newcommand{\dket}[1]{|#1 \rrangle}
\newcommand{\dbraket}[2]{\llangle #1|#2\rrangle}
\newcommand{\ketbra}[2]{| #1\rangle\langle #2|}

\newcommand{\tr}{\mathrm{Tr}}

\newcommand{\norm}[1]{\left\lVert#1\right\rVert}
\newcommand{\expval}[2]{\ensuremath{\langle #1 \rangle_{#2}}}
\makeatletter
\renewcommand{\eqref}[1]{Eq.~(\ref{#1})} 
\newcommand{\figref}[1]{Fig.~\ref{#1}} 
\newcommand{\tabref}[1]{Table~\ref{#1}} 
\newcommand{\secref}[1]{Sec.~\ref{#1}} 
\newcommand{\doublehat}[1]{\hat{\hat{#1}}}
\newcommand\numberthis{\addtocounter{equation}{1}\tag{\theequation}}
\DeclareMathOperator*{\argmax}{arg max}

\newcommand{\printfnsymbol}[1]{%
 \textsuperscript{\@fnsymbol{#1}}%
}
\makeatother

\usepackage[english]{babel}

\newcommand{\utchem}{Chemical Physics Theory Group, Department  of  Chemistry,  University  of  Toronto,  Toronto,  Ontario  M5G 1Z8,  Canada}
\newcommand{\utcomp}{Department  of  Computer Science,  University  of  Toronto,  Toronto,  Ontario  M5S 2E4,  Canada}
\newcommand{\vectorinst}{Vector  Institute  for  Artificial  Intelligence,  Toronto,  Ontario  M5S  1M1,  Canada}
\newcommand{\cifar}{Canadian  Institute  for  Advanced  Research,  Toronto,  Ontario  M5G  1Z8,  Canada}
\newcommand{\cqt}{Centre for Quantum Technologies, National University of Singapore 117543, Singapore}
\newcommand{\mitphys}{Department of Physics, Massachusetts Institute of Technology, Cambridge, MA 02139, USA}
\newcommand{\harvardphys}{Department of Physics, Harvard University, Cambridge, MA 02138, USA}
\newcommand{\harvardchem}{Department of Chemistry and Chemical Biology, Harvard University, Cambridge, MA 02138, USA}
\newcommand{\mitreslab}{Research Laboratory of Electronics, Massachusetts Institute of Technology, Cambridge, MA 02139, USA}
\newcommand{\qols}{QOLS, Blackett Laboratory, Imperial College London SW7 2AZ, UK}
\newcommand{\boehringer}{current address: Boehringer Ingelheim, Amsterdam, Netherlands}
\newcommand{\majulab}{MajuLab, CNRS-UNS-NUS-NTU International Joint Research Unit UMI 3654, Singapore}
\newcommand{\nie}{National Institute of Education and Institute of Advanced Studies, Nanyang Technological University 637616, Singapore}

\begin{document}
\title{Noisy intermediate-scale quantum (NISQ) algorithms
}
\begin{abstract}
A universal fault-tolerant quantum computer that can solve efficiently problems such as integer factorization and unstructured database search requires millions of qubits with low error rates and long coherence times. While the experimental advancement towards realizing such devices will potentially take decades of research, noisy intermediate-scale quantum (NISQ) computers already exist. These computers are composed of hundreds of noisy qubits, i.e. qubits that are not error-corrected, and therefore perform imperfect operations in a limited coherence time. In the search for quantum advantage with these devices, algorithms have been proposed for applications in various disciplines spanning physics, machine learning, quantum chemistry and combinatorial optimization. The goal of such algorithms is to leverage the limited available resources to perform classically challenging tasks. In this review, we provide a thorough summary of NISQ computational paradigms and algorithms. We discuss the key structure of these algorithms, their limitations, and advantages. We additionally provide a comprehensive overview of various benchmarking and software tools useful for programming and testing NISQ devices.
\end{abstract}

\author{Kishor Bharti}
\thanks{These authors contributed equally to this work.\\ 
\urlstyle{same}
\url{kishor.bharti1@gmail.com}\,
\url{a.cervera.lierta@gmail.com}\, \url{thihakyaw.phy@gmail.com}}
\affiliation{\cqt}
\author{Alba Cervera-Lierta}
\thanks{These authors contributed equally to this work.\\ 
\urlstyle{same}
\url{kishor.bharti1@gmail.com}\,
\url{a.cervera.lierta@gmail.com}\, \url{thihakyaw.phy@gmail.com}}
\affiliation{\utcomp}
\affiliation{\utchem}
\author{Thi Ha Kyaw}
\thanks{These authors contributed equally to this work.\\ 
\urlstyle{same}
\url{kishor.bharti1@gmail.com}\,
\url{a.cervera.lierta@gmail.com}\, \url{thihakyaw.phy@gmail.com}}
\affiliation{\utcomp}
\affiliation{\utchem}
\author{Tobias Haug}
\affiliation{\qols}
\author{Sumner Alperin-Lea}
\affiliation{\utchem}
\author{Abhinav Anand}
\affiliation{\utchem}
\author{Matthias Degroote}
\affiliation{\utcomp}
\affiliation{\utchem}
\affiliation{\boehringer}
\author{Hermanni Heimonen}
\affiliation{\cqt}
\author{Jakob S. Kottmann}
\affiliation{\utcomp}
\affiliation{\utchem}
\author{Tim Menke}
\affiliation{\harvardphys}
\affiliation{\mitreslab}
\affiliation{\mitphys}
\author{Wai-Keong Mok}
\affiliation{\cqt}
\author{Sukin Sim}
\affiliation{\harvardchem}
\author{Leong-Chuan Kwek}
\email{cqtklc@gmail.com}
\affiliation{\cqt}
\affiliation{\majulab}
\affiliation{\nie}
\author{Al\'an Aspuru-Guzik}
\email{alan@aspuru.com}
\affiliation{\utcomp}
\affiliation{\utchem}
\affiliation{\vectorinst}
\affiliation{\cifar}

\date{\today}

\maketitle

\tableofcontents

\section{Introduction}\label{ch:intro}

Quantum computing originated in the eighties when physicists started to speculate about computational models that integrate the laws of quantum mechanics \cite{kaiser2011hippies}. Starting with the pioneering works of Benioff and Deutsch, which involved the study of quantum Turing machines and the notion of universal quantum computation \cite{benioff1980computer,deutsch1985quantum}, the field continued to develop towards its natural application: the simulation of quantum systems \cite{feynman1982simulating,lloyd1996universal,manin1980computable}. Arguably, the drive for quantum computing took off in 1994 when Peter Shor provided an efficient quantum algorithm for finding prime factors of composite integers, rendering most classical cryptographic protocols unsafe \cite{shor1994algorithms}. Since then, the study of quantum algorithms has matured as a sub-field of quantum computing with applications in search and optimization, machine learning, simulation of quantum systems and cryptography~\cite{montanaro2016quantum}.

In the last forty years, many scientific disciplines have converged towards the study and development of quantum algorithms and their experimental realization. Quantum computers are, from the computational complexity perspective, fundamentally different tools available to computationally intensive fields. The implementation of quantum algorithms requires that the minimal quantum information units, \textit{qubits}, are as reliable as classical bits. Qubits need to be protected from environmental noise that induces decoherence but, at the same time, their states have to be controlled by external agents. This control includes the interaction that generates entanglement between qubits and the measurement operation that extracts the output of the quantum computation. It is technically possible to tame the effect of noise without compromising the quantum information process by developing quantum error correction (QEC) protocols \cite{shor1995scheme, lidar2013quantum,terhal2015quantum}. Unfortunately, the overhead of QEC in terms of the number of qubits is, at the present day, still far from current experimental capabilities. To achieve the goal of fault-tolerant quantum computation, the challenge is to scale up the number of qubits with sufficiently high qubit quality and fidelity in operations such as quantum gate implementation and measurement~\cite{aharonov2008fault,knill1998resilient,kitaev2003fault}. As the system size grows, it becomes highly challenging to contain the errors associated with cross-talk and measurements below the required error-correction threshold.

Most of the originally proposed quantum algorithms require millions of physical qubits to incorporate these QEC techniques successfully, realizing the daunting goal of building a fault-tolerant quantum computer may take decades. Existing quantum devices contain on the order of 100 phyisical qubits. They are sometimes denoted as ``\textit{Noisy Intermediate-Scale Quantum} (NISQ)'' devices~\cite{preskill2018quantum}, meaning their qubits and quantum operations are not QEC and, therefore, imperfect. One of the goals in the NISQ era is to extract the maximum quantum computational power from current devices while developing techniques that may also be suited for the long-term goal of the fault-tolerant quantum computation~\cite{terhal2015quantum}.

\subsection{Computational complexity theory in a nutshell}\label{ch1:subsec:non-complexity}

The definition of a new computational paradigm opens the window to tackle those problems that are inefficient with the existing ones. New computational complexity classes have been recognized through the study of quantum computing, and proposed algorithms and goals have to be developed within well-known mathematical boundaries.

In this review, we will often use some computational complexity-theoretic ideas to establish the domain and efficiency of the quantum algorithms covered. For this reason, we provide in this subsection a brief synopsis for a general audience and refer to~\cite{arora2009computational} for a more comprehensive treatment.

Complexity classes are groupings of problems by hardness, namely the scaling of the cost of solving the problem with respect to some resource, as a function of the ``size'' of an instance of the problem. The most well-known ones being
described informally here. \textit{i)} $\textrm{P}$: problems that can be solved in time polynomial with respect to input size by a deterministic classical computer. \textit{ii)} $\textrm{NP}$: a problem is said to be in $\textrm{NP}$, if the problem of verifying the correctness of a proposed solution lies in $\textrm{P}$, irrespective of the difficulty of obtaining a correct solution. \textit{iii)} $\textrm{PH}$: stands for Polynomial Hierarchy. This class is a generalization of $\textrm{NP}$ in the sense that it contains all the problems which one gets if one starts with a problem in the class $\textrm{NP}$ and adds additional layers of complexity using quantifiers, i.e. there exists $\left(\exists\right)$ and for all $\left(\forall\right)$. As we add more quantifiers to a problem, it becomes more complex and is placed higher up in the polynomial hierarchy.  Let us denote the classes in $\textrm{PH}$ by $\Sigma_i$ such that $\textrm{PH}=\cup_i \Sigma_i$. We have $\Sigma_1 = \textrm{NP}$. The class  $\Sigma_i$ in $\textrm{PH}$ can be interpreted in the context of two-player games where problems correspond to asking whether there exists a winning strategy in $\frac{i}{2}$ rounds for the player $1$ in a game. Here, one can interpret the quantifiers by asking whether there exists a move $k_1$, such that no matter what move $k_2$ is played, there exists a move $k_3$, and so on for $\frac{i}{2}$ rounds such that player $1$ wins the two-player game. With increasing $i$, one would expect the problem to become more complex and hence  $\Sigma_i \subseteq  \Sigma_{i+1}$. \textit{iv)} $\textrm{BPP}$: stands for Bounded-error Probabilistic Polynomial-time. A problem is said to be in $\textrm{BPP},$ if it can be solved in time polynomial in the input size by a probabilistic classical computer. \textit{v)} $\textrm{BQP}$: stands for Bounded-error Quantum Polynomial-time. Such problems can be solved in time polynomial in the input size by a quantum computer. \textit{vi)} $\textrm{PSPACE}$: stands for Polynomial Space. The problems in $\textrm{PSPACE}$ can be solved in space polynomial in the input size by a deterministic classical computer. Each class in $\textrm{PH}$ is contained in $\textrm{PSPACE}$. However, it is not known whether $\textrm{PH}$ is equal to $\textrm{PSPACE}$. 
\textit{vii)} $\textrm{EXPTIME}$: stands for Exponential Time. The problems in $\textrm{EXPTIME}$ can be solved in time exponential in the input size by a deterministic classical computer.
\textit{viii)} $\textrm{QMA}$: stands for Quantum Merlin Arthur and is the quantum analog of the complexity class $\textrm{NP}.$ A problem is said to be in $\textrm{QMA}$, if given a ``yes" as an answer, the solution can be verified in time polynomial (in the input size) by a quantum computer.

\begin{figure}[t!]
\centering
\includegraphics[width=\columnwidth]{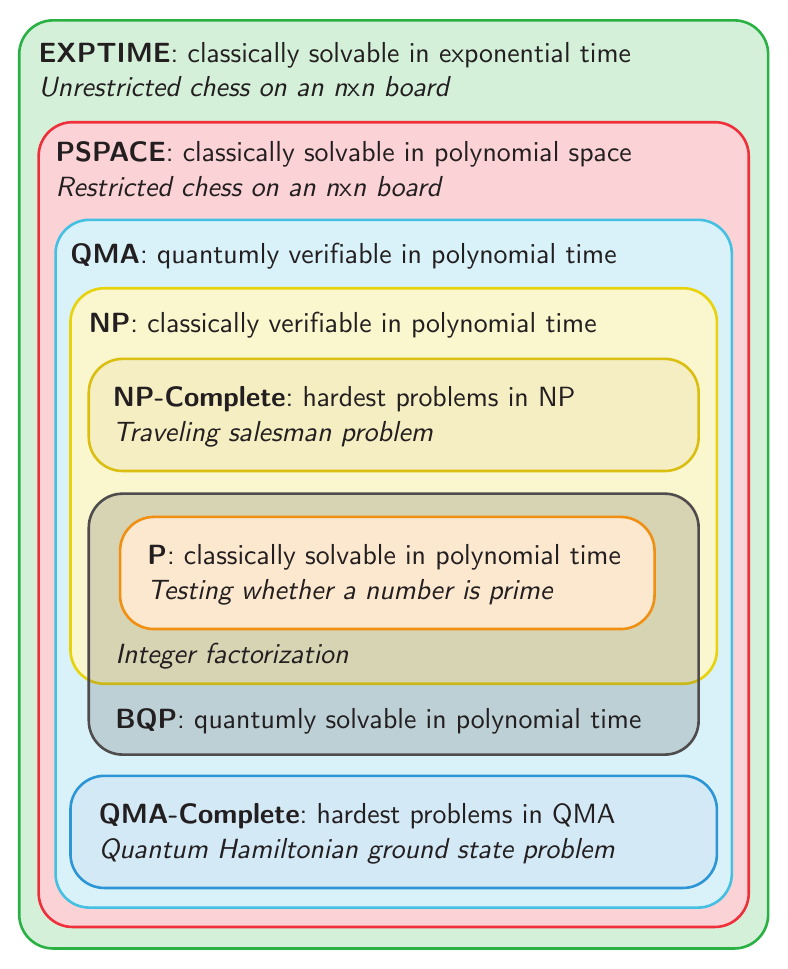}
\caption{An illustrative picture of some relevant complexity classes together with a problem examples. For the chess example, the word ``restricted'' refers to a polynomial upper bound on the number of moves. The containment relations are suggestive. Some of them have not been mathematically proven, being a well-known open problem whether $\textrm{P}$ is equal to $\textrm{NP}$.}
\label{Fig:Complexity}
\end{figure}

Widely believed containment relations for some of the complexity classes are shown in a schematic way in Fig.~\ref{Fig:Complexity}.

To understand the internal structure of complexity classes, the idea of ``reductions'' can be quite useful. One says that problem $A$ is reducible to problem $B$ if a method for solving $B$ implies a method for solving $A$; one denotes the same by $A\leq B$. It is a common practice to assume the reductions as polynomial-time reductions. Intuitively, it could be thought as solving $B$ is at least as difficult as solving $A$. Given a class $C$, a problem $X$ is said to be \textit{$C$-hard} if every problem in class $C$ reduces to $X$. We say a problem $X$ to be \textit{$C$-complete} if $X$ is $C$-hard and also a member of $C$. The $C$-complete problems could be understood as capturing the difficulty of class $C$, since any algorithm which solves one $C$-complete problem can be used to solve any problem in $C$.

A canonical example of a problem in the class $\textrm{BQP}$ is integer factorization, which can be solved in polynomial time by a quantum computer using Shor's factoring algorithm~\cite{shor1994algorithms}. However, no classical polynomial-time algorithm is known for the aforementioned problem. Thus, the integer factorization problem is in BQP, but not believed to be in P~\cite{arora2009computational}. While analyzing the performance of algorithms, it is prudent to perform complexity-theoretic sanity checks. For example, though quantum computers are believed to be powerful, they are not widely expected to be able to solve $\textrm{NP}$-Complete problems, such as the travelling-salesman problem, in polynomial time. The quantum algorithms, however, could provide a speedup with respect to classical algorithms for $\textrm{NP}$-Complete problems.

\subsection{Experimental progress}\label{subsec:experimental_progress}

Here we present a somewhat biased and not exhausted summary of very recent quantum experiments.
Interested readers should consult \cite{acin2018quantum} and references therein for more information about various quantum computing architectures.

Experimental progress in quantum computation can be measured by different figures of merit. The number of physical qubits available must exceed a certain threshold to solve problems beyond the capabilities of a classical computer. However, there exist several classical techniques capable of efficiently simulating certain quantum many-body systems. The success of some of these techniques, such as Tensor Networks \cite{verstraete2008matrix,orus2014practical}, rely on the efficient representation of states that are not highly entangled~\cite{vidal2003efficient,vidal2004efficient}. 
With the advent of universal quantum computers, one would expect to be able to generate and manipulate these highly entangled quantum states.

Hence, one imminent and practical direction towards demonstrating quantum advantage over classical machines consist of focusing on a region of the Hilbert space whose states can not be represented efficiently with classical methods. Alternatively, one might tackle certain computational tasks which are believed to be intractable with any classical computer, as the ones belonging to only quantum complexity classes.

Two recent experimental ventures exhibit this focus.
In 2019, the Google AI Quantum team implemented an experiment with the 53-qubit Sycamore chip \cite{arute2019quantum} in which single-qubit gate fidelities of $99.85\%$ and two-qubit gate fidelities of $99.64\%$ were attained on average. 
Quantum advantage was demonstrated against the best current classical computers in the task of sampling the output of a pseudo-random quantum circuit.

An additional quantum advantage experiment was carried out by Jian-Wei Pan's group using a \textit{Jiuzhang} photonic quantum device performing Gaussian boson sampling (GBS) with $50$ indistinguishable single-mode squeezed states \cite{zhong2020quantum}.
Here, quantum advantage was seen in sampling time complexity of a \textit{Torontonian} of a matrix \cite{quesada2018gaussian}, which scales exponentially with the photon clicks output.  The Torontonian is a matrix function that determines the probability distribution of measurement outcomes, much like the permanent and \textit{Hafnian} in other boson sampling models. Intuitively speaking, while the total number of perfect matchings in a bipartite graph is given by the permanent, the Hafnian corresponds to the total number of perfect matchings in an arbitrarily given graph. Moreover, while the Hafnian is used in experiments counting the number of photons in each mode, the  Torontonian corresponds to the case where one detects whether there are photons in each mode (see~\secref{sec:GBS} for more details about GBS and the related terms).

There are several quantum computing platforms that researchers are actively developing at present in order to achieve scalable and practical universal quantum computers. 
By ``universal'', it is meant that such a quantum computer can perform native gate operations that allow it to easily and accurately approximate any unitary gate (see \secref{sec:compilers} for more details).
Two of the most promising platforms, superconducting circuits and quantum optics, have already been mentioned;
In addition to these, trapped-ion devices are also leading candidates. For instance, major achievements are recent high-fidelity entangling gates reported by the Oxford group~\cite{hughes2020benchmarking}, all-to-all connectivity achieved by IonQ~\cite{nam2020ground}, and transport and reordering capabilities in 2D trap array by the Boulder group~\cite{wan2020ion}.
In the last example, besides facilitating efficient transport of ions and quantum information exchange, the 2D architecture can be viewed towards attaining much more sophisticated quantum error correction code or surface code \cite{lidar2013quantum}, the smallest of it has been realized in superconducting qubit setup \cite{corcoles2015demonstration}. 

Scientists and engineers are also developing hybrid quantum computing platforms trying to achieve similar feats described above. These devices might not necessarily possess universal quantum gate sets, as many are built to solve specific problems.
Notably, coherent Ising machines \cite{utsunomiya2011mapping,wang2013coherent, marandi2014network,mcmahon2016fully,inagaki2016coherent} based on mutually coupled optical parametric oscillators are promising and have shown success in solving instances of hard combinatorial optimization problems.
Recently, it has been shown that the efficiency of these machines can be improved with error detection and correction feedback mechanisms \cite{kako2020coherent}.
The reader is advised to refer to the recent review article \cite{yamamoto2020coherent} for an in-depth discussion about coherent Ising machines. Quantum annealing~\cite{finnila1994quantum,kadowaki1998quantum} has  been another prominent approach towards quantum advantage in the NISQ era~\cite{perdomo2018opportunities,bouland2020prospects,hauke2020perspectives}. Refer to \secref{subsec:quantum-annealing} for more details about quantum annealing.

Lastly, unlike the past decades of academic research in lab-based quantum technologies, we are witnessing the emergence of cloud quantum computers with which anyone with internet access can now control and manipulate delicate qubits and perform quantum computations on the fly. 
Presently, IBM Quantum is leading the effort followed by Rigetti Computing and Xanadu Quantum Cloud.

\subsection{NISQ and near-term} \label{ch1:subsec:NISQ_near-term}

The experimental state-of-the-art and the demand for QEC have encouraged the development of innovative algorithms capable of reaching the long-expected \textit{quantum advantage}. This goal can be defined as a purpose-specific computation that involves a quantum device and that can not be performed classically with a reasonable amount of time and energy resources. 
The term \textit{near-term} quantum computation has been coined to cluster all these quantum algorithms specially developed to be run on current quantum computing hardware or those which could be developed in the next few years,. 
It is important to note that NISQ is a hardware-focused definition, and does not necessarily imply a temporal connotation. NISQ devices can implement the model of quantum circuits, in which all gates adhere to the topology of a specified graph $G$, the nodes of which correspond to qubits. The gates typically operate on one or two qubits. Because each gate operation involves a certain amount of noise, NISQ algorithms are naturally limited to shallow depths~\cite{barak2021classical}.
Near-term algorithms, however, refers to those algorithms designed for quantum devices available in the next few years and carries no explicit reference to the absence of QEC. The phrase ``near-term'' is subjective since different researchers may have other thoughts on how many years can be considered ``near-term''.
Predicting experimental progress is always challenging, and such predictions are influenced by human bias. Algorithms developed for near-term hardware may be unfeasible if hardware advancement does not match the algorithm's experimental requirements.

\subsection{Scope of the review}

This review aims to accomplish three main objectives. The first is to provide a proper compilation of the available algorithms suited for the NISQ era and which can deliver results in the near-term. We present a summary of the crucial tools and techniques that have been proposed and harnessed to design such algorithms. The second objective is to discuss the implications of these algorithms in various applications such as quantum machine learning (QML), quantum chemistry, and combinatorial optimization. Finally, the third objective is to give some perspectives on potential future developments given the recent quantum hardware progress.

Most of the current NISQ algorithms rely on harnessing the power of quantum computers in a hybrid quantum-classical arrangement. Such algorithms delegate the classically difficult part of some computation to the quantum computer and perform the other on some sufficiently powerful classical device. These algorithms update variationally the variables of a parametrized quantum circuit and hence are referred to as \textit{Variational Quantum Algorithms} (VQA) \cite{cao2019quantum,mcardle2020quantum, endo2020hybrid, cerezo2020variational}.

The first proposals of VQA were the Variational Quantum Eigensolver (VQE) \cite{peruzzo2014variational,mcclean2016theory,wecker2015progress}, originally proposed to solve quantum chemistry problems, and the Quantum Approximate Optimization Algorithm (QAOA) \cite{farhi2014quantum}, proposed to solve combinatorial optimization problems. These two algorithms may be thought of as the parents of the whole VQA family. While NISQ devices can arguably achieve quantum advantage for sampling problems, the corresponding question for the optimization problems remains unanswered~\cite{barak2021classical,barak2015beating}. It is important to mention that, as of now, there is no provable quantum advantage for VQA with NISQ devices~\cite{barak2021classical}. We cover the main VQA blocks in \secref{ch:building}.

Other quantum computing paradigms propose different kinds of algorithms. They are inspired and hybridized with analog approaches.These include quantum annealing, digital-analog quantum computation, Gaussian Boson Sampling and analog quantum computation. We present their fundamental properties in \secref{sec:non-variational}.  

In \secref{ch:lemon}, we examine the theoretical and experimental challenges faced by NISQ algorithms and the methods developed to best exploit them. We include the theoretical guarantees that some of these algorithms lay on as well as techniques to mitigate the errors coming from the use of noisy quantum devices. We also cover the possible trainability challenges that VQA have and how to map theoretical NISQ circuits to real hardware.
Section \ref{ch:application} presents the large variety of applications that NISQ algorithms introduce. Techniques to benchmark, compare and quantify current quantum devices performance are presented in \secref{sec:benchmark}. Like any other computational paradigm, quantum computing requires a language to establish human-machine communication. We explain different levels of quantum programming and provide a list of open-source quantum software tools in \secref{sec:software}. Finally, we conclude this review in \secref{ch:outlook} by highlighting the increasing community involvement in this field and by presenting the NISQ, near-term and long-term goals of quantum computational research.

\section{Building blocks of variational quantum algorithms}\label{ch:building}

\begin{figure*}[t!]
\centering
\includegraphics[width=\textwidth]{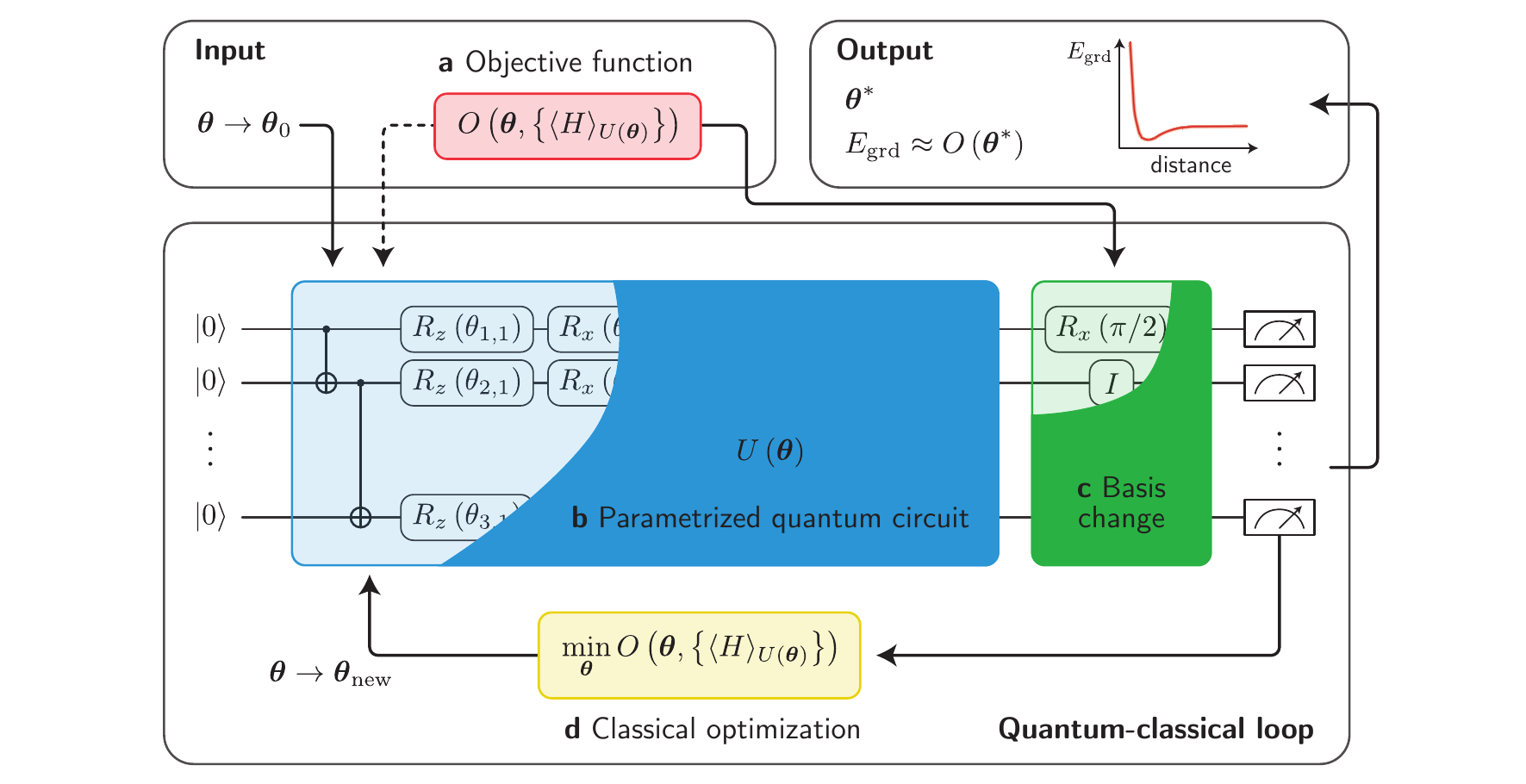}
\caption{Diagrammatic representation of a Variational Quantum Algorithm (VQA). A VQA workflow can be divided into four main components: \textit{a)} the objective function $O$ that encodes the problem to be solved; \textit{b)} the parameterized quantum circuit (PQC) $U$, which variables $\boldsymbol{\theta}$ are tuned to minimize the objective; \textit{c)} the measurement scheme, which performs the basis changes and measurements needed to compute expectation values that are used to evaluate the objective; and \textit{d)} the classical optimizer that minimizes the objective. The PQC can be defined heuristically, following hardware-inspired ans\"atze, or designed from the knowledge about the problem Hamiltonian $H$. 
Inputs of a VQA are the circuit ansatz 
$U(\boldsymbol\theta)$ and the initial parameter values $\boldsymbol\theta_0$. 
Outputs include optimized parameter values $\boldsymbol{\theta}^*$
and the minimum of the objective.}
\label{Fig:VQA_diag}
\end{figure*}

A VQA comprises several modular components that can be readily combined, extended and improved with developments in quantum hardware and algorithms. Chief among these are the objective function, the cost function to be variationally minimized; the parameterized quantum circuit (PQC), those unitaries whose parameters are manipulated in the minimization of objective; the measurement scheme, which extracts the expectation values needed to evaluate the objective; and the classical optimizer, the method used to obtain the optimal circuit parameters that minimize the objective.  
In the following subsections, we will define each of these pieces, presented diagrammatically in \figref{Fig:VQA_diag}.

\subsection{Objective function}\label{sec:objective_function}

The Hamiltonian is a quantum operator that encodes information about a given physical system, such as a molecule or a spin chain. Its expectation value yields the energy of a quantum state, which is often used as the minimization target of a VQA, i.e. obtaining the Hamiltonian ground state. Other problems not related to real physical systems can also be encoded into a Hamiltonian form, thereby opening a path to solve them on a quantum computer. Hamiltonian operators are not all that can be measured on quantum devices; in general, any expectation value of a function written in an operational form (i.e. decomposed or encoded into a quantum operator) can be also be evaluated on a quantum computer.
After the Hamiltonian or operator of a problem has been determined, it must be decomposed into a set of particular operators that can be measured with a quantum processor. Such a decomposition, which is further discussed in \secref{sec:pauli_strings}, is an important step of many quantum algorithms in general and of VQA in particular.

Within a VQA, one has access to measurements on qubits whose outcome probabilities are determined by the prepared quantum state. 
Let us consider only measurements on individual qubits in the standard computational basis and denote the probability to measure qubit $q$ in state $\ket{0}$ by $p_0^q$, where the qubit label $q$ will be omitted whenever possible. 
The central element of a VQA is a parametrized cost or \textit{objective function} $O$ subject to a classical optimization algorithm, 
    $\min_{\boldsymbol{\theta}} O\left(\boldsymbol{\theta}, \left\{\boldsymbol{p}_0\left(\boldsymbol{\theta}\right)\right\}\right)$.
The objective function $O$ and the measurement outcomes $\boldsymbol{p}_0$ of one or many quantum circuit evaluations depend on the set of parameters $\boldsymbol{\theta}$.

In practice it is often inconvenient to work with the probabilities of the measurement outcomes directly when evaluating the objective function. Higher level formulations employ expectation value of the Hamiltonian $H$ of the form
\begin{align}
    \expval{H}{U\left(\boldsymbol{\theta}\right)} \equiv \bra{0}U^\dagger\left(\boldsymbol{\theta}\right) H U\left(\boldsymbol{\theta}\right) \ket{0},
    \label{eq:def_expectationvalue}
\end{align}
describing measurements on the quantum state generated by the unitary $U\left(\boldsymbol{\theta}\right)$, instead of using the probabilities for the individual qubit measurements directly. Arbitrary observables can be decomposed into basic measurements of the so-called Paulis strings, which can be evaluated in the computational basis, as explained below and in \secref{sec:measurement}.  
Restricting ourselves to expectation values instead of pure measurement probabilities, the objective function becomes
\begin{align}
    \min_{\boldsymbol{\theta}} O\left(\boldsymbol{\theta}, \left\{\expval{H}{U\left(\boldsymbol{\theta}\right)}\right\}\right). 
\end{align}

This formulation often allows for more compact definitions of the objective function. For the original VQE~\cite{peruzzo2014variational} and QAOA~\cite{farhi2014quantum} it can, for example, be described as a single expectation value $\min_{\boldsymbol{\theta}} \expval{H}{U\left(\boldsymbol{\theta} \right)}$,

where the differences solely appear in the specific form and construction of the qubit Hamiltonian. 

The choice of the objective function is crucial in a VQA to achieve the desired convergence. Vanishing gradient issues during the optimization, known as barren plateaus, are dependent on the cost function used~\cite{cerezo2020cost} (see \secref{sec:BP} for details).

\subsubsection{Pauli strings}\label{sec:pauli_strings}

To extract the expectation value of the problem Hamiltonian, 
it is sufficient to express it as a linear combination of primitive tensor products of Pauli matrices $\hat{\sigma}_{x},\hat{\sigma}_{y},\hat{\sigma}_{z}$. 
We refer to these tensor products as \textit{Pauli strings}
$\hat{P}=\bigotimes_{j=1}^n \hat{\boldsymbol{\sigma}}$, where $n$ is the number of qubits, $\hat{\boldsymbol{\sigma}}\in\{\hat{I},\hat{\sigma}_{x},\hat{\sigma}_{y},\hat{\sigma}_{z}\}$ and $\hat{I}$ the identity operator. Then, the Hamiltonian can be decomposed as
\begin{equation}
    H = \sum_{k=1}^{M} c_{k}\hat{P}_k,
    \label{eq:Pauli_string}
\end{equation}
where $c_k$ is a complex coefficient of the $k$-th Pauli string and the number of Pauli strings $M$ in the expansion depends on the operator at hand.
An expectation value in the sense of \eqref{eq:def_expectationvalue} then naturally decomposes into a set of expectation values, each defined by a single Pauli string
\begin{equation}
   \langle H \rangle_U = \sum_{k=1}^{M} c_{k}\langle \hat{P}_k\rangle_{U}\,.
   \label{eq:Pauli_string_exp}
\end{equation}

Examples of Hamiltonian objectives include molecules (by means of some fermionic transformation to Pauli strings, as detailed in \secref{sec:chemistry}), condensed matter models written in terms of spin chains, or optimization problems encoded into a Hamiltonian form (see \secref{sec:optimization}).

\subsubsection{Fidelity}

Instead of optimizing in respect to the expectation value of an operator, several VQAs require a subroutine to optimize the state obtained from the PQC $U\left(\boldsymbol{\theta}\right)$, $\ket{\Psi}_{U\left(\boldsymbol{\theta}\right)}$ in respect to a specific target state $\ket{\Psi}$. 
A commonly used cost function is the fidelity between the PQC and the target state
\begin{equation}
    F\left(\Psi, \Psi_{U\left(\boldsymbol{\theta}\right)}\right) \equiv \lvert\langle \Psi |\Psi_{U\left({\boldsymbol{\theta}}\right)}\rangle\rvert^{2},
    \label{eq:fidelity_target}
\end{equation}
which is equivalent to the expectation value over the projector $\hat{\Pi}_\Psi=\ket{\Psi}\bra{\Psi}$. The state preparation objective is then the minimization of the infidelity $1-F\left(\Psi, \Psi_{U\left(\boldsymbol{\theta}\right)}\right)$ or just the negative fidelity
\begin{align}
    \max_{\boldsymbol{\theta}}F\left(\Psi, \Psi_{U\left(\boldsymbol{\theta}\right)}\right) = \min_{\boldsymbol{\theta}} \left(- \expval{\hat{\Pi}_\Psi}{U\left(\boldsymbol{\theta}\right)}\right).
\end{align}
If we know the efficient circuit $U_{\Psi}$ that prepares the target state $|\Psi\rangle$, we can compute the fidelity with the inversion test by preparing the quantum state $U^\dagger_{\Psi}|\Psi_{U\left({\boldsymbol{\theta}}\right)}\rangle$ and measuring the projector into the zero state $\hat{\Pi}_0=|0\rangle^{\otimes n}\langle0|^{\otimes n}$ with the fidelity given by
$F\left(\Psi, \Psi_{U\left(\boldsymbol{\theta}\right)}\right) = \expval{\hat{\Pi}_0}{U_{\Psi}^\dagger U\left(\theta\right)}$~\cite{havlivcek2019supervised}. 
If one wants to avoid optimizing in respect to a projector onto a single state, one can instead use a local observable that also becomes maximal for the target state, namely $\hat{O}=\frac{1}{N}\sum_{k=1}^N\ket{0_k}\bra{0_k}\otimes I_{\bar{k}}$, where $I_{\bar{k}}$ is the identity matrix for all qubits except $k$ and $\ket{0_k}$ is the zero state for qubit $k$ ~\cite{cerezo2020cost,barison2021efficient}.
Alternatively, one can use randomized measurements to measure the fidelity $\text{Tr}(\rho_1\rho_2)$ of two density matrices $\rho_1$, $\rho_2$~\cite{van2012measuring,elben2019statistical,elben2020cross}. First, one selects $m$ unitaries $\{V_k\}_k$, which are chosen as tensor product of Haar random unitaries over the local $d$-dimensional subspace. These unitaries are applied on each quantum state $\rho_i=V_k\rho V_k^\dagger$ and $\rho_i$ is sampled in the computational basis. Then, one estimates the probability $P_{V_k}^{(i)}(\mathbf{s})$ of measuring the computational basis state $\mathbf{s}$ for each quantum state $\rho_i$ and unitary $V_k$. The fidelity is given by
\begin{align}
\tr[{\rho_1\rho_2}]= \label{eq:rand_measl} 
\frac{d^{N}}{m}\sum_{k=1}^m\sum_{\mathbf{s},\mathbf{s}'}(-d)^{-\mathcal{D}[\mathbf{s},\mathbf{s}']}\;P_{V_k}^{(1)}(\mathbf{s})P_{V_k}^{(2)}(\mathbf{s}').
\end{align}
where $\mathcal{D}[\mathbf{s},\mathbf{s}']$ is the Hamming distance between sampled computational basis states $\mathbf{s}$ and $\mathbf{s}'$.
The number of measurements scales exponentially with the number of qubits, however the scaling is far better compared to state tomography. Importance sampling has been proposed to substantially reduce the number of samples necessary~\cite{rath2021importance}.

Objective formulations over fidelities are prominent within state preparation algorithms in quantum optics~\cite{krenn2020computer, krenn2020conceptual, kottmann2020quantum}, excited state algorithms~\cite{lee2018generalized, kottmann2020feasible} and QML~\cite{cheng2018information, benedetti2019generative, perez2020data,huang2021power} (see also \secref{sec:QML} for more references and details).
In these cases, the fidelities are often defined in respect to computational basis states $e_{i}$, such that $F_{e_{i}}=\lvert\langle\Psi\left(\boldsymbol{\theta}\right)\vert{e_{i}}\rangle\rvert^2$.

\subsubsection{Other objective functions}

Hamiltonian expectation values are not the only objective functions that are used in VQAs. Any cost function that is written in an operational form can constitute a good choice. One such example is the conditional value-at-risk (CVaR). Given the set of energy basis measurements $\{ E_1, \ldots E_M \}$ arranged in a non-decreasing order, instead of using the expectation value from \eqref{eq:def_expectationvalue} as the objective function, it was proposed  to use \cite{barkoutsos2020improving}
\begin{equation}
    \text{CVaR}(\alpha) = \frac{1}{\lceil \alpha M \rceil} \sum_{k=1}^{\lceil \alpha M \rceil} E_k\,,
\end{equation}
which measures the expectation value of the $\alpha$-tail of the energy distribution. Here, $\alpha \in (0,1]$ is the confidence level. The CVaR$(\alpha)$ can be thought of as a generalization of the sample mean ($\alpha = 1$) and the sample minimum ($\alpha \to 0$). 

Another proposal \cite{li2020quantum} is to use the Gibbs objective function
\begin{equation}
    G = - \ln \langle e^{-\eta H} \rangle,
\end{equation}
which is the cumulant generating function of the energy. The variable $\eta > 0$ is a hyperparameter to be tuned. For small $\eta$, the Gibbs objective function reduces to the mean energy in \eqref{eq:def_expectationvalue}. Since both the CVaR and the Gibbs objective function can be reduced to the mean energy for suitable limits of the hyperparameters ($\alpha \to 1$ and $\eta \to 0$ respectively), their performances are guaranteed to be at least as good as using the mean energy $\expval{H}{}$. Empirically, by tuning the hyperparameters, both measures have been shown to outperform $\expval{H}{}$ for certain combinatorial optimization problems \cite{barkoutsos2020improving,li2020quantum}. 

\subsection{Parameterized quantum circuits}\label{sec:PQC}

Following the objective function, the next essential constituent of a VQA is the quantum circuit that prepares the state that best meets the objective. It is generated by means of a unitary operation that depends on a series of parameters, the PQC. In this subsection, we describe how this quantum circuit is defined and designed.

We define the state after application of the PQC as
\begin{equation}
\ket{\Psi\left(\boldsymbol{\theta}\right)} = U\left(\boldsymbol{\theta}\right)\ket{\Psi_{0}},
\label{eq:var_qcircuit}
\end{equation}
where $\boldsymbol{\theta}$ are the variational parameters and $\ket{\Psi_{0}}$ is some initial state.
Typically, $\ket{\Psi_{0}}$ is a product state with all qubits in the $|0\rangle$ state, i.e. $\ket{00\cdots 0} = \ket{0}^{\otimes n}$, where $n$ is the number of qubits. In some VQAs, it is convenient to prepare that state in a particular form before applying the PQC. The state preparation operation would then depend on some other unitary operation $P$ that may depend on variational parameters $\boldsymbol{\phi}$,
$\ket{\Psi_{0}} = P\left(\boldsymbol{\phi}\right)\ket{0}^{\otimes n}$.

One example are the quantum feature maps defined in \secref{sec:SL} that encode the data into the PQC.

Any known property about the final state can also be used to obtain the initial guess. For instance, if we expect that the final state solution will contain all elements of the computational basis, or if we want to exploit a superposition state to seed the optimization, an initial state choice may be
$P\ket{0}^{\otimes n} = H_{d}^{\otimes n}\ket{0}^{\otimes n}$,
where $H_{d}$ is the Hadamard gate.
Applied to all qubits, $H_{d}$ generates the even superposition of all basis states, i. e.
\begin{equation}
    |D\rangle = H_{d}^{\otimes n}\ket{0}^{\otimes n} = \frac{1}{\sqrt{{n}}}\sum_{i=1}^{n}|e_{i}\rangle,
    \label{eq:sp_H}
\end{equation}
where $\ket{e_{i}}$ are the computational basis states.

In quantum chemistry algorithms, the initial state usually corresponds to the Hartree-Fock approximation 
(see \secref{sec:chemistry} for details). The choice of a good initial state will allow the VQA to start the search in a region of the parameter space that is closer to the optimum, helping the algorithm converge towards the solution.

The choice of the ansatz $U$ greatly affects the performance of a VQA.
From the perspective of the problem, the ansatz influences both the convergence speed and the closeness of the final state to a state that optimally solves the problem. On the other hand, the quantum hardware on which the VQA is executed has to be taken into account: Deeper circuits are more susceptible to errors, and some ansatz gates are costly to construct from native gates. Accordingly, most of the ans\"atze developed to date are classified either as more \textit{problem-inspired} or more \textit{hardware efficient}, depending on their structure and application. 

\subsubsection{Problem-inspired ans\"atze}\label{sec:HamiltonAnsatz}

An arbitrary unitary operation can be generated by an Hermitian operator $\hat{g}$ which, physically speaking, defines an evolution in terms of the $t$ parameter,
\begin{align}
 G(t) = e^{-i \hat{g} t}.\label{eq:lie_product_formula}
\end{align}
As an example, the generator $\hat{g}$ can be a Pauli matrix $\hat{\sigma}_{i}$ and thus, $G(t)$ becomes a single-qubit rotation of the form
\begin{equation}
    R_k\left(\theta\right) = e^{-i\frac{\theta}{2}\hat{\sigma}_k} = \cos(\theta/2)I - i\sin(\theta/2)\hat{\sigma}_k, \label{eq:def_single_qubit_rotation}
\end{equation}
with $t=\theta$ and $\hat{g}=\frac{1}{2}\hat{\sigma}_k$, corresponding to the spin operator.

From a more abstract viewpoint, those evaluations can always be described as time evolution of the corresponding quantum state, so that the generator $\hat{g}$ is often just referred to as a Hamiltonian. Note, however, that this Hamiltonian does not necessarily need to be the operator that describes the energy of the system of interest. In general, such generators can be decomposed into Pauli strings in the form of \eqref{eq:Pauli_string}. 

Within so-called \textit{problem-inspired} approaches, evolutions in the form of ~\eqref{eq:lie_product_formula}, with generators derived from properties of the system of interest are used to construct the parametrized quantum circuits. The unitary coupled-cluster approach (see below), mostly applied for quantum chemistry problems, is one prominent example. The generators then are elementary fermionic excitations, as shown in \eqref{eq:ucc_excitations}.

The \textit{Suzuki-Trotter} (ST) expansion or decomposition \cite{suzuki1976generalized} is a general method to approximate a general, hard to implement unitary in the form of ~\eqref{eq:lie_product_formula} as a function of the $t$ parameter. 
This can be done by decomposing $\hat{g}$ into a sum of non-commuting operators $\{\hat{o}_k\}_k$, with $\hat{g}=\sum_k c_k \hat{o}_k$ and some coefficients $c_k$. The operators $\hat{o}_k$ are chosen such that the evolution unitary $e^{-i\hat{o}_k t}$ can be easily implemented, for example as Pauli strings $\hat{P}_{k}$. The full evolution over $t$ can now be decomposed into integer $m$ equal-sized steps as
\begin{equation}
    e^{-i\hat{g}t}=\lim_{m\rightarrow \infty}\left(\prod_{k}e^{-i\frac{c_{k}\hat{o}_k t}{m}}\right)^{m}.
    \label{eq:trotter}
\end{equation}
For practical purposes, the time evolution can be approximated by a finite number $m$. 
When Pauli strings are used, this provides a systematic method to decompose an arbitrary unitary, generated by $\hat{g}$, into a product of  multi-qubit rotations $e^{-i\frac{c_k\hat{P}_k t}{m}}$, that can themselves be decomposed into primitive one and two qubit gates. 
Above, we have used the second order ST decomposition to approximate the true unitary at each time step $t$.
The error incurred from the approximation can be bounded by $||U_{\hat{g}}(\Delta_t)-U_{\hat{g}} ^{\textrm{ST}}(\Delta_t)||\leq \sum_{k=1}^m ||[[H_k, H_{>k} ],H_k ]]+ [[H_{>k},H_k ],H_{>k} ]] ||\Delta_t ^3$, where $H_{>k}=\sum_{\beta>k}H_\beta$ and $H_k =c_k \hat{o}_k$~\cite{poulin2014trotter}.

Knowledge about the physics of the particular Hamiltonian to be \textit{trotterized} can reduce substantially the number of gates needed to implement this method. For instance, in \cite{kivlichan2018quantum}, it is shown that by using fermionic swap gates, it is possible to implement a Trotter step for electronic structure Hamiltonians using first-neighbour connectivity circuits with $N^{2}/2$ two-qubit gates width and $N$ depth, where $N$ is the number of spin orbitals. They also show that implementing arbitrary Slater determinants can be done efficiently with $N/2$ gates of circuit depth. 

\paragraph{Unitary Coupled Cluster.}\label{sec:unitary_coupled_cluster}
Historically, problem-inspired ans\"atze were proposed and implemented first.
They arose from the quantum chemistry-specific observation that the unitary coupled cluster (UCC) ansatz \cite{taube2006new}, which adds quantum correlations to the Hartree-Fock approximation, is inefficient to represent on a classical computer \cite{yung2014transistor}.
Leveraging quantum resources, the UCC ansatz was instead realized as a PQC on a photonic processor \cite{peruzzo2014variational}.
It is constructed from the parametrized cluster operator $T(\boldsymbol{\theta})$ and acts on the Hartree-Fock ground state $\ket{\Psi_\text{HF}}$ as
\begin{equation}
    \ket{\Psi(\boldsymbol{\theta})} = e^{T(\boldsymbol{\theta})-T(\boldsymbol{\theta})^\dag} \ket{\Psi_\text{HF}}.\label{eq:ucc_canonical_form}
\end{equation}
The cluster operator is given by $T(\boldsymbol{\theta}) = T_1(\boldsymbol{\theta}) + T_2(\boldsymbol{\theta}) + \cdots$ with
\begin{eqnarray}
    T_1(\boldsymbol{\theta}) &=& \sum_{\substack{i\in\text{occ} \\ j\in\text{virt}}} \theta_i^j \hat{a}^\dag_j \hat{a}_i \\
    T_2(\boldsymbol{\theta}) &=& \sum_{\substack{i_1,i_2\in\text{occ} \nonumber\\ j_1,j_2\in\text{virt}}} \theta_{i_1,i_2}^{j_1,j_2} \hat{a}^\dag_{j_2}\hat{a}_{i_2} \hat{a}^\dag_{j_1}\hat{a}_{i_1},
    \label{eq:ucc_excitations}
\end{eqnarray}
and higher-order terms following accordingly \cite{omalley2016scalable}.
The operator $\hat{a}_k$ is the annihilation operator of the $k$-th Hartree-Fock orbital, and the sets \textit{occ} and \textit{virt} refer to the occupied and unoccupied Hartree-Fock orbitals.

Due to their decreasing importance, the series is usually truncated after the second or third term.
The ansatz is termed \textit{UCCSD} or \textit{UCCSDT}, respectively, referring to the inclusion of single, double, and triple excitations from the Hartree-Fock ground state. The $k$-UpCCGSD approach restricts the double excitations to pairwise excitations but allows $k$ layers of the approach~\cite{lee2018generalized}.
After mapping to Pauli strings as described in \secref{sec:pauli_strings}, the ansatz is converted to a PQC usually via the Trotter expansion in~\eqref{eq:trotter}.

In its original form, the UCC ansatz faces several drawbacks in its application to larger chemistry problems as well as to other applications. For strongly correlated systems, the widely proposed UCCSD ansatz is expected to have insufficient overlap with the true ground state and results typically in large circuit depths~\cite{lee2018generalized, grimsley2019adaptive}. 
Consequently, improvements and alternative ans\"atze are proposed to mitigate these challenges.
 
We restrict our discussion here to provide a short overview of alternative ansatz developments. For more details about the UCC ansatz, see \secref{sec:chemistry}.

\paragraph{Factorized Unitary Coupled-Cluster and Adaptive Approaches.}
The non commuting nature of the fermionic excitation generators, given by the cluster operators~\eqref{eq:ucc_excitations} leads to difficulties in decomposing the canonical UCC ansatz~\eqref{eq:ucc_canonical_form} into primitive one- and two-qubit untiaries. First approaches employed the Trotter decomposition ~\eqref{eq:trotter} using a single step~\cite{romero2018strategies, mcclean2016theory}. The accuracy of the so obtained factorized ansatz depends however on the order of the primitive fermionic excitations~\cite{grimsley2019trotterized, Izmaylov2020order}.

Alternative approaches propose to use factorized unitaries, constructed from primitive fermionic excitations, directly~\cite{evangelista2019exact, Izmaylov2020order}. Adaptive approaches, are a special case of a factorized ansatz, where the unitary is iteratively grown by subsequently screening and adding primitive unitary operators from a predefined operator pool. The types of operator pools can be divided into two classes: Adapt-VQE~\cite{grimsley2019adaptive}, that constructs the operator pool from primitive fermionic excitations, and  Qubit-Coupled-Cluster~\cite{ryabinkin2018qubit} that uses Pauli Strings.
In both original works, the screening process is based on energy gradients with respect to the prospective operator candidate. Since this operator is the trailing part of the circuit, the gradient can be evaluated through the commutator of the Hamiltonian with the generator of that operator.
In contrast to commutator based gradient evaluation, direct differentiation, as proposed in~\cite{kottmann2020reducing} allows gradient evaluations with similar cost as the original objective and generalizes the approach by allowing screening and insertion of operators at arbitrary positions in the circuit. This is, for example, necessary for excited state objectives as discussed in~\secref{sec:var_excited}.

Extended approaches include iterative methods~\cite{ryabinkin2020iterative}, operator pool construction from involutory linear combination of Pauli strings~\cite{lang2020iterative}, Pauli string pools from decomposed fermionic pools~\cite{tang2019qubit}, mutual information based operator pool reduction~\cite{zhang2020mutual}, measurement reduction schemes based on the density matrix reconstruction~\cite{liu2020efficient}, and external perturbative corrections~\cite{ryabinkin2021aposteriori}.

\paragraph{Variational Hamiltonian Ansatz.}
Motivated by adiabatic state preparation, the Variational Hamiltonian Ansatz (VHA) was developed to reduce the number of  parameters and accelerate the convergence \cite{wecker2015progress,mcclean2016theory}. 
Instead of the Hartree-Fock operators, the terms of the fermionic Hamiltonian itself are used to construct the PQC.
For this purpose, the fermionic Hamiltonian $H$ is written as a sum of $M$ terms $H = \sum_i \hat{h}_i$.
Which parts of the Hamiltonian are grouped into each term $\hat{h}_i$ depends on the problem and there is a degree of freedom in the design of the algorithm.
The PQC is then chosen as
\begin{equation}
    U_\text{VHA} = \prod_{i=1}^M e^{\left( i\theta_i\hat{h}_{i} \right)},
\end{equation}
with the operators in the product ordered by decreasing $i$.
The unitary corresponds to $n$ short time evolutions under different parts of the Hamiltonian, where the terms $\hat{h}_i$ of the Hamiltonian can be repeated multiple times.
The initial state is chosen so that it is easy to prepare yet it is related to the Hamiltonian, for example the eigenstate of the diagonal part of $H$.
The Fermi-Hubbard model with its few and simple interaction terms is proposed as the most promising near-term application of the method.
However, it is also shown that the VHA can outperform specific forms of the UCCSD ansatz for strongly correlated model systems in quantum chemistry.
In \secref{sec:chemistry} we discuss some VQE-inspired algorithms that also use adiabatic evolution to improve the performance of the algorithm.

\paragraph{Quantum Approximate Optimization Algorithm.}
One of the canonical NISQ era algorithms, designed to provide approximate solutions to combinatorial optimization problems, is the Quantum Approximate Optimization Algorithm (QAOA) \cite{farhi2014quantum}.
While QAOA can be thought of as a special case of VQA, it has been studied in depth over the years both empirically and theoretically, and it deserves special attention.

The cost function $C$ of a QAOA is designed to encode a combinatorial problem by means of bit strings that form the computational basis. With the computational basis vectors $\ket{e_{i}}$, one can define the problem Hamiltonian $H_{P}$ as (see \secref{sec:MaxCut} for an example)
\begin{equation}
H_{P} \equiv \sum_{i=1}^{n}C(e_{i})|e_{i}\rangle \langle e_{i}|,
\label{eq:HP_QAOA}
\end{equation}
and the mixing Hamiltonian $H_{M}$ as
\begin{equation}
H_{M}\equiv\sum_{i=1}^{n} \hat{\sigma}_{x}^{i}. \label{eq:mxing_operator}
\end{equation}
The initial state in the QAOA algorithm is conventionally chosen to be the uniform superposition state $\ket{D}$ from \eqref{eq:sp_H}.
The final quantum state is given by alternately applying $H_P$ and $H_M$ on the initial state $p$-times,
\begin{equation}
\ket{\Psi(\boldsymbol{\gamma},\boldsymbol{\beta})} 
\equiv e^{-i\beta_{p}H_{M}}e^{-i\gamma_{p}H_{P}}\cdots e^{-i\beta_{1}H_{M}}e^{-i\gamma_{1}H_{P}}\vert D\rangle,\label{eq:QAOA_evolved_state}
\end{equation}
with $\boldsymbol{\gamma}\equiv\left(\gamma_{1},\gamma_{2},\cdots,\gamma_{p}\right)$ and $\boldsymbol{\beta}\equiv\left(\beta_{1},\beta_{2},\cdots,\beta_{p}\right)$.
A quantum computer is used to evaluate the objective function
\begin{equation}
C(\boldsymbol{\gamma},\boldsymbol{\beta}) \equiv 
\bra{\Psi(\boldsymbol{\gamma},\boldsymbol{\beta})} H_{P}(\boldsymbol{\gamma},\boldsymbol{\beta}) 
\ket{\Psi(\boldsymbol{\gamma},\boldsymbol{\beta})}, \label{eq:QAOA_objective}
\end{equation}
and a classical optimizer is used to update the $2p$ angles $\boldsymbol{\gamma}$ and $\boldsymbol{\beta}$ until $C$ is maximized, i.e. $C(\boldsymbol{\gamma^*},\boldsymbol{\beta^*}) \equiv \max_{\boldsymbol{\gamma},\boldsymbol{\beta}}  C(\boldsymbol{\gamma},\boldsymbol{\beta})$. 
Here, $p$ is often referred to as the QAOA level or depth. Since the maximization at level $p-1$ is a constrained version of the maximization at level $p$, the performance of the algorithm improves monotonically with $p$ in the absence of experimental noise and infidelities.

In adiabatic quantum computing (see \secref{subsec:quantum-annealing}), we start from the ground state of $H_{M}$ and slowly move towards the ground state of $H_{P}$ by slowly changing the Hamiltonian. In QAOA, instead, we alternate between $H_{M}$ and $H_{P}.$ One can think of QAOA as a Trotterized version of quantum annealing. Indeed, the adiabatic evolution as used in quantum annealing can be recovered in the limit of $p \to \infty$.

For a combinatorial optimization problem with hard constraints to be satisfied, penalties in the cost function can be added. This might not be an efficient strategy in practice as it is still possible to obtain solutions which violate some of the hard constraints. 
{A variation of the QAOA to deal with these constraints was also discussed in the Sec. VII from the original proposal \cite{farhi2014quantum}.} Building on previous work in quantum annealing~\cite{hen2016quantum,hen2016driver}, it was proposed to encode the hard constraints directly in the mixing Hamiltonian~\cite{hadfield2017quantum}. This approach yields the main advantage of restricting the state evolution to the feasible subspace where no hard constraints are violated, which consequently speeds up the classical optimization routine to find the optimal angles. This framework was later generalized as the \textit{Quantum Alternating Operator Ansatz}   to consider phase-separation and mixing unitary operators ($U_P (\gamma)$ and $U_M(\beta)$ respectively) which need not originate from the time-evolution of a Hamiltonian \cite{hadfield2019from}. The operators $e^{-i\beta H_{M}}$ and $e^{-i\gamma H_{P}}$ from \eqref{eq:QAOA_evolved_state} are replaced by $U_M(\beta)$ and $U_P (\gamma)$ respectively. It is worth noting that both the Quantum Approximate Optimization Algorithm and the Quantum Alternating Operator Ansatz are abbreviated ``QAOA'' in the literature. In this case, we suggest ``QuAltOA'' as an acronym for the Quantum Alternating Operator Ansatz to distinguish the same from the Quantum Approximate Optimization Algorithm. 

{The use of QAOA for combinatorial optimization is presented in \secref{sec:optimization}. Some theoretical guarantees of this ansatz are introduced in \secref{sec:theoretical_garantees}.}

\begin{figure}[tb]
\centering
\includegraphics[width=1.0\columnwidth]{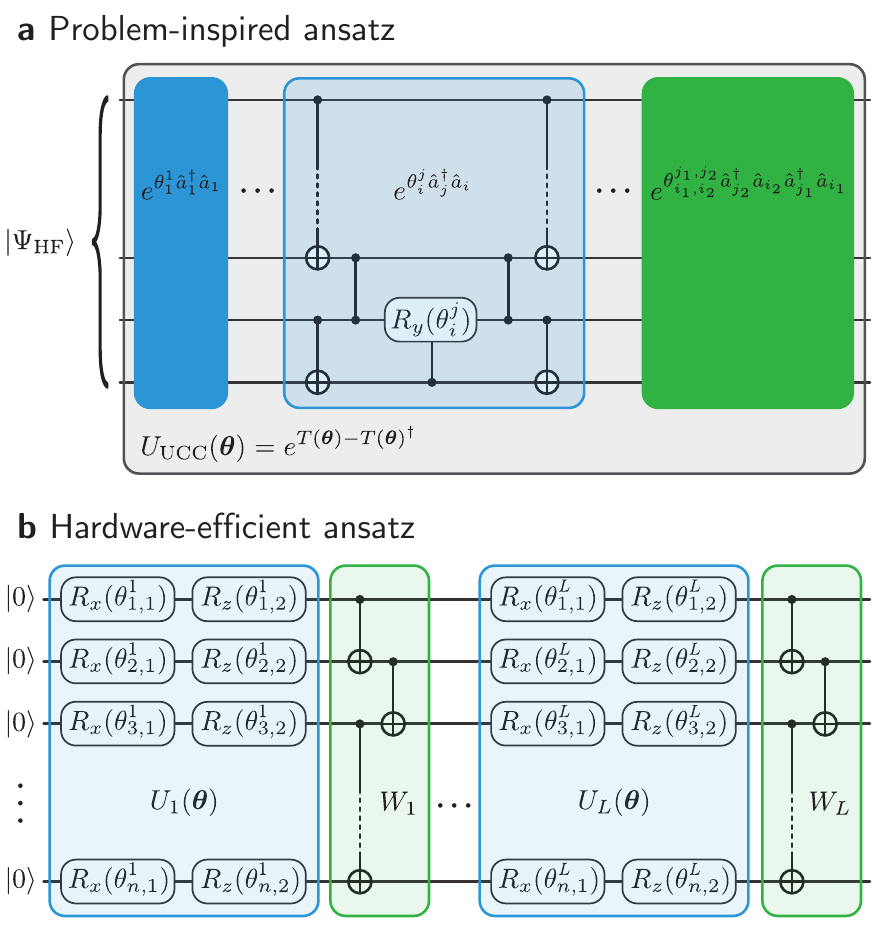}
\caption{Example problem-inspired and hardware-efficient ans\"atze. (a) Circuit of the Unitary Coupled Cluster ansatz with a detailed view of a fermionic excitation as discussed in \cite{yordanov2020efficient}. (b) Hardware-efficient ansatz tailored to a processor that is optimized for single-qubit $x$- and $z$-rotations and nearest-neighbor two-qubit CNOT gates.}
\label{Fig:ansatz_type_comparison}
\end{figure}

\subsubsection{Hardware-efficient ans\"atze}\label{sec:HardwareAnsatz}

Thus far, we have described circuit ans\"atze constructed from the underlying physics of the problem to be solved.
Although it has been shown computationally that such ans\"atze can ensure fast convergence to a satisfying solution state, they can be challenging to realize experimentally.
Quantum computing devices possess a series of experimental limitations that include, among others, a particular qubit connectivity, a restricted gate set, and limited gate fidelities and coherence times.
Therefore, existing quantum hardware is not suited to implement the deep and highly connected circuits required for the UCC and similar ans\"atze for applications beyond basic demonstrations such as the $\text{H}_2$ molecule \cite{moll2018quantum}.

A class of \textit{hardware-efficient ans\"atze} has been proposed to accommodate device constraints \cite{kandala2017hardware}.
The common trait of these circuits is the use of a limited set of quantum gates as well as a particular qubit connection topology.
The gate set usually consists of a two-qubit entangling gate and up to three single-qubit gates.
The circuit is then constructed from blocks of single-qubit gates and entangling gates, which are applied to multiple or all qubits in parallel.
Each of these blocks is usually called \textit{layer}, and the ansatz circuit generally has multiple such layers.

The quantum circuit of a hardware-efficient ansatz with $L$ layers is usually given by
\begin{equation}
\label{eq:HE_ansatz}
U(\boldsymbol{\theta})=\prod_{k=1}^{L}U_{k} \left(\boldsymbol{\theta}_{k}\right)W_{k},
\end{equation}
where $\boldsymbol{\theta} = \left(\boldsymbol{\theta}_{1},,\cdots, \boldsymbol{\theta}_{L}\right)$ are the variational parameters, $U_{k}\left(\boldsymbol{\theta}_{k}\right)=\exp\left(-i\boldsymbol{\theta}_{k}V_{k}\right)$ is a unitary derived from a Hermitian operator $V_{k}$, and $W_{k}$ represents non-parametrized quantum gates.
Typically, the $V_{k}$ operators are single-qubit rotation gates, i.e. $V_{k}$ are Pauli strings acting locally on each qubit. In those cases, $U_{k}$ becomes a product of combinations of single-qubit rotational gates, each one defined as in \eqref{eq:def_single_qubit_rotation}. 
$W_{k}$ is an entangling unitary constructed from gates that are native to the architecture at hand, for example CNOT or CZ gates for superconducting qubits or XX gates for trapped ions~\cite{krantz2019quantum, wright2019benchmarking}. Following this approach, the so-called \textit{Alternating Layered Ansatz} is a particular case of these Hardware-efficient ans\"atze which consists of layers of single qubit rotations, and blocks of entangling gates that entangle only a local set of qubits and are shifted every alternating layer.

The choice of these gates, their connectivity, and their ordering influences the portion of the Hilbert space that the ansatz covers and how fast it converges for a specific problem.
Some of the most relevant properties of hardware-efficient ans\"atze, namely expressibility, entangling capability and number of parameters and layers needed are studied in Refs. \cite{sim2019expressibility, woitzik2020entanglement, nakaji2020expressibility,bravoPrieto2020scalingof} and further discussed in \secref{sec:expressibility}.

Instead of making a choice between the problem-inspired and hardware-efficient modalities, some PQC designers have chosen an intermediate path.
One example is the use of an exchange-type gate, which can be implemented natively in transmons, to construct a PQC that respects the symmetry of the variational problem \cite{ganzhorn2019gate, sagastizabal2019error}.
Such an ansatz leads to particularly small parameter counts for quantum chemistry problems such as the H$_2$ and LiH molecules~\cite{gard2020efficient}.
Another intermediate approach, termed QOCA for its inspiration from quantum optimal control, is to add symmetry-breaking unitaries, akin to a hardware-efficient ansatz, into the conventional VHA circuit \cite{choquette2020quantum}.
This modification enables excursions of the variational state into previously restricted sections of the Hilbert space, which is shown to yield shortcuts in solving fermionic problems.

\subsection{Measurement}\label{sec:measurement}

To gain information about the quantum state that has been prepared on the quantum hardware, one needs to estimate the expectation value of the objective function $\langle\hat{O}\rangle_{U_{\boldsymbol{\theta}}}$.

The most direct approach to estimate expectation values is to apply a unitary transformation on the quantum state to the diagonal basis of the observable $\hat{O}$ and obtaining the probability of measuring specific computational states corresponding to an eigenvalue of $\hat{O}$.
In other words, to determine whether a measured qubit is in the $\ket{0}$ or $\ket{1}$ state. For experimental details on this task we refer to existing reviews, such as for superconducting qubits~\cite{krantz2019quantum} or ion traps~\cite{haffner2008quantum}.
However, on NISQ devices, the tranformation to the diagonal basis mentioned before can be an overly costly one. As a NISQ friendly alternative, most observables of interest can be efficiently parameterized in terms of Pauli strings, as shown above, 
and transformed into their diagonal basis by simple single-qubit rotations, as shown below.

\paragraph{Measurement of Pauli strings.} The expectation value of the $\hat{\sigma}_z$ operator on a particular qubit can be measured by reading out the probabilities of the computational basis state $\left\{\ket{0}, \ket{1}\right\}$ as
\begin{align}
    \langle \psi| \hat{\sigma}_z|\psi\rangle \equiv \langle \hat{\sigma}_z \rangle = |\alpha|^2 - |\beta|^2,
\end{align}
where $|\alpha|^2$ is the probability to measure the qubit in state $\ket{0}$, $|\beta|^2$ is the probability to measure the qubit in state $\ket{1}$ and $\ket{\psi}=\alpha\ket{0}+\beta\ket{1}$.
Measurements defined by $\hat{\sigma}_x$ and $\hat{\sigma}_y$ can be defined similarly by transforming them into the $\hat{\sigma}_z$ basis first. The transformation is given by primitive single-qubit gates

\begin{align}
  &\hat{\sigma}_x =R_y^{\dagger}\left(\pi/2\right)\hat{\sigma}_z R_y\left(\pi/2\right) = H_\text{d}\hat{\sigma}_{z} H_\text{d},\label{eq:measureX}\\
  &\hat{\sigma}_y = R^\dagger_x\left(\pi/2\right) \hat{\sigma}_z R_x\left(\pi/2\right) =  S H_\text{d}\hat{\sigma}_{z} H_\text{d} S^{\dagger},\label{eq:measureY}
\end{align}
where 
$S=\sqrt{\hat{\sigma}_z}$ and $H_\text{d}=(\hat{\sigma}_x+\hat{\sigma}_z)/\sqrt{2}$ is the Hadamard gate. 
Then, to measure  $\hat{\sigma}_x$ on a quantum state $\ket{\psi}$, we rotate $\hat{\sigma}_x$ into the $z$-axis by applying $H_d$ and measure in logical $\hat{\sigma}_z$ basis, i.e. 
\begin{equation}\langle \hat{\sigma}_x \rangle \equiv \bra{\psi}\hat{\sigma}_x \ket{\psi} =\bra{\psi}H_\text{d} \hat{\sigma}_z H_\text{d}\ket{\psi}=\alpha \beta^* + \alpha^* \beta.
\end{equation}
The same applies for $\langle\hat{\sigma}_{y}\rangle$.
Arbitrary Pauli strings $\hat{P}$, with primitive Pauli operations $\hat{\sigma}_{f(k)} \in\left\{\sigma_x, \sigma_y, \sigma_z \right\}$ on qubits $k\in K$, can then be measured by the same procedure on each individual qubit as

\begin{align}
    \expval{\hat{P}}{U} &= \expval{\prod_{k\in K} \sigma_z(k) }{\tilde{U}U}
\end{align}
where $\tilde{U}$ is a product of single qubit rotations according to~\eqref{eq:measureX} and~\eqref{eq:measureY} depending on the Pauli operations $\hat{\sigma}_{f(k)}$ at qubit $k$.

So far we discussed expectation values of a physical observable $\langle \hat{\mathcal{O}}\rangle$, which is the mean value averaged over an infinite number of measurements.
In practice, one can sample only a finite number of single-shot measurements $N_\text{s}$ of the quantum state and thus can estimate the expectation values within some finite error. 
For a Pauli string $\hat{P}$, the number of measurement samples $N_\text{s}$ needed to estimate the expectation value $\expval{\hat{P}}{U}$ with an additive error of at most $\epsilon$ with a failure probability of at most $\delta$ is bounded by Hoeffding's inequality~\cite{huang2019nearterm}
\begin{equation}
N_\text{s}\ge \frac{2}{\epsilon^2}\log\left(\frac{2}{\delta}\right)\,.
\end{equation}
In particular, the error $\epsilon$ decreases with the inverse square-root of the number of measurements $\epsilon\propto 1/\sqrt{N_\text{s}}$.

For many problems, such as quantum chemistry-related tasks, the number of terms in the cost Hamiltonian to be estimated can become very large. A naive way of measuring each Pauli string separately may incur a prohibitively large number of measurements. 
Recently, several more efficient approaches have been proposed~(see \cite{bonet2020nearly} for an overview). The common idea is to group different Pauli strings that can be measured simultaneously such that a minimal number of measurements needs to be performed.

Pauli strings that commute qubit-wise, i.e. the Pauli operators on each qubit commute, can be measured at the same time~\cite{kandala2017hardware,mcclean2016theory}. The problem of finding the minimal number of groups can be mapped to the minimum clique cover problem, which is NP-hard in general, but good heuristics exist~\cite{verteletskyi2020measurement}. One can collect mutually commuting operators and transform them into a shared eigenbasis, which adds an additional unitary transformation to the measurement scheme~\cite{crawford2019efficient,crawford2019efficient,gokhale2019minimizing,yen2020measuring}. Combinations of single qubit and Bell measurements have been proposed as well~\cite{hamamura2020efficient}.

Alternatively, one can use a method called unitary partitioning to linearly combine different operators into a unitary, and use the so-called Hadamard test (see below) to evaluate it~\cite{izmaylov2019unitary,zhao2020measurement}.
In \cite{izmaylov2019revising}, the observables can be decomposed into the so-called mean-field Hamiltonians, which can be measured more efficiently if one measures one qubit after the other, and uses information from previous measurement outcomes. 

For specific problems such as chemistry and condensed matter systems, it is possible to use the structure of the problem to reduce the number of measurements \cite{gokhale2019n,huggins2019efficient,cade2020strategies, cai2020resource}. In particular, in ~\cite{cai2020resource}, where a Fermi-Hubbard model is studied using VQE, the number of measurements is reduced by considering multiple orderings of the qubit operators when applying the Jordan-Wigner transformation. In the context of quantum chemistry, the up-to-date largest reduction could be achieved by the Cartan subalgebra approach of ~\cite{yen2020cartan}. Other approaches use classical shadows~\cite{hadfield2020measurements}, a classical approximation of the quantum state of interest, or neural network estimators~\cite{torlai2020precise} to decrease the number of measurements. All those kind of optimizations require an understanding of the underlying problem and are usually not applicable for every use of the VQE.

\paragraph{Measurement of overlaps.} Several VQA require the measurement of an overlap of a quantum state $\ket{\psi}$ with unitary $U$ in the form of $\bra{\psi}U\ket{\psi}$. This overlap is in general not an observable and has both real and imaginary parts. The Hadamard test can evaluate such a quantity on the quantum computer using a single extra qubit~\cite{miquel2002interpretation}. The idea is to apply a controlled $U$ operation, with control on that qubit, and target $U$ on the quantum state. Then, one can measure from the this single qubit state both real and imaginary part of the overlap. 
A downside of this method is the requirement to be able to implement a controlled unitary, which may require too many resources on current quantum processors.
Alternative methods to measure the overlap without the use of control unitaries have been proposed~\cite{mitarai2019methodology}. One idea is to decompose $U$ into a sum of Pauli strings, and then to measure the expectation value of each Pauli string individually.
Another approach is possible if $U$ can be rewritten into a product of unitaries $U_q$ that act locally on only a few qubits. Then, one can find via classical means the diagonalization of $U_q=V_q^\dagger D V_q$, with diagonal matrix $D$ and $V_q$ being a unitary. The overlap can be found by applying the $V_q$ unitaries on the state $\ket{\psi}$, measure the outcomes in the computational basis and do post-processing of the results with the  classically calculated eigenvalues of $D$.

\paragraph{Classical shadows.} This is a powerful technique to accurately predict $M$ expectation values $\textrm{Tr}(\hat{O}_i \rho)$, $1\leq i\leq M$ of an unknown quantum state $\rho$~\cite{huang2020predicting}. 
The method is based on and inspired from shadow tomography \cite{aaronson2019shadow}.
First, a random unitary $U$ is applied on the state $\rho \rightarrow U\rho U^\dagger$ and then all the qubits are measured in the computational basis. This step is repeated with several random unitaries $U$. Common choices for $U$ are unitaries which can be efficiently computed on a classical computer such as random $n$-qubit Clifford circuits or tensor products of single qubit rotations. 
By post-processing the measurement results, one can gather a classical shadow, which is a classical representation of the quantum state $\rho$. 

There exists performance guarantees that classical shadows with size of order $\log M$ suffice to predict $M$ expectation values simultaneously.
For investigations involving  classical shadow tomography protocols in the presence of noise, refer to \cite{koh2020classical,chen2020robust}. Experimental realizations have been performed recently as well~\cite{struchalin2021experimental,zhang2021experimental}.

\subsection{Parameter optimization}
\label{sec:parameter_optimization}

In principle, the PQC parameter optimization to minimize the objective is not different from any multivariate optimization procedure and standard classical methods can be applied \cite{lavrijsen2020classical}.
However, in the NISQ era, the coherence time is short, which means that high-depth analytical gradient circuits cannot be implemented.
In addition, one of the biggest challenges in parameter optimization is the large number of measurements required for estimating the mean value of an observable to a high precision. Due to this high sampling rate, the measurement process can become a significant bottleneck in the overall algorithm runtime. Thus, an effective optimizer for PQCs should try to minimize the number of measurements or function evaluations. 
As a last criterion, the optimizer should be resilient to noisy data coming from current devices and precision on expectation values that is limited by the number of shots in the measurement.
These three requirements imply that certain existing algorithms are better suited for PQC optimization and are more commonly used, and that new algorithms are being developed specifically for PQC optimization.
Some intuitive concepts of the mechanisms behind optimisation of quantum problems have been investigated in~\cite{mcclean2020low}.
Recently,~\cite{bittel2021training} have shown that the classical optimization corresponding to VQAs is a NP-hard problem.  

In this section, we first review two classes of optimization, gradient-based and gradient-free. 
We also consider resource-aware optimization methods and strategies that additionally minimize quantities associated with the quantum cost of optimization.
While we reserve more detailed descriptions to the respective references and the Supplementary Material, we highlight the main features and advantages for each optimization strategy. 

\subsubsection{Gradient-based approaches}

A common approach to optimise an objective function $f(\boldsymbol{\theta})$ is via its gradient, i.e. the change of the function with respect to a variation of its $M$ parameters $\boldsymbol{\theta}=(\theta_{1},\cdots,\theta_{M})$. The gradient indicates the direction in which the objective function shows the greatest change. This is a local optimization strategy as one uses information starting from some initial parameter value $\boldsymbol{\theta}^{(0)}$ and iteratively updates $\boldsymbol{\theta}^{(t)}$ over multiple discrete steps $t$. A common update rule for each $\theta_{i}$ is
\begin{equation}
\theta^{(t+1)}_i=\theta^{(t)}_i-\eta \ \partial_i f(\boldsymbol{\theta})\,,\label{eq:grad_update}
\end{equation}
or $\boldsymbol{\theta}^{(t+1)}=\boldsymbol{\theta}^{(t)}-\eta \ \boldsymbol{\nabla} f(\boldsymbol{\theta})$, where $\eta$ is a small parameter called learning rate and
\begin{equation}
    \partial_{i}\equiv \frac{\partial}{\partial\theta_{i}}, \ \boldsymbol{\nabla}=\left(\partial_{1},\cdots,\partial_{M}\right)
    \label{eq:gradient}
\end{equation}
is the partial derivative with respect to the parameter $\theta_{i}$ and the gradient vector, respectively, using Einstein notation. 

There are various ways of estimating the gradient on a quantum computer \cite{romero2018strategies}. The most relevant of them are detailed in the Supplementary Material and summarized in the following paragraphs.

\paragraph{Finite difference.} One can compute the gradients using finite differences, i.e. $\partial_{i} f(\boldsymbol{\theta}) \approx (f(\boldsymbol{\theta} + \epsilon \mathbf{e}_i) - f(\boldsymbol{\theta} - \epsilon \mathbf{e}_i))/2\epsilon$, where $\epsilon$ is a small number and $\mathbf{e}_i$ is the unit vector with 1 as its $i$-th element and 0 otherwise. As the objective function $f(\theta)$ is obtained with limited accuracy, a good estimation of the gradient requires smaller $\epsilon$, i.e. more samples taken from the quantum hardware.

\paragraph{Parameter shift rule.} This strategy was proposed in~\cite{romero2018strategies} and developed in~\cite{mitarai2018quantum,schuld2019evaluating}. This method computes the gradients exact and $\epsilon$ can be large (commonly $\epsilon=\pi/2$). This method assumes that the unitary to be optimized can be written as $U(\boldsymbol{\theta}) = V G(\theta_i) W$, where $G=e^{-i \theta_i g}$ is the unitary affected by the parameter $\theta_i$, $g$ is the generator of $G$ and $V, W$ are unitaries independent of $\theta_i$. If $g$ has a spectrum of two eigenvalues $\pm \lambda$ only, the gradient can be calculated by measuring the observable at two shifted parameter values as follows:
    \begin{equation}
        \partial_{i}\langle f(\boldsymbol{\theta}) \rangle = \lambda \left( \langle f(\boldsymbol{\theta}_+) \rangle - \langle f(\boldsymbol{\theta}_-) \rangle \right),
    \end{equation}
where $\boldsymbol{\theta}_{\pm} = \boldsymbol{\theta} \pm (\pi / 4\lambda) \boldsymbol{e}_i$. 
This rule can be generalised to the case where the generator $g$ does not satisfy the eigenspectrum condition (see Supplementary Material for details). It can also be adapted to calculate analytical gradients for fermionic generators of Unitary Coupled-Cluster operators~\cite{kottmann2020feasible} and higher order derivatives~\cite{mari2020estimating}.

\paragraph{L-BFGS.}
It is a quasi-Newton method that efficiently approximates the ``inverse Hessian'' using a limited history of positions and gradients \cite{liu1989limited, fletcher2000mathematics}.
While effective in simulations, recent studies observed BFGS methods do not perform well in experimental demonstrations of VQA due to the level of noise in the cost function and gradient estimates \cite{lavrijsen2020classical}. 
Two heuristics were proposed to find quasioptimal parameters for QAOA using BFGS \cite{zhou2020quantum}, \textsc{INTERP} and \textsc{FOURIER} explained in the supplementary material. Efficient initialization of parameters has also been reported using the Trotterized quantum annealing (TQA) protocol~\cite{sack2021quantum}. These heuristic strategies can be easily extended to gradient-free optimization methods such as Nelder-Mead.
    
\paragraph{Quantum natural gradient.}
The update rule of standard gradient descent assumes that the parameter space is a flat Euclidean space. However, in general this is not the case, which can severely hamper the efficiency of gradient descent methods. In classical machine learning, the natural gradient was proposed that adapts the update rule to the non-Euclidean metric of the parameter space~\cite{amari1998natural}. 
Its extension, the quantum natural gradient (QNG) defines the following update rule \cite{stokes2020quantum}:
    \begin{equation}
\theta^{(t+1)}_i=\theta^{(t)}_i-\eta \  \mathcal{F}^{-1}(\boldsymbol{\theta})\partial_{i} f(\boldsymbol{\theta})\,,
\label{eq:natural_grad_update}
 \end{equation}
where $\mathcal{F}(\boldsymbol{\theta})$ is the Fubini-Study metric tensor or quantum Fisher information metric given by
\begin{equation}
    \mathcal{F}_{ij} = \text{Re}(\braket{\partial_i\psi(\boldsymbol{\theta}) \vert\partial_j \psi(\boldsymbol{\theta})}- \braket{\partial_i\psi(\boldsymbol{\theta} )\vert\psi(\boldsymbol{\theta})} \braket{\psi(\boldsymbol{\theta})\vert\partial_j \psi(\boldsymbol{\theta})})\,.
    \label{eq:Fubiny_Study}
\end{equation}

Superior performance of the QNG compared to other gradient methods has been reported~\cite{yamamoto2019natural,stokes2020quantum} and it has been shown that it can avoid becoming stuck in local minima~\cite{wierichs2020avoiding}. It can be generalized to noisy quantum circuits~\cite{koczor2019quantum}. 
The QNG can be combined with adaptive learning rates $\eta(\theta_i^{t})$ that change for every step of gradient descent to speed up training. For hardware efficient PQCs, one can calculate adaptive learning rates using the quantum Fisher information metric~\cite{haug2021optimal}.
While the full Fubini-Study metric tensor is difficult to estimate on quantum hardware, diagonal and block-diagonal approximations can be efficiently evaluated~\cite{stokes2020quantum} and improved classical techniques to calculate the full tensor exist~\cite{jones2020efficient}. A special type of PQC, the natural PQC, has a euclidean quantum geometry such that the gradient is equivalent to the QNG close to a particular set of parameters~\cite{haug2021natural}.
    
\paragraph{Quantum imaginary time evolution.}
Instead of using the standard gradient descent for optimization, a variational imaginary time evolution method was proposed in \cite{mcardle2019variational} to govern the evolution of parameters. They focused on many-body systems described by a $k$-local Hamiltonian 
and considered a PQC that encodes the state $\ket{\psi(\tau)}$ as a parameterized trial state
$\ket{\psi(\boldsymbol{\theta}(\tau))}$.
The evolution of $\boldsymbol{\theta}(\tau)$ with respect to all the parameters can then be obtained by solving a differential equation (see Supplementary for details). It was later shown in~\cite{stokes2020quantum} that this method is analogous to the gradient descent via the QNG when considering infinitesimal small step sizes.

\paragraph{Hessian-aided gradient descent.}
A recent work~\cite{huembeli2020characterizing} proposed computing the Hessian and its eigenvalues to help analyze the cost function landscapes of QML algorithms.
Tracking the numbers of positive, negative, and zero eigenvalues provides insight whether 
the optimizer is heading towards a stationary point.
The Hessian can be computed by doubly applying the parameter shift rule as shown in  \cite{mitarai2019methodology} and reproduced in the supplementary material.
While a deeper analysis is necessary to compare their performance, both QNG and Hessian-based methods try to accelerate optimization by leveraging local curvature information.
    
\paragraph{Quantum Analytic Descent.}
A method consisting of using a classical model of the local energy landscape to estimate the gradients is proposed in \cite{koczor2020quantum}. 
In this hybrid approach, a quantum device is used to construct an approximate ansatz landscape and the optimization towards the minima of the corresponding approximate surfaces can be carried out efficiently on a classical computer. Using this approximate ansatz landscape, the full energy surface, gradient vector and metric tensor can be expressed in term of the ansatz parameters. The analytic descent has been shown to achieve faster convergence as compared to the QNG. 
    
\paragraph{Stochastic gradient descent.} 
A major drawback of gradient-based methods is the high number of measurements. The stochastic gradient descent (SGD) algorithm addresses this issue by replacing the normal parameter update rule with a modified version
    \begin{equation}
        \boldsymbol{\theta}^{(t+1)} = \boldsymbol{\theta}^{(t)} - \alpha \  \boldsymbol{g}(\boldsymbol{\theta}^{(t)}),
    \end{equation}
where $\alpha$ is the learning rate and $\boldsymbol{g}$ is an unbiased estimator of the gradient of the cost function. As an estimator, one can take the measurement of the gradient with a finite number of shots~\cite{harrow2019low}.
This technique can be combined with sampling of the parameter-shift rule terms~\cite{sweke2020stochastic} or by extending it to doubly stochastic gradient. For the latter, the finite measurements are performed for only a subset of the expectation values of the Hamiltonian terms. This sampling can be performed in the extreme situation where only one Pauli-term is evaluated at a single point in the quadrature.
This is a very powerful method that reduces the number of measurements drastically~\cite{anand2020experimental}.
This method can be extended beyond circuits that allow the parameter-shift rule by expressing the gradient as an integral~\cite{banchi2020measuring}. To accelerate the convergence of SGD for VQA, different strategies are proposed \cite{lyu2020accelerated} and briefly explained in the Supplementary.

\subsubsection{Gradient-free approaches}

In this section, we discuss optimization methods for VQA that do not rely on gradients measured on the quantum computer.

\paragraph{Evolutionary algorithms.}
Evolutionary strategies \cite{rechenberg1978evolutionsstrategien,schwefel1977numerische} are black-box optimization tools for high dimensional problems that use a search distribution, from which they sample data, to estimate the gradient of the expected fitness to update the parameters in the direction of steepest ascent. More recently, natural evolutionary strategies (NES)~\cite{wierstra2014natural} have demonstrated considerable progress in solving these high dimensional optimization problems. They use natural gradient estimates for parameter updates instead of the standard gradients. They have been adapted for optimization of VQA~\cite{zhao2020natural, anand2020natural} and have been shown to have similar performance as the state-of-the-art gradient based method. In \cite{anand2020natural} it is shown that NES, along with techniques like Fitness shaping, local natural coordinates, adaptive sampling and batch optimization, can be used for optimization of deep quantum circuits.
    
\paragraph{Reinforcement learning.}
Several authors have used reinforcement learning (RL) to optimize the QAOA parameters \cite{garcia2019quantum,khairy2019reinforcement,wauters2020reinforcement,yao2020noise,yao2020policy}. This framework consists of a decision-making agent with policy $\pi_{\boldsymbol{\theta}}(a|s)$ parameterized by $\boldsymbol{\theta}$, which is a mapping from the state space $s \in \{S\} $ to an action space $a \in \{A\}$. In response to the action, the environment provides the agent with a reward $r$ from the set of rewards $\{R\}$. The goal of RL is to find a policy which maximizes the expected total discounted reward. For more details, refer to \secref{sec:applications_ML_RL}. In the context of QAOA, for example, $\{S\}$ can be the set of QAOA parameters ($\boldsymbol{\gamma},\boldsymbol{\beta}$) used, $a$ can be the value of $\gamma$ and $\beta$ for the next iteration, and the reward can be the finite difference in the QAOA objective function between two consecutive iterations. The policy can be parameterized by a deep neural network with the weights $\boldsymbol{\theta}$. The policy parameters $\boldsymbol{\theta}$ can be optimized using a variety of algorithms such as Monte-Carlo methods~\cite{hammersley2013monte,sutton2018reinforcement}, Q-Learning~\cite{watkins1992q} and policy gradient methods~\cite{sutton2018reinforcement}.
    
\paragraph{Sequential minimal optimization.}
In machine learning, the sequential minimal optimization (SMO) method~\cite{platt1998sequential} has proven successful in optimizing the high-dimensional parameter landscape of support vector machines.
The method breaks the optimization into smaller components for which the solution can be found analytically.
This method has been applied to variational circuit optimization~\cite{nakanishi2020sequential}, circuit optimization with classical acceleration~\cite{parrish2019jacobi} and circuit optimization and learning with Rotosolve and Rotosolect~\cite{ostaszewski2019quantum}.

\paragraph{Surrogate model-based optimization.} 
When function evaluations are costly, it pays off to not only use the current function value to inform a next parameter value, but to use all previous evaluations to extract information about the search space.
The function values in memory are used to build a surrogate model, an auxiliary function that represents the full expensive cost function based on the current information.
All optimization happens on the surrogate cost landscape, so no explicit derivatives of the cost function are needed.
Through the use of a fitted cost function, these methods are also expected to be more resilient to noise.
Several classical surrogate models have been included in the scikit-quant package~\cite{lavrijsen2020classical,scikit-quant}.
In the Bound optimization by quadratic approximation (BOBYQA) algorithm~\cite{powell2009bobyqa}, a local quadratic model is formulated from the previous function values. It is then minimized in the trust region to obtain a new parameter value.
When the evaluation at this new parameter value does not result in a lower function value, the trust region is altered and the quadratic model is optimized in this new parameter space.
It was shown that this method works well when the PQC is initialized close to the optimal parameters but has more problems with shallow optimization landscapes and gets stuck in local minima~\cite{lavrijsen2020classical}.
The stable noisy optimization by branch and fit (SnobFit)~\cite{huyer2008snobfit} algorithm uses a branching algorithm to explore new areas in parameter space.

\subsubsection{Resource-aware optimizers}

Optimization methods and strategies adopted for early demonstrations of VQA are largely general-purpose and black-box with minimal emphasis on reducing the quantum resources used in the optimization. Therefore, they are more costly and prone to errors than their classical counterparts.
Optimizers developed in more recent years are tailored to additionally minimize quantities associated with the quantum cost of the optimization, e.g. number of measurements or real hardware properties. Additionally, one can use circuit compilation methods as the ones described in \secref{sec:compilers}.
    
\paragraph{ROSALIN.} 
    While VQA leverage low-depth circuits to execute on near-term quantum processors, a significant challenge in implementing these algorithms is the prohibitive number of measurements, or shots, required to estimate each expectation value that is used to compute the objective.
    To address the challenge, \cite{arrasmith2020operator} developed a shot-frugal optimizer called ROSALIN (Random Operator Sampling for Adaptive Learning with Individual Number of shots) that effectively distributes fractions of a predefined number of shots to estimate each term of the Hamiltonian as well as each partial derivative.
    Given the expectation value of the Hamiltonian decomposed into the $h_{i}$ terms as in \eqref{eq:Pauli_string},
    
    the authors note several strategies for allocating shots for estimating each term $\expval{h_i}{}$. 
    While a naive strategy would allocate equal numbers of shots per term, the authors observed lower variance in the energies using weighted approaches in which the number of shots allocated to the $i$-th term $b_i$ is proportional to the corresponding Hamiltonian coefficient $c_i$. 
    
\paragraph{SPSA.}
    In experimental realizations of VQA, the optimizer is often hindered by statistical noise.
    In \cite{kandala2017hardware} this issue is circumvented by applying the simultaneous perturbation stochastic approximation (SPSA) algorithm \cite{spall1992multivariate}, in which the algorithm hyperparameters are determined by experimental data on the level of statistical noise.
    Compared to the finite-difference gradient approximation, which requires $O(p)$ function evaluations for $p$ parameters, SPSA requires only two evaluations, as explained in the supplementary.
    The convergence of SPSA with various types of PQCs has been studied~\cite{woitzik2020entanglement}.

\section{Other NISQ approaches}\label{sec:non-variational}

We proceed to review some of the notable NISQ algorithms, besides VQA. These algorithms do not require tuning the parameters of a PQC in an adaptive feedback manner and often exploit analog or hybrid paradigms that constitute alternatives to the digital quantum computation.

\subsection{Quantum annealing}\label{subsec:quantum-annealing}

Quantum annealing (QA) ~\cite{finnila1994quantum,kadowaki1998quantum} derives its inspiration from simulated annealing (SA), a classical global optimization technique, usually employed to solve combinatorial optimization problems. SA can be valuable in discovering global optima in situations involving many local optima.
The word ``annealing'' comes from metallurgy, which represents heating and slow cooling. In SA, one identifies the objective function with the energy of a statistical-mechanical system. 
The system is assigned an artificially-induced control parameter, called temperature. Like annealing, SA starts with some high temperature $T$, and then the value of $T$ is brought down following some temperature variation function called ``annealing schedule'' such that the final temperature is $T=0.$ The algorithm chooses a candidate state close to the current state randomly. If it improves the solution, it is always accepted with probability $1$. If it does not, then the acceptance is determined based on a temperature-dependent probability function. The idea of tolerating worse solutions can be considered as a virtue of the algorithm. In SA, the probability that a bad solution is accepted is slowly decreased as the solution space is explored. This relates to the notion of ``slow cooling'' in annealing.

In QA, one utilises quantum-mechanical fluctuations, like quantum tunnelling, to explore the solution space. This is analogous to the idea of using thermal fluctuations in SA to explore the solution space. In QA, artificial degrees of freedom of quantum nature via non-commutative operators are introduced, which induces quantum fluctuations.
The strength of these quantum fluctuations is controlled using an annealing schedule (similar to SA, where we decrease the temperature).
The physical idea behind annealing schedule in QA is to move the system from some initial Hamiltonian ground state to the ground state of the problem Hamiltonian. The concept of QA is related to the notion of quantum adiabatic evolution, which is being used for adiabatic quantum computation~\cite{farhi2000quantum,albash2018adiabatic}.

We proceed to a formal discussion now. Adiabatic quantum computation is model of computation based on quantum mechanical processes operating under adiabatic conditions~\cite{farhi2000quantum,albash2018adiabatic}. Before understanding adiabatic quantum computation, one needs to grasp the concept of $k$-local Hamiltonian.
\begin{definition} \label{def:k-loc}
\textbf{k-local Hamiltonian}: A $k$-local Hamiltonian is a Hermitian matrix of the form $H = \sum_{i=1}^r \hat{h}_i$ where each  term is a Hermitian operator acting non-trivially on at-most $k$ qudits, i.e., $\hat{h}_i = h \otimes I$ where $h$ is a Hamiltonian acting on at-most $k$ neighbouring qudits and $I$ is the identity operator. 
\end{definition}

Let us consider a time-dependent hamiltonian $H(s)$, for $s\equiv \frac{t}{T} \in [0,1]$ and a quantum system initialized in the ground state of $H(0)$. We assume that $H(s)$ varies smoothly as a function of $s$ and $H(s)$ has a unique ground state for $s\in[0,1]$. A quantum state initialized in $\ket{\psi(t=0)}$ evolves according to the following  Schr\"odinger equation (setting $\hbar=1$),
\begin{equation}
\label{eq: schro_1}
i \frac{d}{dt} \ket{\psi(t)} = H(t) \ket{\psi(t)}.
\end{equation}
The above equation can be further, equivalently, written as
\begin{equation}
\label{eq: schro_2}
i \frac{d}{ds} \ket{\psi(s)} = TH(s) \ket{\psi(s)}.
\end{equation}
Assuming $\ket{\psi(0)}$ is a ground state of $H(0),$ then in the limit $T\rightarrow \infty$, $\ket{\psi(t)}$ is a ground state of $H(1)$ obtained via evolution \eqref{eq: schro_1}. Such an evolution will be, henceforth, referred as adiabatic evolution according to $H$ for time $T$. 

Now  we proceed to define adiabatic quantum computation. 
\begin{definition}
\textbf{Adiabatic quantum computation (adapted from \cite{aharonov2008adiabatic})}: An adiabatic quantum computation is specified by two $k$-local Hamiltonians $H_0$ and $H_1$  acting on n qudits and a map $s(t): \left[ 0,T\right] \longrightarrow [0,1].$ The input of the computation is the ground state of $H_0$, which is unique and is a product state. The desired output is given by a quantum state  which is $\epsilon-$close in l2-norm to the ground state of $H_1$. Furthermore, $T$ is the smallest time such that the adiabatic evolution generated via $H(s)=(1-s)H_0 + sH_1$ for time $T$ yields the desired output. The running time of the algorithm is given by $T.\text{max}_s \norm{H(s)},$ where $\norm{.}$ denotes the spectral norm.
\end{definition}

QA relaxes the strict requirement of adiabatic evolution, thus allowing diabatic transitions due to the finite temperature of the system, from fast changes of Hamiltonian parameters, and the interaction with the noisy environment~\cite{hauke2020perspectives}. Because of diabatic transitions, QA is prone to getting trapped in excited states.

QA has been investigated for problems in diverse areas including machine
learning~\cite{li2018quantum,benedetti2017quantum,o2015bayesian,benedetti2016estimation,benedetti2018quantum}, protein folding~\cite{perdomo2008construction,perdomo2012finding,babbush2012construction,babej2018coarse}, fault diagnosis \cite{perdomo2015quantum,perdomo2019readiness}, compressive sensing~\cite{ayanzadeh2019quantum}, finance~\cite{orus2019quantum,bouland2020prospects,marzec2016portfolio,rosenberg2016solving,venturelli2019reverse,cohen2020portfolio}, fermionic simulation \cite{babbush2014adiabatic} and high
energy physics~\cite{mott2017solving,das2019track}. The protein folding problem entails calculating a protein's lowest free-energy structure given its amino-acid sequence. The goal is to solve the protein folding problem by mapping it to a Hamiltonian and then using QA to identify low-energy conformations of the protein model. In~\cite{perdomo2012finding}, authors use five and eight qubits for the four-amino-acid sequence to encode and solve the protein folding problem for a short tetrapeptide and hexapeptide chain.  QA has been one of the prominent approaches in the NISQ era in the search for quantum advantage~\cite{perdomo2018opportunities,bouland2020prospects,hauke2020perspectives}.

A major experimental implementation of QA is the D-Wave machine. It attempts to solve problems in
a particular form called Quadratic Unconstrained Binary Optimization
(QUBO)~\cite{lucas2014ising}.  Optimization problems can be cast as a polynomial unconstrained
binary optimization (PUBO) expressed in the form of a $k$-local interaction
with $k\geq3$ over binary variables $x_{i}\in\left\{ 0,1\right\}$ ~\cite{hauke2020perspectives,perdomo2019readiness}.
QUBO is a special case of PUBO with $k=2.$  For a vector of $n$ binary variables
$\textbf{x}\in\left\{ 0,1\right\} ^{n}$ and problem specified values of $Q\in\mathbb{R}^{n\times n}$
and $\textbf{c}\in\mathbb{R}^{n}$, QUBO is defined as
\begin{equation}
\arg\min \textbf{x}^{T}Q\textbf{x}+\textbf{c}^{T}\textbf{x}.\label{eq:QUBO_1}
\end{equation}
Using the map $x_{i}\rightarrow\frac{1-\sigma_{z}^{i}}{2}$, one can convert
the problem in expression \ref{eq:QUBO_1} to ground state finding
problem of the following diagonal $n$-qubit Ising Hamiltonian (up to a constant),
\begin{equation}
H_\text{QUBO}=-\sum_{i,j}J_{i,j}\hat{\sigma}_{z}^{i}\hat{\sigma}_{z}^{j}-\sum_{i}h_{i}\hat{\sigma}_{z}^{i},\label{eq:QUBO_Ham}
\end{equation}
where $\hat{J}_{i,j}=-\frac{Q_{i,j}}{4}$ and $h_{i}=\frac{-c_{i}+\sum_{j}Q_{i,j}}{2}.$

Starting with the ground state of the base Hamiltonian $H_{0}=-\sum_{i}\hat{\sigma}_{x}^{i}$,
solving the QUBO problem on a quantum annealer corresponds to implementing
the annealing schedule $A(t)$ and $B(t)$ for the Hamiltonian
\begin{equation}
H(t)=A(t)H_{0}+B(t)H_\text{QUBO}.\label{eq:Annealing_Schedule}
\end{equation}
Here, $A(0)=B(T)=1$ and $A(T)=B(0)=0$, where $T$ is computation
time. Because annealing does not necessarily satisfy the constraints
of adiabatic evolution, one can end up in excited states as mentioned
earlier. However, one can run the annealing schedule multiple times
and take the best answer i.e, the one corresponding to lowest energy. The qubits in an annealer are not necessarily all-to-all connected, necessitating additional engineering restrictions, such as the minor embedding problem~\cite{choi2008minor,choi2011minor,klymko2014adiabatic}.

The potential of QA has been studied extensively ~\cite{hastings2020power,hauke2020perspectives,farhi2002quantum,denchev2016computational,brady2016spectral}. The performance of D-Wave annealers have also been explored comprehensively~\cite{shin2014quantum,albash2015reexamining,cohen2020portfolio}. 
In particular, an extensive study comparing the performance of quantum annealing with other quantum-inspired and classical optimization state-of-the-art strategies, and in the context of a real- world application, can be found in \cite{perdomo2019readiness}.
For the details of QA, refer to \cite{hauke2020perspectives} and the references therein. A review on Adiabatic Quantum Computation is presented in \cite{albash2018adiabatic}. Refer to~\secref{sec:QML} and Appendix~\secref{sec:finance} for the discussions regarding applications of QA in machine learning and finance respectively.

\subsection{Gaussian boson sampling} \label{sec:GBS}

\begin{figure}[ht!]
    \centering
    \includegraphics[width=\columnwidth]{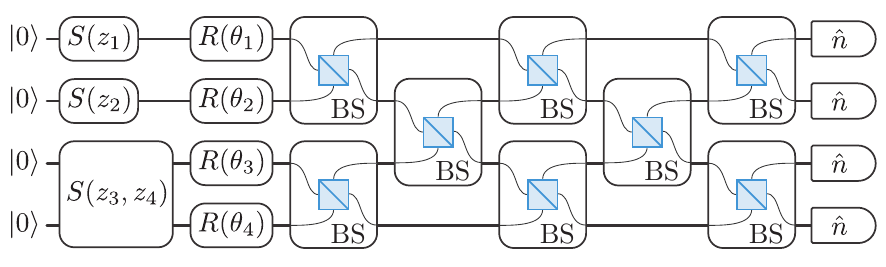}
    \caption{Gaussian boson sampling circuit for a photonic setup. The qumodes are prepared in gaussian states from the vacuum by squeezing operations $S(z_i)$, followed by an interferometer consisting of phaseshifters $R(\theta) = e^{i\theta_j}$ and beam-splitters BS. At the end, photon number resolving measurements are made in each mode.}
    \label{fig:boson_sampling}
\end{figure}

Boson sampling was first proposed as a candidate for quantum computational supremacy by  \cite{aaronson2011computational}. The scenario consists of having $n$ photons that enter an optical circuit consisting of $m$ modes.
This state is then acted upon by a series of phase-shifters and beam-splitters. A phase-shifter adds a phase $R(\theta) = e^{i\theta_j}$ with some angle $\theta_j$ to the amplitude in mode $j$, and acts as the identity in the other $m-1$ modes. A beam-splitter acts on two modes with a rotation $\begin{pmatrix}
\cos\phi & -\sin\phi \\
\sin\phi & \cos\phi \end{pmatrix}$ for some angle $\phi$ and as the identity in the other $m-2$ modes. Finally, a measurement is made where the number of photons in each mode is found. An optical circuit with these elements is shown in Figure \ref{fig:boson_sampling}. Each of these measurement outcomes represent a sample from the the symmetric wavefunction that bosonic systems have. Aaronson and Arkhipov found that the existence of an efficient classical algorithm for sampling from the distribution implies the existence of a classically efficient algorithm for the calculation of the permanent of a related matrix. Such an algorithm would imply the collapse of the polynomial hierarchy (see \ref{ch1:subsec:non-complexity}) to the third order, which is believed to be unlikely~\cite{arora2009computational}. Consequently, such an algorithm is unlikely to exist.

Gaussian boson sampling (GBS) is a variant of boson sampling, where instead of photon states as inputs into the optical circuit, Gaussian states are used as inputs \cite{hamilton2017gaussian}. Gaussian states are those whose Wigner quasi-probability distributions $W(q,p)$ have Gaussian shape. A good introduction to the theory can be found in~\cite{serafini2017quantum}. They have the advantage that they can be created deterministically \cite{hamilton2017gaussian,kruse2019detailed}. They also provide additional degrees of freedom in comparison to boson sampling. Where boson sampling is equivalent to sampling from the permanent of a matrix, GBS is computationally equivalent to sampling from the Hafnian function of a matrix.  Given a graph $G$ with adjacency matrix $E,$ the Hafnian of $E$
is the number of perfect matchings of the graph $G.$ A matching of
a graph $G$ is a subset of edges $M$ such that no two edges in $M$
have a vertex in common. A matching $M$ is perfect if every vertex
is incident to exactly one edge in $M.$ While the Permanent gives the
number of perfect matchings for a bipartite graph; the Hafnian gives perfect
matching for any graph. Thus, the Hafnian can be thought of as a generalization
of the Permanent. Using the adjacency matrix $E,$ the relation between
the Hafnian and the Permanent is given by
\begin{equation}\label{ch2:eq:haf_per}
\mathrm{Haf}\begin{pmatrix} 0 & E \\
E^T & 0
\end{pmatrix} = \mathrm{Per}(E).
\end{equation}
The hardness of simulating a noisy version of GBS has been studied~\cite{qi2020regimes} and GBS has recently become the second platform to show quantum computational supremacy \cite{zhong2020quantum}, and the latest experimental venture towards dynamically programmable GBS nanophotonic chip was carried out by~\cite{arrazola2021quantum}. 

\subsubsection{The protocol}

In GBS we consider $m$ quantum modes (\textit{qumodes}), represented by harmonic oscillators with canonically conjugate variables $q$ and $p$. Gaussian states of the qumodes are those represented by a Wigner-function $W(q,p)$ that has a Gaussian form. These states can be efficiently represented by the complex amplitude $\alpha = \frac{1}{\sqrt{2\hbar}}(q+ip)$ and a covariance matrix $\Sigma \in \mathbb{C}^{2m \times 2m} $. A general pure Gaussian state can be generated from a vacuum with three steps: \textit{i)} Single mode squeezing; \textit{ii)} multi-mode linear interferometry; and \textit{iii)} single-mode displacements. In the GBS protocol the state is then measured in the Fock-basis, performed in practice using photon number resolving detectors. The optical circuit in Figure~\ref{fig:boson_sampling} shows how the system is initialized in the vacuum state, followed by single- and multi-mode squeezing operators $S(z_i)$ and $S(z_i, z_j)$, respectively, and an interferometer with phaseshifters $R(\theta_j)$ and beamsplitters BS. At the end of the protocol, the photon number in each mode is measured. 

For a Gaussian state with zero mean (of the Wigner-function), the probability of detecting $s_i$ photons in the $i$-th qumode is given by:~\cite{hamilton2017gaussian,kruse2019detailed}
\be\label{ch2:eq:GBS_probabilites}
P(s_1,s_2...s_m) = \frac{1}{\textrm{det}(Q)}\frac{\textrm{Haf}(A_s)}{\sqrt{s_1!s_2!\cdots s_m!}}
\ee
where all the matrices are defined in terms of the covariance matrix $\Sigma$:
\begin{align*}
Q = \Sigma + \mathbb{1}/2 \\
A = X(\mathbb{1}-Q^{-1}) \\
X = \begin{bmatrix} 0 & \mathbb{1}\\
\mathbb{1} & 0
\end{bmatrix}
\end{align*}
The $A_s$ matrix is a matrix created from $A$ such that if $s_i = 0$, we delete the rows and colums $i$ and $i+m$ of the matrix, and if $s_i\neq 0$, we repeat the rows and columns $s_i$ times. 

This means that by manipulating the covariance matrix $\Sigma$, we control the matrix from which we sample the Hafnian. For a pure Gaussian state, it can be shown that the $A$ matrix is symmetric~\cite{bromley2020applications}. 

A simpler form of the experiment where instead of counting the number of photons in each mode, we only detect if there are photons or not in each mode, can be used to sample from the so-called Torontonian function of a matrix~\cite{quesada2018gaussian}. If the probability of observing more than one photon per output mode stays low enough, this model has been shown to stay classically intractable to simulate. A more general experiment instead, where the mean of the Gaussian states is non-zero, can be used to sample from the loop Hafnian~\cite{bjorklund2019faster}.

\subsubsection{Applications}

A number of algorithms for applications of GBS have been investigated, and are reviewed by \cite{bromley2020applications}. Here we only briefly summarise that work. Typically GBS algorithms are based on heuristics, and GBS devices are often used to provide a seed for starting points of classical algorithms. GBS can also be viewed as directly giving access to a statistical distribution, such as in the case of point processes~\cite{jahangiri2020point}.

Problems in chemistry have been tackled using GBS. Vibrational spectra of molecules have been computed using GBS by mapping the phononic modes of the molecule to the qumodes of the GBS device~\cite{huh2015boson}, and by extension electron-transfer reactions have been studied~\cite{jahangiri2020quantum}. The technique of sampling high-weight cliques has also been applied to predict molecular docking configurations~\cite{banchi2020molecular}. 

The largest number of GBS algorithms are for graph problems, since the adjacency matrix of a graph is a natural fit as the symmetric $A$ matrix. The Hafnian function computes the number of perfect matchings of a graph, so the samples from the GBS device are with high likelihood from sub-graphs with high density. This is how GBS is used to identify dense subgraphs~\cite{arrazola2018using}, and to get good initial guesses for classical search algorithms to compute the max-clique of a graph~\cite{banchi2020molecular}. 

GBS can also be used to build succinct feature vectors, or ``fingerprints'', of larger graphs via coarse-graining techniques. These feature vectors can then be used as inputs to statistical methods or machine learning to classify graphs. One such problem is to measure the similarity between graphs~\cite{schuld2020measuring}, which has applications in tasks such as checking fingerprint comparison or detecting mutations of molecules.

GBS can also be used as a type of importance sampling device to speed up algorithms requiring randomness. This is how stochastic search algorithms have been sped up by sampling from a GBS device encoding the graph to be searched, instead of sampling uniformly~\cite{arrazola2018quantum}. 

Recently variational methods have been used within the GBS framework~\cite{banchi2020training} and applied to stochastic optimization and unsupervised learning. The method is based on varying the squeezing and interferometer parameters in the device and updating based on the measurement outcomes.

\subsection{Analog quantum simulation}

Simulating a quantum system is a hard problem for classical computers as the Hilbert space increases exponentially with the size of the system. 
As a solution to this long-standing problem, Feynman suggested the ground breaking idea to harness that physical systems given us by nature are quantum-mechanical. He proposed to use quantum systems that are well-controlled in the lab to simulate other quantum systems of interest~\cite{feynman1982simulating}. This concept has spurred the field of analog quantum simulation~\cite{georgescu2014quantum,trabesinger2012quantum}. 

The core idea differs from digital quantum simulation~\cite{lloyd1996universal}. Digital quantum simulators decompose the quantum dynamics to be simulated into a circuit of discrete gate operations that are implemented on a quantum processor. The quantum processor is a well controlled quantum system, that is engineered to be able to efficiently apply a set of specific quantum gates that are universal, i.e. a sequential application of those gates can realize arbitrary unitaries (see \secref{sec:gate_set}).

With this universal approach, a wide range of quantum problems can be simulated to a desired accuracy with a polynomially increase in quantum resources only~\cite{lloyd1996universal}. However, current quantum processors have only limited coherence time and lack the capability to correct errors that inevitably appear during the computation, severely limiting the range of dynamics that can be simulated. 
In contrast, the idea of analog quantum simulators is to map the problem Hamiltonian to be simulated $\hat{H}_\text{sys}$ to the Hamiltonian of the quantum simulator $\hat{H}_\text{sim}$, which can be controlled to some degree, $\hat{H}_\text{sys} \leftrightarrow \hat{H}_\text{sim}$. One then runs the quantum simulator, and maps the results back to the problem. 

The range of problems that can efficiently mapped to the simulator is limited, however as one uses the native quantum dynamics of the simulator, the accessible system size, coherence length and errors is often more favorable compared to current digital quantum simulators. 

\subsubsection{Implementations} 

A wide-range of implementations in various controlled quantum systems has been achieved, ranging from solid state superconducting circuits~\cite{houck2012chip}, quantum dot arrays~\cite{hensgens2017quantum},
nitrogen-vacancy centers~\cite{yao2012scalable}, atomic and molecular physics based platforms such as trapped ions~\cite{blatt2012quantum}, interacting
photons~\cite{chang2014quantum,hartmann2016quantum}, Rydberg atoms~\cite{adams2019rydberg}, and cold
atoms~\cite{bloch2012quantum,gross2017quantum,amico2020roadmap}. 

Concepts of analog quantum simulation have been used within VQA as well, such as problem inspired ans\"atze (see \secref{sec:HamiltonAnsatz}) or protocols inspired by quantum control~\cite{yang2017optimizing,meitei2020gate}. Experimental results for a quantum many-body problem beyond current classical computational capabilities have been reported for 2-D systems~\cite{choi2016exploring}.

\subsubsection{Programmable quantum simulators} 

An analogue quantum system, such as a superconducting circuit, can be adapted to simulate arbitrary dynamics~\cite{bastidas2020fully}. The idea is to drive the parameters of the Hamiltonian $H(t)$ that describes the analog quantum simulator in time $t$. This can be done by adjusting the physical parameters of the quantum simulator in time. 
The driving protocol is engineered via machine learning methods~\cite{haug2020engineering} such that the effective dynamics of the driven system over a time $T$ corresponds to the evolution of a problem Hamiltonian one wants to simulate. 
The effective dynamics that is generated can realize long-range interactions as well as complicated many-body terms, which are natively not supported by the quantum simulator and are often hard to simulate on digital quantum simulators. By periodically driving the analog quantum simulator with the aforementioned driving protocol, various problem Hamiltonians can be simulated~\cite{oka2019floquet}. One can realize complicated many-body dynamics or chemistry problems , as well as solve combinatorial tasks such as SAT-3.
Trapped ion  based analog quantum simulators have been recently used for the implementation of quantum approximate optimization algorithm~\cite{Pagano2020quantum}.

Highly controllable analog quantum simulators have also been proposed for engineering quantum chemistry Hamiltonians by combining different cold atom species embedded within cavity modes, which mediate long-range interactions required to simulate Coulomb repulsion. Optical fields can be used to modify the potential and interaction parameters in order to simulate large scale chemistry problems~\cite{arguello2019analogue} as well as quantum spin model with tunable interactions for system sizes ranging from 64 to 256 qubits \cite{ebadi2020quantum}. 
For ion traps, a programmable quantum simulator can be designed by light fields, that are applied to manipulate the internal degrees of freedom as well as the interaction between different ions. This allows one to  simulate various types of spin Hamiltonians with a high degree of control over the parameters~\cite{monroe2019programmable}.

\subsection{Digital-analog quantum simulation and computation}
\label{sec:digi-ana}

As opposed to analog simulators that are limited by the Hamiltonians they can simulate~\cite{goldman2014periodically,kyriienko2018floquet}, digital quantum simulators can simulate any system's Hamiltonian, but with sometimes costly quantum resources. 
To benefit from a combination of the two approaches, the digital-analog method to quantum computation~\cite{dodd2002universal, parra2020digital} and simulation~\cite{mezzacapo2014digital,yung2014transistor} has been proposed. These schemes combine the application of digital single-qubit gates with the underlying analog Hamiltonian of the quantum processor. This approach allows for universal simulation of quantum dynamics while replacing two qubit gates for an analog Hamiltonian and has been argued to be more resilient against certain types of noise than digital quantum computing~\cite{parra2020digital, martin2020digital, garcia2021noise}.

Digital-analog quantum simulation has been proposed to simulate the Rabi model \cite{mezzacapo2014digital}, Dicke model \cite{mezzacapo2014digital,lamata2017basic}, and fermionic systems \cite{garcia2015fermion, celeri2021digital}. Digital-analog quantum simulation has been reviewed in \cite{lamata2018digital}, whereas digital-analog quantum computing is more recent. The implementation of digital-analog quantum computing has been proposed for superconducting platforms \cite{yu2021superconducting,gonzalez2021digital}. So far the computing framework has been used to simulate Ising models \cite{parra2020digital}, where the analog blocks can be used to enhance the effective connectivity of the qubits to simulate graphs that have a different connectivity from the native connectivity of the quantum device~\cite{galicia2020enhanced}. The analog blocks have also been applied to reduce the operation count required to perform the quantum Fourier transform \cite{martin2020digital}. 

The digital-analog approach has also been combined with VQA (see \secref{ch:building}) resulting in a digital-analog QAOA algorithm, where the two-qubit gates have been replaced by analog blocks \cite{headley2020approximating}. This also has two versions: \textit{i)} where a layer of entangling gates is replaced by an analog block; and \textit{ii)} where a continuous analog block is applied continuously with single-qubit operations overlayed. 

\subsection{Iterative quantum assisted eigensolver}\label{sec:IQAE}

Almost all of the VQAs
update a PQC's parameters in a feedback loop. However, there
exist alternative algorithms that can circumvent this approach with the ansatz given by~\cite{bharti2020iterative,huang2019nearterm,mcclean2017hybrid}
\begin{equation}
\vert\phi\left(\boldsymbol{\alpha}(t),\boldsymbol{\theta}\right)\rangle=\sum_{i=0}^{m-1}\alpha_{i}(t)\vert\psi_{i}\left(\boldsymbol{\theta}_{i}\right)\rangle,\label{eq:IQAE_Ansatz}
\end{equation}
where $\alpha_{i}\in\mathbb{C}$ and $\theta_{i}\in\mathbb{R}^{k_{i}}$
for non-negative integers $k_{i}.$ This ansatz is a
linear combination of quantum states, where the $\alpha_{i}$ parameters
are stored on a classical device. In the special case $m=1$ it corresponds to the usual
PQC, whereas for $m>1$ this ansatz subsumes it. 
This ansatz has been used for finding
the ground state of Hamiltonians~\cite{bharti2020quantumeigensolver,bharti2020iterative}, excited state~\cite{parrish2019quantum,parrish2019quantum2,huggins2020non,stair2020multireference} the simulation of quantum dynamics~\cite{bharti2020quantum,haug2020generalized}, error mitigation~\cite{mcclean2017hybrid},
nonlinear dynamics~\cite{bharti2020quantum,haug2020generalized}, linear systems~\cite{huang2019nearterm} and semidefinite programming~\cite{bharti2021nisq}. 
If one keeps the parameters of the PQC $\theta_{i}$
fixed and only varies the $\alpha_{i}$, the algorithm can be considered a borderline non-VQA algorithm. 
Update of $\theta_{i}$ parameters has been shown to cause trainability
issues in VQA (see \secref{sec:BP}) and thus by fixing $\theta_{i}$ one
can by construction circumvent these issues.
We present here the iterative quantum assisted eigensolver algorithm (IQAE) as an illustration, and in the applications subsection  the quantum assisted simulator for closed systems (see \secref{sec:quantum_assisted_simulator}), open systems (see \secref{sec:OpenQuantumSystem}) and Gibbs state preparation (see \secref{sec:gibbsstate}).

The IQAE algorithm provides an approximation to the ground
state of a Hamiltonian $H$.
Without loss of generality, the $N$-qubit Hamiltonian $H$ is assumed to
be a linear combination of unitaries
\begin{equation}
H=\sum_{i=1}^{m}\beta_{i}U_{i}\,.\label{eq:LCU_Ham}
\end{equation}
Here, $\beta_{i}\in\mathbb{C}$ and $U_{i}\in SU\left(2^{N}\right)$
for $i\in\left\{ 1,2,\cdots,m\right\}.$ The unitaries $U_{i}$ act on at most $\mathcal{O}\left(poly\left(\log N\right)\right)$
qubits. This condition can be relaxed if the unitaries Pauli strings (see \secref{sec:pauli_strings}).
The ansatz state is taken as linear combination of ``cumulative
$K$-moment states'' $\mathbb{CS}_{K}$, which
is generated using some efficiently preparable quantum states and
the unitaries defining the Hamiltonian in \eqref{eq:LCU_Ham}.
For pedagogical reasons, we present the
definition of $K$-moment states and cumulative $K$-moment states.
\begin{defn}  \label{def:cumulant_states} (adapted from \cite{bharti2020iterative})
For a given positive integer $K$, a set of unitaries $\mathbb{U}\equiv\left\{ U_{j}\right\} _{j=1}^{m}$ and
 a quantum state $\vert\psi\rangle,$  
$K$-moment states is the set of quantum states of the form $\left\{ U_{j_{K}}\cdots U_{j_{2}}U_{j_{1}}\vert\psi\rangle\right\} _{j}$
for  $U_{j_{l}}\in\mathbb{U}.$ Let us denote the aforementioned set by $\mathbb{S}_{K}$. We define the singleton set $\left\{ \vert\psi\rangle\right\} $
 as the $0$-moment state (denoted by $\mathbb{S}_{0}$). Finally, we define the  cumulative
$K$-moment states $\mathbb{CS}_{K}$ as $\mathbb{CS}_{K}\equiv\cup_{i=0}^{K}\mathbb{S}_{i}$.
\end{defn}
As instructive example, note that the set of $1$-moment states is $\left\{ U_{j}\vert\psi\rangle\right\} _{j=1}^{m}$, 
where the unitaries $\left\{ U_{j}\right\} _{j=1}^{m}$ make up the Hamiltonian $H$. The set of cumulative $1$-moment states is $\mathbb{CS}_{1}=\{\vert \psi \rangle\} \cup \left\{ U_{j}\vert\psi\rangle\right\} _{j=1}^{m},$ and the set of cumulative $K$-moment states is $\mathbb{CS}_{K}=\{\vert \psi \rangle\} \cup \left\{ U_{j_1}\vert\psi\rangle\right\} _{j_1=1}^{m} \cup \dots \cup \left\{ U_{j_K}\dots U_{j_1}\vert\psi\rangle\right\} _{j_1=1,\dots,j_K=1}^{r}$.

Now, the ansatz is is given by
$\vert\xi\left(\alpha\right)\rangle^{\left(K\right)}=\sum_{\vert\chi_{j}\rangle\in\mathbb{CS}_{K}}\alpha_{j}\vert\chi_{j}\rangle$. The ground state problem reduces to the following optimization program
\begin{align*}
\min_{\alpha}&\,\alpha^{\dagger}\mathcal{D}^{\left(K\right)}\alpha\\
 \text{subject to }&\alpha^{\dagger}\mathcal{E}^{\left(K\right)}\alpha=1\,.\numberthis\label{eq:QCQP_1}
\end{align*}
Here, the overlap matrices $\mathcal{D}^{\left(K\right)}$ and $\mathcal{E}^{\left(K\right)}$
are given by
$\mathcal{D}_{nm}^{\left(K\right)}=\sum_{i}\beta_{i}\langle\chi_{n}\vert U_{i}\vert\chi_{m}\rangle$ and
$\mathcal{E}_{nm}^{\left(K\right)}=\langle\chi_{n}\vert\chi_{m}\rangle$.
These overlap matrices can be computed on a quantum computer without
the requirement of any complicated measurement involving multi-qubit
controlled unitaries. For example, for a Hamiltonian composed of Pauli strings, the product of Pauli strings is a Pauli string up to a phase
factor $\pm$1 or $\pm\iota$. Thus, the overlap matrices are simply expectation values $\langle\psi\vert \hat{P}\vert\psi\rangle$ of some Pauli string $\hat{P}$, which can be easily measured (see \secref{sec:measurement}).

The optimization program \ref{eq:QCQP_1}
is a quadratically constrained quadratic program (QCQP) with single
equality constraint. The Algorithms proceeds in three serial and disjoint steps.
\begin{enumerate}
    \item Select ansatz, which can be done on paper
    \item Estimate overlap matrices on quantum computer, which can be done efficiently in a parallel fashion.
    \item Post-processing on a classical computer to solve the QCQP based on the overlap matrices from step $2$.
\end{enumerate}

As a major speedup compared to standard VQA, there is no feedback loop between classical and quantum computer such that the calculations can be easily parallelized. The ansatz can be improved by changing $K$ to $K+1.$  The ansatz construction
is systematic and there is no trainability issue such
as the barren plateau problem (see \secref{sec:BP}). For the QCQP, there  exist conditions which tell whether
a local minima is a global minima as
a stopping criteria for the classical solver. Moreover, the Lagrangian
relaxation of the program \ref{eq:QCQP_1} is a semidefinite program and efficiently solvable. 
\section{Theoretical challenges}

\subsection{Barren plateaus} \label{sec:BP}

It was recently shown that the expectation
value of the gradient of the objective function corresponding to randomly initialized PQCs (RPQC) decays exponentially to zero as a function of the number of qubits~\cite{mcclean2018barren}. 
The mathematical basis of this result hinges
on the fact that the PQC from \eqref{eq:HE_ansatz} becomes a
unitary $2$-design as the circuit depth increases polynomially with the circuit width i.e, the number of qubits. The notion of unitary $2$-design
has been used extensively in the recent proofs of barren plateau in
RPQCs, which necessitates a small discussion about their mathematical
structure.

Using the notion of $2$-design, the appearance of barren plateaus
in the training landscape has been established for various kind of ans\"atze.
Barren Plateaus can be thought of as a consequence of the exponentially
large dimension of the Hilbert space when the number of qubits increases
and the fact that the variational circuit unitary, when the parameters
are initialized at random, is a 2-design. Consequently, the strategies
proposed to tackle this problem focus on reducing the space dimension
of this unitary or breaking the randomness properties related to the
2-designs. Another way to think of the origin of the barren plateau
issue could be the problem-agnostic nature of the ansatz, faced with exponentially
large parameter space. Thus, one could attempt to devise ans\"atze as well
as the optimization methodology in a problem-aware manner by using physically-inspired or problem-specific ans\"atze as the ones presented in \secref{sec:PQC} or those proposed in~\secref{sec:IQAE}.

Besides the exponential parameter space that induced barren plateaus, other physical phenomena can also generate them. In particular, the noise and decoherence present in the quantum computing experiments also generates this problem in VQAs \cite{wang2020noise}. Entanglement-induced barren plateaus have also been reported recently~\cite{marrero2020entanglement}.

While certain ans\"atze can be assumed or proven to form (approximate) 2-designs, such proofs are challenging for the general ones.
To numerically verify the presence of barren plateaus, past studies often considered computing the gradients and variances of a local observable using a particular ansatz over increasing system sizes~\cite{mcclean2018barren, skolik2020layerwise}.

Another attempt to avoid a barren plateau is to initialize the variational circuit with a particular state choice. Intuitively, the algorithm will start in a particular region of the Hilbert space allowing the optimization subroutine to potentially find the minima in a closer region. This strategy include all physically inspired methods mentioned in \secref{sec:PQC}. The use of clever encodings for the algorithm parameters can also be understood as a initialization strategy \cite{cervera2020meta} (see \secref{sec:QML}). Classical algorithms such as neural networks can also be used to learn the proper circuit encodings~\cite{verdon2019learning}.

A good choice for the initial state is often not enough to reduce the size of the Hilbert space. Although expressive circuit ans\"atze are usually a requirement for the success of a VQA (see \secref{sec:expressibility} for more details), it can expand the parameter space that the optimizer has to explore. Several works propose circuit structures that reduce that space by introducing correlations between the variational parameters of the circuit \cite{volkoff2020large}, block-wise initialization of those parameters~\cite{grant2019initialization} or exploring particular ansatz structures~\cite{sharma2020trainability}.

The mentioned works require a circuit design that is not necessarily hardware efficient. Other ideas focus on the classical parts of the VQA instead of the quantum circuit designs. One example is using local instead of global cost functions for the optimization. It has been shown~\cite{cerezo2020cost} that barren plateaus also emerge in shallow depth circuits, and that the use of local cost functions reduces the exponential decay tendency to a polynomial one. The optimization strategy may also reduce the effect of the vanishing gradients, for instance by training the circuit layer by layer~\cite{skolik2020layerwise,lyu2020accelerated} or by measuring low depth gradients~\cite{harrow2019low}. 
Certain variational quantum algorithms for quantum simulation can be free of barren plateaus when at every training step the state to be learned is close to the state of the circuit~\cite{haug2021optimal}.

Barren plateaus are a roadblock in the trainability and hence any
PQC ansatz which suffers from this phenomena will fail to properly
train the parameters in its search for the near-optimal (or optimal) performance.
As shown in Ref.~\cite{arrasmith2020effect}, even the family of gradient-free approaches which perform local search, therefore mimicking gradient-based optimization, seem to face similar challenges. However, one can circumvent this issue by using hybrid quantum states of the form of equation \eqref{eq:IQAE_Ansatz} or hybrid density matrices as introduced in Ref.~\cite{haug2020generalized}. The idea is to
write the overall ansatz as a classical combination of quantum states
i.e, $\vert\phi\left(\boldsymbol{\alpha}(t),\boldsymbol{\theta}\right)\rangle=\sum_{i=0}^{m-1}\alpha_{i}(t)\vert\psi_{i}\left(\boldsymbol{\theta}_{i}\right)\rangle.$
Tuning the $\boldsymbol{\theta}_{i}$ can often lead to barren plateaus. One can
avoid such issues by fixing $\boldsymbol{\theta}_{i}$ by harnessing the structure
of the problem to find the basis states of the ansatz, i.e, $\left\{ \vert\psi_{i}\left(\boldsymbol{\theta}_{i}\right)\rangle\right\} $
(see \secref{sec:IQAE} for more details). Interestingly, quantum convolutional neural networks also do not exhibit barren plateaus~\cite{pesah2020absence}.

\subsection{Expressibility of variational ans\"atze}\label{sec:expressibility}

A cornerstone in the success of VQA is choosing the proper ansatz for the problem. In addition to trainability, i.e. how well the ansatz can be optimized, another major quality is expressibility. This concerns whether a given PQC is able to generate a rich class of quantum states. The number of PQC layers, parameters or entangling gates required to achieve a given accuracy is also linked to the expressibility of the circuit.

\paragraph{Expressibility.} 
Sampling states from a PQC $\ket{\psi_{\boldsymbol{\theta}}}$ for randomly chosen $\boldsymbol{\theta}$ generates a distribution of states.
Expressibility is defined as the deviation of this distribution from the Haar measure, which samples uniformly from the full Hilbert space
\begin{equation}
A^{(t)}=\vert\vert \int_\text{Haar}(\ket{\psi}\bra{\psi})^{\otimes t} \text{d}\psi-\int_{\boldsymbol{\theta}}(\ket{\psi_{\boldsymbol{\theta}}}(\bra{\psi_{\boldsymbol{\theta}}})^{\otimes t} \text{d}\psi_{\boldsymbol{\theta}} \vert\vert_\text{HS}^2\,,
\end{equation}
where $\int_\text{Haar}\text{d}\psi$ denotes the integration over a state $\ket{\psi}$ distributed according to the Haar measure and $\vert\vert A \vert\vert_\text{HS}^2=\text{Tr}(A^\dagger A)$ the Hilbert-Schmidt norm. An ansatz circuit $U$ with small $A^{(t)}_U$ is more expressive, with $A^{(t)}_U=0$ corresponding to being maximally expressive, as it generates quantum states with a distribution closer to the Haar measure. The PQC samples uniformly from the full Hilbert space and thus is able to approximate any possible state. This is especially important in the case where one wants to train the PQC to represent a particular quantum state while having little prior information about the state. A highly expressive PQC is more likely to be able to represent the target state. 

\paragraph{Entangling capability.} This measure denotes the power of a PQC to create entangled states and can be used as another quantifier of the expressiveness of an ansatz. In \cite{sim2019expressibility} the Meyer-Wallach $Q$ measure~\cite{meyer2002global} has been proposed to estimate the number and types of entangled states a particular PQC can generate.
One defines a linear mapping $\iota_j(e)$ that acts on the computational basis $
\iota_j(b) \ket{b_1 \cdots b_n} =  \delta_{bb_j} \ket{b_1 \cdots \tilde{b}_j \cdots b_n}$,
where $b_j \in \{0, 1\}$ and $\tilde{b}_{j}$ denotes absence of the $j$-th qubit. The entanglement measure $Q$ is then defined as
\begin{equation} \label{eq:meyer_wallach1}
Q(\ket{\psi}) \equiv \frac{4}{n} \sum_{j=1}^n D \big( \iota_j(0) \ket{\psi}, \iota_j(1) \ket{\psi} \big),
\end{equation}
where $D$ is the generalized distance defined by the coefficients of two states $\ket{u} = \sum u_i \ket{e_{i}}$ and $\ket{v} = \sum v_i \ket{e_{i}}$,
\begin{equation} \label{eq:meyer_wallach_distance}
D(\ket{u}, \ket{v}) = \frac{1}{2} \sum_{i,j} \vert u_i v_j - u_j v_i \vert^2.
\end{equation}
It can be rewritten as the average of the purity of each qubit~\cite{brennen2003observable}
\begin{equation} \label{eq:meyer_wallach2}
Q(\ket{\psi}) = 2(1-\frac{1}{n}\sum_{k=1}^n\text{Tr}[\rho_k^2])\,,
\end{equation}
where $\rho_k$ is the density matrix of the $k$-th qubit.
Thus, $Q(\ket{\psi})$ is an entangling monotone~\cite{scott2004multipartite} and can be interpreted as the average of the entanglement of each qubit with the rest of the system.

Only if the state is a product state we find $Q=0$, whereas $Q=1$ is reached for certain entangled states such as the GHZ state. 
The entangling capability of a PQC is then defined as the average $Q$ of states randomly sampled from the circuit.
\begin{equation} \label{eq:entangling_capability_score}
\text{Ent} = \frac{1}{\vert S\vert} \sum_{\boldsymbol\theta_i \in S} Q \big( \ket{\psi_{\boldsymbol\theta_i}} \big),
\end{equation}
where $S= \{ \boldsymbol\theta_i \}_i $ is the set of sampled circuit parameters. 

\paragraph{Parameter dimension} The parameter dimension $D_\text{C}$ is the number of independent parameters of the quantum state that is generated by the PQC~\cite{haug2021capacity}. From this measure, one can calculate the redundancy of a PQC, i.e. the fraction of parameters that can be removed without loss of expressive power. 
A further local measure of expressibility is the effective quantum dimension $G_\text{C}(\boldsymbol{\theta})$, which can be used to calculate the expressive power of initialization strategies for the PQC. Under a small variation of the PQC parameter $\boldsymbol{\theta}$, it measures how many independent directions in the parameter space exist for the quantum state. Both measures can be calculated as the number of non-zero eigenvalues of the Fubini-Study metric tensor defined in \eqref{eq:Fubiny_Study}. \\

In \cite{sim2019expressibility,nakaji2020expressibility,haug2021capacity}, a wide class of circuits have been investigated with the aforementioned expressibility measures. It has been found that certain types of ans\"atze are more expressive, e.g. layered PQCs consisting of CNOT or $\sqrt{\text{iSWAP}}$ gates are more expressive than CZ. 
There is a trade-off between an ansatz being expressive and trainable. Making an ansatz more expressive most likely will result in reducing the gradient of the objective function. In \cite{holmes2021connecting}, the authors suggest several strategies for reducing expressibility and improving trainability, including correlating parameters or restricting rotation angles of parameterized gates. Interpolating the PQC parameters between fixed and random angles has been proposed as another method~\cite{haug2021capacity}. Expressibility of PQCs have been further explored using classical Fisher information~\cite{abbas2020power} and memory capacity~\cite{wright2020capacity}.

It has been shown that alternating layered ansatz (see \secref{sec:HardwareAnsatz}) is both relatively expressive as well as does not exhibit barren plateaus in certain regimes~\cite{nakaji2020expressibility}. In VQE algorithms, there is a trade-off between the number of layers in this ansatz and the correlation length of critical Hamiltonians.  However, in the critical phase, the number of layers must exceed a certain threshold dictated by the system size to show an exponential improvement. The circuit depth unravels an effective correlation length that can be used as an estimation of the number of free parameters in the ansatz \cite{bravoPrieto2020scalingof}.

\subsection{Reachability} \label{sec:reachability}

Reachability discusses the question whether a given PQC $\ket{\Psi(\boldsymbol{\theta})}$ with parameters $\boldsymbol{\theta}$ is capable of representing a quantum state that minimizes some objective function.

This can be quantified by the reachability deficit over finding the minimum of an objective function $\hat{O}$~\cite{akshay2020reachability} as
\begin{equation}
f_\text{R}=\text{min}_{\psi\in\mathcal{H}}\bra{\psi}O\ket{\psi}-\text{min}_{\boldsymbol{\theta}}\bra{\Psi(\boldsymbol{\theta})}O\ket{\Psi(\boldsymbol{\theta})},
\end{equation}
where the first term on the right side is the minimum over all states $\ket{\psi}$ of the Hilbert space, whereas the second term is the minimum over all states that can be represented by the PQC. The reachability deficit is equal or greater than zero $f_\text{R}\ge0$, with $f_\text{R}=0$ when the PQC can generate a state $\ket{\Psi(\boldsymbol{\theta}^*)}$, where $\boldsymbol{\theta}^*$ are the parameters that minimizes the objective function.

Reachability has been studied in-depth for QAOA.
Although QAOA has been shown to exhibit quantum computational universality \cite{lloyd2018quantum,morales2019on}, which implies that any unitary operator is reachable under the QAOA ansatz, this statement does not hold true for finite fixed depths $p$. 
In fact, it was shown that QAOA exhibits reachability deficits for the MAX-2-SAT and MAX-3-SAT problems, where the optimal value of the objective function cannot be found using a fixed circuit depth $p$ beyond a critical clause density (defined as the ratio between the number of clauses and the number of variables in the problem)~\cite{akshay2020reachability}. 
In other words, for problems with a certain clause density, there is a critical depth $p^*$ for which the optimal solution can only be found (up to a threshold) if $p \geq p^*$. As $p^*$ grows with the clause density, this limits the performance of QAOA for problem instances with high clause density.

Similar reachability deficits have also been found in the variational Grover search problem~\cite{akshay2020reachability}.
Moreover, by re-analyzing the experimental data from Google's Sycamore quantum processor on the application of QAOA to various graph optimization problems \cite{harrigan2021quantum}, authors from \cite{akshay2020reachability2} also discovered reachability deficits in this case, where the graph density (defined as the ratio between the number of graph edges to the number of graph nodes) replaces the clause density as the order parameter.

Note that the reachability deficits are distinct from the barren plateau problem, where the gradients of the objective function concentrate to zero for many choices of initial variational parameters, thus slowing down the optimization process. On the other hand, the reachability deficit for $p < p^*$ is independent of the initial parameters.

\subsection{Theoretical guarantees of the QAOA algorithm} \label{sec:theoretical_garantees}

The QAOA has several key analytical results which have contributed to its considerable interest in recent years.
The quantum advantage of QAOA algorithm has been studied in \cite{farhi2016quantum}, where they showed that the efficient sampling of the output distribution of QAOA, even for the lowest depth case of $p = 1$, implies the collapse of the polynomial hierarchy (see \secref{ch1:subsec:non-complexity} ). 
Following the conjecture from complexity theory that the polynomial hierarchy does not collapse, this result propels QAOA as a possible candidate for establishing some quantum advantage. {In particular, it has been shown that for $p = 1$, 420 qubits would suffice to demonstrate quantum advantage~\cite{dalzell2020how}.}

The power of QAOA compared to classical algorithms is an ongoing topic of research. For specific instances of the Max-Cut problem, QAOA for $p=1$ was shown to perform equally well or worse than classical algorithms~\cite{hastings2019classical,bravyi2019obstacles}. For more discussion on QAOA for Max-Cut, refer to ~\secref{sec:MaxCut}. {For QAOA of depth $p$, the measurement outcomes of a qubit depend on the $p$-neighbourhood of that qubit. Thus if $p$ is too small, it does not `see' the whole graph  \cite{farhi2020seetypical,farhi2020seeworst}. For large $p$, the QAOA algorithms can `see' the whole graph with no known indications regarding the performance limitations.} 
    
For the case where the problem Hamiltonian $H_P$ takes the form
\begin{equation}
    H_P = \sum_i \omega_A \hat{\sigma}_{z}^{2i} + \omega_B \hat{\sigma}_{z}^{2i+1} + \gamma_{AB} \hat{\sigma}_{z}^{2i} \hat{\sigma}_{z}^{2i+1} + \gamma_{BA} \hat{\sigma}_{z}^{2i+1} \hat{\sigma}_{z}^{2i+2},
\end{equation}
where $\omega_{A(B)}$ are the coefficients for the even (odd) sites and $\gamma_{AB(BA)}$ are the interaction strength between first (second) neighbour spins. Taking $H_M$ as defined in \eqref{eq:mxing_operator} for a 1D lattice, Ref. \cite{lloyd2018quantum} showed that QAOA can be used to implement universal quantum computation. This result was proven and generalized in a later work \cite{morales2019on} to include a larger class of problem and mixing Hamiltonians that can provide computational universality for QAOA.

By connecting VQA with optimal control theory, Pontryagin's minimum principle of optimal control is used to show that the bang-bang protocol (in which the evolution switches abruptly between two Hamiltonians) is optimal for a fixed total time $T$ \cite{yang2017optimizing}. Since QAOA can be regarded as a bang-bang ansatz by switching between unitary evolution under $H_P$ and $H_M$ respectively, this suggests the optimality of QAOA as a VQA. However, recent works have challenged this claim. By generalizing the argument in  Ref. \cite{yang2017optimizing}, it has been shown that the optimal protocol actually possesses the `bang-anneal-bang' structure \cite{brady2020optimal}. Such protocols begin and end with a bang, with regions of smoothly varying control function (akin to quantum annealing) in between. It was also shown that when the total time $T$ is large, bang-bang QAOA suffers from the proliferation of local minima in the control parameters, rendering it difficult to find optimal (or near-optimal) QAOA parameters.
\section{Programming and Maximizing NISQ utility}\label{ch:lemon}

Current NISQ devices have a limited number of qubits ($\sim50-100$) available.
In addition, due to their noisy nature and short coherence time, one can only perform a restricted number of gate operations. 
In order to make maximal use of the currently available quantum resources, there are two approaches from the operational point-of-view: the bottom-up and the top-down. 
The bottom-up approach refers to the scenario where one has full control over the design of any quantum computing platform to keep pushing the performance quality such as gate fidelity and coherence time for a given hardware design constraints.
The top-down approach means implies that one does not get involved in hardware design and simply makes use of what has already been made or fabricated in the experimental labs.
In this section, we focus on the latter approach, i.e. extending or maximizing the utility of current and near-term quantum devices from an algorithmical perspective.
Finally, we also present a summary of software tools to control, program and maximize the utility of NISQ algorithms.

\subsection{Quantum error mitigation (QEM)}

Sensitivity to errors and noise are the two most prominent roadblocks towards scalable universal quantum computers. 
Fault-tolerant quantum computing can be attained by encoding non-Abelian anyons in topological materials~\cite{kitaev2003fault} or applying quantum error correction codes~\cite{raussendorf2007fault}.
While the former is still in its infancy, the latter mandates physical resources exceeding our current experimental capabilities.
In the NISQ era of running hybrid quantum/classical algorithms, it is desirable to use all the restricted and available qubits as logical qubits without applying QEC techniques.
As we discuss throughout this review article, the hybrid quantum/classical algorithms rely on computing the expectation value of some physical observables using quantum processors.
Quantum error mitigation (QEM) techniques discussed in this subsection need no extra qubit and can suppress errors in finding expectation values with simple classical post-processing and different runs of quantum circuits.
To be precise, with QEM, we are not interested to recover the ideal quantum output state $\hat{\rho}^{(0)}$, but to estimate the ideal observables $\hat{A}$ expectation value: $E[\mu^{(0)}]=\braket{\hat{A}^{(0)}}=\tr{(\hat{\rho}^{(0)}\hat{A})}$ \cite{li2017efficient,temme2017error,kandala2019error}, sometimes surpassing the break-even point, where the effective gates are superior to their physical building blocks, at an affordable cost with respect to near-term quantum hardware \cite{zhang2020error}.
Here, $\mu$ is the outcome of a measurement and we use superscript $(0)$ to denote an ideal noise-free realization of a state, operation or observable quantity.
Recently, it was also shown how to achieve stochastic error mitigation for a continuous time evolution \cite{sun2020mitigating}. For a comprehensive treatment on quantum error mitigation, refer to~\cite{endo2020hybrid}.

\subsubsection{Zero-noise extrapolation}
\label{ch4:subsubsec:ZNE}
\cite{li2017efficient}, and  \cite{temme2017error} independently and concurrently proposed the Richardson extrapolation QEM, namely zero-noise extrapolation (ZNE), where a quantum program is to operate at various effective noise levels of a quantum processor.
It is then extrapolated to an estimated value at a noiseless level.

Formally, a quantum circuit/system in the presence of noise can be modelled as an open quantum system \cite{breuer2002theory} using the Gorini-Kossakowski-Sudarshan-Lindblad equation or in short the Lindblad master equation (setting $\hbar=1$):
\begin{equation}
\frac{d}{d t}\hat{\rho}(t)=-i\left[\hat{K}\left(t\right),\hat{\rho}\left(t\right)\right]+\doublehat{\mathcal{L}}\left[\hat{\rho}\left(t\right)\right],
\label{eq:ch4:Lindblad_master_eq}
\end{equation}
where $\hat{K}(t)$ acts as time-dependent driving Hamiltonian, and $\doublehat{\mathcal{L}}[.]=\sum_k \Gamma_k ( \hat{\mathcal{O}}_k[.]\hat{\mathcal{O}}_k^\dagger - \frac{1}{2}\{\hat{\mathcal{O}}_k \hat{\mathcal{O}}_k ^\dagger,[.] \})$ is a superoperator.
The above equation in general describes Markovian dynamics for $\Gamma_k \ge 0$. 
Whenever loss rate $\Gamma_k$ become negative \cite{fleming2012non,rivas2010entanglement}, the above equation would also describe non-Markovian dynamics \cite{tan2010non,bastidas2018floquet,kyaw2020dynamical}.
To ensure the complete positivity, we require $\int_0 ^t \Gamma(t') dt' >0$, $\forall t$.
In general, $\Gamma_k$ are fixed by the nature of the noise experienced by a quantum system. 
Mathematically, one can parametrize $\Gamma_k$ with a dimensionless scalar $\lambda$, i.e., $\Gamma_k \rightarrow \lambda\Gamma_k$.
When $\lambda=0$, there is no noise and the second term (loss term) in \eqref{eq:ch4:Lindblad_master_eq} is zero, resulting in pure unitary dynamics. 
When $\lambda=1$, the actual quantum device loss rate is matched.
In summary, ZNE involves two steps.
\begin{enumerate}
    \item Noise-scaling: we make a number of measurements $E[\mu^{(\lambda_j)}]$ for $\lambda \ge 1.$
    \item Extrapolation: we estimate $E[\mu^{(0)}]$ from the previous step.
\end{enumerate}

\paragraph{Noise-scaling} can be accomplished in three ways. Firstly, in Ref. \cite{temme2017error} it was proposed to use a \textit{time-scaling approach} by taking $\lambda > 1$, which means that the time-dependent driving Hamiltonian $\hat{K}(t)$ is now rescaled by $\frac{1}{\lambda}\hat{K}(t/\lambda)$.  
This approach is only possible if the user has full control over back-end quantum processor.
Control pulses for each quantum gate have to be recalibrated and be applied for longer duration.
Secondly, one can apply a technique called \textit{circuit folding} \cite{giurgica2020digital}. 
Suppose that a quantum circuit is composed of $d$ unitary layers such that $U=L_d\cdots L_2 L_1$ where $d$ refers to the circuit depth and each $L_j$ either represents a single layer of gate operations or just a single quantum gate.
The circuit folding is then achieved by
\begin{equation}
    U\rightarrow U(U^\dagger U)^n,
\end{equation}
where $n$ is some positive integer. 
Since $U^\dagger U$ is an identity, this action has no effect on an ideal circuit.
However, in noisy circuit, $U$ is imperfect and the above $1+2n$ circuit operations would increase the noise level.
Thirdly, instead of entire circuit folding, one can also use \textit{gate folding} technique \cite{giurgica2020digital}:
\begin{equation}
    L_j \rightarrow L_j (L_j^\dagger L_j)^n.
\end{equation}
The second and third techniques do not require users to have full control of quantum computer back-end and thus we expect to be of greater use in software level control of quantum circuits.

\paragraph{Extrapolation step} of the ZNE method can be considered as a regression problem if we choose to consider a generic model for calculating the expectation value $E_{\textrm{model}}[\mu^{(\lambda;\Upsilon)}]$, where the meaning of \textit{model} would become clear shortly and $\Upsilon$ corresponds to the model parameters. 
We note that the expectation value $E$ is a real number that can only be obtained in the infinite measurement limit.
With limited number of measurement samples $N$, statistical estimation is $\hat{E}[\mu^{(\lambda)}]=E[\mu^{(\lambda)}]+\hat{\delta}$~\footnote{The hat notation used is in accordance with statistics notation and should not be confused with a quantum operator.}, where $\hat{\delta}$ is a random variable with zero mean and variance $\sigma^2 = \mathbb{E}(\hat{\delta}^2)=\sigma_0 ^2 /N$.
Here, $\sigma_0 ^2$ is the single-shot variance. 
Given a set of $m$ scaling parameters ${\lambda}=\{ \lambda_1, \lambda_2,\cdots,\lambda_m \}$ with $\lambda_j \ge 1$, and the corresponding measurement outcomes ${\mu}=\{\mu_1, \mu_2,\cdots,\mu_m \}$, the ZNE is nothing but to build a good estimator $\hat{E}[\mu^{(0)}]$ for $E[\mu^{(0)}]$ such that its bias $\mathbb{E}(\hat{E}[\mu^{(0)}]-E[\mu^{(0)}])$, and its variance $\mathbb{E}(\hat{E}[\mu^{(0)}]^2)-\mathbb{E}(\hat{E}[\mu^{(0)}])^2$ are both reasonably small.
Onwards, let us adopt a simplified notation of $E[\mu^{(\lambda)}]=E(\lambda)$.
Now let us mention briefly the statistical models.
The expectation value $E(\lambda)$ cannot be of any arbitrary function, which would make ZNE impossible to extrapolate back to $E(0)$.
Depending on some underlying noise model assumption, one can apply various statistical models.

\paragraph{The polynomial extrapolation} is based on the polynomial model of degree $d$ such that 
\begin{equation}
    E^{(d)}_{\textrm{poly}}(\lambda)=c_0 + c_1 \lambda+\cdots + c_d \lambda^d,
\end{equation}
where $c_j$ are $d+1$ unknown real parameters.
This extrapolation is justified in weak noise limit and we need the number of data points $m$ to be equal or larger than $d+1$.
Consequently, we can obtain two other variants: \textit{the linear extrapolation} $(d=1)$ and \textit{the Richardson extrapolation} $(d=m-1)$ \cite{temme2017error}.
By construction, the error with respect to the true expectation value is $O(m)$ when we have large sample size $N\rightarrow \infty$.
By using the interpolating \textit{Lagrange polynomial}, the estimator is explicitly given by:
\begin{equation}
    \hat{E}_{\textrm{Rich}}(0)= \hat{c}_0 = \sum_{k=1}^m \mu_k \prod_{i\neq k} \frac{\lambda_i}{\lambda_i -\lambda_k},
\end{equation}
with the assumption that all $\lambda_j$ are different.
One important observation is that the Richardson model based ZNE is dictated by a statistical uncertainty which is exponentially scaling with the number of data points. 
There are also other statistical models such as \textit{poly-exponential extrapolation} \cite{giurgica2020digital} and \textit{exponential extrapolation}~\cite{endo2018practical}.
Various exponential extrapolation methods have been proposed and investigated in~\cite{cai2020multi} and applied to depolarizing noise in~\cite{vovrosh2021efficient}.
In fact, \cite{cai2020multi} shows that the ZNE, quasi-probability and stabiliser-based approach can be combined by exploiting novel aspects of the individual technique.

The ZNE scheme suffers from a few limitations.
The scheme works by extrapolation, and hence it is challenging to obtain result guarantees in general.
The number of measurement shots required to obtain the mitigated expectation value can be relatively high compared to the unmitigated case as seen above.
More importantly, the fundamental drawback of both ZNE and probabilistic error cancellation (PEC) \cite{temme2017error}, or quasi-probability method (which is discussed next) is that one needs to know the precise physical noise model in advance, which in itself is a difficult problem.
Experimentalists in the lab will have imperfect knowledge about the real noise, which will typically differ from the canonical ones.
We will also discuss a more practical approach based on gate set tomography proposed in Ref. \cite{endo2018practical}, which does not require explicit knowledge of the noise model and mitigates any localized Markovian errors, so that the error in the final output is only due to unbiased statistical fluctuation.

\begin{figure*}[t]
\centering
\includegraphics[scale=1.5]{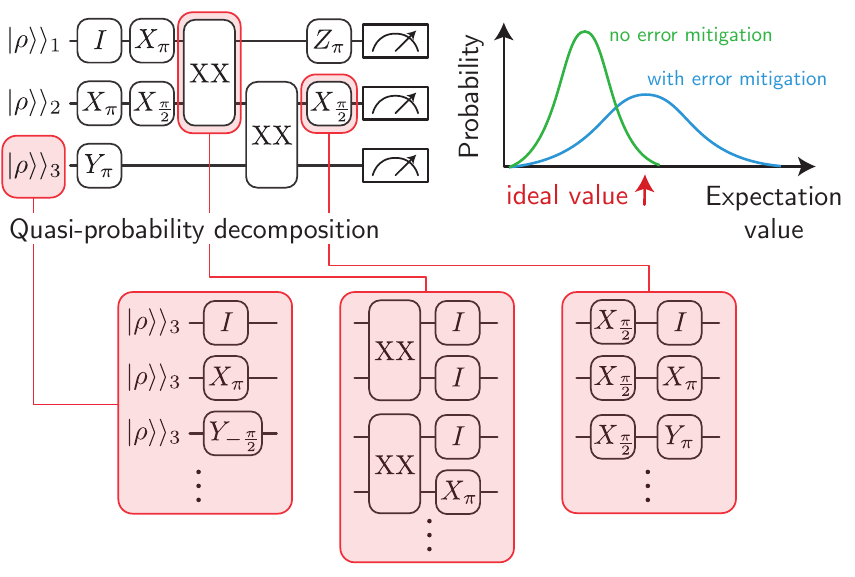}
\caption{Quantum computing of the expected value of an observable using gate set tomography-based PEC. 
Quasiprobability decomposition of initial state preparation, examplary single- and two-qubit processes are computed.
Implementing the resulting decomposition is done using the Monte Carlo approach.
With QEM, the probability distribution of expected value of the physical observable is now centered around an ideal value with larger variance as compared to the one without QEM. 
Inspired by \cite{zhang2020error}.
}
\label{fig:ch4:QEM_quasiprobability_decomposition}
\end{figure*}

\subsubsection{Probabilistic error cancellation}
\label{ch4:subsubsec:PEC}

Let us familiarize ourselves with the notations used in quantum tomography \cite{greenbaum2015introduction,merkel2013self}, which we adopt here.
A quantum state is represented by a density matrix $\hat{\rho}$, and a physical observable is denoted by a Hermitian $\hat{A}$ operator.
An operation is a map on the states space such that one can use the Kraus representation to denote it as: $\doublehat{\mathcal{L}}[\hat{\rho}]=\sum_j \hat{K}_j \hat{\rho} \hat{K}^\dagger _j.  $
We note that this equivalence with \eqref{eq:ch4:Lindblad_master_eq} is only valid when we have Markovian dynamics. 
Here, $\hat{K}_j$ are Kraus operators.
In terms of the Pauli transfer matrix representation, $\hat{\rho}$ in \eqref{eq:ch4:Lindblad_master_eq} can be written as a column vector \cite{navarrete2015open}, denoted as $\dket{\rho}$.
Similarly, the Lindblad superoperator $\doublehat{\mathcal{L}}$ can be recast as a square matrix, i.e., using the Pauli transfer matrix representation, since it is a linear map.
For simplicity and without loss of generality, we may absorb the unitary dynamics (the first term in \eqref{eq:ch4:Lindblad_master_eq}) into $\doublehat{\mathcal{L}}$ onwards.
A physical observable $\hat{A}$ is now written as a row vector $\dbra{A}$.
Consequently, the expectation value is $\braket{\hat{A}}=\tr{[\hat{A}\hat{\rho}]}= \dbraket{A}{\rho}$.
Likewise, the expectation of $\hat{A}$ after the state $\hat{\rho}$ passing through a series of linear maps is read as: $\tr{[\hat{A}\doublehat{\mathcal{L}}_N \circ\cdots \circ \doublehat{\mathcal{L}}_1 (\hat{\rho})]}=\dbra{A}\mathcal{L}_N \cdots \mathcal{L}_1 \dket{\rho}$.

The central theme of probabilistic error cancellation (PEC) or quasiprobability decomposition introduced by the IBM team in Ref. \cite{temme2017error} is that one can estimate the expectation value of an observable by sampling from a set of erroneous circuits, labelled by $\mathcal{L}^{(l)}_{\textrm{tot}}$ for $l=1,2,\cdots,$ such that 
\begin{equation}\label{eq:ch4:PEC_expect}
    \braket{\hat{A}^{(0)}}= \sum_l q_l \dbra{A^{(l)}}\mathcal{L}^{(l)}_{\textrm{tot}}\dket{\rho^{(l)}}.
\end{equation}
The expectation of an observable is going to be far-off (without QEM) from the ideal value due to the presence of noise\footnote{The superscript $(0)$ to denote the ideal noise-free realisation of a state, operation or observable quantity.}.
Given specific error models (assuming experimentalist has full and correct knowledge about them), the real numbers ${q_l}$, which represent quasiprobabilities, can be efficiently derived.
Here, each $\mathcal{L}_{\textrm{tot}}^{(l)}$ represents the total sequence of noisy gates in the $l$th circuit. 
Monte Carlo sampling could be used to compute $\braket{\hat{A}^{(0)}}$ by randomly choosing the $l$th circuit with the probability $p_l =|q_l|/C$, where $C=\sum_l |q_l|$. 
Lastly, the computed result is given by the expected value of effective measurement outcomes $\braket{A^{(0)}}=CE[\mu_{\textrm{eff}}]$, where the effective outcome is $\mu_{\textrm{eff}}=\textrm{sgn}(q_l)\mu^{(l)}$ if the $l$th circuit is chosen and $\mu^{(l)}$ is the outcome from the $l$th circuit.

As a consequence, the mean value of the PEC outcome centers around the ideal one with larger variance due to $C$ (see \figref{fig:ch4:QEM_quasiprobability_decomposition} right corner).

The above PEC method relies on the correct knowledge of error model $\mathcal{L}^{(l)}_{\textrm{tot}}$ as is apparent from \eqref{eq:ch4:PEC_expect}.
To enable practical implementations, \cite{endo2018practical} proposes to combine linearly independent basis set operations and gate set tomography to fully remove impact of localized Markovian errors by systematically measuring the effect of errors to design efficient QEM circuits.
The set of operations including measurement and single-qubit Clifford gates is universal in computing expected values of observables. 
For the single-qubit case, any operation $\mathcal{L}$ which is a $4\times 4$ real matrix in the Pauli transfer matrix representation, can be expressed as a linear combination of 16 basic operations, i.e., $\mathcal{L}=\sum_{i=1}^{16} q_i \mathcal{B}_i ^{(0)}$, which are composed of $\{\pi, H,S, R_x, R_y \}$ gates \cite{endo2018practical}.
Similarly, the same decomposition can be applied to the two-qubit case. 
See \figref{fig:ch4:QEM_quasiprobability_decomposition} for example decompositions.

A way to systematically measure errors is through gate set tomography (GST), with which one can even mitigate state preparation and measurement errors.
In short, the purpose of GST is to measure noisy individual quantum circuit performance a priori.
For a single-qubit gate, one prepares initial states $\ket{0}, \ket{1}, \ket{+_x}$, and $\ket{+_y}$, where $\ket{+_x}$ and $\ket{+_y}$ are the eigenstates of Pauli operators $\hat{\sigma}_x$ and $\hat{\sigma}_y$ with $+1$ eigenvalue, respectively.
For noisy devices, these four states are denoted as $\Bar{\rho}_1$, $\Bar{\rho}_2$, $\Bar{\rho}_3$ and $\Bar{\rho}_4$, accordingly. 
We also use $\Bar{\mathcal{L}}$ (superoperator) to denote a noisy/imperfect gate to be measured.
Since what we care about are expectation value of physical observables, for single-qubit case, we have observables $\hat{{I}},\hat{\sigma}_x, \hat{\sigma}_y, \hat{\sigma}_z$, denoted as $\Bar{A}_1, \Bar{A}_2, \Bar{A}_3, \Bar{A}_4$. 
The mean value of observables, the $4\times4$ matrix $\Tilde{A}$, is nothing but $\Tilde{A}_{j,k}=\tr{[\Bar{A}_j \Bar{\mathcal{L}}\Bar{\rho}_k]}$.
Similarly, we can also construct the $4\times4$ matrix $g$ without applying any gate to the initial states as $g_{j,k}=\tr{[\Bar{A}_j \Bar{\rho}_k]}$.
This is repeated for each qubit and each single-qubit gate.
Statistical estimation of the initial states $\Bar{\rho}_k$ and the observables $\Bar{A}_j$ are then given by 
\begin{align}
    \dket{\hat{\rho}_k} &= T_{\bullet,k},\\
    \dbra{\hat{A}_j} &= (gT^{-1})_{j,\bullet},
\end{align}
where we note that the hat symbol represents the statistical estimate and $T_{\bullet,k} (T_{j,\bullet})$ denotes the $k$th column ($j$th row) of the matrix $T$, where $T$ is an invertible $4\times 4$ matrix with the following relationship $\hat{\mathcal{L}}=Tg^{-1}\Tilde{\mathcal{L}}T^{-1}$.
The same procedure applies for the two-qubit case with the only difference being that there are total of $16$ initial states: $\Bar{\rho}_{k_1}\otimes \Bar{\rho}_{k_2}$ and $16$ observables: $\Bar{A}_{j_1}\otimes \Bar{A}_{j_2}$ to be measured. 
Similarly, we have $g=g_1\otimes g_2$ and $T=T_1 \otimes T_2$.
We have to implement two-qubit gate GST for each qubit pair involved in a quantum program run.

{Quasiprobability decomposition} is then computed based on GST results above.
From GST, we have estimation of initial states $\dket{\hat{\rho}_k}$, observables to be measured $\dbra{\hat{A}_j}$, and gates $\hat{\mathcal{L}}$.
Let's denote $\mathcal{L}^{(0)}$ as the Pauli transfer matrix of the ideal gate with no error.
The main idea of decomposition comes from a very simple idea that a noisy gate operation comes from an ideal operation followed by a noise operation, i.e., $\mathcal{L}=\mathcal{N}\mathcal{L}^{(0)}$. 
Hence, the inverse of the noise is given by
$
    \mathcal{N}^{-1}=\mathcal{L}^{(0)}\hat{\mathcal{L}}^{-1}=\sum_i q_{\mathcal{L},i} \hat{\mathcal{B}}_i.
$
And, by applying the inverse of the noise after the operation, we can obtain the operation without error: $\mathcal{L}^{(0)}=\mathcal{N}^{-1}\mathcal{L}$.
Notice that the matrices in the above equation are obtained from the first GST step.
The remaining task is to determine quasiprobabilities $q_{\mathcal{L},i}$ for each qubit and gate involved by solving the above equation.
We note that instead of quasiprobabilistic decomposition of quantum gates, one could in principle use randomized compiling technique proposed in Ref. \cite{wallman2016noise}.

GST-based PEC experiments have recently been done in trapped-ion systems~\cite{zhang2020error} and superconducting circuits~\cite{song2019quantum}.
Lastly, a similar strategy was recently applied to mitigating errors in measurement readout \cite{kwon2020hybrid}.

\subsubsection{Other QEM strategies}\label{sec:other_QEM}
We have seen that the quantum error mitigation techniques discussed so far do not require any ancilla or extra qubits with the caveat that one needs to perform more measurements. At the same time, one is only interested in information about the expectation value.
Along this line of thought, there exist several proposals, which we will outline below.
However, some of the methods might require ancilla qubits.

\paragraph{Subspace expansion method} \cite{mcclean2017hybrid,colless2018computation,mcardle2019error,sagastizabal2019experimental,mcclean2020decoding,barron2020preserving} are designed to mitigate errors in the VQE routine, where we often tend to find an approximate ground state $\ket{\psi_a}$ of a system Hamiltonian $H$.
However, such state may differ from the true ground state $\ket{\psi_g}$ due to noisy processes.
In general, we do not know which error occurred to the state. 
The subspace expansion method works by resolving the action of $H$ on the linear combination of quantum states ansatz~\eqref{eq:IQAE_Ansatz}. The subspace is spanned by a set of operators $\hat{\mathcal{O}}_i$, i.e., $\{\ket{\hat{\mathcal{O}}_i \psi_a}\}$. Now, one proceeds to evaluate $H_{ij}=\bra{\psi_a}\hat{\mathcal{O}}_i H \hat{\mathcal{O}}_j \ket{\psi_a}$, and $S_{ij}=\bra{\psi_a}\hat{\mathcal{O}}_i \hat{\mathcal{O}}_j \ket{\psi_a}$.
The latter is needed since the subspace states are in general not orthogonal to each other.
By solving the generalized eigenvalue problem $HC=SCE$, with eigenvectors $C$ and diagonal matrix of eigenvalues $E$, we can obtain the Hamiltonian spectra including the excited states (see \secref{sec:var_excited}).

This method requires an appropriate choice of subspace operators to mitigate errors due to external noise. 
In general, without knowing the noise models of quantum device, it would require an exponential number of expansion operators to obtain the optimal groundstate.

\paragraph{Stabilizer based approach} \cite{bonet2018low,mcardle2019error,sagastizabal2019experimental,cai2021quantum} relies on the information associated with conserved quantities such as spin and particle number conserving ansatz. 
If any change in such quantities is detected, one can pinpoint an error in the circuit, which is akin to stabilizer measurement in quantum error correction schemes.
We can implement the stabilizer measurements by adding ancilla qubits to the qubit registers or taking additional measurements and post-processing.

\paragraph{Individual error reduction} method was proposed in Ref. \cite{otten2019accounting}.
As we have seen earlier, Markovian noise can be modelled using the Lindblad master equation, \eqref{eq:ch4:Lindblad_master_eq}, where we have $\frac{d\hat{\rho}}{dt}=\doublehat{\mathcal{L}}(\hat{\rho})=\sum_i \mathcal{L}_i (\hat{\rho})$, where each $\mathcal{L}_i$ denotes a noise channel present. 
Here, we have absorbed the unitary component into $\doublehat{\mathcal{L}}$.
It was shown that 
\begin{align}
    \Tilde{\rho}(T)&= \hat{\rho}(T)- \sum_{j=1}^m \frac{1}{g_j}\left(\hat{\rho}(T) -\hat{\rho}_j (T) \right),\label{eq:ch4:individual_error}\\
    &= \hat{\rho}^{(0)}(T) + \mathcal{O}(\tau^2).
\end{align}
Notational explanations are as follows.
$\hat{\rho}(T)$ is the density matrix after applying quantum gates with the presence of all associated noise channels at the final evolution time $T$. In contrast, $\hat{\rho}_j (T)$ is the state under the influence of all the noise channels but one less $\mathcal{L}_j$ according to the ratio $g_j$.
Notice that if $g_j=1$, we have fully removal of the entire channel $\mathcal{L}_j$.
$\hat{\rho}^{(0)}(T)$ is the ideal output state without any error, while $\tau$ is the evolution time for each noise process after the gate application.
We note that the first-order error $\mathcal{O}(\tau)$ is removed. 
As usual, what we want to obtain is $\braket{\hat{A}}=\tr[\hat{\rho}^{(0)}(T)\hat{A}]$ for a physical observable $\hat{A}$.
We can arrive at it by using \eqref{eq:ch4:individual_error}.
Though its result is neat and beautiful, this method assumes a perfect removal of individual noise channel.
Hence, it is relatively unrealistic on current quantum hardware as compared to other strategies.

\paragraph{Dynamic error suppression/robust control techniques} concern suppression of experimental gate errors at the pulse control level, which can be passive as well as active one. 
Pulse shaping technique is a strategy for passive cancellation of system-bath interaction.
Traditionally, this method stands on the shoulder of a mean to obtain high-fidelity quantum gate in nonlinear qubits such as transmons, commonly known as derivative removal of adiabatic gate (DRAG) scheme \cite{motzoi2009simple,gambetta2011analytic,de2015fast}.
On the other hand, \textit{dynamical decoupling} (DD) \cite{viola1999dynamical,santos2005dynamical,viola2005random} is a very well-known and widely used quantum control technique in the literature, which is designed to suppress decoherence via fancy pulses to the system so that it cancels the system-bath interaction to a given order in time-dependent perturbation theory \cite{lidar2014review} in an active manner. 
Recently, DD experiments were performed on the 16-qubit IBMQX5, 5-qubit IBMQX4, and the 19-qubit Rigetti Acorn chips \cite{pokharel2018demonstration}, where the gain in substantial gate fidelity relative to unprotected, free evolution of individual transmon qubits was demonstrated.
One may combine DD and pulse shaping technique to obtain dynamically corrected gates \cite{khodjasteh2009dynamically,edmunds2020dynamically} composing of shaped pulses which actively drive state evolution within a Hilbert space in order to cancel certain system-bath couplings.
With the availability of Qiskit Pulse \cite{alexander2020qiskit} that allows users to control backend pulse shapes and sequences of a quantum processor on the fly, a recent study \cite{carvalho2020error}, based on the Qiskit Pulse and robust control techniques, demonstrates enhancement up to: $\sim10\times$ single-qubit gate coherent-error reduction; $\sim5\times$ average coherent-error reduction; $\sim10\times$ increase in calibration window to one week of valid pulse calibration; $\sim12\times$ reduction gate-error variability across qubits and over time; and up to $\sim 9\times$ reduction in single-qubit gate error (including crosstalk). The improvements rendered by  ~\cite{carvalho2020error} have implications on the performance of multiqubit gates in trapped ions~\cite{milne2020phase}.

In light of these recent developments, together with IBM Qiskit Pulse \cite{alexander2020qiskit}, we envisage a possibility to realize/encode holonomic quantum gates \cite{zanardi1999holonomic,zhang2015fast} which are robust against parameter fluctuations and attain even better gate fidelity and performance.

\paragraph{Lanczos-inspired approach} \cite{suchsland2020algorithmic} estimates the expectation value of a physical observable $\tr{[\hat{\rho}^{(0)}\hat{A}]}$ by constructing a basis of the order-$m$ Krylov subspace $\mathcal{K}^{(m)}$ spanned by $\{\ket{\Psi},H\ket{\Psi},H^2 \ket{\Psi},...,H^m \ket{\Psi} \}$.
A way to look at this is to systematically construct the objective function to be minimised.
For the order-$m$, the objective function is 
\begin{equation}
    E_{L,k,n,m}=\min_{\substack{a\in \mathbb{R}^m}}\sqrt[k]{\frac{\bra{\Psi}H^k (\sum_{i=0}^{m-1}a_i H^i)^n \ket{\Psi}}{\bra{\Psi}(\sum_{i=0}^{m-1}a_i H^i)^n \ket{\Psi}}}.
\end{equation}
Due to the Krylov expansion, this technique can reduce the impact of different sources
of noise with cost of an increase in the number of measurements to be performed, without additional experimental overhead.
Calculating dynamic quantities such as Hamiltonian moments \cite{vallury2020quantum} and quantum power method based on higher-order Suzuki-Trotter expansion \cite{seki2020quantum} on near-term quantum computers are two recent examples that fall under the same approach here.

\paragraph{Learning-based and AI-inspired methods} employ machine learning techniques such as regression for error mitigation. The process consists of training different candidate circuit variants with non-Clifford gates substituted with gates with efficient classical simulability~\cite{strikis2020learning,czarnik2020error}. A recent approach suggests merging zero noise extrapolation with learning-based methods for near-Clifford circuits~\cite{lowe2020unified}.
There are also genetic algorithms to mitigate errors in quantum simulations \cite{las2016genetic,spagnolo2017learning}.

\subsection{Circuit compilation} \label{sec:compilers}

As it will be discussed in \secref{sec:software}, a quantum computer is composed of its hardware (quantum) and software (classical). The software translates a quantum algorithm into a set of instructions that implement the desired quantum operations and read out the qubit states. This process can be understood as \textit{quantum compilation} \cite{chong2017programming}, but the term is not limited to this particular application. When mapping a quantum circuit to a specific device architecture, one needs to consider the available quantum gates, the qubit connectivity that allows two-qubit gates implementation, and experimental limitations such as decoherence time, which imposes a certain maximum circuit depth in terms of the number of gates. For these reasons, it has become indispensable to develop tools that allow for circuit simplifications and efficient mappings of the general algorithm to specific hardware. These tools are also known as quantum compilers since they translate the theoretical circuit to the realistic simulator or device. In the following lines, we describe some of these tools. Many of them are suited both for NISQ and fault-tolerant quantum computation.

\subsubsection{Native and universal gate sets}\label{sec:gate_set}

The available gates that can be implemented experimentally on a particular hardware platform are sometimes referred to as the \textit{native gate set}. With a universal gate set $\mathcal{G}\in SU(d)$ (also called \textit{instruction gate set}), any unitary operation can be constructed efficiently. More formally, the \textit{Solovay-Kitaev theorem} \cite{dawson2006solovay} states that given this universal set $\mathcal{G}$, any unitary operation $U\in SU(d)$ can be approximated with $\epsilon$ accuracy with a finite sequence $S$ of gates from $\mathcal{G}$. This sequence scales logarithmically as $\mathcal{O}(\log^{c}(1/\epsilon))$, where $c$ is a constant that depends on the theorem proof. For $d=2^n$ this theorem guarantees that qubit quantum circuits can be decomposed using a finite gate sequence. Although this is one of the most important theorems in quantum computation, it is an existence theorem, i.e. it does not provide the  decomposition that it predicts. It also requires that the gate set contains the inverse of all gates. Further developments presented in \cite{bouland2018trading} tried to remove this assumption.

The Clifford group is an important object in quantum information science because of its applications in quantum error correction, randomized benchmarking and investigations for quantum advantage. The generalized Pauli operators in prime dimension $p$ are given by
\begin{equation}
T_{(a,b)}=
\begin{cases}
    \omega^{-\frac{ab}{2}}{Z}^{a}\mathcal{X}^{b} & \left(a,b\right)\in\mathbb{Z}_{p}\times\mathbb{Z}_{p},p\neq2\\
\iota^{ab}\mathcal{\mathcal{Z}}^{a}\mathcal{X}^{b} & \left(a,b\right)\in\mathbb{Z}_{2}\times\mathbb{Z}_{2},p=2
\end{cases}
\label{eq:Generalized_Pauli}
\end{equation}
where $\omega=\exp\left(\frac{2\pi i}{p}\right)$ and $\mathbb{Z}_{p}$ denotes an integer modulo $p$. The $\mathcal{Z}$ and $\mathcal{X}$ operators are defined via their action on computational basis states $\left\{ \vert k\rangle\right\} _{k}$, with $\mathcal{\mathcal{X}}\vert k\rangle=\vert k+1\mod p\rangle,\label{eq:Generalized_X}$ and $\mathcal{Z}\vert k\rangle=\omega^{k}\vert k\rangle.$
The unitaries which map the set of generalized Pauli operators to themselves up to a phase are called \textit{Clifford unitaries}. Let us denote the set of $p$ dimensional Clifford unitaries by $\mathcal{C}_{p}$.
Mathematically speaking,
\begin{equation}
U\in\mathcal{C}_{p}\iff\exists\phi:UT_{\left(a_{1},b_{1}\right)}U^{\dagger}=\exp\left(i\phi\right)T_{\left(a_{2},b_{2}\right)}\label{eq:Clifford_Operators}
\end{equation}
where $T_{\left(a_{1},b_{1}\right)}$ and $T_{\left(a_{2},b_{2}\right)}$ are generalized Pauli operators. The set of Clifford unitaries $\mathcal{C}_{p}$ forms a group, called the \textit{Clifford group}. In this review, we focus on $p=2$, i.e. the qubit Clifford group.

There are infinitely many universal gate sets, but the Clifford group's gates $H$, $S$ and $CNOT$, together with the $T$ gate, compose the most commonly used set. The Clifford group alone can be simulated efficiently as stated by the \textit{Gottesman–Knill theorem} \cite{Aaronson2004improved}. The theorem states that no quantum advantage can be found without the use of the $T$ gate. For this reason, many algorithms try to simplify and reduce quantum circuits to the minimal number of $T$ gates, giving an estimation of the classical efficiency of that particular circuit \cite{amy2013meet,gosset2013algorithm,heyfron2018efficient, kissinger2019reducing,amy2019t}. 

Besides these minimal reduction algorithms, other basic decompositions are useful. Even if only a native gate set is available experimentally, other basic gates can be constructed and used in algorithms. As an example, $S$ and $T$ gates are particular cases of the single-qubit rotational gate $R_{z}$, and the $H$ gate can be obtained from $R_{y}$ and $R_{x}$ gates as $H=R_{y}(-\pi/2)R_{x}(\pi)$. Any single-qubit gate can be decomposed into the gate sequence
$U(\theta,\phi,\lambda)=R_{z}(\phi)R_{y}(\theta)R_{z}(\lambda)$.
This motivates using single-qubit rotational gates and at least one entangling gate (e.g. $CNOT$ or $CZ$ gate) as native gate sets. Any two-qubit gate can be obtained from this minimal set by using circuit decompositions \cite{barenco1995elementary,blaauboer2008analytical, watts2013metric,deguise2018simple,peterson2020two}. The particular choice of the entangling gate can be motivated from the experimental platform. Depending on the technology used to construct the quantum device, a natural 2-qubit gate implementation can be more suited. Some examples are the use of $CZ$ gates in tunable superconducting circuits~\cite{krantz2019quantum}, cross-resonance gates in fixed frequency superconducting qubits~\cite{krantz2019quantum,kjaergaard2020superconducting}, or the $XX$ gates in trapped ions~\cite{haffner2008quantum}. More expressive gate sets with continuous gate parameters or long-range interactions can be achieved by further control over the hardware parameters in time \cite{lacroix2020improving,foxen2020demonstrating,krinner2020demonstration,bastidas2020fully}. The complexity of the circuit decomposition into CNOT and $R_{z}$ gates is analyzed in \cite{amy2018controlled}.

\subsubsection{Circuit decompositions}

Once the native gate set is fixed, the next step consists of decomposing the theoretical unitary circuit into this basic set. A raw translation of all single and two-qubit gates into the native set might imply a large circuit depth, reducing the effectiveness of that decomposition. Moreover, finding the decomposition of gates acting on more than one qubit might prove challenging in general. Besides common circuit decompositions mentioned before, one may need mathematical tools to understand and derive general circuit reductions to particular smaller pieces.

One of these mathematical tools is the so-called \textit{ZX-Calculus}. It is a graphical language that maps quantum circuits to particular graph representations and derives a set of rules to manipulate these graphs. Its application range goes from measurement-based quantum computation to quantum error correction. For a complete review about ZX-calculus and its variety of applications, see Ref. \cite{wetering2020zx}. For the purpose of this review, we are interested in the quantum circuit simplification applications \cite{cowtan2019phase, de2019techniques,duncan2020graph,kissinger2020reducing, hanks2020effective}. 

Other approaches use well-known artificial intelligence algorithms to find optimal circuit decompositions, for instance, the use of reinforcement learning \cite{zhang2020topological,pirhooshyaran2020quantum}. Evolutionary algorithms such as genetic algorithms have been widely studied \cite{colin1999quantum, massey2004evolving, massey2006human, bang2014genetic, heras2016genetic, spagnolo2017learning, li2017approximate, lamata2018quantum, potovcek2018multi}. In these approaches, multiple random circuits composed by the native gate set are generated and evolved later on. The evolution strategy includes the definition of possible \textit{mutations} such as introducing a new gate on a particular qubit, the swap between circuit gates or the deletion of a particular gate. Then, a multi-objective loss function is used to estimate the success of each circuit family until a given convergence criterion, after which the circuit with the best performance is selected. These works have to be added to those focusing finding the optimal PQC for a given VQA, as discussed in \secref{sec:PQC}. A VQA for circuit compilation using a genetic algorithm as optimization subroutine is presented in Ref. \cite{khatri2019quantum}, called Quantum Assisted Quantum Compiler.

\subsubsection{The qubit mapping problem}

After decomposing and simplifying the quantum circuit into the native gates, a hardware-specific task remains: mapping the resultant circuit to the particular qubit connectivity or topology, a task also known as the \textit{qubit routing problem}. In general, due to experimental limitations, not all qubits are connected, which means that two-qubit operations are not always possible. A naive approach to circumvent this limitation consist of swapping each qubit state with its neighbour (by using $SWAP$ gates) until we find a connected pair, perform the desired two-qubit operation and swap back the states of the qubits involved, returning to the original state with the intended two-qubit gate applied to it.
This translates into a significant growth of the circuit depth for circuit topologies with a sparse qubit connection graph.

Some NISQ algorithms presented in this review may include the qubits' connectivity by means of the loss function or the available rules used to decompose the unitaries. However, quantum compilation is a hardware-specific transformation; it might be more useful to apply this step independently of the quantum circuit and depending on the chip architecture. Unfortunately, the qubit mapping problem is NP-complete \cite{Botea2018OnTC}. Several heuristic approaches based on dynamic programming and depth partitioning have been explored \cite{zulehner2017exact,siraichi2018qubit,zulehner2018efficient, li2019tackling,zulehner2019compiling,cowtan2019qubit} as well as using reinforcement learning \cite{pozzi2020using}. Exact methodologies based on reasoning engines such as Boolean satisfiability solvers have also been proposed \cite{wille2019mapping,tan2020optimal}. The so-called LHZ architecture is an approach that solves the connectivity issue at the cost of increasing the number of qubits
~\cite{lechner2020quantum}. The same framework can be applied to a quantum annealing system as well~\cite{lechner2015quantum}. Encoding this problem into a QUBO (see \secref{subsec:quantum-annealing}) to solve it using classical simulated annealing has also been proposed in Ref. \cite{dury2020qubo}. Approaches for circuit compilation based on commutation algebra of quantum gates have been been suggested in ~\cite{itoko2020optimization,itoko2019quantum}. 

There can be many possible qubit mappings of a given algorithm into a particular device if not all qubits are required. In those cases, one can put some extra effort to find the best performing qubits in terms of error rates and coherence times \cite{nishio2020extracting,niu2020hardware}. In that direction, finding the mapping with the lowest circuit depth may prove valuable to reduce the errors due to decoherence \cite{zhang2020slackq}.

Finally, the use of circuit synthesis with connectivity constraints have also been proposed. Some of these works are based on Gaussian elimination processes where you take the matrix representation of the circuit transformation and manipulate it to extract the basic transformations (in particular, the CNOT gates that respect the connectivity) \cite{nash2020quantum,kissinger2019cnot}. In \cite{gheorghiu2020reducing}, the strategy consist on slicing the circuit into smaller parts that can be adapted and transformed to fit into the particular topology. One can also adapt this problem to the syndrome decoding problem \cite{de2020quantum}. Reference \cite{de2020architecture} solves the qubit routing for phase polynomial circuits, i.e. circuits that only contain CNOTs and $R_{z}$ gates.

\subsubsection{Resource-aware circuit design}

As described in \secref{sec:PQC}, there are different strategies to design a circuit ansatz. Unfortunately, many of them require circuit depths, qubit connectivity and a number of parameters beyond the capabilities of current quantum hardware. In the following paragraphs, we discuss strategies to design and adapt PQC and VQA to the devices characteristics.

\paragraph{ADAPT-VQE.}
    Early VQA employed a fixed ansatz design with its parameters tuned using a classical optimizer.
    The Adaptive Derivative-Assembled Pseudo-Trotter ansatz Variational Quantum Eigensolver (ADAPT-VQE) was introduced as a more scalable and efficient way to simultaneously design and optimize a parameterized ansatz \cite{grimsley2019adaptive}.
    At each iteration, ADAPT-VQE constructs an ansatz by adding an operator corresponding to the largest gradient from a carefully designed operator pool. That is, given an operator $\hat{\tau}_i$ from the operator pool, the gradient of the energy with respect to the corresponding parameter $\theta_i$ is defined as
    \begin{equation}
        \partial_{i}E = \bra{\psi}[H, \hat{\tau}_i]\ket{\psi},
    \end{equation}
    where $\ket{\psi}$ is the ansatz at the current iteration to be updated.
  After computing the gradient components and choosing the operator corresponding to the largest gradient, the gate operation implementing $\hat{\tau}_i$ is added to the ansatz with its parameter value initialized at $0$.
    The ansatz is then optimized before adding the next operator.
    The ADAPT-VQE algorithm terminates when the norm of the gradient vector falls below a predefined threshold.
    
    In the case of fermionic ADAPT-VQE, the operator pool consists of fermionic operators that are transformed into quantum gate operations through, e.g., the Jordan-Wigner mapping. 
    A more hardware-efficient variant of the ADAPT-VQE algorithm is the qubit ADAPT-VQE, in which the pool consists of gate operators acting directly on qubits \cite{tang2019qubit}. 
    Both versions of ADAPT-VQEs were able to to generate optimized circuits with reduced depths and CNOT counts compared to previous ansatz construction and optimization methods.

\paragraph{MI-ADAPT-VQE.}
    The mutual information-assisted ADAPT-VQE (MI-ADAPT-VQE), introduced by \cite{zhang2020mutual}, leverages the density matrix renormalization group (DMRG) \cite{white1992density,hallberg2006new} method to accelerate the circuit constructions for the ADAPT-VQE routine. Instead of gradients, it uses the mutual information to guide circuit constructions.
    At the beginning of the algorithm, the pair-wise quantum mutual information is approximated using DMRG, which is then applied to construct a reduced pool of entangling gates. 
    In each iteration of the method, new circuits are generated in which quantum gates are mainly distributed among pairs of qubits corresponding to large mutual information. This avoids allocating quantum resources on pairs of qubits that are less important to entangle. 
    Numerical experiments suggest that the number of new circuits needed in each step of the adaptive construction can be significantly reduced using MI-ADAPT-VQE, saving both time and quantum resources. The number of trial circuits in certain cases can be reduced to about $5\%$ for $H_2$ and $10\%$ for $H_2O$ as compared to ADAPT-VQE, using an operation pool based on the qubit coupled-cluster method \cite{ryabinkin2020iterative}.
    
\paragraph{MoG-VQE.}
    To reduce two-qubit gate counts for near-term experiments, the multiobjective genetic variational quantum eigensolver (MoG-VQE) optimizes for both the energy and the number of CNOTs in the quantum circuit \cite{chivilikhin2020mogvqe}.
    The MoG-VQE algorithm combines two evolutionary strategies: \textit{i)} NSGA-II \cite{deb2000fast}, a multiobjective genetic algorithm, to propose a circuit structure to minimize both the energy and CNOT count, and \textit{ii)} CMA-ES \cite{hansen2003reducing} to tune parameters and evaluate optimized energies for the qubit topologies suggested by the NSGA-II algorithm. 
    MoG-VQE initializes a diverse population by sampling a checkerboard pattern of two-qubit circuit blocks. 
    To vary the populations over different generations, the three possible mutation operators are: \textit{i)} inserting a two-qubit circuit block in a random position; \textit{ii)} removing a two-qubit circuit block in a random position; and \textit{iii)} adding or removing 10 circuit blocks to help escape from local minima. 
    The authors note that \textit{iii)} is selected with a lower probability than mutation operators \textit{i)} and \textit{ii)}.
    Parents are selected using the tournament selection method.
    For each circuit topology, the corresponding energy is evaluated using the CMA-ES optimizer.
    These steps repeat until some termination criteria are satisfied. 
    Using MoG-VQE, the authors reported significant reductions in the CNOT counts compared to those of other hardware-efficient circuits when estimating ground state energies of several molecules.
    For example, for a 12-qubit LiH Hamiltonian, MoG-VQE generated a circuit corresponding to estimating the ground state energy to within chemical precision using only 12 (non-nearest-neighbor) CNOTs. 
    
\paragraph{PECT.} An alternative approach for adaptively constructing and optimizing an ansatz was introduced by the Parameter-efficient circuit training (PECT) scheme \cite{sim2020adaptive}.
PECT enables optimizations of predefined ans\"atze, such as unitary coupled-cluster or the low-depth circuit ansatz (LDCA) \cite{dallaire-demers2018lowdepth}, by dynamically pruning and adding back parameterized gates during an optimization.
After selecting an ansatz $U$, a subset of gate operations from $A$ is chosen while other parameterized gate operations are tuned to identity operations.
This results in an ansatz substructure $A'$ with reduced circuit depth and gate count.
Parameters of $A'$ are then optimized, following what the authors call a ``local optimization'' step.
After local optimization, to refine or reparameterize the ansatz substructure, parameters with small magnitudes are pruned or removed.
A heuristic growth rule is used to grow back the same number of parameters that was pruned. 
Steps of local optimization and ansatz reparameterization are repeated until termination criteria are met.
Because PECT optimizes parameter subsets at any iteration, circuits that are executed on the quantum computer have reduced depths and CNOT counts compared to the original ansatz. 
Using PECT, the authors were able to optimize 12-qubit LDCA circuits, naively equipped with hundreds to low thousands of parameters, to estimate ground state energies of LiH and $\text{H}_2\text{O}$.
Previous optimizations of LDCA were limited to 8 qubits.

\subsection{Quantum software tools} \label{sec:software}

\begin{figure*}[ht!]
    \centering
    \includegraphics[width=\linewidth]{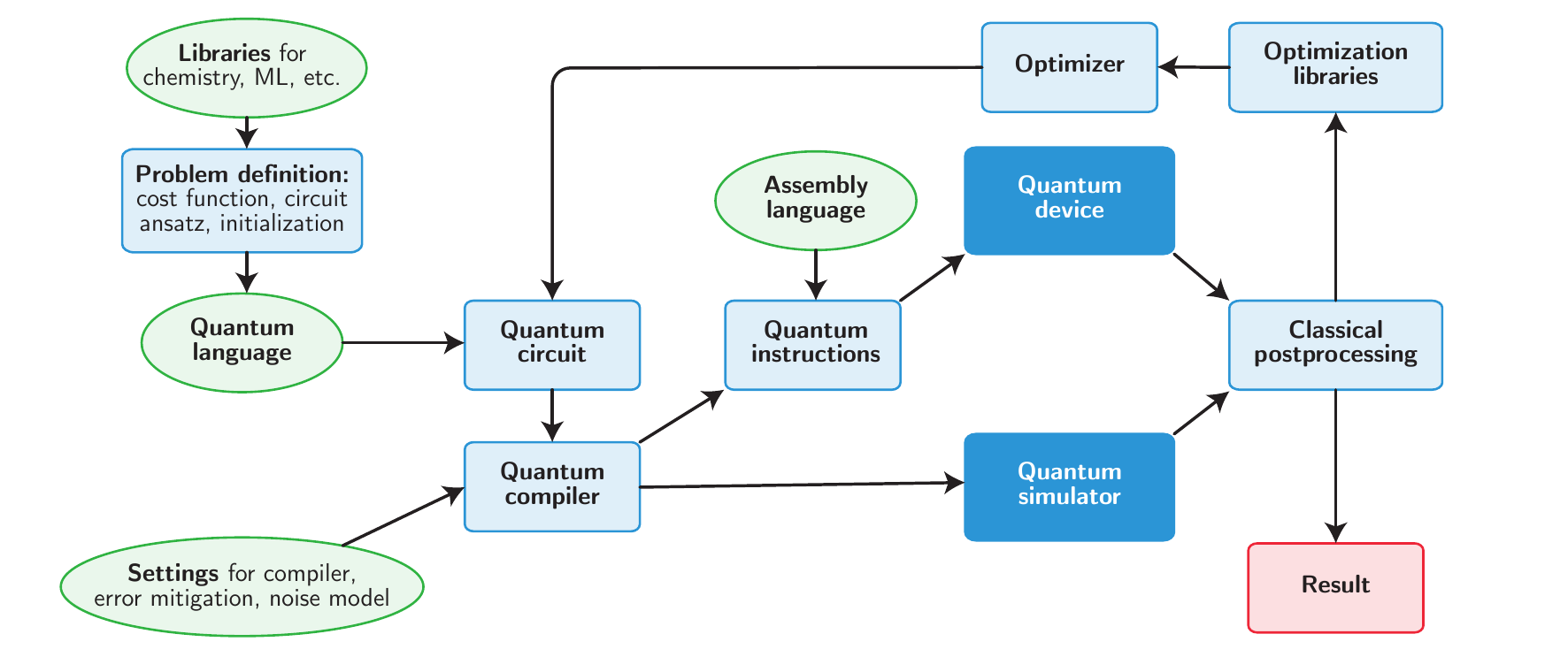}
    \caption{Schematic representation of a standard NISQ programming workflow (color online). Green circular boxes represent the libraries and languages used for designing, optimizing and running a quantum algorithm in a real quantum device or in a simulator. External libraries can be used to define the problem or to improve the performance of the algorithm by simplifying the circuit or using error-mitigation techniques. An assembly language will be needed to translate the theoretical algorithm to a set of physical operations on the quantum hardware. Classical post-processing is necessary to manipulate the result of the computation and to either obtain the final result or send the provisional one to a classical optimizer (VQA).}
    \label{fig:software}
\end{figure*}

A quantum computer is a hybrid device composed of quantum hardware and classical software that controls it by sending a list of instructions and processing the results of the computation. This hybrid nature is accentuated in the NISQ era, as explained in the current review. Thus, the classical subroutines are part of the core in state-of-the-art NISQ algorithms and a language to communicate with the quantum device is a bare necessity. On top of that, almost all progress in quantum algorithms is tested in quantum simulators making them essential to perform proof-of-concept simulations, before or until the algorithm is applicable on real devices.

Figure \ref{fig:software} represents diagrammatically the typical workflow of a NISQ algorithm. The individual parts of the problem, such as the objective function to optimize, the quantum circuit design or the initialization parameters, are translated into quantum circuits by a classical pre-computation step. The syntax of this language includes the quantum gates, qubit initialization, objective function definition, etc. The theoretical circuit is then compiled to fulfill experimental limitations such as qubit connectivity, native quantum gate set or circuit depth. To accomplish this task, compilers that allow for circuit simplification (see \secref{sec:compilers}), or noise models (for simulation purposes) might be useful. After this pre-processing step, the algorithm is ready to enter into the quantum-classical loop. The quantum circuit can be run in a quantum simulator or real hardware. In the latter case, an assembly language \cite{cQASM,OpenQASM,pyquil,strawberry} will translate the quantum circuit into a set of instructions for the device. After the qubits are measured, the result can be post-processed and techniques such as error mitigation might be used. Either the algorithm finishes or the result is sent to the classical optimizer that computes the next loop variables (e.g. for VQA). 

We define a \textit{quantum software library} as a library or a set of libraries written in a classical programming language (e.g. python or C++) that allows writing a quantum program. 
In some cases, these libraries are open-source and can be used directly on real hardware or on a quantum simulator. The proliferation of all these libraries, simulators and devices has also created a necessity for some multi-platform languages. These are those that can use multiple quantum software libraries as a backend, reducing the programming efforts substantially by unifying the language syntax. Some of these packages include built-in sub-libraries suited for particular applications, from chemistry to QML, or particular well-known algorithms such as VQE or QAOA.

We provide a list of some open-source libraries suited for NISQ computation in Tab.~\ref{tab:software} from App.~\ref{app:tables}. This list represents just a snapshot of the state-of-the-art of the quantum software ecosystem as new tools are being developed and some projects are being abandoned. An updated list of quantum software resources can be found in Ref. \cite{qcreports,awesomeqosf} and a detailed comparison analysis between some of these languages and libraries in Ref. \cite{gay2006quantum,nguyen2020extending,heim2020quantum,garhwal2019quantum}. Due to the broad applications of NISQ algorithms, specific libraries used in other fields beyond quantum computation can also be required, e.g. quantum chemistry and machine learning libraries or external compilers and simulators. These libraries are used for other applications besides purely quantum computation, so we consider and list them as external libraries in Tab.~\ref{tab:ext_soft} from App.~\ref{app:tables}, although most of them integrated in the quantum software libraries.

\section{Applications}\label{ch:application}

\subsection{Many-body physics and chemistry}\label{sec:chemistry}

Understanding the static and dynamic properties of quantum mechanical systems is a core challenge at the heart of many fields such as chemistry and physics. 
Classical numerical methods often struggle in solving these problems, due to the exponential increase of resources needed with growing number of particles to simulate. Owing to their quantum mechanical nature, quantum computers offer a way to simulate even large-scale many-body systems~\cite{feynman1982simulating, vonburg2020quantum}.
The initial application for chemistry was to obtain molecular energies via quantum phase estimation on a quantum computer \cite{aspuru2005simulated}. Besides the molecular energy, properties that can be extracted from a successfully prepared ground state, such as energy derivatives with respect to the nuclear framework, are of similar interest ~\cite{kassal2009quantum, obrien2019calculating}. Fault tolerant quantum algorithms have the potential to become \textit{killer applications} in the computational discovery of chemical reaction mechanisms ~\cite{reiher2017elucidating} and NISQ algorithms could play a major role in their realization.
Here we review various NISQ algorithms that have been proposed to tackle quantum chemistry and many-body physics related problems. We start by introducing concepts on mapping physical problems onto the quantum computer. Then, we introduce algorithms for common challenges, such as finding the static as well as dynamic properties of quantum systems in various settings. All NISQ algorithms discussed in this section are listed in \tabref{tab:many_body}.

\subsubsection{Qubit encodings}

In general, any physical system can be written in terms of a Hamiltonian which is the sum of its kinetic and potential energy. 
In quantum theory, each physical system is associated with a language of operators and an algebra establishing  such language. 
Depending on the system constituents, there are three types of particles (operators) in Nature: fermions, bosons and anyons.
The first two are elementary particles obeying Fermi-Dirac (FD) and Bose-Einstein (BE) statistics, respectively.
The latter being quasiparticles obeying continuous or anyonic statistics, and existing only in two-dimensional confinement.
Quantum computers (QC) operate in a language of qubits (a distinguishable set of spin-$1/2$ particles).
Hence, the quantum simulation of a physical system refers to performing a one-to-one mapping from the system operator to the QC language, preserving the underlying statistics. For a recent review on hardware-dependent
mappings of spin Hamiltonians into their corresponding quantum
circuit, refer to ~\cite{tacchino2020quantumreview}.

In the standard model of quantum computation, a two-level system or spin-$1/2$ particle is denoted by its spin orientation $\ket{\uparrow}=\ket{0}=(1,0)^T$ and $\ket{\downarrow}=\ket{1}=(0,1)^T$. 
An $N$-qubit system is then constructed from the standard Pauli matrices $\hat{\sigma}_x^i,\hat{\sigma}_y^i,\hat{\sigma}_z^i$, where the superscript $i$ refers to the $i^\text{th}$ local qubit site.
These operators satisfy the commutation relations of an $\bigoplus_{i=1}^N su(2)_i$ algebra
    $[\hat{\sigma}_\mu^l,\hat{\sigma}_\nu^m]=2i\delta_{lm}\epsilon_{\mu\nu\lambda}\hat{\sigma}_\lambda^l$,
where $\epsilon_{\mu\nu\lambda}$ is the totally anti-symmetric Levi-Civita symbol with $\mu,\nu,\lambda\in \{x,y,z\}$. 

\paragraph{Fermions. }
In the second quantized notation, $N$ fermions are denoted by fermionic operators $\hat{f}_i ^\dagger (\hat{f}_i)$, the creation (annihilation) operators of a fermion in the $i^\text{th}$ mode/site $(i=1,\cdots,N).$
The fermionic operators obey Pauli's exclusion principle and the anti-symmetric nature of the fermion wave function.
Hence, the fermionic algebra is defined by the  anti-commutators $\{\hat{f}_i,\hat{f}_j \}=0, \{\hat{f}_i^\dagger,\hat{f}_j \}=\delta_{ij}.$
There are a number of well-known mappings that allow the description of a fermionic system by the standard model of QC.
They are the Jordan-Wigner transformation \cite{jordan1928pauli}, Bravyi-Kitaev transformation \cite{bravyi2002fermionic} and Ball-Verstraete-Cirac transformation \cite{ball2005fermions,verstraete2005mapping}.
In~\cite{steudtner2019quantum}, the two-dimensional topology of most proposed qubit architectures is taken explicitly into account and compared to some of the aforementioned one-dimensional mappings.
More advanced mappings, using the interaction graph of the Hamiltonian~\cite{setia2018bravyikitaevsuperfast, setia2019superfast} or customized quasi-local and local encodings~\cite{havlicheck2017operator, chien2020custom, derby2020compact, jiang2020optimal} have been introduced as well. Other approaches try to lower the qubit requirements of the mapped fermionic operators by taking inspiration from classical error correction codes and the internal symmetries of the system~\cite{bravyi2017tapering, steudtner2018fermiontoqubit}, other examples are mappings based on point-group symmetries of molecular Hamiltonians~\cite{setia2020reducing}. Recently, mappings of SU($N$) fermions to qubits have been proposed~\cite{consiglio2021variational}.\\ 

In the following, we will briefly outline the oldest and most intuitive mapping: the Jordan-Wigner transformation. In this mapping, the qubit states are equivalent to the second-quantized occupation number vectors, and fermionic creation and annihilation operators are transformed to qubit raising and lowering operators $\hat{\sigma}_{\pm}^j =( \hat{\sigma}_x ^j \pm i \hat{\sigma}_y ^j)/{2}$ combined with strings of $\hat{\sigma}_z$ operators that ensure the correct anti-commutation properties:
\begin{equation}
    \hat{f}_j \rightarrow \left(\prod_{l=1}^{j-1} -\hat{\sigma}_z^l \right)\hat{\sigma}_- ^j, \quad f_j^\dagger \rightarrow \left(\prod_{l=1}^{j-1} -\hat{\sigma}_z^l \right)\hat{\sigma}_+ ^j.
\end{equation}
In this new transformation, one can verify that $\hat{f}_j^\dagger, \hat{f}_j$ satisfy the above anticommutation relations, while $\hat{\sigma}_\mu^j$ satisfy the commutation relations showed above. 
The reader is referred to the literature~\cite{aspuru2005simulated,seeley2012bravyikitaev,tranter2015bravyikitaev,tranter2018comparison,somma2003quantum} and the respective original references for details and comparisons regarding the other transformations. 

\paragraph{Bosons. }
Bosonic operators satisfy the commutation relations
$[\hat{\tilde{b}}_i,\hat{\tilde{b}}_j]=0, [\hat{\tilde{b}}_i,\hat{\tilde{b}}_j ^\dagger]=\delta_{ij}$
in an infinite-dimensional Hilbert space. 
At first, it seems it is impossible to simulate bosonic systems due to the nature of infinite dimensions.
However, sometimes we are interested in studying some finite modes of excitations above the ground state.
Hence, the use of the entire infinite dimensional Hilbert space is unnecessary. 
In a finite dimensional basis, the bosons $\hat{b}_i^\dagger, \hat{b}_i$ obey the following commutation relations \cite{batista2004algebraic}
\begin{equation}
    [\hat{b}_i,\hat{b}_j]=0, \ [\hat{b}_i,\hat{b}_j^\dagger]=\delta_{ij}\left[1-\frac{N_b +1}{N_b !}(\hat{b}_i^\dagger)^{N_b} (\hat{b}_i)^{N_b} \right],
    \label{eq:ch4:truncated_boson_commutator}
\end{equation}
with $\hat{b}_i^\dagger \hat{b}_i \ket{n_i}=n_i \ket{n_i}$ with $n_i = 0,\cdots,N_b$, where $N_b$ is the maximum truncated excitation number, corresponding to the $i^\text{th}$ bosonic site/mode.
A direct consequence is one can then write down the creation and annihilation operators as 
\begin{equation}
    \hat{b}_i ^\dagger = \sum_{n=0}^{N_b -1} \sqrt{n+1} \ketbra{n+1}{n},
\end{equation}
and $\hat{b}_i$ is complex conjugate of $\hat{b}_i ^\dagger$.
There are infinite means to translate such truncated operators into the QC language, the so-called Pauli words.
A Commonly used one is known as standard binary or compact encoding \cite{somma2003quantum,veis2016quantum,sawaya2019quantum,mcardle2019digital,sawaya2020resource}, where $\{\alpha,\beta\in \mathbb{W}\}$ in $\ketbra{\alpha}{\beta}$ are now written in terms of binary strings. 
Using the following identities:
$\ketbra{0}{1}\equiv \hat{\sigma}_-;\ketbra{1}{0}\equiv \hat{\sigma}_+;\ketbra{0}{0}\equiv(I+\hat{\sigma}_z)/2;\ketbra{1}{1}\equiv(I-\hat{\sigma}_z)/2$, Pauli words translation can be accomplished.
Recently, detailed studies on various encodings (binary, Gray, Unary, block Unary), have been studied and Gray code in particular is found to be resource efficient (in terms of number of qubits and two-qubit entangling gates) in simulating some specific bosonic and spin Hamiltonians \cite{sawaya2020resource}. 

\paragraph{Anyons. }
As seen above, we can now proceed to simulate more general particle statistics, in particular, hard-core anyons.
With ``hard-core'', we refer to the Pauli's exclusion principle where only zero or one particle can occupy a single mode.
The anyonic operators $\hat{a}_i, \hat{a}_i ^\dagger$ obey the commutation relations
    $[\hat{a}_i,\hat{a}_j]_\theta = [\hat{a}_i ^\dagger ,\hat{a}_j ^\dagger]_\theta=0$, $[\hat{a}_i,\hat{a}_j ^\dagger]_{-\theta} =\delta_{ij}(1-(e^{-i\theta +1 })\hat{n}_j)$ and $[\hat{n}_i,\hat{a}_j ^\dagger] = \delta_{ij} \hat{a}_j^\dagger$,
where $\hat{n}_j =\hat{a}_j ^\dagger \hat{a}_j, [\hat{A},\hat{B}]_\theta = \hat{A}\hat{B}-e^{i\theta}\hat{B}\hat{A}$, with $(i\leq j)$ and $0\leq \theta < 2\pi$. 
Specifically, $\theta=\pi \textrm{ mod}(2\pi)$ gives rise to canonical fermions, and $\theta=0 \textrm{ mod}(2\pi)$ would recover hard-core bosons.
By simply applying the following isomorphic mapping between algebras \cite{somma2003quantum}:
\begin{eqnarray}
    \hat{a}_j ^\dagger  &=& \prod_{i<j}\left(\frac{e^{-i\theta}+1}{2}+\frac{e^{-i\theta}-1}{2}\hat{\sigma}_z ^i \right) \hat{\sigma}_+ ^j, \nonumber \\
    \hat{a}_j  &=& \prod_{i<j}\left(\frac{e^{i\theta}+1}{2}+\frac{e^{i\theta}-1}{2}\hat{\sigma}_z ^i \right) \hat{\sigma}_- ^j, \quad
    \hat{n}_j = \frac{1+\hat{\sigma}_z ^j}{2},
\end{eqnarray}
we would obtain Pauli words for the QC. 
The above mapping would also ensure the anyonic algebra shown above. 

\subsubsection{Constructing electronic Hamiltonians} 
\label{sec:constructing_electronic_hamiltonians}

The electronic structure problem is one of the most prominent task within VQA (see for example the reviews \cite{cao2019quantum, mcardle2020quantum}) and was the pioneering task for the variational quantum eigensolver \cite{peruzzo2014variational, mcclean2016theory}. In this section, we will illustrate how the original continuous many-electron problem can be discretized to a second-quantized formulation that can itself be encoded into qubits by the techniques introduced in the beginning of \secref{sec:chemistry}. This encoded qubit systems define then the central problem of the VQAs further described in \secref{sec:VQE}.

The electronic structure problem aims to approximate eigenfunctions of an electronic Hamiltonians
\begin{align}
    H_\text{e} = \mathcal{T}_\text{e} + \mathcal{V}_\text{ee} + \mathcal{V}_\text{ext},\label{eq:electronic_hamiltonian}
\end{align}
describing a system of $N_\text{e}$ electrons through their accumulated kinetic energies $\mathcal{T}_\text{e}=-\frac{1}{2}\sum_{k=1}^{N_\text{e}} \Delta_{\boldsymbol{r}_k} $, the electronic Coulomb repulsion $\mathcal{V}_\text{ee} = \sum_{k\neq l} V_\text{ee}\left(\boldsymbol{r}_k-\boldsymbol{r}_l\right) =  \sum_{k\neq l}\frac{1}{\lvert \boldsymbol{r}_k - \boldsymbol{r}_l \rvert}$, and an external potential $\mathcal{V}_\text{ext} = \sum_{k=1}^{N_\text{e}} V_\text{ext}\left(\boldsymbol{r}_k\right)$ that is usually given by the accumulated Coulomb potential of nuclear point charges. If the external potential is not explicitly spin dependent, the electronic Hamiltonian only acts on the spatial coordinates $\boldsymbol{r}_k \in \mathbb{R}^3$ of the electrons and, to ensure proper electronic wave functions, the fermionic anti-symmetry is achieved via restrictions in the Hilbert-space. 
We refer to \cite{herbst2018thesis, kottmann2018thesis, rhowedder2010thesis} and the textbook \cite{yserentant2010regularity} for the direct construction and discretization of this continous Hilbert spaces.

A more compact, but formally equivalent, definition is offered through second quantization by introducing the abstract anti-commuting field operators $\hat{\psi}^\dagger\left(x\right)$ and $\hat{\psi}\left(x\right)$ that create and annihilate electrons at spin-coordinate $x_k\in\mathbb{R}^3 \times \left\{\pm\frac{1}{2}\right\}$~\cite{jordan1927mehrkorperproblem,surjan2012second, jorgensen2012second}. The electronic Hamiltonian can then be written as
\begin{align}
    H_\text{e} =& \int \operatorname{d}x\;
    \hat{\psi}^\dagger\left(x\right)
    \left(T\left(x\right) + V_\text{ext}\left(x\right)\right) \hat{\psi}\left(x\right) \label{eq:electronic_hamiltonian_second_quantized}\\
    &+ \int \operatorname{d}x\operatorname{d}y\;
    \hat{\psi}^\dagger\left(x\right) \hat{\psi}^\dagger\left(y\right)
    V_\text{ee}\left(x-y\right)  \hat{\psi}\left(y\right) \hat{\psi}\left(x\right)\nonumber
\end{align}
where the potential operators still only act on the spatial part of the spin components. Although direct approaches on real-space grids are possible~\cite{kottmann2018coupledGS, kottmann2018thesis, kunitsa2020grid, kivlichan2017bounding} the majority of VQA employs a fixed set of three dimensional functions (so-called orbitals) to capture the spatial part of the electronic Hilbert space. The orbitals are usually determined by solving a mean-field problem (Hartree{\textendash}Fock) within a set of globally defined atomic orbitals.
Alternatives to the standard representation are, for example, direct determination of system adapted orbitals~\cite{kottmann2020reducing}, compactification of basis sets through intrinsic atomic orbitals~\cite{barison2020quantum} and optimized virtual orbitals represented by plane-waves~\cite{bylaska2020quantum}. 

For the formal description of the discretized second-quantized electronic Hamiltonian, the origin of the orbitals is not important as long as they form an orthonormal set of $H^1\left(\mathbb{R}^3\right)$ functions. Using such a set of spatial orbitals we can formally expand the field operators in the corresponding spin-orbitals
\begin{equation}
    \hat{\psi}^\dagger\left(x\right) = \sum_k \phi_k^\ast\left(x\right) f_k^\dagger, \quad
    \hat{\psi}\left(x\right) = \sum_k \phi_k\left(x\right) f_k,
\end{equation}
where $f_k^\dagger$ and $f_k$ are fermionic creation and annihilation operators obeying the anticommutation relations shown in the previous subsection. Using the expansion from \eqref{eq:electronic_hamiltonian_second_quantized} leads to the common discretized second-quantized Hamiltonian,
\begin{align}
    H_\text{e} = \sum_{kl} h_{kl} f_k^\dagger f_l + \sum_{klmn} g_{klmn} f_k^\dagger f_l^\dagger f_n f_m,
    \label{eq:electronic_hamiltonian_second_quantized_discretized}
\end{align}
with the molecular integrals~\cite{fermann2020fundamentals}
\begin{align}
    &h_{kl} = \int \phi_k^\ast\left(x\right)
    \left(T\left(x\right) + V_\text{ext}\left(x\right)\right)
    \phi_l\left(x\right)
    \operatorname{d}x,\label{eq:molecular_integrals} \\
    &g_{klmn} \int
    \phi_k^\ast\left(x\right)
    \phi_l^\ast\left(y\right)
    V_\text{ee}\left(x-y\right)
    \phi_m^\ast\left(x\right)
    \phi_n^\ast\left(y\right)
    \operatorname{d}x\operatorname{d}y.\nonumber
\end{align}
Note that the indices of the two body integrals are denoted in the standard Dirac notation $g_{klmn} \equiv \bra{kl}V_\text{ee}\ket{mn}$ but other notations, such as Mulliken $\left(km\vert lm\right) = \bra{kl}V_\text{ee}\ket{mn}$  are sometimes used.
Generally speaking, an arbitrary set of spatial orbitals, that can in principle be any set of orthonormal $H^1\left(\mathbb{R}^3\right)$ functions, defines a discretized second-quantized Hamiltonian as in \eqref{eq:electronic_hamiltonian_second_quantized_discretized} over the corresponding molecular integrals \eqref{eq:molecular_integrals}. This discretized Hamiltonian can then be encoded into a qubit Hamiltonian by corresponding fermion to qubit mappings discussed in \secref{sec:chemistry}.

\subsubsection{Variational quantum eigensolver}\label{sec:VQE}

Estimating the ground state and its energy of Hamiltonians is an important
problem in physics, which has numerous applications ranging from solid-state
physics to combinatorial optimization (see \secref{sec:optimization}). While this problem is in general QMA-hard and even quantum computers are not expected to be able to efficiently solve it in general~\cite{kempe2006complexity}, there is hope that approximate solutions of the ground state could be found faster and for larger system sizes compared to what is possible with classical computers. 

To this end, VQE \cite{peruzzo2014variational,mcclean2016theory} has been proposed, to find the ground state of a Hamiltonian $H$ in a manner that is suited for NISQ devices~\cite{wecker2015progress}. Following the concept introduced in \secref{sec:objective_function} and \secref{sec:PQC}, a parameterized circuit $U(\boldsymbol{\theta})$ is minimized with respect to the objective function, which in general is the expectation value of the energy of the Hamiltonian from \eqref{eq:def_expectationvalue}.

The approximated ground state is given by the quantum state $\ket{\psi_\text{min}}=U(\boldsymbol{\theta}_\text{min})\ket{0}$ which minimizes the energy $\text{min}_{\boldsymbol{\theta}} \langle H_{\boldsymbol{\theta}}\rangle\ge E_\text{g}$ upper bounded by the true ground state energy $E_\text{g}$ as guaranteed by the Rayleigh-Ritz variational principle \cite{gould2012variational}. VQE has been intensively studied in both theory and experiments, and various adaptions and extensions have been proposed, which we discuss in the following paragraphs.

\paragraph{Self-verification. } Whether the variational quantum simulator has converged to an actual eigenstate of the Hamiltonian, can be checked directly on the quantum processor by verifying that the variance of the energy $\text{var}=\langle(H-\langle H\rangle)^2\rangle$ is zero. This has been demonstrated for solving a many-body Hamiltonian on 8 qubits on a ion-trap \cite{kokail2019self} (see also \secref{sec:nuclear_physics}).

\paragraph{Accelerated VQE. } A key computational effort in VQE lies in estimating the cost function, which is achieved by repeatedly running the circuit and taking measurements of the Pauli strings (see \secref{sec:measurement}). For a given desired additive error bounded by $\epsilon$, it takes $O(1/\epsilon^2)$ number of samples. This can be improved by using the Quantum Phase estimation algorithm to estimate the expectation value, which takes only $O(\text{log}(1/\epsilon))$ samples, however at the cost of additional computation which may be hard in the NISQ era. To leverage a trade-off between the advantages and disadvantages of both methods, an accelerated versions of VQE that interpolates between regular measurements and quantum phase estimation has been proposed~\cite{wang2019accelerated}.

\paragraph{Measurement-based VQE. } In \cite{ferguson2020measurement}, the authors present two strategies to implement the VQE algorithm on a measurement-based quantum computer, an alternative quantum computing paradigm that uses entanglement as a resource and achieves the the desired computation by performing particular sets of local measurements (see \cite{briegel2009measurement} for a review). They propose a way to generate the needed variational state families using measurements on a highly entangled state and provide an equivalence between the measurement- and gate-based schemes.

\paragraph{Reusing qubits in VQE. } A recent proposal suggested a VQE method that relies on fewer qubits by re-using some of them~\cite{liu2019variational}. The core idea is to represent a virtual $N$ qubit state by $R+V<N$ physical qubits, where $R$ qubits have to be reusable qubits, e.g. they can be measured and re-initialized during the circuit runtime. These intermittent measurements are possible on current ion trap hardware~\cite{pino2020demonstration}. The $R+V$ qubits are entangled by a PQC, then $R$ qubits are measured and the outcome is recorded. The $R$ qubits are re-initialized to the $\ket{0}^{\otimes R}$ state, and again entangled with the $V$ other qubits by another PQC. This procedure is repeated until in total $N$ qubits have been measured. The concept and expressiveness of this type of ansatz is the same as Tensor networks methods such as MPS, which have been highly useful for the classical calculation of many-body problems, and open up a way to perform quantum computing of many qubits on devices with limited number of qubits.

\paragraph{Adiabatically assisted VQE. }

The ground state of more challenging Hamiltonians can be difficult to find for standard VQE due to convergence to local minima instead of the global minima of the energy. 
To alleviate this, quantum annealing  (see \secref{subsec:quantum-annealing}) can be used to adiabatically assist the optimisation procedure, as proposed in the adiabatically assisted VQE \cite{garcia2018addressing}. 
This approach uses an objective function $O(s)=\bra{0}U^\dagger(\boldsymbol{\theta}) H(s) U(\theta)\ket{0}$, where $H(s)=(1-s)H_0+sH_1$. Here, $H_0$ is a Hamiltonian with easily preparable groundstate and the goal is to find the ground state of a Hamiltonian $H_1$. 
In this algorithm, VQE is run for multiple discrete steps $s_n$. One starts with $s_0=0$ and finds the minimal parameters $\boldsymbol{\theta}_0^*$ of the objective function $O(s_0)$. Then, $\boldsymbol{\theta}_0^*$ is used as initial guess for VQE for the next increasing step $s_{1}=s_0+\Delta s$ with objective function $O(s_1)$. This procedure is repeated until $s=1$ is reached. 
This approach eases the optimization task, as the initial Hamiltonian $H_0$ is a simple Hamiltonian with a ground state that can be easily found via optimization. For small steps $\Delta s$, the ground state of the Hamiltonian $H(s)$ and $H(s+\Delta s)$ will not differ too much, making the optimization task at every step less challenging compared to directly solving for $H(1)$. Previous works \cite{mcclean2016theory} also suggest to use adiabatically prepared states as initial states of a VQE algorithm (see \secref{sec:PQC}).

\subsubsection{Variational quantum eigensolver for excited states}\label{sec:var_excited}

The methods of VQE have been extended to obtain the excited states of a given Hamiltonian. Finding excited states or the spectrum of a Hamiltonian is an important problem in quantum chemistry and many-body physics. Various proposals have been put forward.

\paragraph{Folded spectrum method. }  A straightforward way of calculating excited states is the folded spectrum method proposed by ~\cite{peruzzo2014variational}. To find an excited state of a Hamiltonian $H$ with approximate energy $\lambda$, the above defined VQE method is here applied to the objective function $C(\boldsymbol{\theta})=\expval{\left(H-\lambda\right)^2}{U_{\left(\boldsymbol{\theta}\right)}}$. VQE will target the eigenstate with an energy that is closest to $\lambda$. This method requires an approximate knowledge of the energy of the excited state that one wants to find, as well as estimating $\langle H^2\rangle$, which may require a excessively large number of measurements to be performed.

An extension of this method can also be used to find states that are constrained to a specific value of the conserved quantity of the problem, such as total particle number or magnetization~\cite{ryabinkin2018constrained}. Here, one defines the objective function $C(\boldsymbol{\theta})=\expval{H}{\mathcal{U}\left(\boldsymbol{\theta}\right)}+\sum_i\mu_i(\expval{S_i}{\mathcal{U}(\boldsymbol{\theta})}-s_i)^2$, where $S_i$ is the operator corresponding to the conserved quantity, and $s_i$ is the target value of that quantity. Note, that this does not restrict the target space to be an eigenstate of $S_i$.

\paragraph{Orthogonally constrained VQE. }
Excited states can be found by constraining the VQE objective function such that it penalizes the ground state \cite{higgott2019variational}.
First, one finds an approximation to the ground state of Hamiltonian $H$ via VQE with $\boldsymbol{\theta}_0=\text{arg min}_{\boldsymbol{\theta}} \expval{H}{\mathcal{U}(\boldsymbol{\theta})}$ and approximated ground state $\ket{\psi(\boldsymbol{\theta}_0)}=\mathcal{U}(\boldsymbol{\theta}_0)\ket{0}$. Then, one uses this information to formally project out the approximate ground state to find the next highest excited state. One defines the Hamiltonian $H_1=H+a\ket{\psi(\boldsymbol{\theta}_0)}\bra{\psi(\boldsymbol{\theta}_0)}$, with some sufficiently large positive parameter $a$. The ground state of $H_1$ then corresponds to the first excited state of $H$ and can be found with a VQE. This procedure can be repeated to find higher excited states up to any order by sequentially accumulating the projector terms of all states found. The Hamiltonian for the $k$-th excited state is then given by $H_k = H+ \sum_{i}^{k-1}a_{i}\ket{\psi(\boldsymbol{\theta}_{i})}\bra{\psi(\boldsymbol{\theta}_i)}$.
Combined with the unitary coupled cluster ansatz, the orthogonally constrained VQE can find excited states of small molecules~\cite{higgott2019variational,lee2018generalized}. It was further extended for adaptive circuit construction~\cite{kottmann2020feasible} and imaginary time evolution~\cite{jones2019variational}.
 
The projector term requires calculating the overlap $\vert \langle \psi(\theta)|\psi(\theta_0)\rangle\vert$, which can be achieved for example by the SWAP test, by applying the inverse of the circuit that generated the ground state $\vert\bra{0}U^\dagger(\theta)U(\theta_0)\ket{0}\vert^2$, or randomized measurements~\cite{elben2020cross}. 
An alternative approach that relies on a discriminator circuit that is trained in parallel to distinguish between the excited state to be learned and previously found lower-lying states has been proposed~\cite{tilly2020computing} and demonstrated on a small model system. Scalable proposals still remain an open research question.
Since the projector term does not require the overlap itself, but the absolute square of it, it can be computed with the help of ~\eqref{eq:fidelity_target} by computing the fidelity of the current trial state with the previously found states~\cite{lee2018generalized, kottmann2020reducing}. 

\paragraph{Subspace expansion. }
The subspace expansion method was introduced in \secref{sec:other_QEM} for error mitigation. This method can be also used to find excited states~\cite{mcclean2017hybrid} and it was demonstrated for a small molecule in~\cite{colless2018computation}. After finding the ground state of a Hamiltonian $H$ with VQE, one follows the steps that were detailed in \secref{sec:other_QEM}. One expands the prepared quantum state with different appropriate operators that match the low-energy excitations of $H$  and generates a set of states that span the low-energy subspace. Then, overlaps between the states are measured, which are then used to solve a generalized eigenvalue problem on a classical computer. The eigenvalues and eigenstates give the excited states of the Hamiltonian.
For quantum chemistry problems, the subspace expansion method was also proposed for including dynamical correlations to ground states over external corrections~\cite{takeshita2020increasing}, in the spirit of classical quantum chemistry methods, like for example \textit{CAS-CI}~\cite{roos1980complete}.

As alternative approach, the expansion in the subspace can also be accomplished by real-time evolving a reference state, and picking states at different evolution times as basis for expansion~\cite{stair2020multireference}. This is motivated by the fact that the time evolution can be seen as an approximate Krylov expansion of the quantum state. Then, one proceeds to solve the generalized eigenvalue problem to find eigenstates and eigenvalues of the Hamiltonian.

\paragraph{Subspace-search VQE/State-averaged VQE. }
The core idea of a subspace-search VQE (SSVQE)~\cite{nakanishi2019subspace} or state-averaged VQE (SAVQE)~\cite{yalouz2021stateaveraged,arimitsu2021analytic} is to minimize the energy of a PQC $U(\boldsymbol{\theta})$ over a set of orthogonal quantum states. The goal is to find the $k$-th eigenstates with lowest eigenenergy of a Hamiltonian $H$. 
In the weighted SSVQE the cost function is 
\begin{equation}
L(\boldsymbol{\theta})=\sum_{j=1}^k w_j \bra{\varphi_j}U^\dagger(\theta)HU(\boldsymbol{\theta})\ket{\varphi_j}\,, \label{eq:SSVQE_weighted}
\end{equation}
where $\{\ket{\varphi_j}\}_{j=0}^k$ is a set of $k$ easily preparable mutually orthogonal quantum states (with $\braket{\varphi_i|\varphi_j}=\delta_{i,j}$) and $\{w_j\}_j$ are positive real numbers with $w_i>w_j$ for $i<j$. Minimizing $\boldsymbol{\theta}^*=\text{arg min}_{\boldsymbol{\theta}} L(\boldsymbol{\theta})$ to its global minimum gives us the ground state and excited states $\ket{\psi_j}=U(\boldsymbol{\theta})\ket{\varphi_j}$, where $j=1$ is the ground state and $j>1$ the excited states sorted in ascending order. This algorithm gives all $k$ eigenstates in a single optimization routine. Note however that the more states to be optimized, the more complex the optimization landscape and the effort to minimize becomes.
An alternative formulation of the algorithm to find specifically the $k$-th lowest eigenstate is the unweighted SSVQE. Here, one minimizes 
    $L_1(\boldsymbol{\theta})=\sum_{j=1}^k \bra{\varphi_j}U^\dagger(\boldsymbol{\theta})H U(\boldsymbol{\theta})\ket{\varphi_j}$.
However, due to the absence of weights, the found states $\ket{\psi'_j}=U(\boldsymbol{\theta}^*)\ket{\varphi_j}$ for minimal $\boldsymbol{\theta}^*$  are not proper eigenstates of $H$, but are superposition states that span the subspace of the $k$ lowest energies. As final step to find the $k$-th eigenstate, one fixes $\boldsymbol{\boldsymbol{\theta}}=\theta^*$ to its minimized value, and then maximizes $\boldsymbol{\phi}^*=\text{max}_{\boldsymbol{\phi}} L_2(\boldsymbol{\phi})$, with $L_2(\boldsymbol{\phi})=\sum_{j=1}^k \bra{\varphi_j}V^\dagger(\boldsymbol{\phi}) U^\dagger(\theta^*)H U(\boldsymbol{\theta}^*)V(\boldsymbol{\phi})\ket{\varphi_j}$ and $V(\boldsymbol{\phi})$ being a unitary that acts only on the Hilbert space of the $k$ lowest eigenstates. Then, for the maximized $\boldsymbol{\phi}^*$, the $k$-th lowest eigenstate is given by $\ket{\psi_k}=U(\boldsymbol{\theta}^*)V(\boldsymbol{\phi}^*)\ket{\varphi_k}$. Besides general applications that involve excited states, state-averaged orbital-optimized VQEs (SA-OO-VQE) were proposed to treat chemical systems that require a ``democratic description of multiple states'' as for example necessary in the vicinity of conical intersections~\cite{yalouz2021stateaveraged}.  Here, ``democratic description'' corresponds to treating degenerate or quasi-degenerate states at the same footing.

\paragraph{Multistate contracted VQE. }
This algorithm combines the non-weighted SSVQE with the subspace expansion to find the ground state and excited states~\cite{parrish2019quantum}. First, one runs the non-weighted SSVQE routine to find the unitary $U(\boldsymbol{\theta}^*)$ to find $k$ states that span the subspace of the $k$ smallest eigenvalues $\ket{\psi'_j}=U(\boldsymbol{\theta}^*)\ket{\varphi_j}$. Then, to find the correct eigenstates, one runs the subspace expansion and measures the overlap matrix $H_{ij}=\bra{\psi'_i}H\ket{\psi'_j}$, and diagonalizes it to find estimates of the $k$ lowest eigenergies and eigenstates.

\paragraph{Fourier transform of evolution. } Recent experiments have determined the spectra of molecular and many-body Hamiltonians using superconducting processors~\cite{aleiner2020accurately,roushan2017spectroscopic,google2020hartree}. A particular method to determine the eigenergies of Hamiltonians via Fourier transforming the dynamics of observables has been applied in~\cite{aleiner2020accurately,roushan2017spectroscopic}. The idea is to prepare a Fock state that has overlap with the eigenstates whose eigenvalues one wants to calculate. The Fock state is then evolved in time with the Hamiltonian and specific observables are measured over a range of time. The Fourier transform of the time evolution of the observables can be used to deduce the eigenenergies of the Hamiltonian.

\paragraph{Witness-assisted variational eigenspectra solver (WAVES). }
WAVES core idea is to use a single reference qubit as an eigenstate witness to variationally find the ground state and excited states~\cite{santagati2018witnessing}. A variational ansatz applied to a reference state is chosen. Then, the time evolution operator $U(t)=\exp(-iHt)$ is evolved on the ansatz state as a control unitary $CU(t)$, with the control being the single qubit in a superposition state. Then, full tomography is performed on the single qubit to read out its von-Neumann entropy. If the variational state is an eigenstate of the Hamiltonian $H$, then the entropy is zero. Further, the energy of the state can be estimated from the state of the qubit as well. The ansatz is variationally updated using the information from the qubit in a iterative fashion until the ground state is found. Excited states can be found by applying an appropriate excitation operator on the found ground state, and then variationally minimizing the von-Neumann entropy of the qubit. As last step, the authors suggest to use the iterative phase estimation algorithm to further improve the accuracy of the excited state as well determine its eigenvalue. 
This method requires to implement a controlled time evolution operator, similar to non-variational proposals~\cite{jensen2020quantum}, which are considered to be challenging for larger systems on NISQ devices.

\subsubsection{Hamiltonian simulation}\label{sec:hamilton_simulation}

A major application for quantum computers is the simulation of the dynamics of Hamiltonians for problems such as many-body physics and chemistry. 
One standard approach for quantum simulation of Hamiltonians is based on the Trotter-Suzuki expansion from \eqref{eq:trotter}, where the evolving unitary is split up into small discrete timesteps of efficiently implementable unitaries, which can be run on the quantum computer. 
 Naturally, the depth of the quantum circuit increases polynomially with the desired time to be evolved and target accuracy, which may not be feasible on NISQ devices without access to error correction.
The relevant algorithms are reviewed in the following.
We remark that some necessary tools to simulate many-body interaction Hamiltonian ~\cite{menke2019automated,bravyi2008quantum} have also been proposed.

\paragraph{Variational quantum simulator. }
VQA have been proposed to solve dynamical problems in the NISQ era~\cite{li2017efficient}. 
The core idea is to iteratively update an efficiently implementable variational quantum state $\ket{\psi(\boldsymbol{\theta})}$ with a new set of parameters $\boldsymbol{\theta} \rightarrow\boldsymbol{\theta}'$ such that it minimizes the error between the actual time evolution $\exp(-iH\delta t)\ket{\psi(\boldsymbol{\theta})}$ for a timestep $\delta t$ and the updated variational state $\ket{\psi(\boldsymbol{\theta}')}$. 
The rules to update the parameters $\boldsymbol{\theta}$ to solve the Schr\"odinger equation $i\text{d}/\text{d}t \ket{\psi(t)}=H\ket{\psi(t)}$ can be found by the variational McLachlan’s principle $\delta\vert\vert (\text{d}/\text{d}t +iH)\ket{\psi(\boldsymbol{\theta})}\vert \vert =0$ with $\vert \vert\ket{\psi}\vert \vert=\sqrt{\braket{\psi\vert\psi}}$ and demanding that $\boldsymbol{\theta}$ remains real-valued. One finds a set of linear equations of motion $\boldsymbol{A}\dot{\boldsymbol{\theta}}=\boldsymbol{C}$ with
\begin{eqnarray}
    A_{i,j}&=&\text{Re}(\partial_{\theta_i}\bra{\psi(\boldsymbol{\theta})}\partial_{\theta_j}\ket{\psi(\boldsymbol{\theta})}),\nonumber\\
    C_{i}&=&\text{Im}(\partial_{\theta_i}\bra{\psi(\boldsymbol{\theta})}H\ket{\psi(\boldsymbol{\theta})})\,.
\end{eqnarray}

At a given step of the iteration, one needs to measure the elements of $\boldsymbol{A}$ and $\boldsymbol{C}$ using the Hadamard test or methods from \cite{mitarai2019methodology} (see \secref{sec:measurement}), and then update $\boldsymbol{\theta}$ with the solution of the linear equation of motion by a small timestep $\delta t$. The solver can be combined with adaptive strategies to reduce the complexity of the  Ansatz circuit~\cite{zhang2020low,yao2020adaptive}. 

VQS has been applied on the IBM quantum processor to simulate energy transfer in molecules~\cite{lee2021simulating} as well as to simulate a time-dependent Hamiltonian~\cite{lau2021quantum}.
A straightforward extension of the variational quantum simulator can be applied to solve the Schrödinger equation in imaginary time~\cite{mcardle2019variational}, for time-dependent problems~\cite{yuan2019theory} or for general linear differential equations~\cite{endo2020variational,kubo2020variational}.
Its implementation to open quantum systems~\cite{endo2020variational,yuan2019theory}is discussed in \secref{sec:OpenQuantumSystem}. Using the hardware-efficient structure of the PQC, it is possible to reduce the cost of measuring the $\boldsymbol{A}$ and $\boldsymbol{C}$ matrices~\cite{benedetti2020hardware}. 
Alternatively, the projected - Variational Quantum Dynamics method (p-VQD) has been proposed to bypass the measurement of aforementioned matrices~\cite{barison2021efficient,otten2019noise}. Here, one variationally maximizes the fidelity between the PQC $\ket{\psi(\boldsymbol{\theta})}$ and the Trotter evolved state $\exp(-iH\delta t)\ket{\psi(\boldsymbol{\theta}')}$. The optimized PQC yields the state evolved by a time $\delta t$. This algorithm is then repeated to gain evolution for longer times. By appropriately choosing the evolution time $\delta t$, barren plateaus can be avoided~\cite{haug2021optimal}.

\paragraph{Subspace variational quantum simulator. }
The subspace variational quantum simulator (SVQS) \cite{heya2019subspace} builds upon the idea of the SSVQE \cite{nakanishi2019subspace} introduced earlier in \secref{sec:var_excited}. The core idea is to rotate the initial state to be evolved onto the low-energy subspace found by the weighted SSVQE, then evolve it in time within the subspace, and then apply the reverse mapping.
First, run the weighted SSVQE by preparing $k$ initial states $\left\{\ket{\varphi_j}=\sigma_j^x\ket{0}\right\}_{j=0}^k$ which are orthogonal with each other ($\braket{\varphi_i|\varphi_j}=\delta_{i,j}$) and lie in the computational subspace, as well as a PQC $U(\boldsymbol{\theta})$. Now as in the weighted SSVQE minimize \eqref{eq:SSVQE_weighted}.

Then, prepare an initial state $\ket{\psi_{\mathrm{in}}}$ to be evolved, which is encoded into the computational subspace by applying the Hermitian conjugate of the obtained circuit $U^\dagger(\boldsymbol{\theta})$.
Here, the evolution of the state in time is performed by applying single-qubit rotations on each qubit $\mathcal{T}(t)=\bigotimes_{j} R_Z (-E_j t)$, where $\{E_j\}_{j}$ are the eigenenergies of the eigenstates $\{\ket{E_j}\}_{j}$ obtained by SSVQE earlier.
Finally, the state $\mathcal{T}(t)U^\dagger (\boldsymbol{\theta})\ket{\psi_{\mathrm{in}}}$ in the computational subspace is reverse mapped by applying $U(\boldsymbol{\theta})$, giving the evolved state
\begin{equation}
\ket{\psi(t)}=U(\boldsymbol{\theta})\mathcal{T}(t)U^\dagger (\boldsymbol{\theta})\ket{\psi_{\mathrm{in}}}\,.
\end{equation}
This method has the key advantage that since the evolution is directly implemented as simple rotations in the computational subspace, the circuit depth is independent of the evolution time to be simulated. However, the initial SSVQE optimization can be difficult, especially when one considers many eigenstates $k$.

\paragraph{Variational fast forwarding. }
Similar to the idea of the SVQS, variational fast forwarding (VFF) relies on the idea of evolving a quantum state in time $\exp(-iHt)$ within a diagonal subspace, such that an enhanced evolution time can be achieved
\cite{cirstoiu2020variational}. First, a circuit that implements a small timestep of the desired evolution is implemented as $V(\delta t)=\exp(-iH\delta t)$. Then, an approximate diagonal factorization of $V(\delta t)$ is trained for a particularly structured  variational circuit \begin{equation}
    U(\boldsymbol{\theta},\boldsymbol{\gamma},\delta t)=W(\boldsymbol{\theta})D(\boldsymbol{\gamma},\delta t)W^\dagger(\boldsymbol{\theta})\,.
\end{equation}
Here, $D(\boldsymbol{\gamma},\delta t)$ is composed of commuting unitaries and chosen to parameterize the eigenvalues of unitary $V(\delta t)$, whereas $W(\boldsymbol{\theta})$ represents its eigenvectors. Then, the evolution to an arbitrary time $T=N\delta t$, where $N$ is some integer, is found by fast forwarding with $U(\boldsymbol{\theta},\boldsymbol{\gamma},N\delta t)=W(\boldsymbol{\theta})D^N(\boldsymbol{\gamma},\delta t)W^\dagger(\boldsymbol{\theta})$. 
For the training of the variational Ansatz, the fidelity between $V(\delta t)$ and $U(\boldsymbol{\theta},\boldsymbol{\gamma},\delta t)$ is maximized by a quantum-classical feedback loop with a cost function that uses the local Hilbert-Schmidt test~\cite{khatri2019quantum}.
As alternative approach, it was proposed to diagonalize the Hamiltonian $H$ instead of the unitary $V(\delta t)$, and fast forward via $U(\boldsymbol{\theta},\boldsymbol{\gamma},T)=W(\boldsymbol{\theta})\exp(-iD(\boldsymbol{\gamma})T)W^\dagger(\boldsymbol{\theta})$ ~\cite{commeau2020variational}.
Fast-forwarding can also be performed without requiring to train via a feedback loop using the linear combination of states approach (\eqref{eq:IQAE_Ansatz})~\cite{lim2021fast}.

\paragraph{Quantum Assisted Simulator. }\label{sec:quantum_assisted_simulator}
The VQS algorithm employs a classical-quantum feedback loop to update
the parameters of the PQC. Until the
classical processor has calculated its output, the classical-quantum
feedback loop delays any use of the quantum device, slowing the algorithm on current cloud computing framework.
The VQS algorithm, as well as its VQE based variant, i.e. SVQS share similarities and most of
the concerns faced by VQE, such as the barren plateau issue (see \secref{sec:BP}). 
Further, the VQS algorithm requires controlled-unitaries, which make it difficult to realize for current-term devices. 
To tackle the issues faced by VQS, the quantum assisted simulator (QAS) was suggested recently~\cite{bharti2020quantum}. The QAS algorithm does not need any classical-quantum feedback loop, can be parallelized, evades the barren plateau problem by construction, supplies a systematic
approach to constructing the ansatz and does not require any complicated
unitaries.

The QAS algorithm shares its approach with IQAE (see \secref{sec:IQAE}). The ansatz is given as a linear combination of states $\vert\phi\left(\boldsymbol{\alpha}(t)\right)\rangle=\sum_{\vert\psi_{i}\rangle\in\mathbb{CS_{K}}}\alpha_{i}(t)\vert\psi_{i}\rangle$ (see \eqref{eq:IQAE_Ansatz}), with classical coefficients $\boldsymbol{\alpha}(t)$ for ansatz state $\vert\psi_{i}\rangle$, which can be systematicly constructed (see Definition \ref{def:cumulant_states}). The Hamiltonian $H$ is given as a linear combination of unitaries (see \eqref{eq:LCU_Ham}).
The QAS algorithm employs Dirac-Frenkel principle to obtain the following classical evolution equation for $\boldsymbol{\alpha}(t)$
\begin{equation}
\mathcal{E}\frac{\partial\boldsymbol{\alpha}(t)}{\partial t}=-\iota\mathcal{D}\boldsymbol{\alpha}(t).\label{eq:Dirac_Frenkel_QAS}
\end{equation}
Here, $\mathcal{E}_{i,j}=\langle\psi_{i}\vert\psi_{j}\rangle$ and
$\mathcal{D}_{i,j}=\sum_{k}\beta_{k}\langle\psi_{i}\vert U_{k}\vert\psi_{j}\rangle$ are overlap matrices that can be efficiently measured on a quantum computer, i.e. for $H$ given as combination of Pauli strings, the overlaps are measurement of Pauli strings.

Recently, QAS was run on the IBM quantum computer and showed superior performance compared to Trotter and VQS for a time-dependent Hamiltonian~\cite{lau2021quantum}. A novel Hamiltonian simulation algorithm based on truncated Taylor series was proposed recently~\cite{lau2021nisq}. The classical post-processing in the aforementioned algorithm corresponds to a QCQP. 

\subsubsection{Quantum information scrambling and thermalization}

Quantum information scrambling is a quantum phenomena occurring when initially local states become increasingly non-local with the time-evolution of the system. It can be analyzed by computing the so-called out-of-time-ordered correlation function (OTOC) and has strong implications in thermalization in closed quantum systems dynamics. Recent experiments have been carried out to study this phenomena in a few qubits trapped-ion devices and simulators \cite{landsman2019verified,joshi2020quantum}, and in a 53 superconducting qubit processor \cite{mi2021information}. The algorithms proposed are based on the well-known teleportation algorithm and use single and two-qubit gates to reproduce the the scrambling process. 

In the context of VQAs, a variation of the VQE algorithm has been proposed to obtain the thermal evolution of quantum systems \cite{verdon2019quantum}. The authors present the Quantum Hamiltonian-Based Models (QHBM), an extension of the VQA's PQC to mixed states instead of pure states. Within this approach, the QHBM are classically trained to learn a mixed state distribution as a function of the optimization parameters. A direct application of such a model is the Variational Quantum Thermalizer (VQT), an algorithm which goal is to prepare a fixed-temperature thermal state of a given Hamiltonian.

The limitations of using variational QML algorithms to learn a scrambling unitary have also been studied in
\cite{holmes2020barren}, where it is found trainability issues related with barren plateaus (see \secref{sec:BP}).

\subsubsection{Simulating open quantum systems}\label{sec:OpenQuantumSystem}

In the following, we deal with the physics of open quantum systems \cite{huh2014linear} which are well-described by the Lindblad master equation from \eqref{eq:ch4:Lindblad_master_eq}.
By sampling from a mixture of pure state trajectories evolved by a non-Hermitian Hamiltonian and random quantum jumps,
one recovers the Lindblad dynamics.

\paragraph{Trotter simulation of open systems. } NISQ quantum hardware can be used to directly simulate the dynamics of small-scale open systems by using ancillas combined with measurements in the spirit of the quantum jump method~\cite{koppenhofer2020quantum,hu2020quantum}. Here, the unitary part of the dynamics is implemented via a Suzuki-Trotter decomposition (see \secref{sec:PQC}). The non-unitary part of the dynamics that encodes the interaction with the external degrees of freedom is simulated by entangling the circuit with ancillas and subsequently measuring them.  For every time step of the dynamics a new set of ancilla qubits has to be provided. Current quantum computers based on superconducting circuits do not allow one to measure and re-use qubits, thus requiring a linear increase in the number qubits with every timestep. Further, in general the circuit depth scales polynomially with simulation time.

\paragraph{Generalized variational quantum simulator. } In Ref.~\cite{endo2020variational} the VQS algorithm is extended to simulate
the method of quantum jumps in a variational setting. They implemented
the algorithm for 2D Ising Hamiltonians for $6$ qubits and observed
a dissipation induced phase transition. In another work \cite{yuan2019theory},
VQS is extended to mixed states and simulate the Lindblad dynamics fully without the need of stochastic sampling.
The idea is to write the density matrix as $\rho=\rho\left(\boldsymbol{\theta}\left(t\right)\right)$
and simulate the evolution of $\rho$ via evolution of the parameters
$\boldsymbol{\theta}(t).$  One can re-express  \eqref{eq:ch4:Lindblad_master_eq} as 
$\frac{d}{dt}\rho=\sum_{i}g_{i}S_{i}\rho T_{i}^{\dagger}$,
where $S_{i}$ and $T_{i}$ are unitaries and $g_{i}$ are coefficients.
Using Dirac and Frenkel equation, the evolution of parameters is given
by
\begin{gather}
\sum_{j}M_{i,j}\dot{\theta_{j}}=V_{i},\label{eq:Frenkel_evolution_1}\\
M_{i,j}=\text{Tr}\left[\left(\partial_{i}\rho\left(\boldsymbol{\theta}(t)\right)\right)^{\dagger}\partial_{j}\rho\left(\boldsymbol{\theta}(t)\right)\right]\\
V_{i}=\text{Tr}\left[\left(\partial_{i}\rho\left(\boldsymbol{\theta}(t)\right)\right)^{\dagger}\sum_{j}g_{j}S_{j}\rho T_{j}^{\dagger}\right]\,.
\end{gather}
This method can also be extended to deep quantum neural network type Ansatzes~\cite{liu2020solving}.

These algorithms, however, suffer from the canonical drawbacks of the VQS algorithm, such as the requirement of feedback loop, trainability
issues and necessity of controlled unitaries.

\paragraph{Generalized quantum assisted simulators. }
Recently, the generalized quantum assisted simulator (GQAS)~\cite{haug2020generalized}
was proposed as extension of the quantum assisted simulator to tackle above issues (see \secref{sec:quantum_assisted_simulator}). Instead of using a density matrix, the GQAS
algorithm introduced the concept of ``hybrid density matrix''
\begin{equation}
\hat{\rho}=\sum_{k,l}\beta_{k,l}\vert\psi_{k}\rangle\langle\psi_{l}\vert\label{eq:Hybrid_density_matrix}
\end{equation}
for $\beta_{k,l}\in\mathbb{C}$ and $\vert\psi_{l}\rangle$ are chosen from the set of cumulative $K$ moment states (see Definition~\ref{def:cumulant_states}). A classical device stores the coefficients $\boldsymbol{\beta}$ and the quantum states correspond to some quantum register. A hybrid
density matrix is a valid density matrix, if $\text{Tr}\left(\hat{\rho}\right)=1$
and $\hat{\rho}\succcurlyeq0.$ Note that the normalization condition is
fulfilled when $\text{Tr}\left(\hat{\rho}\right)=\text{Tr}\left(\boldsymbol{\beta}\mathcal{E}\right)=1,$
where $\mathcal{E}_{k,l}=\langle\psi_{k}\vert\psi_{l}\rangle.$ Using
Dirac-Frenkel principle, the simulation of open system dynamics for
the hybrid density matrix is given by
\begin{multline}
\mathcal{E}\frac{\text{d}}{\text{d}t}\boldsymbol{\beta}(t)\mathcal{E}=-\iota(\mathcal{D}\boldsymbol{\beta}(t)\mathcal{E}-\mathcal{E}\boldsymbol{\beta}(t)\mathcal{D})+\\
\sum_{n=1}^{K}\gamma_{n}(\mathcal{R}_{n}\boldsymbol{\beta}(t)\mathcal{R}_{n}^{\dagger}-\frac{1}{2}\mathcal{F}_{n}\boldsymbol{\beta}(t)\mathcal{E}-\frac{1}{2}\mathcal{E}\boldsymbol{\beta}(t)\mathcal{F}_{n}),\label{eq:GQAS_Dirac_Frenkel}
\end{multline}
where $\mathcal{D}_{k,l}=\langle\psi_{k}\vert H\vert\psi_{l}\rangle$, $\mathcal{R}_{k,l}^{n}=\langle\psi_{k}\vert L_{n}\vert\psi_{l}\rangle$ and $\mathcal{F}_{k,l}^{n}=\langle\psi_{k}\vert L_n^\dagger L_n\vert\psi_{l}\rangle$. 
For a given choice of ansatz, the quantum computers only have to compute the overlap matrices as measurements of Pauli strings. Then, the classical
computer uses this information to simulate the dynamics. There is
no quantum-classical feedback loop, which on the currently available quantum computers can speed up computations substantially.

\subsubsection{Nonequilibrium steady state}\label{sec:nonequilibrium}
Unlike the previous \secref{sec:OpenQuantumSystem}, we concern the physics of open quantum system that is out-of-equilibrium in nature, which is common in designing devices for molecular-scale electronics \cite{xiang2016molecular}, excitonic transport \cite{kyaw2017parity} as well as quantum thermodynamics \cite{vinjanampathy2016quantum}.
By ``out-of-equilibrium'', we mean that a quantum system and bath/s are constantly driven by external forces such as voltage differences, during which the composite particles of the system and bath are also interacting each other.

Notice that the method used in the previous \secref{sec:OpenQuantumSystem} would also lead to extremely high dimensional matrices in the Lindblad like master equation approach $d{\hat{\rho}}/dt=\doublehat{\mathcal{L}}\hat{\rho}$ (see \secref{ch4:subsubsec:ZNE}, \eqref{eq:ch4:Lindblad_master_eq}), and it deems impossible to capture all the degrees-of-freedom involved.
However, one may relax some of the constraints involved in the problem setup, say time-independent dissipation and non-interaction among particles with small system size. 
The steady state density matrix of a quantum system $\hat{\rho}_{\textrm{SS}}$ at the limit $t\rightarrow\infty$ is then given by solving 
\begin{equation}
    \doublehat{\mathcal{L}}\ket{\hat{\rho}_{\textrm{SS}}}=0,
\end{equation}
or equivalently $\doublehat{\mathcal{L}}^\dagger\doublehat{\mathcal{L}}\ket{\hat{\rho}_{\textrm{SS}}}=0$.
A recent study \cite{yoshioka2020variational} has shown that with ancilla qubits, the above non-Hermitian superoperator $\doublehat{\mathcal{L}}$ can be simulated. The main idea is to map the density matrix of $N$ qubits onto a vector of twice the number of qubits $2N$
\begin{equation}
\hat{\rho}=\sum_{ij}\rho_{ij}\ket{i}\bra{j}\rightarrow\ket{\hat{\rho}}=\sum_{ij}\frac{\rho_{ij}}{C}\ket{i}_P\ket{j}_A,
\end{equation}
where $C=\sqrt{\sum_{ij}\vert\rho_{ij}\vert^2}$. By using a digital quantum computer and the variational approach one iteratively minimize the expectation value of a parameterized density matrix $\ket{\rho_{\boldsymbol{\theta}}}=U(\boldsymbol{\theta})$ with $\text{min}_{\boldsymbol{\theta}} \bra{0}^{\otimes 2N}U^\dagger(\boldsymbol{\theta})\doublehat{\mathcal{L}}^\dagger\doublehat{\mathcal{L}}U(\boldsymbol{\theta})\ket{0}^{\otimes 2N}$. A drawback of this approach is that measuring expectation values from the parameterized density matrix directly is difficult and thus requires an additional transformation.

Beyond the Lindblad master equation, to capture and describe truly ``out-of-equilibrium'' processes, the nonequilibrium Green's function (NEGF) formalism \cite{stefanucci2013nonequilibrium,dalla2013keldysh,sieberer2016keldysh} is commonly used. 
These existing Green's function techniques are very complicated to be solved.
Many assumptions need to be made in order to have some closed form and do some calculations.
In particular, it requires that the interaction among particles are weak such that one does not need to find higher-order Feynman diagrams in finding the self-energy functional.

Since some of the existing quantum algorithms provide promising speedup over classical ones, one may wonder to use quantum algorithms to solve the NEGF, with a strategy of leaving classically hard computational tasks to the quantum processor and feeding its output back to classical computer, which could be done in a variational fashion. 
There exists a number of proposals \cite{kreula2016non,jaderberg2020minimum,endo2020calculation} in the literature that undertake such hybrid quantum-classical approach.
However, these methods assume no interaction among composite particles.
In generic open quantum system in which many-body effects cannot be neglected, one would like to go beyond those assumptions.
It is yet to see any quantum advantage of those near-term quantum algorithms over existing methods \cite{hartle2008multimode,li2016resummation,fitzpatrick2017observation} for solving nonequilibrium steady state solution of an extremely complex physical setup such as vibrationally-coupled electron transport with multiple electronic levels \cite{hartle2008multimode}.

\subsubsection{Gibbs state preparation}\label{sec:gibbsstate}
Finding the ground state of quantum Hamiltonians is known to be QMA-hard.
Under reasonable assumptions, preparing Gibbs state corresponding
to arbitrarily small temperatures is as challenging as the Hamiltonian
ground state problem. Gibbs state preparation has applications in
many areas including quantum annealing, quantum SDP solvers, Boltzmann
training and simulation of equilibrium physics. For a Hamiltonian
$H,$ the Gibbs state at temperature $T$ (with $k_\text{B}=1$) is given by
\begin{equation}
\rho(T)=\frac{\exp\left(-\frac{H}{T}\right)}{\text{Tr}\left(\exp\left(-\frac{H}{T}\right)\right)}.\label{eq:Gibbs_state}
\end{equation}
Some of the approaches to prepare Gibbs state are mentioned in the following
\begin{enumerate}
\item Starting with $d$-dimensional maximally mixed state $\frac{I_{d}}{d}$,
under imaginary time evolution for time $\tau,$ one gets Gibbs state
corresponding to temperature $T=\frac{1}{2\tau}$~\cite{verstraete2004matrix}.
\item One can start with maximally entangled state $\vert\xi\rangle_{d}=\frac{1}{\sqrt{d}}\sum_{j}\vert j,j\rangle_{AB}$
of a system combined of two equally sized subsystems A and B, and evolve it under imaginary time evolution using Hamiltonian
$H\otimes I.$ After tracing out system $B$, the state of system $A$ at time $\tau$ is given
by Gibbs state corresponding to temperature $T=\frac{1}{2\tau}.$
\item The Gibbs state of a system is the density matrix which corresponds
to minimum of its free energy. Thus, one can variationally tune the
parameters of a parametrized density matrix such that it leads to
minimization of free energy.
\end{enumerate}
Recently, a few NISQ algorithms for Gibbs state preparation have been
proposed, which apply the aforementioned ideas. In \cite{yuan2019theory}, authors
used VQS based imaginary time evolution to prepare Gibbs state following
the second approach. The first approach does not work in VQS based
imaginary time evolution. In another work \cite{chowdhury2020variational}, the third approach was used to
prepare Gibbs states. The aforementioned works require complicated
controlled unitaries and classical-quantum feedback loop. 
In \cite{haug2020generalized},
QAS based imaginary time evolution (see \secref{sec:quantum_assisted_simulator}) was suggested to prepare the Gibbs
state with either first and second
approach. The QAS approach does not require any
classical quantum feedback loop or complicated controlled
unitaries. Using random circuits as initial state, ~\cite{richter2020simulating} suggested an approach based on imaginary time evolution prepare Gibbs state.

\subsubsection{Simulation of topological phases and phase transitions}
NISQ devices can be used to study the ground states of quantum Hamiltonians for understanding topological phases and phase transitions. An important example is  the one-dimensional cluster Ising Hamiltonian, describing a symmetry-protected topological phase of matter. The ground state of this Hamiltonian is the one-dimensional cluster state, which can be created by applying Hadamard gates to all qubits, followed by control-Z gates on each pair of neighboring qubits. State tomography and symmetry arguments were used to study the entanglement measures of this state and to highlight its topological nature~\cite{choo2018measurement,azses2020identification}. A modified algorithm was implemented to simulate an enlarged family of Hamiltonians and study the quantum phase transition between a topological and a topologically-trivial phases of matter~\cite{smith2019crossing}. NISQ devices were also used to simulate the dynamics of fundamental models of quantum magnetism~\cite{bassman2020towards,smith2019simulating} and topological phases in one and two dimensions~\cite{mei2020digital}.

\subsubsection{Many-body ground state preparation}

The preparation of non-trivial many-body quantum states is crucial for many applications in quantum metrology and quantum information processing \cite{kyaw2014measurement}. QAOA has been used as a resource-efficient scheme for many-body quantum state preparation. In this context, the state $\ket{\psi}$ for a system with linear dimension $L$ (e.g. $L$ can refer to the number of spins in a 1D spin chain) is non-trivial if there is no local unitary circuit $U$ with depth $\mathcal{O}(1)$ which can generate $\ket{\psi}$ from a product state $\ket{\phi}$: $\ket{\psi} = U \ket{\phi}$ \cite{ho2019efficient}. The Greenberger-Horne-Zeilinger (GHZ) state, which is an essential resource in several quantum metrology proposals \cite{dur2014improved,geza2014quantum}, is an example of a non-trivial quantum state due to its highly-entangled nature, and is the ground state of the 1D Ising Hamiltonian with periodic boundary conditions, i.e. $H_P = - \sum_{i=1}^{L} \hat{\sigma}_{z}^{i} \hat{\sigma}_{z}^{i+1}$

Using QAOA, it has been shown that the GHZ state can be prepared efficiently with perfect fidelity using $p = L/2$, where $p$ is the QAOA depth \cite{ho2019efficient}. The authors conjectured that the ground state of the 1D transverse-field Ising model

with $L$ even and periodic boundary conditions, can be prepared perfectly at any point in the phase diagram using QAOA with $p = L/2$. The ground state of the antiferromagnetic Heisenberg model with open boundary conditions $H_P = \sum_{i=1}^{L-1} \hat{\boldsymbol{\sigma}}^{i} \cdot \hat{\boldsymbol{\sigma}}^{i+1}$, 

where $\hat{\boldsymbol{\sigma}}^{i} \equiv (\hat{\sigma}_{x}^{i},\hat{\sigma}_{y}^{i},\hat{\sigma}_{z}^{i})$, has also been prepared with near perfect fidelity using QAOA. Using a long-range 1D Ising Hamiltonian $H_P = -\sum_{i<j} J_{ij} \hat{\sigma_{z}}^{i} \hat{\sigma}_{z}^{j}$, 

where $J_{ij} = J_0 / |i-j|^{\alpha}$, QAOA can achieve the ultrafast preparation of a GHZ state with a circuit depth of $\mathcal{O}(1)$ (for $\alpha = 0$) \cite{ho2019ultrafast}. This result was generalized by \cite{wauters2020polynomial}, which showed that QAOA can prepare the ground states of the fully-connected ferromagnetic $q$-spin model (note that $q$ is used here instead of the conventional $p$ in order to avoid confusion with the QAOA depth $p$)
\begin{equation}
    H = -\frac{1}{N^{q-1}} \bigg( \sum_{i=1}^{N} \hat{\sigma}_{z}^{i} \bigg)^{q} - h \bigg( \sum_{i=1}^N \hat{\sigma}_{x}^{i} \bigg)
\end{equation}
with resources scaling polynomially with the number of spins $N$. Since the system can encounter a first-order phase transition where the spectral gap becomes very small, QAOA greatly outperforms quantum annealing in this instance since an exponentially long annealing time is needed.

\subsubsection{Quantum autoencoder}

The quantum autoencoder~\cite{romero2017quantum} (QAE)  is a VQA for the compression of data on a quantum computer. It finds a new data state representation which requires fewer qubits than the data was originally defined upon. This new encoding is said to be a representation in the \textit{latent space}. The process of transforming the data into the latent space is referred to as \textit{encoding}, and the converse, transformation of states in the latent space back onto the original, is known as \textit{decoding}. 

Training a QAE requires the minimization of an objective defined over several related quantum states. For a set of $n$-qubit states $\{ \ket{\psi}_{i}\}$, the goal of the QAE is to find a unitary circuit $\mathcal{E}(\boldsymbol{\theta})$ which accomplishes the following transformation
\begin{equation}
    E:  \mathcal{H}^{n} \rightarrow \mathcal{H}^{k}\otimes\mathcal{H}^{n-k}| E\ket{\psi}_i = \ket{\phi}_i \otimes \ket{0}^{\otimes (n-k)},
\end{equation}
where $k$ is the dimension of the latent space. 
Thus, the application of a perfectly trained autoencoder to any state of the relevant set yields a product state that consists of the transformed state on $k$ qubits with a $(n-k)$-qubit ``trash'' state. In principle, the trash state could be any state, but the all-zero state is chosen for simplicity.

The loss function of the QAE may be defined in several ways. It is a fidelity loss function (see \secref{sec:objective_function}), in which minimization is performed by increasing the overlap between a (partial) measurement of the state resulting from the application of the encoder and a known state. 
The most practical definition for training the autoencoder, called ``trash training'', uses as its objective the overlap between the ``trash'' qubits and the $\ket{0}^{\otimes(n-k)}$ state. Formulated in the density matrix picture, the objective of minimization is
\begin{equation}
O = - \text{Tr}(I^{\otimes k}\otimes \ketbra{0}{0}^{\otimes (n-k)} \rho_i),
\end{equation}
where $\rho_i = \sum_i p_i \ketbra{\psi_{i}}{\psi_{i}}$ with, in general, all states in the set equally weighted. 

The QAE can be trained by training only the encoding circuit, due to the unitarity of the encoder; the decoding operation is achieved by the complex conjugate of the encoder circuit. Improvements in the encoding results translates to improvements in the decoder, a boon not possessed by classical autoencoders.

After the successful training of a QAE, the encoder and decoder circuits may be used for data transformation in further algorithms, action upon the data in the latent space, in which the data is represented more densely, may prove powerful in further applications.

A data re-uploading strategy to construct a QAE encoder is presented in \cite{bravo2020quantum2}, where the so-called enhanced feature quantum autoencoder (EF-QAE) is trained to compress the ground state of the 1-D Ising model as a function of the external field and samples of handwritten digits. The QAE has also been deployed experimentally in the compression of qutrits on a photonic device~\cite{pepper2019experimental}. Small states have been experimentally compressed losslessly on photonic devices by \cite{huang2020realization}. \cite{bondarenko2020quantum} designed a QAE capable of denoising entangled quantum states, such as GHZ or W states, subject to spin flip errors and random unitary noise. 

\subsubsection{Quantum computer-aided design}\label{sec:quantum_aided_design}

Two recent proposals focus the computing power of NISQ devices back on the processors themselves:
Techniques were developed to simulate quantum hardware on a quantum computer \cite{kottmann2020quantum, kyaw2020quantum}.
They establish the paradigm of ``quantum computer-aided design'', indicating that classically intractable simulations of quantum hardware properties can be performed on a quantum computer, thereby improving the prediction of device performance and reducing experimental testing cycles.

In the first approach, optical path modes are mapped to sets of qubits, and quantum optical elements are mapped to digital quantum circuits that act on the qubits \cite{kottmann2020quantum}.
Photonic setups can then be flexibly simulated.
The framework is used to simulate both a Boson sampling experiment and the optimization of a setup to prepare a high-dimensional multipartite entangled state.

The second proposal introduces quantum simulation techniques for superconducting circuit hardware \cite{kyaw2020quantum}.
A circuit module consisting of coupled transmon qubits is designed.
The corresponding superconducting circuit Hamiltonian, which is written in a basis of multi-level operators, is efficiently mapped to a set of data qubits \cite{sawaya2020resource}.
Simulations of a multi-level extension to the VQE algorithm \cite{higgott2019variational} are used to determine the spectrum of the superconducting circuit.
The resulting states and eigenenergies are directly related to experimentally relevant device characteristics and can be used to seed simulations of time dynamics.

Device and setup design is a key challenge in improving and scaling quantum systems. 
Therefore, digital quantum simulation of quantum processors will be a relevant application for NISQ quantum computers as classical resources become too small to capture the relevant Hilbert space of the hardware.

\subsection{Machine learning} \label{sec:QML}

The goal of machine learning is to facilitate a computer to act without being explicitly programmed to. As per Tom Mitchell~\cite{mitchell1997machine}, given some class of tasks $\mathcal{T}$ and performance metric $\mathcal{P}$, a computer program is said to learn from experience $\mathcal{E}$ if
\begin{equation}
 \mathcal{P}(\mathcal{T})\propto\mathcal{E},
\end{equation}
i.e. its performance measured by $\mathcal{P}$ for task \textbf{$\mathcal{T}$} increases with $\mathcal{E}$. 

Depending on the kind of experience $\mathcal{E}$ permitted to have during the learning process, the machine learning algorithms are classified into three categories:
\begin{enumerate}
\item \textit{Supervised learning.} Given a function $y=f(x),$ the goal 
is to learn $f$ so that it returns the label $y$ for the unlabelled data $x.$ A canonical example would be pictures of cats and monkeys, with the task to recognize the correct animal. Given training examples from the joint distribution $P(Y, X)$, the task of supervised learning is to infer the probability of a label $y$ given
example data, $x$, i.e., $P\left(Y=y\vert X=x\right)$.
\item \textit{Unsupervised learning.} 
The data is provided without any label. The task is to recognize an underlying pattern in this data. Given access to several examples $x\in X$ 
the algorithm goal is to learn the probability distribution $P(X)$ or some important properties of the aforementioned distribution.
\item \textit{Reinforcement learning.} In this case, neither data nor label is provided. The machine has to generate data and improve the aforementioned data generation process via optimizing a given reward function. This is similar to how a human child learns to walk. If it fails, the output acts as a negative reward.  
\end{enumerate}

Machine learning has uncovered applications in physics such as 
Monte Carlo Simulation~\cite{huang2017accelerated,liu2017self}, many-body physics~\cite{carleo2017solving},  
phase transition~\cite{wang2016discovering}, 
quantum foundations~\cite{bharti2019teach}, 
and state tomography~\cite{torlai2018neural}.

For a meticulous
review on machine learning for physics, refer to~\cite{carleo2019machine, dunjko2018machine,bharti2020machine}.

Most of the success in machine learning come from the use of artificial neural networks, structures capable of learning sophisticated distributions and that encompass multiple features that can be fine-tuned depending on the problem to tackle. In that direction, there are several proposals to define a model for quantum neural networks with different kind of activation functions~\cite{schuld2014quest,wan2017quantum,torrontegui2019unitary}.  For  implementations of artificial neuron and artificial neural network on the NISQ hardware, refer to~\cite{tacchino2019artificial,tacchino2020quantum}. 

The merger of quantum theory and machine learning has recently led birth to a new discipline, known as quantum machine learning (QML). Both algorithms that deal classically with data from a quantum origin and quantum algorithms that process quantum and classical data are usually known as QML applications. However, in this review, we will focus only on those algorithms that process data quantum-mechanically, in particular, those that use quantum algorithms that can be run in NISQ computers. For QML review that mainly focus on fault-tolerant quantum algorithms check~\cite{biamonte2017quantum}. For a survey of quantum computational learning theory, refer to~\cite{arunachalam2017survey}. An analysis of QML from a classical ML perspective can be found at \cite{ciliberto2018quantum,dunjko2018machine}, and for near-term devices in \cite{perdomo2018opportunities,benedetti2019parameterized,li2021recent}.

It might be surprising that a linear theory as quantum physics can generate the non-linearities that a machine learning model needs. However, the linearity of quantum mechanics comes from the dynamical part (quantum states evolution) and one can encounter multiple sources of non-linearities
arising from measurement, post-selection or coupling the system with environment. Quantum operations in the Hilbert space can also encode non-linear behaviour, as it will be shown with Kernel methods.  
 
In the following subsections, we will present the quantum mechanical analogs of the three machine learning categories defined above. The algorithms discussed will be listed in \tabref{tab:ml_algorithms}.

\subsubsection{Supervised learning} \label{sec:SL}

The two prominent methods to perform a supervised learning classification task using a NISQ computer are  \textit{quantum Kernel estimation} \cite{schuld2019quantum,havlivcek2019supervised, kusumoto2019experimental,huang2021power} and \textit{Variational Quantum Classifier} (VQC)~\cite{farhi2018classification, mitarai2018quantum}.

Classical Kernel methods include well-known machine learning algorithms such as Support Vector Machines (SVM) \cite{cortes1995support}, Principal Component Analysis (PCA) or Gaussian Processes, among others. The rich theoretical structure of Kernel methods can be expanded to the quantum world by defining and working in the Hilbert space with the quantum equivalent of feature vectors \cite{schuld2019quantum}. To that aim, one needs to modify and adapt the well-known theorems to work in a quantum feature space. For more details about classical Kernel methods we refer to \cite{hofmann2008kernel}. A review on Kernel methods in the context of QML can be found in \cite{mengoni2019kernel}. In the following lines, we will directly describe the quantum versions of them. The basics of supervise learning with quantum computers are presented in~\cite{schuld2018supervised}.

Given an input set $\mathcal{X}$ and quantum Hilbert space $\mathcal{H}$, data $\boldsymbol{x}\in\mathcal{X}$ is encoded into a quantum state (\textit{quantum feature vector}) $|\Phi(\boldsymbol{x})\rangle$ by means of the \textit{quantum feature map}, i.e.  $\Phi:\mathcal{X}\rightarrow\mathcal{H}$. The inner product of two quantum feature vectors defines a kernel
\begin{equation}
\kappa\left(\boldsymbol{x}_{i},\boldsymbol{x}_{j}\right)\equiv\left\langle \Phi\left(\boldsymbol{x}_{i}\right)|\Phi\left(\boldsymbol{x}_{j}\right)\right\rangle _{\mathcal{H}},
\end{equation}
for $\boldsymbol{x}_{i},\boldsymbol{x}_{j}\in\mathcal{X}$. In comparison with classical kernels, the inner product is defined in a Hilbert space by replacing the standard definition $\langle\cdot,\cdot\rangle$ by the Dirac brackets $\langle \cdot |\cdot\rangle$. For a map $\Phi$, the \textit{reproducing kernel Hilbert space} takes the form
\begin{equation}
\mathcal{R}_{\phi}=\left\{ f:\mathcal{X}\rightarrow\mathbb{C}\vert \ f(\boldsymbol{x})=\left\langle w|\Phi(\boldsymbol{x})\right\rangle _{\mathcal{H}},\forall \boldsymbol{x}\in\mathcal{X},|w\rangle\in\mathcal{H}\right\} \label{eq:RKHS}.
\end{equation}
The orthogonality of $|w\rangle$ w.r.t. $|\Phi(\boldsymbol{x})\rangle$ defines a decision boundary, i.e. depending on the sign of the inner product, $\boldsymbol{x}$ lies in one side of the hyperplane. The function $f$ is thus a linear function in $\mathcal{H}$. The \textit{representer theorem} \cite{Scholkopf2001generalized} states that this function can be approximated by the linear function $f^{\star}$ by using the kernel defined above, i.e.
\begin{equation}
f^{\star}(\boldsymbol{x})=\sum_{i=1}^{\mathcal{D}}\alpha_{i}\kappa\left(\boldsymbol{x},\boldsymbol{x}_{i}\right)
\label{eq:representer_theorem}
\end{equation}
for an input dataset $\mathcal{D}$. Using \eqref{eq:representer_theorem}, one can solve a convex optimization problem to get the coefficients $\alpha_{i}$. The analysis so far entails the connection between linear models in reproducing kernel Hilbert space with kernelized models in the input space. 

One can use a quantum computer to calculate the inner product of feature mapped quantum states to obtain the kernel $\kappa$. This kernel can be fed to a classical device, which can use \eqref{eq:representer_theorem} to obtain the coefficients $\alpha_{i}$, for instance, by maximizing a cost function of the form \cite{havlivcek2019supervised}
\begin{equation}
    C(\boldsymbol{\alpha}) =  \sum_{i=1}^{\mathcal{D}}\alpha_{i}- \frac{1}{2} \sum_{i,j}^{\mathcal{D}}y_{i}y_{j}\alpha_{i}\alpha_{j}\kappa\left(\boldsymbol{x}_{i},\boldsymbol{x}_{j}\right),
\end{equation}
where $y_{i}$ are the labels of the training points and constrained to $\sum_{i=1}^{\mathcal{D}}\alpha_{i}y_{i}=0$. Ideas based on connections between kernel methods and quantum circuit based machine learning has been used to justify that the models for QML can be framed as kernel methods~\cite{schuld2021quantum}. For some of the other relevant works on quantum kernel methods, refer to~\cite{park2020theory,blank2020quantum}. A high-dimensional data classification experiment with quantum kernel methods was carried out recently~\cite{peters2021machine}.
The encoding of data into quantum circuits is characterized by the quantum Fisher information metric~\cite{haug2021largescale}. For hardware efficient PQCs, the kernel can be related to radial basis function kernels~\cite{haug2021optimal,haug2021largescale}. Measuring the quantum kernel using the SWAP or inversion test scales as $\mathcal{D}^2$. Using randomized measurements~\cite{elben2020cross}, the kernel can be computed in a time that scales linearly with the dataset size $\mathcal{D}$, which allows for processing large datasets with quantum computers~\cite{haug2021largescale}.

Also, a Gaussian Boson Sampling device (see \secref{sec:GBS}) can be used for computing kernel functions \cite{schuld2020measuring}.

Another approach is to use a variational circuit $U(\boldsymbol{\theta})$ and directly perform the classification task in the reproducing kernel Hilbert space, without using \eqref{eq:representer_theorem}. This approach is sometimes referred as a variational quantum classification. Data is also embedded into the state $|\Phi(\boldsymbol{x})\rangle$ and then processed with a PQC $U(\boldsymbol{\theta})$. The resultant state becomes
\begin{equation}
|\Psi(\boldsymbol{x},\boldsymbol{\theta})\rangle=U(\boldsymbol{\theta})|\Phi(\boldsymbol{x})\rangle,
\label{eq:VQC_state}
\end{equation}
which parameters are estimated by training it to match the target states $|y_{i}\rangle$ that represent the $y_{i}$ labels of the training points, i.e. by minimizing the infidelity
\begin{equation}
C(\boldsymbol{\theta}) = \sum_{i=1}^{\mathcal{D}}\left(1-|\langle y_{i}|\Psi(\boldsymbol{x}_{i},\boldsymbol{\theta})\rangle|^2\right).
\label{eq:VQE_cost}
\end{equation}

Both methods require a way to encode the data into a quantum state. There are several strategies to define the quantum feature map. It is a key step in the success of the classification task, as the needed non-linearities must come from it. Furthermore, to eventually obtain any quantum advantage, one should search from the set of classically intractable feature maps. One of the first proposed approaches was the amplitude encoding ~\cite{schuld2016prediction} also required in other quantum algorithms \cite{harrow2009quantum}. This approach encodes the classical data points into the amplitudes of a quantum state, i.e. $|\Phi(\boldsymbol{x})\rangle=\sum_{i}x_{i}|e_{i}\rangle$, where $|e_{i}\rangle$ are the basis states. However, this raw encoding requires \textit{i)} knowing which gates can be used to perform this operation for general data points and \textit{ii)} having an efficient way to extract and process these amplitudes. Although the first point can eventually be overcome by using similar approaches as the ones used to define a PQC, the second one requires tools as QRAM~\cite{Giovannetti2008quantum}, experimentally challenging for the NISQ era. The studies towards QRAM in~\cite{park2019circuit} proposed an approach to update classical data consisting of $M$ entries of $n$ bits each using $O(n)$ qubits and $O(Mn)$ steps. A \textit{forking-based} sampling scheme was suggested in~\cite{park2019parallel} to reduce the resource requirements for state preparation for tasks involving repeated state preparation and sampling. At the moment of writing, building a QRAM remains challenging and further investigations are required.

In general, the encoding strategies used in state-of-the-art algorithms consist on introducing the classical data points into the parameters of the quantum circuit gates. As briefly mentioned in \secref{sec:PQC}, one designs a state preparation circuit $E$ that encodes the data points,
\begin{equation}
|\Phi(\boldsymbol{x})\rangle = E(\boldsymbol{x},\boldsymbol{\phi})|0\rangle.
\label{eq:encoding}
\end{equation}
The use of $\boldsymbol{\phi}$ parameters is optional and they can be subject to the optimization subroutine too. 

Typically, the encoding gate is designed using the same structure of a layer-wise PQC from \eqref{eq:HE_ansatz}. Data points are introduced in layers of single-qubit rotational gates $R$, as defined in \eqref{eq:def_single_qubit_rotation}, followed by an entangling gate unitary $W$, e.g.
\begin{equation}
 E(\boldsymbol{x})=\prod_{k=1}^{L_{E}}\left(\bigotimes_{i=1}^{n}R_{k}(\boldsymbol{x}_{i})\right) W_{k},
\end{equation}
with $L_{E}$ being the total number of encoding layers. Then, the whole Variational Quantum Classifier (VQC) is composed by this encoding circuit and the processing one to be optimized, i.e. $U_{VQC}(\boldsymbol{\theta},\boldsymbol{x})= E(\boldsymbol{x}) U(\boldsymbol{\theta})$.

Alternatively, some works propose to remove the distinction between the encoding $E$ and processing $U$ circuits and introduce the data values along the circuit \cite{vidal2019input,perez2020data,lloyd2020quantum,schuld2020effect,nghiem2020unified}. This strategy, sometimes called \textit{input-redundancy} or \textit{data-reuploading}, introduce the data in all circuit layers, e.g. 
\begin{equation}
 U_{VQC}(\boldsymbol{\theta},\boldsymbol{x})=\prod_{k=1}^{L}\left(\bigotimes_{i=1}^{n}R_{k}(\boldsymbol{x}_{i},\boldsymbol{\theta})\right) W_{k},
\end{equation}
where $L$ is now the total number of circuit layers. This strategy has proved the universality when applied to one qubit \cite{perez2020data} and can reconstruct the coefficients of the Fourier series \cite{vidal2019input,schuld2020effect}.

The inclusion of encoding strategies and, in particular, the data re-uploading, can help well-known VQA such as the VQE. In general, one of the final goals of a VQE can be the identification of interesting points on a potential energy surface generated by a parametrized Hamiltonian. Often, one is interested in the ground state energy as a function of some Hamiltonian parameter $\lambda$, e.g. the interatomic distance, but other properties, like the energy gap between ground state and first excited state can be interesting as well.~\cite{kyaw2020quantum} To do so, one often needs to scan discretely over $\lambda$ for some particular interval and run a VQE to obtain the ground state energy on each of these points. This becomes an extra computational cost, especially if we are interested only in a particular region of this ground state profile, e.g. to extract the $\lambda_{min}$ whose ground state has minimal energy. In that direction, some proposals suggest to encode the parameters of the Hamiltonian into the PQC and learn the energy profiles \cite{mitarai2019generalization}. In particular, the \textit{Meta-VQE} algorithm \cite{cervera2020meta} proposes to encode the $\lambda$ into the PQC gates together with the optimization parameters. Then optimize an objective function that corresponds with the sum of expectation values for some $M$ training $\lambda$ parameters, i.e. $\expval{\hat{O}}=\sum_{i=1}^{M}
\langle H\rangle_{U (\boldsymbol{\theta}, \lambda_{i})}$. Once the circuit has been optimized, one can run it again with the new $\lambda_{i}$ to directly extract an estimation of the ground state, without having to optimize the full circuit again.
An extension of this approach is the \textit{optimized Meta-VQE} (\textit{opt-meta-VQE}), which consist of using the optimized parameters from the Meta-VQE as starting points of a standard VQE. This tries to avoid vanishing gradients issues (discussed in \secref{sec:BP}) by starting in a particular region of the parameter space instead of random initialization. 

Some VQC also need an extra piece, the definition of the target state $|y_{i}\rangle$ to construct the objective function to be optimized using the fidelity with respect to these states. The goal of the quantum circuit is to divide and push the quantum states that encode the data points into two or more regions of the Hilbert space. To that aim, the parameters of the circuit are trained to match every encoded state into a particular representative of one of these regions. Therefore, the more separated these regions are, the lesser misclassified points are expected. As discussed in \secref{sec:measurement}, qubits measurement implies a certain computational cost. For that reason, many proposals suggest to use the state of only one qubit to train the whole circuit \cite{farhi2018classification,schuld2020circuit}. The cost function estimation reduces to measuring the probability distribution of one qubit. Other works use a more sophisticated definition of these target states by selecting the most orthogonal states of the qubits space \cite{perez2020data,lloyd2020quantum}. This strategy is inspired from optimal state discrimination \cite{helstrom1969quantum}.

Using the nonlinear character of quantum mechanical processes as ``reservoir'',
the notion of \textit{quantum reservoir computing} has been suggested. The reservoir
is a highly nonlinear system whose parameters are arbitrary but fixed.
One can perform reservoir computing by employing a basic training algorithm such as linear regression
at the readout stage. Since the reservoir
parameters are fixed, only training of the readout stage parameters
is required. The aforementioned idea helps utilize the high nonlinearities
of the reservoir without the high computational cost
of training. The concept of employing quantum systems as quantum reservoirs
was first introduced in ~\cite{fujii2017harnessing,nakajima2019boosting}. Quantum reservoir computing has been
proposed for many experimental platforms such as Gaussian states in
the optical set-up~\cite{nokkala2020gaussian}, two-dimensional fermionic lattices~\cite{ghosh2019quantum} and nuclear
spins~\cite{negoro2018machine}. Quantum gate based implementation of quantum reservoir computing
for NISQ devices has also been discussed~\cite{chen2020temporal}. A Gaussian Boson Sampler  (see \secref{sec:GBS}) can also be used for quantum reservoir computing as suggested in ~\cite{wright2020capacity} to perform machine learning tasks such as classification. NISQ devices have also been used for regression~\cite{mitarai2018quantum}.  Distance-based classifier using quantum interference circuits has been proposed in  ~\cite{schuld2017implementing}.

Quantum annealing has been also applied  to supervised learning to predict biological data~\cite{li2018quantum}. Here, the quantum annealer is used to train the parameters of the classification model, which is done by mapping the problem of finding the optimal parameters to a minimization of a QUBO.

\subsubsection{Unsupervised learning}\label{sec:unsupervised_learning}

The use of quantum devices to speed up different unsupervised learning tasks has been investigated thoroughly, leading to different algorithms for generative modelling \cite{benedetti2016estimation,benedetti2017quantum,benedetti2019generative}, clustering~\cite{otterbach2017unsupervised}, among others~\cite{lloyd2013quantum}. An analysis of quantum speedup in unsupervised learning for Fault-Tolerance algorithms is presented in \cite{aimeur2013quantum}.
The task of learning probabilistic generative models in particular has been of interest to the QML community, because of the potential advantage quantum computers may exhibit over their classical counterparts in the near future \cite{perdomo2018opportunities}. For the advantages rendered by quantum correlations such as \textit{contextuality} and \textit{Bell non-locality} for
generative modelling, refer to ~\cite{gao2021enhancing}.

Generative Modelling involves learning the underlying probability distribution from a finite set of samples from a data set, and generating new samples from the distribution. 
There have been several proposals for using parameterized quantum circuits as models for generative learning~\cite{benedetti2019generative,benedetti2018quantum,amin2018quantum}, including quantum Boltzmann machines, quantum circuit Born machines, quantum assisted Helmholtz machines, quantum generative adversarial networks, amongst others~\cite{benedetti2019parameterized,amin2018quantum,benedetti2018quantum,benedetti2019generative}. We discuss some of these proposals in detail hereafter.

\paragraph{Quantum Boltzmann Machines. }
The quantum Boltzmann machine \cite{amin2018quantum} (QBM) extends the classical Boltzmann machine~\cite{ackley1985learning}, a neural architecture capable of several tasks including generative modeling of data. Such models take their name from their physical inspiration, namely, the Boltzmann distribution over the Ising model in the classical case, and the Boltzmann distribution over the transverse-field Ising model, for the quantum case. Such a network consists of a mixture of \textit{visible} and \textit{hidden} vertices, connected by weighted edges. The visible vertices function as both input and outputs to the network, whilst the hidden vertices add extra degrees of freedom to the network.

The QBM can be modeled with the Hamiltonian
\begin{equation}
    H = -\sum_{a}^{N} (b_a \hat{\sigma}_{z}^{a} + \Gamma_a \hat{\sigma}_{x}^{a} ) -\sum_{a,b}\omega_{ab}\hat{\sigma}_{z}^{a}\hat{\sigma}_{z}^{b},
    \label{eq:QBM}
\end{equation}
where $b_a$, $\Gamma_a$, and $\omega_{ab}$ are the parameters to be fine-tuned to generate the training data.
Defining the density matrix $\rho = \frac{e^{-H}}{Z}$ with $Z$ the usual partition function, $Z=\text{Tr}\left(e^{-H}\right)$, the marginal probability that the visible variables are in some state $\boldsymbol{v}$ is given by $P_v=\text{Tr}\left(\Lambda_{v}\rho\right)$, with $\Lambda_{v} = \left(\bigotimes_{\nu} \frac{1+v_{\nu}\hat{\sigma}_{z}^{\nu}}{2}\right) 
\otimes I_{h}$, a projector onto the subspace spanned by the visible variables tensor the identity acting on the hidden variables. The objective of training the QBM, then, is to get the family of probability distributions $P_{v}$ to match the family inherent to the data, $P_{v}^{\text{data}}$, for arbitrary $\boldsymbol{v}$. This is achieved by minimizing the negative log-likelihood measure shown below
 \begin{equation}
     \mathcal{L} = - \sum_{v}P_{v}^{\text{data}}\text{log} \frac{\text{Tr} \Lambda_v e^{-H}}{\text{Tr} e^{-H}}\,.
 \end{equation}
The gradients of $\mathcal{L}$ with respect to the Hamiltonian parameters are difficult to calculate by sampling the Boltzmann machine, both classically and in the quantum variant. Methodologies of approximating these gradients are necessary to advance the deployment of QBM's.

 The QBM may be trained both to be a generator, or a discriminator, with respect to the distribution it is trained to mimic. Consider the joint distribution of input and output variables $x$ and $y$ respectively. In the discriminative case, the objective is to minimize negative log-likelihood with respect to $P_{y|x}$,  For generative learning, the goal is to learn the joint distribution $P_{x,y}$ directly.

 The implementation of the QBM designed by \cite{amin2018quantum} found that a ten qubit QBM with only visible vertices is able to learn a mixture of randomly generated Bernoulli distributions more effectively than a classical Boltzmann machine, and performed better in generative applications. \cite{kieferova2017tomography} found that a QBM outperformed classical Boltzmann machines in generative training to reproduce small Haar-random states.
 Extensions to the QBM, such as the Variational Quantum Boltzmann Machine (VQBM) \cite{zoufal2020variational}, have improved upon trainability. Using ideas similar to ~\cite{zoufal2020variational}, VQBM were also proposed in ~\cite{shingu2020boltzmann}. In addition to its generative capacities, QBMs have shown potential in reinforcement learning \cite{crawford2019reinforcement}, in which they have been shown to achieve better fidelity to data distributions than do restricted Boltzmann machines or deep Boltzmann machines (classical boltzman machines with layers of hidden vertices) of similar sizes.
To suit NISQ devices,~\cite{verdon2017quantum} suggested that QBM can be approximated using QAOA as a subroutine in \cite{anschuetz2019realizing}, an efficient method for training QBMs in NISQ devices based on the eigenstate thermalization hypothesis has been proposed.

\paragraph{Quantum Circuit Born Machines. }
Parametrized quantum circuits can function as generative models to sample from probability distribution.
The Quantum Circuit Born Machine (QCBM) \cite{benedetti2019generative} outputs bitstrings $\boldsymbol{x}$ sampled from measurements in the computational basis of a quantum circuit $U(\boldsymbol{\theta})$, with the probability of each bit string given by the Born rule $p_{\boldsymbol{\theta}}(x)\sim\vert\bra{\boldsymbol{x}}U(\boldsymbol{\theta})\ket{0}\vert^2$. The goal is that the distribution of the QCBM matches the one from a given target distribution.

QCBMs can prepare classical probability distributions as well as entangled quantum states by training them to match the probability distribution corresponding to the desired quantum state~\cite{benedetti2019generative}.
In \cite{liu2018differentiable}, training of QCBMs using the gradients of the parameterized quantum circuit was proposed using the maximum mean discrepancy loss, which calculates the difference of the sampled output from the quantum circuit and the desired distribution in a kernel feature space.

QCBMs are well suited to be run on current NISQ hardware and can serve as benchmarks~\cite{zhu2019training,leyton2019robust,hamilton2019generative} and have been applied to tasks such as generating images~\cite{rudolph2020generation} or financial data~\cite{coyle2020quantum,alcazar2020classical}.
It has been shown that QCBMs can potentially outperform classical computers as they are able to sample from probability distributions that are difficult for classical computers~\cite{du2020expressive,coyle2020born}.

\paragraph{Quantum Generative Adversarial Networks. }
Generative adversarial learning~\cite{goodfellow2014generative} has been one of the most recent breakthrough in machine learning, and have become very powerful tool in the machine learning community, for image and video generation, and materials discovery. 
The GAN consists of two networks, a generator, $F_G(z;\theta_g)$ and a discriminator, $F_D(x;\theta_d)$ - with parameters $\theta_g$ and $\theta_d$ respectively, playing an adversarial game, which can be summarized as follows:
\begin{equation}
\begin{split}
    \min_{\theta_g} \max_{\theta_d} (E_{x\sim p_\text{data}(x)}[\log (F_D(x)] \\
    + E_{z\sim p_z(z)}[\log(1-F_D(F_G(z)))]
\end{split}
\end{equation}
where $p_z(z)$ is a fixed prior distribution, $p_\text{data}(x)$ is the target distribution, $x$ is the data sampled from $p_\text{data}(x)$, and $z$ is the noise sampled from $p_z(z)$. The training of GAN is carried iteratively, until the generator produces a distribution indistinguishable from the target distribution. 

A quantum version of generative adversarial networks (GANs) was proposed theoretically in Refs.~\cite{dallaire2018quantum,lloyd2018quantumgenerative} and further developed for near term quantum devices in Refs.~\cite{ zeng2019learning, romero2019variational,situ2020quantum}, where parameterized quantum circuits are used for adversarial learning instead of classical neural networks. 

The different adaptions of quantum GANs can be divided into different categories, based on the data and networks used being classical and quantum~\cite{romero2019variational}. There have been different studies with hybrid models of GANs using both classical and quantum data, and it has been shown that the training of these networks are robust to moderate level of noise~\cite{anand2020experimental}.

The training of quantum GANs has been demonstrated experimentally on various quantum processing units, for a variety of tasks including, quantum state estimation~\cite{hu2019quantum}, image generation~\cite{huang2020experimental,huang2020realizing}, generating continuous distributions~\cite{anand2020experimental}, learning distribution~\cite{zoufal2019quantum}, among others~\cite{nakaji2020quantum}.

\subsubsection{Reinforcement learning}
\label{sec:applications_ML_RL}

The general framework of reinforcement learning (RL) involves an agent interacting with an environment attempting to maximize an underlying reward function. 

The mathematics of RL can be captured using Markov decision process (MDP)~\cite{sutton2018reinforcement}. An MDP is a $4$-tuple $\left(S,A,R,P\right),$ where $S$ is the set of all possible valid states; $A$ is the set of all possible actions; $R$ is the reward function, i.e. a map $R: \ S\times A\times S\rightarrow\mathbb{R}$; and $P$ is the transition probability, i.e. a map $P: \ S\times A\rightarrow[0,1].$ Specifically, the transition probability $P(\tilde{s}\vert s,a)$ represents the probability of transition to state $\tilde{s}$ given the present state is $s$ and the action $a$ has been taken. The term ``Markov'' in MDP means that transitions are memory-less and depend only on the current state and action. The agents in reinforcement learning learn via trial and error. For a successful training, a proper balance between exploration of unknown strategies and exploitation of prior experience is required.

The training happens via agent-environment interaction. At the beginning of time step $t$, the environment state is $s_{t}.$ From the set $A,$ the agent selects an action $a_{t}.$ The transition probability dictates the next state of the environment $s_{t+1}$ and the agent gets reward $r_{t+1}$ based on the reward function $R.$ 
The agent-environment interaction yields a series of states and actions of the form $\tau = \left(s_{1},a_{1},s_{2},a_{2},\cdots,s_{H},a_{H}\right).$ The aforementioned series is called a trajectory and the number of interactions ($H$) in an episode is called horizon. Suppose the probability of a trajectory is $P\left(\tau\right)$ and the corresponding cumulative reward is $R_{tot}\left(\tau\right).$ Then, the expected reward is $\sum_{\tau}P\left(\tau\right)R_{tot}\left(\tau\right).$

By harnessing quantum mechanical phenomena such as superposition and entanglement, one can expect to achieve speedups in the reinforcement learning tasks \cite{dong2008quantum,dunjko2016quantum,dunjko2017advances,paparo2014quantum}. The aforementioned intuition has led to recent works towards quantum reinforcement learning \cite{dunjko2017advances,cornelissen2018quantum}

We discuss the essence of quantum reinforcement lerning by
providing a brief synopsis of quantum agent environment (AE) paradigm. For details, refer to ~\cite{dunjko2017advances}. In the AE paradigm,
agent and environment are modelled via sequences of unitary maps $\left\{ \mathcal{E}_{A}^{j}\right\} _{j}$
and $\left\{ \mathcal{E}_{E}^{j}\right\} _{j}$ respectively. The
agent and environment have access to memory registers belonging to
Hilbert spaces $\mathcal{H}_{A}$ and $\mathcal{H}_{E}$. The communication
register between the agent and the environment belongs to Hilbert
space $\mathcal{H}_{C}.$ The agent maps $\left\{ \mathcal{E}_{A}^{j}\right\} _{j}$
act on $\mathcal{H}_{A}\otimes\mathcal{H}_{C}$ and the environment
maps $\left\{ \mathcal{E}_{E}^{j}\right\} _{j}$ act on $\mathcal{H}_{E}\otimes\mathcal{H}_{C}$.
The agent and environment interact with each other by applying their
maps sequentially. The set of actions and states correspond to orthonormal
set of vectors $\left\{ \vert a\rangle\vert a\in A\right\} $ and
$\left\{ \vert s\rangle\vert s\in S\right\} $ respectively. The Hilbert
space corresponding to the communication register is given by $\mathcal{H}_{C}=\mathcal{\text{span}}\left(\vert y\rangle\vert y\in S\cup A\right).$
The classical AE paradigm corresponds to the case where the agent
and environment maps are classical.

Quantum reinforcement learning has been studied for algorithm such as 
SARSA, and Q Learning ~\cite{jerbi2019quantum}, which are some of the elementary reinforcement learning algorithms~\cite{sutton2018reinforcement}.

In the set-up of variational quantum circuits, reinforcement learning has been explored for small input sizes~\cite{chen2020variational}. This work revealed a possibility of quadratic advantage in parameter space complexity. Using better encoding schemes, \cite{lockwood2020reinforcement} showed the case of reinforcement learning with variational quantum circuits for larger input sizes. In a follow-up work, \cite{lockwoodplaying} demonstrated the possibility of dealing with the relatively complicated example of playing Atari games.

Reinforcement learning with quantum annealers has also been investigated 
by ~\cite{crawford2016reinforcement}. In their framework, they explore reinforcement learning with quantum Boltzmann machines. A detailed study of basic reinforcement learning protocols with superconducting circuits is provided in ~\cite{lamata2017basic}. Some exciting proposals of reinforcement learning with trapped ions and superconducting circuits have also been proposed recently~\cite{cardenas2018multiqubit}. For quantum eigensolvers, reinforcement learning study has been carried out recently~\cite{albarran2020reinforcement}. Reinforcement learning with optical set-up has been discussed in ~\cite{yu2019reconstruction}.

\subsection{Combinatorial optimization} \label{sec:optimization}

Given a finite set of objects, say $S$, combinatorial optimization
deals with finding an optimal object from the set $S.$ It is a sub-discipline
of mathematical optimization theory, with applications in diverse
fields such as artificial intelligence, logistics, supply chain and
theoretical computer science. Some typical examples of combinatorial
optimization problems are the traveling salesman problem~\cite{lenstra1975some}, job-shop
scheduling~\cite{manne1960job}, max-cut~\cite{festa2002randomized} and Boolean satisfiability~\cite{tovey1984simplified}. 

To understand combinatorial optimization, let us consider the canonical
problem of Boolean satisfiability. Boolean variables admit two truth
values, TRUE and FALSE. 
These  can be combined
together using operators AND or conjunction (denoted by $\land$), NOT or negation (denoted by $\lnot$), and OR or disjunction (denoted by $\lor$). These combinations are called Boolean expressions.

A Boolean expression is said to be satisfiable if it can be TRUE for
appropriate assignment of logical values to its constituent Boolean
variables. Given a Boolean expression $E$, the Boolean satisfiability
problem (SAT) consist of checking if $E$ is satisfiable. The well-known Cook-Levin
theorem showed that SAT is NP-complete~\cite{arora2009computational}. 

Every
combinatorial optimization problem can be expressed as $m$ clauses
over $n$ Boolean variables. A Boolean variable is known as positive
literal, while its negation is known as a negative literal. 
A disjunction of
literals is known as clause or constraint. For every constraint
$C_{\alpha}$ for $\alpha\in\left\{ 1,2,\cdots,m\right\} $ and every
string $z\in\left\{ 0,1\right\} ^{n},$ let define
\begin{align*}
C_{\alpha}(z)=
\begin{cases}
1 & \text{ if \ensuremath{z} satisfies }C_{\alpha}(z)\\
0 & \text{ if }z\text{ does not satisfy}
\end{cases}.
\end{align*}
The goal of a combinatorial optimization problem, framed as such,
is to find a string which maximizes the following objective function,
\begin{equation}
C(z)=\sum_{\alpha=1}^{m}C_{\alpha}\left(z\right),
\label{eq:Objective_Cluase}
\end{equation}
which counts the
number of satisfied constraints. 

Approximate optimization algorithms such as QAOA seeks to find a solution $z$ (usually a bit-string) with a desired approximation ratio $r^*\leq C(z)/C_{\text{max}}$, where $C_{\text{max}}$ is the maximum value of $C(z)$. Using $C(z)$ and computational basis vectors $\vert e_{i}\rangle\in\mathbb{C}^{2^{n}}$ for $i=1,\ldots,2^n$,
one can construct the problem Hamiltonian as the one in \eqref{eq:HP_QAOA}, 
and thus mapping the combinatorial optimization problem to a Hamiltonian ground state problem.

The list of the NISQ algorithms for combinatorial optimization discussed in the following lines are listed in \tabref{tab:combinatorial}.

\subsubsection{Max-Cut}
\label{sec:MaxCut}

Max-Cut is an important combinatorial optimization problem with applications
in diverse fields such as theoretical physics and circuit design.
In theoretical physics, the Max-Cut problem is equivalent to finding
the ground state and its energy of a spin glass Hamiltonian.
Given a graph $G=\left(V,E\right)$ with a vertex set $V$ and edge set
$E$, a cut is a partition of the elements of $V$ into two disjoint
subsets. Given a weight function $w:E\rightarrow\mathbb{R}^{+}$ such
that the edge $\left(i,j\right)\in E$ has weight $E_{ij}$, the Max
Cut problem consist of finding a cut $K\cup\bar{K}=V$ that maximizes
\begin{equation}
\sum_{i\in K,j\in\bar{K},(i,j)\in E}w_{ij}.\label{eq:max_cut_objective_1}
\end{equation}
For every vertex $v_{i}\in V$, let us associate a variable $x_{i}$
which takes values $\pm1.$ Given an arbitrary cut $K\cup\bar{K}=V$,
let us define $x_{i}=1$ if $v_{i}\in K$ and $-1$ otherwise. Then, the
Max-Cut problem is equivalent to the following quadratic program,
\begin{equation}
\max\text{ }\sum_{\left(v_{i},v_{j}\right)\in E}w_{ij}\frac{\left(1-x_{i}x_{j}\right)}{2},\\
\label{eq:max_cut_objective_2}
\end{equation}
subject to $x_{i}\in\left\{ -1,+1\right\} \forall v_{i}\in V$.

Considering $n$ vertices as $n$ qubits in the computational basis,
we can classify qubits by assigning quantum states $\vert0\rangle$
or $\vert1\rangle.$ For the classical objective function in the optimization
program from \eqref{eq:max_cut_objective_2}, we can use the following
Hamiltonian as the problem Hamiltonian,
\begin{equation}
H_{P}=\sum_{\left(i,j\right)\in E}\frac{1}{2}\left(I-\hat{\sigma}_{z}^{i}\otimes \hat{\sigma}_{z}^{j}\right) \equiv \sum_{\left(i,j\right)\in E} C_{ij}.
\end{equation}

It has been shown that it is NP-hard to achieve an approximation ratio of $r^* \geq 16/17 \approx 0.9412$ for Max-Cut on all graphs \cite{hastad2001optimal}. For the QAOA with $p = 1$, it has been shown that for a general graph,
\begin{multline}
\langle C_{ij} \rangle = \frac{1}{2} + \frac{1}{4} (\sin 4\beta \sin \gamma)(\cos^{d_i} \gamma + \cos^{d_j} \gamma) \\
- \frac{1}{4} (\sin^2 \beta \cos^{d_i + d_j - 2 \lambda_{ij}} \gamma)(1 - \cos^{\lambda{ij}} 2\gamma),
\end{multline}
where $d_{i} + 1$ and $d_{j} + 1$ denote the degrees of vertices $i$ and $j$ respectively, and $\lambda_{ij}$ is the number of triangles containing the edge $(i,j)$ in the graph \cite{wang2018quantum}. Here, $\gamma$ and $\beta$ refer to the QAOA parameters from~\eqref{eq:QAOA_evolved_state}. {Analytical results for general Ising optimization problems with $p = 1$ have also been found \cite{ozaeta2020expectation}.}

In the case of unweighted 3-regular (u3R) graphs, the above result gives the approximation ratio of $0.692$, which is consistent with the pioneering result by Farhi, Goldstone and Gutman \cite{farhi2014quantum}. In comparison, the best classical algorithms to date give the approximation ratio of $r^* \approx 0.8786$ for general graphs 
\cite{goemans1995improved}, and $r^* \approx 0.9326$ for u3R graphs \cite{halperin2004max} using semidefinite programming. While QAOA for $p = 1$ does not outperform its classical counterparts for the Max-Cut problem, QAOA has been found to surpass the Goemans-Williamson bound for larger values of $p$ \cite{crooks2018performance}. 

QAOA has also been applied to the clustering problem (from unsupervised learning) by mapping it to Max-Cut problem \cite{otterbach2017unsupervised}. {Remarkably, it was shown that by fixing the QAOA parameters and selecting the typical problem instances from a reasonable distribution, the objective function value concentrates, i.e. the objective function value is almost independent on the instance \cite{brandao2018for}. This implies that the parameters optimized for one instance can be used for other typical instances, which would drastically reduce the optimization cost. Similar concentration behavior was also reported for the Sherrington-Kirkpatrick model in the infinite size limit $(n \to \infty)$ \cite{farhi2019quantum}.}

Recently, a non-local version of QAOA called recursive QAOA (RQAOA) was proposed ~\cite{bravyi2019obstacles}.  It consist of running a QAOA as a subroutine on a specific problem with $N$ qubits and measuring the expectation values of the correlations between the all qubit pairs $(i,j)$ with $M_{ij}=\langle \sigma^i_z\sigma^j_z\rangle$. Then, one picks out the pair of qubits $(n,m)$ that have maximal absolute value of correlation $n,m=\text{arg max}_{(i,j)}\vert M_{ij}\vert$. For $M_{nm}>0$, the selected qubit pair $(n,m)$ are positively correlated  and very likely to be in the same state, whereas for $M_{nm}<0$ they are anti-correlated and likely to be in opposite state. 
Now, this correlation is fixed as a constraint on the problem by fixing the state of the qubit $\sigma^m_z=\text{sign}(M_{nm})\sigma^n_z$. With this constraint, one of the two qubits can be removed as its state is completely determined by the other, reducing the total qubit number by one. Now, the above procedure is repeated for the now reduced problem of size $N-1$ qubits, i.e. one runs the QAOA subroutine, measures the correlations and fixes the qubit pairs with maximal correlation. The RQAOA algorithm is run recursively until the size of the problem is reduced to a small number of qubits such that it can be solved easily classically.
When RQAOA is run with the QAOA subroutine of depth $p=1$, it can efficiently simulated on a classical computer, which can serve as an important benchmark with classical algorithms~\cite{bravyi2019obstacles}.
Numerical experiments with higher $p$ suggest similar or better performance on combinatorial problems compared to other classical algorithms~\cite{bravyi2020hybrid,egger2020warm}.

Finally, QAOA with depth $p=1$ has been investigated in comparison with quantum annealing~\cite{streif2020forbidden}. QAOA is connected to quantum annealing in the sense that in the limit of infinite depth $p$, QAOA is equivalent to quantum annealing (refer to \secref{sec:HamiltonAnsatz} for QAOA, as well as \secref{subsec:quantum-annealing} for quantum annealing). However, QAOA can outperform quantum annealing on specific problems even at depth $p=1$. In fact, QAOA can solve specific problems perfectly for $p=1$, arriving at the correct solution with unit probability, whereas quantum annealing struggles here to find the solution~\cite{streif2020forbidden}. This shows that QAOA is strictly more powerful than quantum annealing.

\subsubsection{Other combinatorial optimization problems}

While the usage of QAOA on Max-Cut has been studied extensively, QAOA has also applications in other important combinatorial optimization problems such as Max-$k$ Vertex Cover, which seeks to find the set of $k$ vertices on a graph that maximizes the number of edges incident on the vertices \cite{cook2019quantum}; Exact-cover problem (given a set $X$ and several subsets $S_i$, find the combination of subsets which contains all elements just once) with applications to the tail-assignment problem \cite{bengtsson2020improved,vikstaal2020applying}; lattice protein folding \cite{fingerhuth2018quantum,robert2021resource}; knapsack problem as applied to battery revenue optimization \cite{grandrive2019knapsack}; multi-coloring graph problems \cite{oh2019solving}; maximum independent set problems \cite{choi2020quantum,saleem2020max} with applications to scheduling; and the vehicle routing problem \cite{utkarsh2020solving}. An adiabatically assisted approach was suggested in ~\cite{garcia2018addressing} to tackle combinatorial optimization problems. Investigations involving variational Grover search could be helpful to solve combinatorial optimization problems~\cite{morales2018variational,zhang2021implementation}. Gaussian Boson Sampling (see \secref{sec:GBS}) has been used to assist in a wide variety of combinatorial optimization problems \cite{bromley2020applications, arrazola2018quantum}, most prominently to solve Max-Clique \cite{arrazola2018using, banchi2020training}. This has applications in predicting molecular docking configurations \cite{banchi2020molecular}, computing vibrational spectra of molecules \cite{huh2015boson}, and electron-transfer reactions \cite{jahangiri2020quantum}. Using NISQ devices, an approach was suggested in ~\cite{metwalli2020finding} for the triangle finding problem and its $k$-clique generalization.

Quantum Annealing, which has been the inspiration of QAOA, is a prominent platform that has been applied to various combinatorial optimization problems and its applications, such as protein folding~\cite{perdomo2012finding}, reviewed in \cite{hauke2020perspectives}. As gate-based devices mature, it will open the possibility for experimental benchmarking of QAOA against state-of-the- art solvers for suitable real-world applications, as performed in \cite{perdomo2019readiness} in the context of quantum annealing machines and including proposals beyond the capabilities of current D-wave devices.

\subsection{Numerical solvers}\label{sec:numerical_solvers}

We proceed to discuss NISQ algorithms used to solve numerical problems such as factoring, singular value decomposition, linear equations and non-linear differential equations, all of them listed in \tabref{tab:numerical_solvers}.

\subsubsection{Variational quantum factoring}\label{sec:factoring}

The factoring problem accepts a composite positive integer $N$ as
input and returns its prime factors as output. There is no known efficient
classical algorithm for prime factorization and the hardness of factoring
is used to provide the security in the RSA public-key cryptosystems.
The famous Shor's factoring algorithm is a polynomial time quantum
algorithm for the factoring problem~\cite{shor1999polynomial} (implying prime factorization is in BQP)  and hence has been extensively
investigated by quantum computing researchers (for details refer to~\cite{anschuetz2019variational} and references therein). The resource estimates
for implementing the Shor's algorithm is, however, way beyond the
capabilities of the NISQ era. A detailed analysis has shown that factoring
a $2048$-bit RSA number would necessitate a quantum processor with
$10^{5}$ logical qubits and circuit depth on the order of $10^{9}$
to run for roughly $10$ days~\cite{van2010distributed,jones2012layered}. On a photonic architecture, using $1.9$
billion photonic modules, factoring a $1024$-bit RSA number is expected
to tale around $2.3$ years~\cite{devitt2013requirements}. To tackle the factoring problem in the
near-term quantum devices, it is imperative to develop NISQ-era compatible alternatives
to Shor's factoring algorithm. 

The factoring
problem can be mapped to the ground state problem of an Ising Hamiltonian~\cite{burges2002factoring,dattani2014quantum}.
To understand the aforesaid mapping, let us consider the factoring
of $m=p\times q$. Suppose the binary representations of $m,p$ and
$q$ are
$m=\sum_{k=0}^{n_{m}-1}2^{i}m_{k}$, 
$p=\sum_{k=0}^{n_{p}-1}2^{i}p_{k}$ and
$q=\sum_{k=0}^{n_{q}-1}2^{i}q_{k}$.
Here, $m_{k}\in\left\{ 0,1\right\} $is the $k$th bit of $m$ and
the total number of bits for $m$ has been denoted by $n_{m}.$ Similar
notation has been employed for $p$ and $q.$ Since $m=p\times q,$
it induces $n_{c}=n_{p}+n_{q}-1$ constraints on the
individual bits of $m,p$ and $q,$
\begin{equation}
\sum_{j=0}^{i}q_{i}p_{i-j}+\sum_{j=0}^{i}z_{j,i}-m_{i}-\sum_{j=1}^{n_{c}}2^{j}z_{i,i+j}=0\,,\label{eq:Constraints_factoring_1}
\end{equation}
for $i\in[0,n_{c})$ and the carry bit from position $i$ to position
$j$ has been represented by $z_{i,j}.$ The constraint $i$ in
\eqref{eq:Constraints_factoring_1} induces clause $C_{i}\equiv\sum_{j=0}^{i}q_{i}p_{i-j}+\sum_{j=0}^{i}z_{j,i}-m_{i}-\sum_{j=1}^{n_{c}}2^{j}z_{i,i+j}$
over $\mathbb{Z}$ such that factoring can be modelled as assignment
of binary variables $\left\{ m_{i}\right\} ,$ $\left\{ p_{i}\right\} $
and $\left\{ q_{i}\right\} $ which solves $\sum_{i=0}^{n_{c}-1}C_{i}^{2}=0.$

One can map the binary variables to quantum observables to quantize
the clause $C_{i}$ to $\hat{C_{i}}$ using the  mapping
$b_{k}\rightarrow\frac{1}{2}\left(1-\sigma_{b,k}^{z}\right)$ 
and obtain the Hamiltonian
$H_{P}=\sum_{i=0}^{n_{c}-1}\hat{C_{i}}^{2}$,
which we refer as
factoring Hamiltonian. Note that the factoring Hamiltonian is a 4-local
Ising Hamiltonian. 

By using the aforementioned ideas, one can use NISQ
algorithms for the ground state problem to tackle the factoring problem (see \secref{sec:VQE} and \secref{sec:optimization}).
In Ref.~\cite{anschuetz2019variational}, authors employ QAOA to find the ground state
of the factoring Hamiltonian and refer to their Algorithm as variational
quantum factoring (VQF) algorithm. Numerical simulations were provided
for numbers as high as $291311$. For a recent experimental realization and detailed analysis of VQF, refer to ~\cite{karamlou2020analyzing}.

\subsubsection{Singular value decomposition} \label{subsec: SVD}

Given a matrix $M\in\mathbb{C}^{m\times n}$, the Singular Value Decomposition (SVD) provides a factorization of the form 
$M=U\Sigma V^{\dagger}$, 
where $U\in\mathbb{C}^{m\times m}$  is a unitary matrix, $\Sigma\in\mathbb{R}_{+}^{m\times n}$ is a rectangular diagonal matrix with non-negative real diagonal entries and $V\in\mathbb{C}^{n\times n}$ is a unitary matrix. 
The diagonal entries of $\Sigma$  are called the singular values of matrix $M.$ The columns of the unitary matrices $U$ and $V$ are called left-singular and right-singular vectors of $M.$ Using Dirac notation, one can write
\begin{equation}
M=\sum_{j=1}^{r}d_{j}\vert u_{j}\rangle\langle v_{j}\vert.
\label{eq:SVD_Dirac}
\end{equation}
where $d_{j},\vert u_{j}\rangle,\vert v_{j}\rangle$ are singular values, left-singular vectors and right-singular vectors. The rank of matrix $M$ is $r$ and is equal to the number of non-zero singular values.

SVD finds applications in  calculating pseudoinverse ~\cite{gregorcic2001singular},  solving homogeneous linear equations ~\cite{klema1980singular}, signal processing ~\cite{vandewalle1991use} and recommendation systems ~\cite{koren2009matrix}. Moreover, the notion of Schmidt decomposition which is used to study entanglement of bipartite quantum states, is related to SVD.

In the quantum information context, the SVD can be used to compute
the Schmidt decomposition of  bipartite quantum states. For a 
quantum state $\vert\psi\rangle\in\mathcal{H}_{A}\otimes\mathcal{H}_{B}$,
the Schmidt decomposition is given by 
\begin{equation}
\vert\psi\rangle=\sum_{i}d_{i}\vert u_{i}\rangle\vert v_{i}\rangle,\label{eq:Schmidt}
\end{equation}
where $d_{i}$ are non-negative real numbers such that $\sum_{i} d_{i}^{2}=1$.
Moreover, $\left\{ \vert u_{i}\rangle\right\} _{i}$ and $\left\{ \vert v_{i}\rangle\right\} _{i}$
correspond to orthonormal basis sets for $\mathcal{H}_{A}$ and $\mathcal{H}_{B}$
respectively. The number of non-zero $d_{i}$, say $\chi$,
is called the Schmidt rank of the quantum state $\vert\psi\rangle$
and is used to quantify the bipartite entanglement. To calculate the
Schmidt decomposition, one can write the bipartite quantum state as
a matrix
$\vert\psi\rangle=\sum_{i,j}A_{ij}\vert i\rangle\vert j\rangle$,
where $\ket{i}$ and $\ket{j}$ are the computational basis states of each qubit, and perform SVD of the matrix $A.$

Ref.~\cite{bravo2020quantum} provide a NISQ algorithm to perform SVD of pure bipartite states. Starting with two unitary circuits, which act on different bipartitions of the system, the authors variationally determine the singular values and singular vectors by training the circuits on exact coincidence of outputs. The central ideas of their method is to variationally find circuits that provides the following transformation of the initial quantum state $\vert\psi\rangle_{AB}$ with Schmidt rank $\chi$,
\begin{equation}
U_{A}\otimes V_{B}\vert\psi\rangle_{AB}=\sum_{i=1}^{\chi}\lambda_{i}e^{i\gamma_{i}}\vert e_{i}\rangle_{A}\vert e_{i}\rangle_{B},\label{eq:SVD_Bravo_Circuits}
\end{equation}
where $U_{A}\vert v_{i}\rangle_{A}=e^{i\alpha_{i}}\vert e_{i}\rangle_{A}$, $V_{B}\vert v_{i}\rangle_{B}=e^{i\beta_{i}}\vert e_{i}\rangle_{B}$ such that $\alpha_{i}=\beta_{i}+\gamma_{i}\in[0,2\pi)$ and $\left\{ \vert e_{k}\rangle_{A,B}\right\} _{k}$ are the compuational basis states in $\mathcal{H}_{A,B}.$ 
Using their algorithm, authors also suggest the possibility
to implement SWAP gate between parties $A$ and $B$ without the requirement of any gate connecting the two subsystems. 

Using variational principles for singular values and Ky Fan theorem \cite{fan1951maximum}, \cite{wang2020variational} provide an alternative NISQ algorithm for SVD. The authors provide proof of principle application of their algorithm in image compression of handwritten digits. They also discuss the applications of their algorithm in recommendation systems and polar decomposition.

\subsubsection{Linear system problem}\label{sec:linear_system_problem}

Systems of linear equations play a crucial role in various areas of science, engineering and finance. 
Given a matrix $A\in\mathbb{C}^{N\times M}$ and $\boldsymbol{b}\in\mathbb{C}^{N}$, the task of the linear system problem (LSP) consists of finding $\boldsymbol{x}\in\mathbb{C}^{M}$ such that 
\begin{equation}
A\boldsymbol{x}=\boldsymbol{b}\,.\label{eq:LSP}
\end{equation}
Depending on the dimensions $M$ and $N$, the LSP takes various forms. If $M=N$ and $A$ is invertible, $\boldsymbol{x}=A^{-1}\boldsymbol{b}$ is unique. If $M\neq N,$ the LSP can be under-determined or over-determined. For the sake of simplicity, it is natural to assume the matrix $A$ to be square i.e. $M=N.$ If the matrix $A$ has at most $s$ non-zero elements per row or column, the LSP is called $s$-sparse.

The quantum version of the LSP, known as the quantum linear system
problem (QLSP), assumes $A$ to be $N\times N$ Hermitian matrix and $\boldsymbol{b}$ to be a unit vector, i.e. it can be represented
as a quantum state $\ket{b}=\sum_{i=1}^{N}b_{i}\vert e_{i}\rangle.$ The QLSP problem thus is formulated as
\begin{equation}
    A\ket{x}=\ket{b} \rightarrow \ket{x}=A^\dagger|b\rangle.
    \label{eq:QLSP}
\end{equation}

The first quantum algorithm proposed for solving the QLSP was the famous Harrow-Hassidim-Lloyd (HHL) algorithm~\cite{harrow2009quantum}. Apart from the size of the matrix $A$, i.e. $N$, and its sparsity $s$, two other dominant factors determining the running time of a LSP or QLSP algorithm are the condition number $(\kappa)$ of the matrix $A$ and the additive error $(\epsilon)$ corresponding to the solution. The condition number of a matrix $A$ is given by ratio of maximal and minimal singular values of $A.$ The best classical algorithm for LSP is the conjugate gradient method with runtime complexity $O\left(Ns\kappa\log\left(\frac{1}{\epsilon}\right)\right).$ On the other hand, the HHL algorithm for QLSP, as originally proposed, has runtime complexity $O\left(\log\left(N\right)s^{2}\frac{\kappa^{2}}{\epsilon}\right).$ Further works on the HHL algorithm has improved $\kappa$ scaling to linear~\cite{ambainis2012variable} and error dependence to $poly\left(\log\left(\frac{1}{\epsilon}\right)\right)$~\cite{childs2017quantum}. Implementation of HHL algorithm, however, requires the fault-tolerant architecture and hence its guarantees can not be leveraged on the NISQ architecture. The largest QLSP solved on a gate based quantum computer corresponds to its implementation on an nuclear magnetic resonance (NMR) processor for $N=8$~\cite{wen2019experimental}. 

Recently, a few VQA based implementations of the QLSP were proposed~\cite{huang2019nearterm,bravo2019variational,xu2019variational2}. Given a QLSP with input $A$ and $\vert b\rangle,$ 
the idea is to find the ground state of the following Hamiltonian,
\begin{equation}
H(u)=A(u)P_{+,b}^{\perp}A(u),\label{eq:Hamiltonian_Subasi}
\end{equation}
where $A(u)$ and $P_{+,b}^{\perp}$ are defined as
\begin{align}
A(u)&\equiv\left(1-u\right)\sigma_{z}\otimes I+u \ \sigma_{x}\otimes A, \\ P_{+,b}^{\perp}&=I-\vert+,b\rangle\langle b,+\vert.
\end{align}
Both $A$ and $|b\rangle$ are assumed to be constructed efficiently with a quantum circuit, i.e. $A=\sum_{k=1}^{K_{A}}\beta_{k}U_{k}$ and $\vert b\rangle=U_{b}\vert0\rangle$, with $K_{A}=O\left(poly\left(\log N\right)\right)$. The phase in $\beta_{k}$ can be absorbed in $U_{k}$ and hence one can assume $\beta_{k}>0.$

The Hamiltonian in \eqref{eq:Hamiltonian_Subasi} for $u=1,$
has a unique ground state, $\vert+\rangle\vert x^{\star}\rangle=\vert+\rangle\frac{A^{-1}\vert b\rangle}{\left\Vert A^{-1}\vert b\rangle\right\Vert _{2}},$ with zero ground state energy. After removing the ancilla, the ground state can be seen to be proportional to $A^{-1}\vert b\rangle.$ 
Thus, one can define the following loss function,
\begin{equation}
L_{H}\left(\vert x\rangle\right)=\langle+,x\vert H(1)\vert+,x\rangle.\label{eq:Loss_Subasi}
\end{equation}
Without the ancilla, the above loss function can be written as $L_{H}\left(\vert x\rangle\right)=\langle x\vert A^{2}\vert x\rangle-\langle x\vert A\vert b\rangle\langle b\vert A\vert x\rangle.$

In \cite{huang2019nearterm}, authors analyze the optimization landscape for VQA based optimization for the loss function of \eqref{eq:Loss_Subasi} and show the presence of barren plateaus which persist independent of the architecture of the quantum circuit for generating $\vert x\left(\theta\right)\rangle.$
Even techniques based on adiabatic morphing~\cite{garcia2018addressing} fail to evade the effect of the barren plateaus. To circumvent the barren plateau problem, ~\cite{huang2019nearterm} proposed a classical-quantum hybrid state (see also \secref{sec:IQAE} and \eqref{eq:IQAE_Ansatz})
$\boldsymbol{x}=\sum_{i=1}^{r}\alpha_{i}\vert\psi_{i}\left(\theta_{i}\right)\rangle$, 
where $\alpha_{i}\in\mathbb{C}$ and $\theta_{i}\in\mathbb{R}^{k_{i}}$ for $i\in\left\{ 1,2,\cdots,r\right\}.$ Note that $\theta_{i}$ are the usual variational parameters and $\alpha_{i}$ are the combination parameters. These parameters are stored on a classical device and the state $x$ is not explicitly created on a quantum processor. Moreover, $x$ may not be normalized. To solve the QLSP, one minimizes the following loss function,
\begin{equation}
L_{R}\left(x\right)=\left\Vert Ax-\vert b\rangle\right\Vert _{2}^{2} =x^{\dagger}A^{\dagger}Ax-2\text{Re}\left\{ \langle b\vert Ax\right\} +1.
\end{equation}
Since optimization with respect to $\theta_{i}$ suffers from the barren plateau problem, one can fix and subsequently drops the variational parameter $\theta_{i}$.

The optimization landscape is convex in $\boldsymbol{\alpha}=\left(\alpha_{1},\alpha_{2},\cdots,\alpha_{r}\right).$ Starting from $\vert\psi_{1}\rangle=\vert b\rangle,$ other quantum states can be generated using the Ansatz tree approach in \cite{huang2019nearterm}. It was proved that finding the combination parameters of $\vert\psi_{1}\rangle,\vert\psi_{2}\rangle,\cdots,\vert\psi_{r}\rangle$ to minimize $L_{R}\left(\sum_{i=1}^{r}\alpha_{i}\vert\psi_{i}\rangle\right)$ is BQP complete. Moreover, using $O \left(K_{A}^{2}\frac{r^{2}}{\epsilon}\right)$ measurements, one can find $\epsilon$-suboptimal solution.
With this approach, linear systems as high as $2^{300}\times2^{300}$ can be solved by considering cases which are also classically tractable.

\subsubsection{Non-linear differential equations}\label{sec:nonlinear_differential}

Nonlinear differential equations (NLDE) are a system of differential equations (DE) that cannot be expressed as a linear system.
The numerical approaches
to tackle DE can be local or global. Local methods
employ numerical differentiation techniques~\cite{butcher1987numerical} such as the Runge-Kutta or discretization of the space of variables. Global methods, on
the other hand, represent the solution via a suitable basis set, and
the goal remains to find optimal coefficients~\cite{gottlieb1977numerical}. In many cases,
as the number of variables or nonlinearity in the differential equations
increase, finding solutions becomes challenging. To achieve higher accuracy,
local methods require a fine grid, which renders high computational
cost. In the case of global methods, high accuracy necessitates a
large number of elements in the basis set, leading to more extensive
resource requirements. To tackle resource challenges, quantum algorithms are proposed. 

Linear DE can be re-expressed as a system of linear
equations using the finite difference method, and one can employ NISQ linear
system algorithms to tackle the problem (see \secref{sec:linear_system_problem}). For a recent theoretical proposal with experimental work on linear differential equations, refer to ~\cite{xin2020quantum}. However NLDE defy this approach for large nonlinearities.

A canonical example of a NLDE appearing
in quantum theory is the $1$-D nonlinear Schr\"odinger
equation
$\left[-\frac{1}{2}\frac{\text{d}^2}{\text{d}x^2}+V(x)+g\left|f(x)\right|^{2}\right]f(x)=Ef(x)$.
Here, $E$ denotes energy, $g$ quantifies nonlinearity, and $V$ is
the external potential. Recently,
NISQ algorithms for NLDE have been proposed. 
Ref.~\cite{lubasch2020variational} use ancillary quantum registers and controlled-multiqubit
operations to implement nonlinearities to simulate the nonlinear Schrodinger equation. 
Ref.~\cite{haug2020generalized} propose the nonlinear quantum assisted simulator (NLQAS) to tackle NLDE without any controlled unitaries. Using NLQAS, they simulate this equation for $8$ qubit system. NLDE have also
been studied in \cite{gaitan2020finding} for fluid dynamics problems. Using differentiable
quantum circuits, \cite{kyriienko2020solving} have also proposed an interesting
approach to solving NLDE via global methods.

\subsection{Other applications}\label{sec:other_app}

In this subsection, we cover other applications for which NISQ algorithms can provide promising improvements. They are listed in \tabref{tab:other_app}.

\subsubsection{Quantum foundations}

One of the first experiments in digital quantum computers were the Bell nonlocality tests known as \textit{Bell inequalities} \cite{brunner2014bell}. Those experiments computed a type of Bell inequalities known as Mermin inequalities in up to five superconducting quantum qubits. The experiment consisted in preparing the GHZ state \cite{greenberger1990bell}, measure it in a particular basis state and obtain the expectation value of the Mermin operator \cite{alsina2016experimental}. These nonlocality tests can be extended to higher dimensions by controlling quantum levels beyond the $|0\rangle$ and $|1\rangle$. As example, \cite{cervera2021experimental} experimentally generate a qutrit GHZ state using a programmable device controlled with Qiskit Pulse software \cite{alexander2020qiskit}, the first step towards performing a GHZ test.

In the context of VQA, the non-classicality in VQEs is examined using contextuality, which is a nonclassical feature of quantum theory~\cite{amaral2018graph}. Using the notion of ``strong contextuality'',
 \cite{kirby2019contextuality} categorized VQE experiments into two categories:
contextual and non-contextual. Such foundational works could be utilized
to comprehend the possible sources of quantum advantage in NISQ algorithms.
Using novel concepts from this field, contextual subspace VQE (CS-VQE) was recently proposed~\cite{kirby2020contextual}. 

In another work, the
variational consistent history (VCH) algorithm was suggested to investigate foundational questions~\cite{arrasmith2019variational}. The consistent
history approach has been used to examine topics from quantum
cosmology and quantum-classical transition. In the VCH algorithm,
the quantum computer is used to compute the ``decoherence functional'',
which is challenging to calculate classically. The classical
computer is employed to tune the history parameter so that the consistency
is improved.

\subsubsection{Quantum optimal control}\label{sec:quantum_optimal_control}

Quantum optimal control is a topic of paramount
importance in the pursuit to harness the potential of Near-Term quantum devices. For a given quantum control system and a cost function
that measures the quality of control, it aims to find a control that can achieve optimal performance.

Some recent works have investigated quantum optimal control
in the NISQ framework. Recent detailed perspective in this direction can be found in Ref. \cite{magann2020pulses}.
Ref.~\cite{li2017hybrid} provides a hybrid quantum-classical approach to quantum
optimal control. To remedy some of the difficulties of classical approaches
to optimal control related to scaling of resources, \cite{dive2018situ} proposed
another NISQ framework. Experimental demonstration
of quantum control for a $12$-qubit system has also been carried on~\cite{lu2017enhancing}.
The aforementioned approaches, however, restrict their target states
to be sparse matrices. For dense target states, \cite{policharla2020algorithmic}
recently proposed a  NISQ algorithm. Along with their algorithm, they also suggested a few  algorithmic primitives to calculate overlap of quantum states and transition matrix elements. Hybrid quantum-classical algorithm have also been implemneted for computing quantum optimal control pulses, in particular for controlling molecular systems~\cite{magann2020digital,castaldo2020quantum}.

\subsubsection{Quantum metrology}\label{sec:quantum_metrology}

Quantum metrology harnesses non-classical features of quantum theory
for parameter estimation tasks. A canonical example could be estimating
the parameter $\phi$ of a unitary map under the action of Hamiltonian
$\hat{H}$, given by
$\hat{\rho}\left(\phi\right)=e^{-i \hat{H}\phi}\hat{\rho}_{0}e^{+i \hat{H}\phi}$
where the density matrix $\hat{\rho}_{0}$ refers to  the initial state of the system. The goal
is to estimate $\phi$ via measurements on $\hat{\rho}(\phi).$ The quantum
Cramér-Rao bound provides a lower bound to the achievable precision,
\begin{equation}
\left(\Delta\phi\right)^{2}\geq\frac{1}{nF_{Q}\left(\hat{\rho}\left(\phi\right)\right)}.\label{eq:QFI_metrology}
\end{equation}
Here, $n$ represents number of samples, $F_{Q}\left(\hat{\rho}\left(\phi\right)\right)$
is quantum Fisher information and $\left(\Delta\phi\right)^{2}$ is
the variance in the estimation of $\phi$. In most of the experiments,
the parameter of interest is either temperature or magnetic field.

Notice that the precision of the estimation procedure increases as the
quantum Fisher information increases. Using it
as a cost function, a few works have recently explored quantum metrology to prepare a better probe state in a VQA set-up~\cite{koczor2020variational,kaubruegger2019variational,ma2020adaptive,beckey2020variational}. 
In addition, \cite{meyer2020variational} provided a toolbox for multi-parameter estimation and \cite{haug2021natural} the natural PQC with the lowest possible quantum
Cramér-Rao bound for a general class of circuits.

\subsubsection{Fidelity estimation}\label{sec:fidelity_estimation}

In \secref{sec:objective_function}, we discussed how to use the fidelity as an objective function, a quantity which is useful to train some VQA algorithms. In addition, estimating the fidelity of a quantum state with respect to another state has a general interest in the context of quantum computing. For this reason, algorithms that can estimate this property may become useful in the NISQ era.

Given the density matrices of two quantum states $\rho_{1}$ and $\rho_{2}$, their fidelity is given by
\begin{equation}
F\left(\rho_{1},\rho_{2}\right)=\left(\text{Tr}\sqrt{\sqrt{\rho_{1}}\rho_{2}\sqrt{\rho_{1}}}\right)^2.\label{eq:Fidelity_quantum}
\end{equation}
Due to the large dimensionality of the Hilbert spaces, computing fidelity can be challenging. 

Recently, variational quantum fidelity estimation
(VQFE) algorithm was proposed to tackle slightly modified version
of the fidelity estimation task which works efficiently when $\rho_{1}$ has low-rank. Ref.~\cite{cerezo2020fidelity} provide lower and upper
bounds on $F\left(\rho_{1},\rho_{2}\right)$ via VQFE. The algorithm
calculate fidelity between $\rho_{1}^{n}$, which is a truncated version
of $\rho_{1}$ obtained by projecting $\rho_{1}$ to subspace spanned
by $n$ largest eigenvalue eigenvectors of $\rho_{1}.$ The bounds
improve monotonically with $n$ and is exact for $n=rank\left(\rho_{1}\right).$
The VQFE algorithm proceeds in three steps: \textit{i)} a variational diagonalization of $\rho_{1}$; \textit{ii)} the matrix elements of $\rho_{2}$ are computed in the eigenbasis of $\rho_{1}$; and \textit{iii)} using the matrix elements from \textit{ii)}, the fidelity is estimated.

\subsubsection{Quantum error correction}\label{sec:variational_error_correction}

The leading error correction schemes carry high resource overheads, which renders them impractical for near-term devices~\cite{fowler2012surface,johnson2017qvector}. Moreover, many of the schemes mandate knowledge of the underlying noise model~\cite{fletcher2008channel,kosut2008robust,kosut2009quantum}. For an encoding process $\mathcal{E}$, decoding process $\mathcal{D}$
and noise model $\mathcal{N}$, the quality of a quantum error correction
scheme can be characterized by how close $\mathcal{D}\text{\textopenbullet}\mathcal{N}\text{\textopenbullet}\mathcal{E}$
is close to identity. The range of $\mathcal{E}$ is called code space
$\mathcal{C}.$ 

In ~\cite{johnson2017qvector}, a  variational error-correcting scheme
i.e, quantum variational error corrector (QVECTOR) was proposed by defining an objective function over the code space $\mathcal{C}$. The authors employ two trainable parametric quantum circuits $V\left(\boldsymbol{p}\right)$
and $W\left(\boldsymbol{q}\right)$ for encoding and decoding respectively,
with tunable parameter vectors $\boldsymbol{p}$ and $\boldsymbol{q}$.
For a given encoding-decoding pair, characterized by $\left(\boldsymbol{p},\boldsymbol{q}\right),$
the authors calculate a quantity called ``average code fidelity'' with respect to Haar
distribution of states over the code space $\mathcal{C}.$ The algorithm
is model-free, i.e. no assumption of the noise model is involved. The
goal of the QVECTOR algorithm is to maximize average code fidelity in a variational set-up. 

In the context of VQA, error correction has also been explored in~\cite{xu2019variational2} where the target logical states are encoded as ground state  of appropriate Hamiltonian. \cite{xu2019variational2} employ imaginary time evolution to find the ground state. The authors implement there scheme for five and seven qubit codes. For a brief discussion on error correction and quantum fault-tolerance, refer to \secref{subsec:fault-tolerance}. 

\subsubsection{Nuclear physics}\label{sec:nuclear_physics}

The Standard Model of particle physics is the theory that describes the nature of the electromagnetic and nuclear interactions. Its current formulation consist of describing the forces as quantum fields, i.e. by using quantum field theory (QFT) formalism. Perturbative calculations of QFT provide with the dynamics of the physical processes at a given energy scale. However, in some cases as in quantum chromodynamics (QCD), perturbation theory can not be applied because the impossibility of observe a free quark or gluon (the fundamental particles affected by QCD interaction) due to confinement. For this reason, QCD calculations are obtained by means of numerical methods such as Monte Carlo simulations in a discretized version of QFT on a lattice structure (LQFT). The high computational cost of LQFT has motivated the study of using quantum computation or simulation to obtain the desired QCD predictions \cite{joo2019status}.

The Schwinger model describes the dynamics of the quantum electromagnetic (QED) interaction in one spatial and temporal dimensions. It is used as a toy model to study QCD since it shows fermion confinement but it is simple enough to be solved analytically. The first experimental quantum simulations of this model were carried out in trapped ions \cite{hauke2013quantum} and later on a superconducting circuit quantum computer \cite{martinez2016real}. A first proposal to use a quantum-classical algorithm to simulate this model was presented in \cite{klco2018quantum}, where the quantum computer simulates the dynamics of the symmetry sectors suggested by a classical computation. In Ref. \cite{kokail2019self} a VQS is used in an analog setup to reduce the number of variational parameters and thus, reduce the computational cost of the algorithm. Their proposal is experimentally implemented in a trapped-ion analog simulator. A significant reduction of the computational cost of LQFT is proposed in \cite{avkhadiev2020accelerating} by using a VQA approach to compute the \textit{optimized interpolating operators} (approximators of the quantum state wavefunction).

Adaptations of the UCC quantum chemistry ansatz, introduced in \secref{sec:HamiltonAnsatz}, to study quantum-variationally QCD are presented in \cite{liu2020quantum,dumitrescu2018cloud}, and for the study neutrino-nucleus scattering in \cite{roggero2020quantum}. A 10-qubit VQC is used in \cite{wu2020application} to study Higgs boson decays and production processes and in \cite{chen2020quantum} a QCNN model is proposed to study basic high-energy processes.
Recently, a PQC is used to learn the parton distribution function of protons~\cite{perez2020determining}.

\subsubsection{Entanglement properties}

Entanglement is a resource for numerous quantum information tasks. A bipartite quantum state $\rho_{AB}\in\mathcal{H}_{A}\otimes\mathcal{H}_{B}$
is called separable if it admits the form
$\rho_{AB}=\sum_{i}p_{i}\rho_{i}^{A}\otimes\rho_{i}^{B}$,
where $p_{i}$ are non-negative and $\sum_{i}p_{i}=1.$ If a state is not separable, then it is called entangled. The problem of detecting whether a state is separable or entangled is known as the separability problem and has been shown to be NP-hard~\cite{gurvits2003classical}. 

As mentioned in \secref{subsec: SVD}, computing the Schmidt rank of $\rho_{AB}$ gives a measure of the bipartite entanglement. Thus, those algorithms that tackle the SVD problem can also be used to extract entanglement properties~\cite{bravo2020quantum}. In~\cite{wang2020detecting}, authors propose a NISQ algorithm for the separability problem by providing a variational approach to employ \textit{the positive map criterion}.
This criterion establishes that the quantum state $\rho_{AB}$ is separable if and only if for arbitrary quantum system $R$ and arbitrary
positive map $\mathcal{N}_{B\rightarrow R}$ from $B$ to $R$, we
have $\mathcal{N}_{B\rightarrow R}\left(\rho_{AB}\right)\geq0.$ The authors start with a positive map and decompose it into a linear combination of NISQ implementable operations. 
These operations are executed on the target state, and the minimal eigenvalue of the final state is variationally estimated. The target state is deemed entangled if the optimized minimal eigenvalue is negative.  

Exploring a similar strategy as the one presented in \cite{bravo2020quantum}, \cite{perez2020measuring} propose a VQA to compute the tangle, a measure of tripartite-entanglement.

VQA have also been employed for extracting the entanglement spectrum of quantum systems in ~\cite{cerezo2020variational,larose2019variational}.

\section{Benchmarking} \label{sec:benchmark}

One of the central questions at the intersection of software and hardware
for NISQ devices is evaluating devices' performance and capabilities. This is where benchmarking concepts come in, to provide
various metrics that attempt to measure different machines' capabilities and compare them across time and other devices. A benchmarking
protocol can be characterized by its inherent assumptions, resource
costs and the information gain. The goal is to build benchmarking
protocols that make minimal and practical assumptions, have low resource
costs, and have high information-gain. 

Benchmarking protocols have been
developed for NISQ as well as fault-tolerant devices. For a pedagogical
summary, refer to~\cite{eisert2019quantum}. In this review, we focus on quantum benchmarking
protocols for NISQ devices. Some of the leading NISQ benchmarking
schemes are randomized benchmarking, quantum volume, cross-entropy benchmarking and application-based benchmarks. 

\subsection{Randomized benchmarking}

The most straightforward way of comparing devices is by simply counting qubits. To really compare different qubits, we must also have a sense of how many operations we can do with them before the noise arising from errors drowns out the signal. Randomized benchmarking (RB) is a convenient method for finding average error rates for quantum operations, in particular for single and two-qubit gates~\cite{magesan2011scalable,magesan2012characterzing}. RB is robust against state preparation and measurement (SPAM) errors
and, unlike tomography, admits an efficient and practical implementation.

RB involves the following assumptions: \textit{i)} for every gate, the incurred
noise is independent of other Clifford gates; \textit{ii)} the involved unitaries
should constitute a $2$-design (see \secref{sec:BP}) and should not be universal. In other
words, no $T$ gate is allowed; \textit{iii)} during the experiment, there is
no drifting in the noise processes; and \textit{iv)} one can describe noise processes
using completely positive trace-preserving (CPTP) maps. 

A RB protocol
starts by sampling a sequence of $m$ \textit{Clifford} gates (see \secref{sec:gate_set}). The sequence
is applied to the initial state, followed by its inverse.
Finally a two-outcome POVM measurement is done to calculate the fidelity between initial state and the output state, followed by
classical post-processing. The RB protocol discretizes time so that it is measured in the number of gates and it then averages over many sequences of each length $m$.
More formally, a $4$-step RB protocol consist of
\begin{enumerate}
\item Generate $K_m$ sequences of $m$ quantum operations $C_{i_j}$ with $i\in [1,m]$ and $j\in[1,K_m]$, where $i$ indexes over the sequence of operations, and $j$ over the statistical samples. These operations are randomly chosen from the Clifford group, and a $m+1$-th operation is chosen that cancels the first $m$ operations such that the net operation is the identity. The operations can be chosen from the $2$-,$4$- or $2^n$-dimensional Clifford groups, depending on whether we are benchmarking single-, two- or $n$-qubit operations \cite{mckay2019three}. These operations come with some error, which is modeled with linear operators $\Lambda_{i_j,j}$, so that the full sequence of $m$ operations is given by
\begin{equation}
    \label{eq:ch5:gate_error}
    S_{K_m} = \bigcirc^{m+1}_{j=1} (\Lambda_{i_j,j} \circ C_{i_j})
\end{equation}
    Here, $\circ$ denotes composition and $\bigcirc$ represents composition of the terms defined with index $j$. 
    \item For each sequence we find the fidelity with the initial state by measuring $\textrm{Tr}[E_{\psi} S_{K_m}( \rho(\psi))]$, where $\rho(\psi)$ is the initial state (with preparation errors) and $E_{\psi}$ is a POVM measurement operator corresponding to the measurement including noise. Without noise, this would be the projector $E_{\psi}=\ketbra{\psi}{\psi}$.
    \item Average over the $K_m$ statistical samples to find the sequence fidelity $\label{eq:ch5:sequence_fidelity}
    F(m,\psi) = \textrm{Tr}[E_{\psi} S_m(\rho(\psi))]$
    where $S_m$ is the mean over the operations $S_{K_m}$.
    \item Fit the data with the function
    \begin{equation}
        \label{eq:ch5:RB_fit}
    F_{fit}(m,\psi) = A_0p^m+B_0,
    \end{equation}
    where we have assumed the errors are independent of gate and time. This is not a fundamental assumption, but can be relaxed~\cite{magesan2011scalable,magesan2012characterzing}. The average gate error is here given by $\epsilon_{RB} = 1-p-(1-p)/2^n$, and the constants $A_0$ and $B_0$ absorb the SPAM errors.
\end{enumerate}

The operations $C_{i_j}$ are chosen from the Clifford group, because these are relatively easy to perform on quantum hardware, and because the final $m+1$-th operation that undoes the sequence can easily be pre-computed on a classical computer. Averaging over the Clifford group (or any other finite group) also has the property that even though the real noise-channel would be more complicated than the purely depolarizing one, the average over the group will still give rise to an exponential decay.

These gate errors extracted from randomized benchmarking can be used to compare the quality of quantum gates, and to estimate that an algorithm of depth $\sim 1/\epsilon_{RB}$ gates can be run on the device before only statistical noise is output. The intuition behind the RB protocol is that a (purely) depolarizing channel will cause exponential decay of an excited state over time.

Simultaneous randomized benchmarking  (SRB) has been proposed to acquire information about crosstalk and undesired coupling between the neighbouring qubits~\cite{gambetta2012characterization}. RB has also been extended for gatesets that do not form a Clifford group~\cite{cross2016scalable,brown2018randomized,hashagen2018real,francca2018approximate,carignan2015characterizing, gambetta2012characterization,harper2017estimating}. In such cases, the expression for $F_{fit}(m,\psi)$ does not follow equation \eqref{eq:ch5:RB_fit}~\cite{helsen2019new}. Employing concepts from representation theory, an extension of RB has been proposed to extract the fidelity for a broad category of gatesets, including T-gate \cite{helsen2019new}. A practically scalable protocol called cycle benchmarking was developed lately to characterize local and global errors for multi-qubit quantum computers~\cite{erhard2019characterizing}.

\subsection{Quantum volume}
\label{subsec:quantum_volume}
To further refine the concept of the computational power of a quantum computer from just qubit count and gate-errors, the IBM Quantum team introduced the ``quantum volume''~\cite{moll2018quantum,cross2019validating}. It is one of the widely accepted metrics for benchmarking
NISQ-era quantum computers. As mentioned earlier, one can not rank
quantum computers based on the number of qubits alone. Quantum
volume gives a rough estimate of the number of effective qubits a
quantum computer has based on their performance on the ``heavy output
generation problem''. The heavy output generation problem is related
to the random circuit sampling task used in Google's quantum
supremacy experiment. Quantum volume treats the depth and width of a quantum
circuit at the equal footing. Hence, its estimation depends on the largest square-sized circuit, which can successfully implement the
heavy output generation problem. The quantum computer's performance
also depends on its software stack, for example, compiler, and thus
quantum volume can increase with the improvements in the software
stack. 

The quantum volume benchmark can be thought analogous to the classical LINPACK benchmark~\cite{dongarra1987linpack}. Like the LINPACK benchmark, it is architecture-agnostic and provides a single real number metric based on the quantum computer's performance for a model problem, i.e., heavy output generation problem. 

More formally, quantum volume can be defined in the following terms.
Given an $n$ qubit quantum computer with the largest achievable \textit{model
circuit} depth $d(m)$ for \textit{model circuit} width $m\in\left\{ 1,2,\cdots,n\right\} $
such that the probability of observing a \textit{heavy output} for
a random selection of model circuit  is strictly greater than $2/3$, the quantum volume $V_{Q}$   is defined as~\cite{cross2019validating} 
\begin{equation}
\log_{2}V_{Q}=\argmax_{m} \min \left(m,d(m)\right).
\label{eq:quantum_volume}
\end{equation}
Intuitively speaking, quantum volume estimates the largest square
random quantum circuit which the quantum computer can successfully
implement the so-called heavy output generation problem. To conclude the
discussion, it remains to describe the ``model circuit'' and the
``heavy output generation problem''.

\begin{figure}[ht!]
    \centering
    \includegraphics[width=\columnwidth]{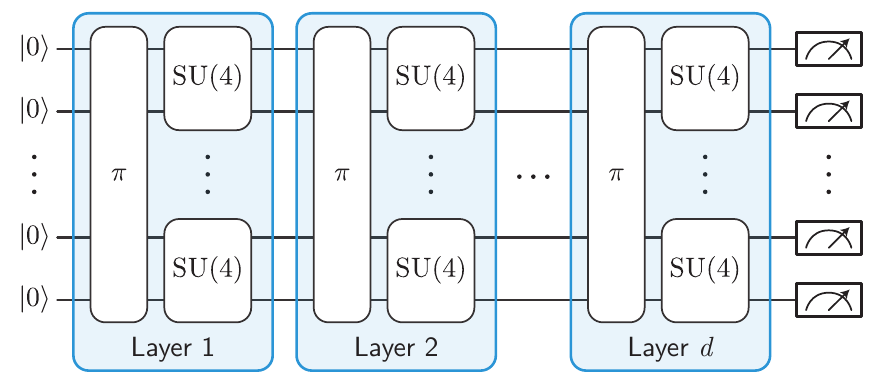}
    \caption{Model circuit for the quantum volume benchmark. Each layer consists of random permutations of qubit labels, followed by application of two-qubit haar-random unitaries. Inspired by \cite{cross2019validating}.}
    \label{fig:QV_model_circuit}
\end{figure}

The model circuit with depth $d$ and width $m$ for estimating quantum
volume is given by $d$-layered sequence $U=U^{(d)}U^{(d-1)}\cdots U^{(1)}$
where layer $t$ consists of random permutations $\pi_{t}\in S_{m}$
applied to qubit labels, followed by the tensor product of Haar-random
two-qubit unitaries from $SU(4)$. If the model circuit width $m$
is odd, one of the qubits is left idle in every layer. See Figure
\ref{fig:QV_model_circuit}  for a pictorial description.

Given a model circuit $U$ with width $m$, the ideal output distribution
over bit strings $x\in\left\{ 0,1\right\} ^{m}$ is given by$
P_{U}(x)=\vert\langle x\vert U\vert0\rangle\vert^{2}.$

One can arrange the probabilities for various bitstrings in ascending
order in a set $\mathbb{P}=\left\{ p_{0}\leq p\leq\cdots\leq p_{2^{m}-1}\right\}.$
The median of the set $\mathbb{P}$is given by $p_{med}=\frac{p_{2^{m-1}}+p_{2^{m-1}-1}}{2}.$
The Heavy outputs are defined as
$
H_{U}=\left\{ x\in\left\{ 0,1\right\} ^{m}\vert p_{U}(x)>p_{med}\right\}.$
The goal of the heavy output problem is to sample a set of strings
such that at least $2/3$ are heavy output. For an ideal quantum circuit,
the expected heavy output probability asymoptotically tends to $\sim0.85$.
For a completely depolarized device, it is $\sim0.5.$ 

On the target
system, one implements $\tilde{U}$ by using a circuit compiler with
native gate set such that $1-F_\text{avg}\left(U,\tilde{U}\right)\leq\epsilon\leq1$
for some approximation error $\epsilon$, where $F_\text{avg}$ is average
gate fidelity, as defined in Ref.~\cite{horodecki1999general}. The role of circuit compiler
is crucial in the aforementioned step. Suppose the observed distribution
for the implemented circuit $\tilde{U}$ of the model circuit $U$
is $g_{u}(x).$ The probability of sampling heavy output is given
by
\begin{equation}
h_{U}=\sum_{x\in H_{U}}q_{U}\left(x\right).
\label{eq:heavy_output_probability}
\end{equation}
For a randomly selected circuit of depth $d$, the probability of sampling
a heavy output is given by
\begin{equation}
h_{d}=\int_{U}h_{U}dU.\label{eq:depth_d_sampling}
\end{equation}
The term $d(m)$ in \eqref{eq:quantum_volume} is equal to
the largest depth $d$ for model circuit of width $m\in\left\{ 1,2,\cdots,n\right\} $
such that $h_{d}>\frac{2}{3}.$ 

The quantum volume benchmark requires simulation of the model circuit's heavy output generation problem on a classical computer. It, hence, is not a scalable method as the quantum volume increases. Moreover, the special treatment for square circuits is not entirely justified. Investigations are needed to devise other interesting benchmarks. A benchmark for rectangular circuits has also been proposed in the literature~\cite{blume2020volumetric}.

At the time of writing, Honeywell's system model H1  has achieved $\log_{2}V_{Q}=9$~\cite{Honeywel2021online}, and IBM quantum device named ``IBM Montreal'' has demonstrated $\log_{2}V_{Q}=6$~\cite{ibmnews}.

\subsection{Cross-entropy benchmarking}\label{subsec:XEB}

The linear cross-entropy benchmarking is a statistical test used by Google in their quantum supremacy experiment~\cite{arute2019quantum,neill2018blueprint}. It measures
how often high-probability bitstrings are sampled in an experimental
scenario. Suppose we perform a sampling task and obtain bitstrings
$\left\{ x_{j}\right\} _{j}$ via measurement on a given $m$-qubit
circuit $\mathcal{C}_{E}.$ The linear cross-entropy benchmarking
fidelity is given by
\begin{equation}
\mathcal{F}_{XEB}=2^{m}\left\langle P\left(x_{j}\right)\right\rangle _{j}-1.\label{eq:Linear_XEB}
\end{equation}
Here, the average $\left\langle .\right\rangle _{j}$ is over the
experimentally observed bitstrings $\left\{ x_{j}\right\} _{j}$ and
$P\left(x_{j}\right)$ denotes the probability of observing bitstring
$x_{j}$ for the ideal circuit version of $\mathcal{C}_{E}$. In other
words, $P\left(x_{j}\right)$ denotes the ideal probability of the
generated sample $x_{j}.$ Since one can not have an ideal circuit in
practice, $P\left(x_{j}\right)$ are calculated using a classical
computer simulation of the ideal circuit. $\mathcal{F}_{XEB}$ compares
how often a bitstring $x_{j}$ is observed experimentally with its
classically simulated ideal probability. For the ideal case, $\mathcal{F}_{XEB}$
approaches unity for a large number of qubits. On the other hand, it
is equal to zero for uniform distribution. As the circuit's nose
grows, $F_{XEB}$ decreases and approaches zero. Since the probabilities
$P\left(x_{j}\right)$ are calculated via classical simulation; it
renders the computation of $\mathcal{F}_{XEB}$ intractable in the
supremacy regime. The classical hardness of spoofing linear cross-entropy
benchmarking was studied by Aaranson and Gunn~\cite{aaronson2019classical}, where they suggested
the absence of any efficient classical algorithm for the aforementioned
task.

\subsection{Application benchmarks}

While hardware benchmarks, such as randomized benchmarking or quantum volume, provide valuable insight into the performance of quantum devices, they may not well represent or predict the performance of VQAs which employ structured circuits. Application benchmarks were developed to complement hardware benchmarks and provide a more complete picture of both the performance and (near-term) utility of quantum devices.
These benchmarks consist executing experimental demonstrations of VQA instances that can be compared to classically computed exact results. 
Examples of application benchmarks can be found in Refs.~\cite{dallaire2018quantum, benedetti2019generative, arute2020hartree, karamlou2020analyzing}. In particular, \cite{arute2020hartree} demonstrated VQE experiments for hydrogen chain binding energy and diazene isomerization mechanism with PQC sizes as big as $12$ qubits and $72$ two-qubit gates.

As a specific example of an application benchmark, quantum circuits that diagonalize spin Hamiltonians have been proposed in recent years \cite{verstraete2009quantum,schmoll2017kitaev,cervera2018exact}. By comparing the results obtained from the quantum device with the analytical solution, one can discern the performance of the computation for a specific purpose experiment. Small experiments have shown that gate fidelities and decoherence times alone do not provide a complete picture of the noise model \cite{cervera2018exact}. 

In that direction, authors of \cite{dallaire2020application} proposed a figure-of-merit called the \emph{effective fermionic length} to quantify the performance of a NISQ device in which the application-at-hand is estimating the energy density of the one-dimensional Fermi-Hubbard model over increasing chain lengths. Theoretically, as the chain length increases, the energy density should approach the infinite chain limit. In practice, the NISQ device will accrue some level of noise and decoherence, which will cause the computed energy density to diverge past some chain length. The maximum chain length after which noise and decoherence start degrading the algorithm performance reveals the ``limit'' of the quantum device in carrying out related algorithms. Ref.~\cite{dallaire2020application} abstracts this idea to redefine an application benchmark as a way to systematically test the limits of a quantum processor using exactly solvable VQA instances that can also be scaled to larger system sizes (e.g chain length in Ref.~\cite{dallaire2020application} or number of preprocessing steps in Ref.~\cite{karamlou2020analyzing}).

Generative models such as the QCBM (see \secref{sec:unsupervised_learning}) can serve as benchmarks for NISQ devices~\cite{leyton2019robust,hamilton2019generative,zhu2019training}. Here, the measurement output of hardware efficient variational ans\"ate are used to represent different types of distributions and study the effect of noise and hardware limitations on the result.

In addition to VQAs, one can analyze more fundamental benchmarks, such as the ability of NISQ devices to violate local-realism by means of Mermin inequalities \cite{alsina2016experimental} or the entanglement power of the devices by trying to construct maximal entangled states \cite{wang201816, cervera2019quantum}. 

\section{Outlook}\label{ch:outlook}

In the last decade, quantum computing has experienced notable progress in applications, experimental demonstrations, and theoretical results. The number of papers in quantum computation, particularly in NISQ applications, is growing almost exponentially. Several reasons explain this community drive, one of those being tremendous improvements in quantum hardware. 
Quantum computing is a relatively young field in science and, as such, there is plenty of room for pioneering research and discoveries. This fact, together with the theoretical, practical, and experimental challenges (several of them covered in this review), has strengthened the motivation for an open-source strategy in the field. Nowadays, many universities and research centers subscribe to an open-access policy that pushes towards the free and open-source publication of all computational tools, data, and programs used in their research. These policies have proved valuable for rapid scientific development as well as for democratizing community knowledge. This way of thinking has percolated through academia walls and it has been introduced into several private companies, not just for its advantage, but also because it facilitates the continuous healthy flow of quantum computing researchers to themselves (and, in some cases, resulting in foundation of startups). Consequently, there is a rich open quantum computing ecosystem composed of universities, institutes, big corporations, startups, and uncountable individual enthusiasts. 
Another product of the symbiosis between academia and the private sector is \textit{cloud quantum computing}. Companies 
are offering access to their hardware remotely, in some cases at zero-cost for their small prototypes and simulators. On the one hand, scientists and quantum computing enthusiasts around the world have the opportunity to experience  real quantum devices from their homes. On the other hand, this increases the chances of finding real-world applications in quantum computation and solving the current challenges of this field. The proliferation of open-source quantum computing languages, simulators, and tools (detailed in \secref{sec:software}) have burgeoned many user communities. Various international initiatives have been set up to attract quantum computing talent, and the private sector's involvement is ramping up.  Several non-profit initiatives are also encouraging the use and development of these tools \cite{UnitaryFund,QOSF}.

Experimental realizations of quantum computation, although in the early stages, have interested many communities in this quantum information subfield. Healthy competition has also arisen between the classical and quantum computing branches. Classical computational scientists have put their efforts into moving the quantum advantage frontier further, raising the bar to claim that a quantum algorithm shows a significant speed-up. Along that direction, an off-shoot is an effort in \textit{dequantization}, first exhibited in the case of recommendation-systems, to devise quantum-inspired classical algorithms that are nearly as fast as their quantum counterparts~\cite{tang2019quantum}. Such attempts have eliminated examples of speed-up for some problems in linear algebra. So far, dequantized machine learning algorithms have been developed for recommendation systems~\cite{tang2019quantum}, principal component analysis and supervised clustering~\cite{tang2018quantum}, stochastic regression~\cite{gilyen2018quantum} and low-rank linear systems \cite{chia2018quantum,arrazola2019quantum}. 

Since NISQ devices are inherently noisy, analysis similar to~\cite{napp2019efficient,zlokapa2020boundaries,zhou2020limits} will be required to find out how much noise a NISQ algorithm can endure until its classical simulation becomes efficient. This is crucial in order to understand the boundary where quantum computers provide an advantage. Investigating the potential of NISQ algorithms using ideas from quantum foundations such as contextuality and entanglement are helpful in that respect \cite{deutsch2020harnessing,bharti2020machine}. More theoretical results as the ones presented in \cite{farhi2016quantum,lloyd2018quantum,biamonte2019universal, bravyi2020quantum,bravyi2021classical,bouland2021noise,movassagh2019quantum} may also prove valuable. It is also imperative to develop strategies that help us bypass complicated measurements involving controlled multi-qubit unitaries~\cite{mitarai2019methodology}.
For machine learning tasks, ideas similar to~\cite{harrow2020small} would be valuable.

Another fascinating frontier that needs to be investigated in the next few years, we believe, is quantum and classical certification schemes for quantum devices and quantum computation ~\cite{eisert2020quantum}. The intractability of quantum computation by classical devices poses the challenge to verify the correct functioning of the quantum devices as well as the correctness of the final output~\cite{eisert2020quantum}. The existence of multiple quantum computing platforms requires new methodologies and figures of merit to benchmark and compare these devices. Other works are being proposed in that direction \cite{kyaw2020quantum,kottmann2020quantum}, as well as the development of benchmarking measures discussed in \secref{sec:benchmark}. Ideas from complexity theory~\cite{mahadev2018classical,metger2020self} and quantum foundations~\cite{bharti2019local,bharti2019robust} could be valuable in this direction.

At the moment of documenting this review, there is no known demonstration of industrially relevant quantum advantage. Quantum computing is still in its early days, and so far a useful quantum computer is missing. The potential of NISQ devices is not fully understood, and a lot of rigorous research is required to release the power of the early quantum computers. However, a number of experiments overcoming classical computational resources have been performed and many theoretical and practical tools are being used and developed, as explained in \secref{ch:lemon}.

\subsection{NISQ goals}

We expect experimental pursuit in the NISQ era would focus on the design of quantum hardware with a larger number of qubits, and gates with lower error rates capable of executing deeper circuits. Along the way, one of the goals is to demonstrate quantum advantage for practical use cases. If the NISQ paradigm is not powerful enough to exhibit any quantum advantage, theoretical pursuits would be required to understand its limitations. The prime direction of the NISQ and near-term era is to engineer the best possible solution with the limited quantum resources available. The tools and techniques invented during this period could be valuable in the fault-tolerant era as well.

To conduct a successful demonstration of quantum advantage, the right blend of the following three crucial components is required:
\begin{enumerate}
\item \textit{Hardware development:} The design of quantum computers with more qubits, lesser error rates, longer coherence times, and more connectivity between the qubits will be one of the top priorities in the NISQ era. Intensive research in new qubits developments, quantum optimal control and material discovery will be indispensable for both universal programmable quantum computers or special-purpose ones.
A way to scale up the number of qubits present in a quantum platform is to design a novel qubit which has built-in autonomous quantum error correction down to the hardware level \cite{paz1998continuous,chamberland2020building} or protected novel qubit \cite{douccot2012physical,nataf2011protected,kyaw2019towards} which is robust against specific noises in the hardware.
As a quantum processor size grows, there is a tremendous need to store quantum information during quantum information processing \cite{kyaw2014z_2,kyaw2015scalable,kyaw2015creation}. 
Even miniaturizing microwave circulator onto the superconducting chip \cite{mahoney2017chip,chapman2017widely} can be seen as a mean to scale up the quantum platform although it has nothing to do with novel qubit design.

\item \textit{Algorithm design:} To harness the potential of noisy but powerful quantum devices, we expect breakthroughs on the algorithm frontier. Algorithms with realistic assumptions, as the ones mentioned in \secref{sec:compilers}, regarding device capabilities will be favored. To lessen the effect of noise, progress towards the design of error mitigation algorithms is expected. Efforts have to be made to develop algorithms that harness the problem's structure in the best possible manner and map it to the given hardware in efficient ways, such as in \secref{sec:digi-ana}. VQA with better expressibility and trainability will also be helpful. 
\item \textit{Application problem:} We have discussed the existing applications of NISQ devices in  many areas in \secref{ch:application}. Collaborations between experts with domain knowledge from these fields and quantum algorithm researchers will be required more and more to develop the field and integrate quantum computation into industrial workflows. New collaborations might reveal difficult problems for classical computers that are well suited for NISQ devices. It is not clear yet which applications will be the first ones to witness quantum advantage, though there is plenty of speculation and opinions. 
\end{enumerate}

\subsection{Long-term goal: fault-tolerant quantum computing} \label{subsec:fault-tolerance}

Noise is regarded as one of the most prominent threats to a quantum computer's practical realization. In 1995, Peter Shor established that by encoding quantum information redundantly using extra qubits, one could circumvent the effect of noise~\cite{shor1995scheme}. The quantum information is spread over multiple physical qubits to generate a logical qubit~\cite{calderbank1996good,knill1997theory, gottesman1997stabilizer,shor1995scheme}. Most of the transformative algorithms such as Shor's factoring algorithm, Grover search algorithm, and HHL require error-corrected qubits for their execution. Soon after Shor's error-correcting code, many others were developed.
Some of the famous error-correcting codes are stabilizer and topological error-correcting codes \cite{terhal2015quantum,fowler2012surface}. While the stabilizer code utilizes extra qubits to protect the logical qubit, topological codes employ a set of qubits positioned on a surface, such as a torus, in a lattice structure.

Over the years, quantum error correction has evolved as a subfield of quantum computation and has transformed from a theoretical pursuit to a practical possibility. The process of detecting and correcting errors can be, itself, prone to noise. Thus error correction alone does not guarantee the prospect of storing or processing quantum information for an arbitrarily long period. The aforesaid issue can be tackled by utilizing the \textit{Quantum Fault-Tolerant threshold theorem}. Informally speaking, it is possible to execute arbitrarily large quantum computation by arbitrarily suppressing the quantum error rate, given the noise in the individual quantum gates are below a certain threshold~\cite{aharonov2008fault}.  If one wants to simulate an ideal circuit of size $N$, the size of the noisy quantum circuit for fault-tolerant quantum computation scales $O\left(N\left(\log N\right)^{c}\right)$, for some constant $c$, given the noisy circuit is subjected to stochastic noise strength $p<p_{c}$ for some noise threshold $p_{c}$ \cite{terhal2015quantum}. This theorem rises some practically relevant questions such as \textit{i)} How high is $p_{c}$; \textit{ii)} what is the value of the constant $c$; and \textit{iii)} what is the value of the multiplicative constant in $O\left(.\right)$. These questions determine the practicality of any fault-tolerant quantum computation scheme~\cite{terhal2015quantum}.

Looking forward, lowering the noise level will be a critical challenge. Though the problem is demanding, significant progress has been made recently at the algorithmic as well as hardware frontier~\cite{lidar2013quantum,terhal2015quantum,campagne2020quantum, noh2020fault}. Quantum error-correcting codes amenable to architectures with limited qubit connectivity have also been proposed~\cite{chamberland2020topological}. As we transition towards fault-tolerant quantum computing, partial quantum error correction demonstrations such as exponential suppression of bit or phase errors~\cite{ai2021exponential} and approximate quantum error correction schemes~\cite{leung1997approximate, faist2020continuous} become highly relevant.   Recently, Monroe and Brown's groups have confirmed the first-ever fault-tolerant operation on a logical qubit~\cite{egan2020fault}.

\vspace{1cm}

We are at an exciting juncture in the history of computing. Completely new kinds of computers that were once only figments of imagination are rapidly becoming a reality. The NISQ era offers fantastic opportunities to current and future researchers to explore the theoretical limits of these devices and discover practical and exciting applications in the near-term. Theoretical investigations and experimental challenges will help us to comprehend quantum devices power and build better algorithms. The success of the field lies in the hands of the researchers and practitioners of the area, so we encourage everyone with interest to join the effort.

\section*{Acknowledgements}

A.A.-G. acknowledges the generous support from Google, Inc. in the form of a Google Focused Award. This work was supported by the U.S. Department of Energy under Award No. DESC0019374 and the U.S. Office of Naval Research (ONS506661). A.A.-G. also acknowledges support from the Canada Industrial Research Chairs Program and the Canada 150 Research Chairs Program.
T.H. is supported by a Samsung GRC project and the UK Hub in Quantum Computing and Simulation, part of the UK National Quantum Technologies Programme with funding from UKRI EPSRC grant EP/T001062/1. 
L.-C.K and K.B  acknowledge the financial support from the National Research  Foundation  and  the  Ministry  of  Education, Singapore. 
We thank Michael Biercuk, Naresh Boddu, Zhenyu Cai, Sam Gutmann, Edward Farhi, Rahul Jain, Dax Koh, Alejandro Perdomo-Ortiz and Mark Steudtner for interesting discussions, feedback and comments.

\bibliographystyle{apsrmp4-1}
\bibliography{NISQ_Review}

\clearpage
\onecolumngrid

\begin{appendices}

\section{NISQ algorithms and tools tables} \label{app:tables}

\subsection{Tables of applications}

\begin{table*}[!h]
    \centering
     \begin{tabularx}{1\linewidth}{L{0.5}|L{1}}
    \hline
    \hline
    \textbf{Algorithm/Application} & \textbf{Proposed implementations} \\
    \hline
    \multicolumn{2}{c}{Variational quantum eigensolver (VQE) and related solvers } \\
    \hline
    VQE & \cite{mcclean2016theory,peruzzo2014variational,wecker2015progress}\\
    \hline
    Adaptive VQE & \cite{grimsley2019adaptive, ryabinkin2018qubit, sim2020adaptive, kottmann2020feasible, zhang2020mutual, tang2019qubit, gomes2021adaptive, stair2021simulating} \\
    \hline
    IQAE & \cite{bharti2020iterative,bharti2020quantumeigensolver} \\
    \hline
    Krylov approaches & \cite{stair2020multireference, huggins2020non, jouzdani2020hybrid} \\
    \hline
    Imaginary time evolution & \cite{motta2020determining, sun2020quantum, mcardle2019variational,bharti2020quantum}\\
    \hline
    Full quantum eigensolver (FQE) & \cite{wei2020full} \\
    \hline
        \multicolumn{2}{c}{VQE for excited states} \\
    \hline
    Folded spectrum & \cite{peruzzo2014variational,ryabinkin2018constrained}\\
        \hline
    Orthogonally constrained VQE & \cite{higgott2019variational, lee2018generalized, kottmann2020feasible}\\
        \hline
    Subspace expansion and linear-response based& \cite{mcclean2017hybrid, takeshita2020increasing, ollitrault2020quantum} \\
            \hline
    Subspace-search VQE & \cite{nakanishi2019subspace}\\
                \hline
    Multistate contracted VQE & \cite{parrish2019quantum}\\
                \hline
    Fourier transform of evolutions & \cite{aleiner2020accurately,roushan2017spectroscopic}\\
                \hline
    WAVES & \cite{santagati2018witnessing}\\
    \hline
    Adiabatically-Assisted & \cite{mcclean2016theory,garcia2018addressing} \\
    \hline
    Projected VQE & \cite{stair2021simulating} \\
    \hline
        \multicolumn{2}{c}{Hamiltonian simulation} \\
        \hline
        Variational quantum simulator (VQS) & \cite{li2017efficient,mcardle2019variational,yuan2019theory,endo2020variational,kubo2020variational,benedetti2020hardware}\\
            \hline
        Subspace VQS & \cite{heya2019subspace}\\
            \hline
         projected-Variational Quantum Dynamics (p-VQD) & \cite{barison2021efficient,otten2019noise}\\
            \hline
        Variational fast forwarding & \cite{cirstoiu2020variational,commeau2020variational}\\
                    \hline
        Quantum assisted simulator & \cite{bharti2020quantum} \\
            \hline
        \multicolumn{2}{c}{Quantum information scrambling and thermalization} \\
    \hline
    Scrambling & \cite{joshi2020quantum,landsman2019verified,holmes2020barren,mi2021information} \\
    \hline
    Thermal state & \cite{verdon2019quantum} \\
    \hline
        \multicolumn{2}{c}{Open quantum systems} \\
            \hline
    Generalized VQS &  \cite{endo2020variational,yuan2019theory,liu2020solving}\\
        \hline
    Generalized quantum assisted simulator & \cite{haug2020generalized} \\
            \hline
     Trotter simulation & \cite{koppenhofer2020quantum,hu2020quantum}\\
    \hline
        \multicolumn{2}{c}{State preparation} \\
    \hline
    Non-equilibrium steady state & \cite{yoshioka2020variational}\cite{kreula2016non,jaderberg2020minimum,endo2020calculation}\\
    \hline
    Gibbs-state & \cite{endo2020variational,chowdhury2020variational,haug2020generalized}\\
    \hline
    Many-body ground state & \cite{ho2019efficient,ho2019ultrafast,wauters2020polynomial}\\
    \hline
        \multicolumn{2}{c}{Quantum autoencoder} \\
    \hline
    Quantum autoencoder & \cite{romero2017quantum,bravo2020quantum2,pepper2019experimental,huang2020experimental,huang2020realization,bondarenko2020quantum}\\
    \hline
        \multicolumn{2}{c}{Quantum computer-aided design} \\
    \hline
    Optical setups &  \cite{kottmann2020quantum} \\
    \hline
    Superconducting circuits & \cite{kyaw2020quantum} \\
    \hline
    \hline
    \end{tabularx}
    \caption{NISQ algorithms for Many-body physics and chemistry applications from \secref{sec:chemistry}.}
    \label{tab:many_body}
\end{table*}

\begin{table*}[h!]
    \centering
     \begin{tabularx}{1\linewidth}{L{0.5}|L{1}}
    \hline
    \hline
    \textbf{Algorithm/Application} & \textbf{Proposed implementations} \\
    \hline
    \multicolumn{2}{c}{Supervised learning} \\
    \hline
    Quantum kernel methods & \cite{havlivcek2019supervised, kusumoto2019experimental, schuld2019quantum, schuld2020measuring,peters2021machine}\\
    \hline
    Variational quantum classifiers (VQC) & \cite{farhi2018classification, mitarai2018quantum, vidal2019input, lloyd2020quantum, perez2020data, schuld2020circuit, schuld2020effect}\\
    \hline
    Encoding strategies in VQA & \cite{mitarai2019generalization,cervera2020meta}\\
    \hline
    Quantum reservoir computing & \cite{fujii2017harnessing, nakajima2019boosting, nokkala2020gaussian, ghosh2019quantum, negoro2018machine, chien2020custom, mitarai2018quantum}\\
    \hline
    Supervised QUBO classifier & \cite{li2018quantum} \\
    \hline
    \multicolumn{2}{c}{Unsupervised learning} \\
    \hline
    Quantum Boltzmann machines (QBM) &  \cite{amin2018quantum,kieferova2017tomography,zoufal2020variational}\\
    \hline
    Energy-based models (e.g., RBMs) &
    \cite{benedetti2016estimation,benedetti2017quantum,benedetti2018quantum}\\
    \hline
    Quantum circuit Born machines (QCBM) & \cite{zhu2019training,benedetti2019generative,liu2018differentiable,leyton2019robust,hamilton2019generative,rudolph2020generation,coyle2020quantum,alcazar2020classical}\\
    \hline
    Quantum generative adversarial networks (QGAN) & \cite{dallaire2018quantum,lloyd2018quantumgenerative, zeng2019learning,romero2019variational,situ2020quantum,hu2019quantum}\\
    \hline
    \multicolumn{2}{c}{Reinforcement learning} \\
    \hline
    Reinforcement learning & \cite{chen2020variational,lockwood2020reinforcement,lockwoodplaying,crawford2016reinforcement,lamata2017basic,cardenas2018multiqubit,albarran2020reinforcement,yu2019reconstruction,jerbi2019quantum}\\
    \hline
    \end{tabularx}
    \caption{NISQ algorithms for machine learning applications from \secref{sec:QML}.}
    \label{tab:ml_algorithms}
\end{table*}
\begin{table*}[h!]
    \centering
     \begin{tabularx}{1\linewidth}{L{0.5}|L{1}}
    \hline
    \hline
    \textbf{Algorithm/Application} & \textbf{Proposed implementations} \\
    \hline
    Max cut & \cite{farhi2014quantum,otterbach2017unsupervised,hastings2019classical,bravyi2019obstacles, headley2020approximating,satoh2020subdivided} \\
    \hline
    Max clique & \cite{banchi2020molecular,arrazola2018using}\\
    \hline
     Triangle finding & \cite{metwalli2020finding}\\
    \hline
    Maximum independent set & \cite{choi2020quantum,saleem2020max,utkarsh2020solving} \\
    \hline
    Max hafnian & \cite{arrazola2018quantum} \\
    \hline
    Vertex cover & \cite{cook2019quantum} \\
    \hline 
    Exact cover & \cite{bengtsson2020improved,vikstaal2020applying,garcia2018addressing} \\
    \hline 
    Knapsack & \cite{grandrive2019knapsack} \\
    \hline
    Graph multi-coloring & \cite{oh2019solving} \\
    \hline
    Fault diagnosis & \cite{perdomo2015quantum,perdomo2019readiness} \\
    \hline
    Bayesian networks & \cite{o2015bayesian} \\
    \hline
    Protein folding & \cite{benedetti2016estimation,perdomo2008construction,babbush2014adiabatic,fingerhuth2018quantum,robert2021resource,babej2018coarse,perdomo2012finding}  \\
    \hline
    \end{tabularx}
    \caption{NISQ algorithms for combinatorial optimization from \secref{sec:optimization}.
    }
    \label{tab:combinatorial}
\end{table*}

\begin{table*}[h!]
    \centering
     \begin{tabularx}{1\linewidth}{L{0.5}|L{1}}
    \hline
    \hline
    \textbf{Algorithm/Application} & \textbf{Proposed implementations} \\
    \hline
    Factoring &  \cite{anschuetz2019variational,karamlou2020analyzing} \\
    \hline
    SVD & \cite{bravo2020quantum,wang2020variational} \\
    \hline
    Linear systems & \cite{bravo2019variational,huang2019nearterm,xu2019variational2} \\
    \hline
    Non-linear differential equations & \cite{lubasch2020variational,haug2020generalized,gaitan2020finding,kyriienko2020solving} \\
    \hline
    Semidefinite programming & \cite{bharti2021nisq} \\
    \hline
    \end{tabularx}
    \caption{NISQ algorithms for numerical solvers applications from \secref{sec:numerical_solvers}.}
    \label{tab:numerical_solvers}
\end{table*}
\begin{table*}[h!]
    \centering
     \begin{tabularx}{1\linewidth}{L{0.5}|L{1}}
    \hline
    \hline
    \textbf{Algorithm/Application} & \textbf{Proposed implementations} \\
    \hline
    Portfolio optimization & \cite{alcazar2021enhancing,bouland2020prospects,cohen2020portfolio,marzec2016portfolio,rosenberg2016solving,venturelli2019reverse,egger2020quantum} \\
    \hline
    Fraud detection & \cite{egger2020quantum,egger2020warm,zoufal2020variational} \\
    \hline
    Option pricing & \cite{kubo2020variational} \\
    \hline
    \end{tabularx}
    \caption{NISQ algorithms for finance applications from  \secref{sec:finance}}
    \label{tab:finance}
\end{table*}
\begin{table*}[h!]
    \centering
     \begin{tabularx}{1\linewidth}{L{0.5}|L{1}}
    \hline
    \hline
    \textbf{Algorithm/Application} & \textbf{Proposed implementations} \\
    \hline
    \multicolumn{2}{c}{Quantum foundations} \\
    \hline
    Bell inequalities & \cite{alsina2016experimental} \\
    \hline
    Contextuality & \cite{kirby2019contextuality,kirby2020contextual} \\
    \hline
    Variational consistent history (VCH) & \cite{arrasmith2019variational} \\
    \hline
    \multicolumn{2}{c}{Quantum optimal control} \\
    \hline
    Quantum optimal control & \cite{magann2020pulses, li2017hybrid, dive2018situ, lu2017enhancing, policharla2020algorithmic} \\
    \hline
    \multicolumn{2}{c}{Quantum metrology} \\
    \hline
    Quantum metrology & \cite{beckey2020variational, kaubruegger2019variational, koczor2020variational, ma2020adaptive, meyer2020variational} \\
    \hline
    \multicolumn{2}{c}{Fidelity estimation} \\
    \hline
    Fidelity estimation & \cite{cerezo2020fidelity} \\
    \hline
    \multicolumn{2}{c}{Quantum error correction (QEC)} \\
    \hline
    Quantum variational error corrector (QVECTOR) & \cite{johnson2017qvector}\\
    \hline
    Variational circuit compiler for QEC & \cite{xu2019variational} \\
    \hline
    \multicolumn{2}{c}{Nuclear physics} \\
    \hline
    Schwinger model & \cite{hauke2013quantum,martinez2016real, klco2018quantum,kokail2019self,avkhadiev2020accelerating} \\
    \hline
    High-energy processes & \cite{dumitrescu2018cloud,liu2020quantum,roggero2020quantum,wu2020application,chen2020quantum,perez2020determining} \\
    \hline
    \multicolumn{2}{c}{Entanglement properties} \\
    \hline
    Schmidt decomposition & \cite{bravo2019variational,wang2020detecting} \\
    \hline
    Multipartite entanglement & \cite{perez2020measuring} \\
    \hline
    Entanglement spectrum &  \cite{larose2019variational,cerezo2020variational} \\
    \hline
    \end{tabularx}
    \caption{NISQ algorithm for other applications listed in  \secref{sec:other_app}.}
    \label{tab:other_app}
\end{table*}

\clearpage
\subsection{Table of software packages}

\begin{table*}[th!]
    \centering
 \begin{tabularx}{1\linewidth}{L{1.8}|C{0.6}|C{0.5}|C{0.8}|C{1.3}}
 \hline
 \hline
\multicolumn{1}{c|}{\textbf{Name}} & \textbf{Language} &  \textbf{Hardware} & \textbf{Multi-platform} & \textbf{Built-in applications} \\
 \hline
 Cirq \cite{cirq} &  python  & Yes & No & Chemistry, ML, Noise characterization, Optimization \\
 \hline
  DQCSim \cite{DQCsim} & python, C++, Rust & No & No & -- \\
  \hline
 IQS \cite{intel} & C++ & No & No & QAOA \\
  \hline
 $|Lib\rangle $\cite{Lib} & python, C++  & Yes** & Yes & -- \\
 \hline
 Pennylane \cite{pennylane} & python  & Yes** & Yes & ML, Optimization \\
 \hline
 ProjectQ \cite{projectq} & python  & Yes** & Yes & Fermionic simulation, Optimization \\
 \hline
 pyquil/Forest \cite{pyquil} & python, Lisp & Yes & No & VQE, QAOA, Noise characterization, Optimization\\
 \hline
 QDK \cite{qsharp} & python, C\#,  Q\#  & Yes** & No & Chemistry, Optimization \\
 \hline
  Qibo \cite{qibo} &  python  & Yes* & Yes & VQE, QAOA, Adiabatic evolution, Optimization \\
 \hline
 Qiskit \cite{Qiskit} &  python  & Yes & No & Chemistry, ML, Optimization, Finance, Noise characterization\\
 \hline
 QTensor~\cite{qtensor} & python, C++ & No & No & Tensor-Network simulator, quantum circuit simulator, QAOA\\
 \hline
 QuEST \cite{quest,jones2020questlink} & C++, Mathematica &  No & Yes &  -- \\
 \hline
 Quimb~\cite{quimb} & python & No & No & Tensor-Network simulator, quantum circuit simulator\\
 \hline
 Qulacs~\cite{qulacs} & python, C++ & No & No & Simulator, noise characterization\\
 \hline
  StrawberryFields \cite{strawberry} & python & Yes* & No & GBS \\
 \hline
 Tequila \cite{tequila} & python  & Yes** & Yes & Chemistry, ML, Noise characterization, Optimization\\
 \hline
XACC \cite{xacc} & python, C++ &  Yes** & Yes & Optimization, VQE, QAOA, RBM and other algorithms \\
 \hline
 Yao \cite{yao} & Julia & No & No & --\\
 \hline
 \end{tabularx}
    \caption{List of open-source quantum software libraries. These packages are designed using common computing languages such as python, C++ or Julia. Some of them can be used in real quantum hardware, either because the developers are also building these devices or because the package include other quantum libraries. Some of them can translate their code to other quantum libraries or simulators. 
    *Hardware not publicly available yet. **Not on own hardware but can be run in a hardware backend. }
    \label{tab:software}
\end{table*}

\clearpage
\subsection{Table of external libraries}
\begin{table*}[h!]
    \centering
 \begin{tabularx}{\linewidth}{L{1}|C{0.5}|C{0.5}}
 \hline
 \hline
 \multicolumn{1}{c}{\textbf{Name}} & \textbf{Language} & \textbf{Application} \\ 
 \hline
OpenFermion \cite{OpenFermion} & python & Chemistry \\ 
\hline
psi4 \cite{psi4} & python & Chemistry \\ 
\hline
PySCF \cite{pyscf1, pyscf2} & python & Chemistry \\ 
\hline
NWChem \cite{nwchem} & Fortran 77, C & Chemistry \\ 
\hline
EntropicaQAOA \cite{EntropicaQAOA} & python & QAOA \\
\hline
TensorFlowQ \cite{TFQ} & python & QML \\ 
\hline
TensorFlow \cite{TF} & python & ML \\ 
\hline
Mitiq \cite{larose2020mitiq} & python & Error Mitigation \\ 
\hline
pyzx \cite{pyzx} & python & Compiler \\ 
\hline
quilc \cite{quilc} & quil & Compiler \\
\hline
Q-Convert \cite{qconvert,qconvert_js} & JavaScript, python & quantum language converter \\
\hline
iTensor \cite{itensor} & Julia,C++ & Tensor Networks \\
\hline
OpenQL \cite{OpenQL} & python, C++ & Compiler \\ 
\hline
JKQ \cite{JKQ} & C++ & Simulator and Compiler \\ 
\hline
ScaffCC \cite{scaffcc} & Scaffold & Compiler \\ 
 \hline
 staq \cite{staq} & C++ & Compiler \\ 
\hline
 Silq \cite{silq} & D & Compiler \\ 
 \hline
QX simulator \cite{QXsim} & python, C++ & Simulator \\ 
\hline
QRACK \cite{qrack} & C++ & Simulator \\ 
\hline
quantum-circuit \cite{qsimulator} & JavaScript & Simulator \\
\hline
QuTip \cite{Qutip} & python & Quantum Info. SDK\\ 
 \hline
 Q-Ctrl \cite{QCtrl} & python & Quantum Control \\ 
\hline
 \end{tabularx}
    \caption{External open-source libraries useful for the NISQ era. These libraries have applications in chemistry, machine learning, circuit compilation and quantum control. Some of them are integrated in the quantum software libraries listed in Table \ref{tab:software}.}
    \label{tab:ext_soft}
\end{table*}

\section{Classical optimization strategies}
\twocolumngrid
In this section, we detail the algorithms and strategies used to optimize the parameters of the PQC. For completeness, we reproduce part of the text shown in the main article and add the corresponding details on the methods.

\subsection{Gradient-based approaches}

\paragraph{Finite difference.}
    This method approximates the gradient of a function $f(\theta)$ as follows:
    \begin{equation}
        \partial_{i} f(\boldsymbol{\theta}) \approx \frac{f(\boldsymbol{\theta} + \epsilon \mathbf{e}_i) - f(\boldsymbol{\theta} - \epsilon \mathbf{e}_i)}{2\epsilon},
    \end{equation}
    where $\epsilon$ is a small number and $\mathbf{e}_i$ is the unit vector with 1 as its $i$-th element and 0 otherwise. The smaller $\epsilon$, the closer the right-hand side of above formula is to the true value of the gradient. However, for small $\epsilon$ the difference of the numerator 
    becomes small as well. As the objective function is an expectation value sampled from the quantum device and it is only estimated with limited accuracy, smaller $\epsilon$ require more samples taken from the quantum hardware to achieve a good estimation of the gradient.

\paragraph{Parameter-shift rule.}
    The analytical gradient can be calculated on quantum hardware using the parameter-shift rule, which was originally proposed in \cite{romero2018strategies} and developed in~\cite{mitarai2018quantum,schuld2019evaluating}. A key advantage is that the gradient is exact even if the difference parameter $\epsilon$ is chosen to be a large number (commonly $\epsilon=\pi/2$), avoiding the issues of the finite difference method.
    We assume that the unitary to be optimized can be written as $U(\boldsymbol{\theta}) = V G(\theta_i) W$, where $G=e^{-i \theta_i g}$ is the unitary affected by the parameter $\theta_i$, $g$ is the generator of $G$ and $V, W$ are unitaries independent of $\theta_i$. If $g$ has a spectrum of two eigenvalues $\pm \lambda$ only, the gradient can be calculated by measuring the observable at two shifted parameter values as follows:
    \begin{equation}
        \partial_{i}\langle f(\boldsymbol{\theta}) \rangle = \lambda \left( \langle f(\boldsymbol{\theta}_+) \rangle - \langle f(\boldsymbol{\theta}_-) \rangle \right),
    \end{equation}
    where $\boldsymbol{\theta}_{\pm} = \boldsymbol{\theta} \pm (\pi / 4\lambda) \boldsymbol{e}_i$. 
    
    This rule can be generalised to the case where the generator $g$ does not satisfy the eigenspectrum condition by decomposing the unitary into commuting terms as $G =  G_1 G_2 .. G_n = e^{-i \theta_i (g_1 + g_2 + .. + g_n)}$, where the generator $g_m$ of $G_m = e^{-i \theta_i g_m}$ satisfies that condition. We can then use the parameter-shift rule on each $G_m$ and calculate the analytical gradient using the product rule. This has been further developed for calculating analytical gradients for fermionic generators of Unitary Coupled-Cluster operators~\cite{kottmann2020feasible} and higher order derivatives~\cite{mari2020estimating}. 
    
    One can also use an auxiliary qubit and controlled unitaries to evaluate the gradient of multi-qubit unitaries where the parameter-shift rule does not apply. This was originally proposed in the context of unitary coupled-cluster~\cite{romero2018strategies} and later generalized for arbitrary gradients~\cite{schuld2019evaluating, yuan2019theory}. A further alternative is the stochastic parameter-shift rule~\cite{banchi2020measuring}, which relies on stochastically sampling scaled evolutions of the generator.
    
\paragraph{L-BFGS.}
    It is a quasi-Newton method that efficiently approximates the ``inverse Hessian" using a limited history of positions and gradients \cite{liu1989limited}. ``Inverse Hessian'' refers to the inverse of the Hessian matrix, where Hessian is a square matrix of second-order partial derivatives of the loss function. The inverse Hessian is employed to adjust gradient updates to the current loss landscape. L-BFGS is a memory-efficient variant of the BFGS method, which stores dense approximations of the inverse Hessian \cite{fletcher2000mathematics}.
    While effective in simulations, recent studies observed BFGS methods do not perform well in experimental demonstrations of VQA due to the level of noise in the cost function and gradient estimates \cite{lavrijsen2020classical}. 
    
    Two heuristics were proposed to find quasioptimal parameters for QAOA using BFGS \cite{zhou2020quantum}: \textit{i)} \textsc{INTERP}, where the optimized parameters at QAOA level $p$ are linearly interpolated and used as initial parameters for the level $p+1$ optimization; and \textit{ii)} \textsc{FOURIER}, where instead of optimizing the $2p$ QAOA parameters $\boldsymbol{\gamma}$ and $\boldsymbol{\beta}$ in \eqref{eq:QAOA_evolved_state}, one can instead optimize $2q$ new parameters $\boldsymbol{u} \equiv (u_1, u_2, \ldots u_q)$ and $\boldsymbol{v} \equiv (v_1, v_2, \ldots v_q)$ defined via the discrete sine and cosine transformations
    \begin{equation}
    \begin{split}
        \gamma_i(\beta_i) = \sum_{j=1}^{q} u_j(v_{j}) \sin(\cos) \left[ \left(i-\frac{1}{2}\right) \left(j-\frac{1}{2}\right) \frac{\pi}{p} \right].
    \end{split}
    \end{equation}
    Similarly to \textsc{INTERP}, the optimal parameters found at level $p$ are used to initialize the parameters for level $p+1$. Efficient initialization of parameters has also been reported using the Trotterized quantum annealing (TQA) protocol~\cite{sack2021quantum}. Note that these heuristic strategies can be easily extended to gradient-free optimization methods such as Nelder-Mead.
    
\paragraph{Quantum natural gradient.}
    The update rule of standard gradient descent \eqref{eq:grad_update} has the implicit assumption that the underlying parameter space is a flat Euclidean space. However, in general this is not the case, which can severely hamper the efficiency of gradient descent methods. In classical machine learning, the natural gradient was proposed that adapts the update rule to the non-Euclidean metric of the parameter space~\cite{amari1998natural}. 
    As an extension to the realm of parameterized quantum circuits, the quantum natural gradient (QNG) has been proposed~\cite{stokes2020quantum}. The update rule for this method is
    \begin{equation}
\theta^{(t+1)}_i=\theta^{(t)}_i-\eta \  \mathcal{F}^{-1}(\boldsymbol{\theta})\partial_{i} f(\boldsymbol{\theta})\,,
\label{eq:natural_grad_update_sup}
 \end{equation}
where $\mathcal{F}(\boldsymbol{\theta})$ is the Fubini-Study metric tensor or quantum Fisher information metric given by
\begin{equation}
    \mathcal{F}_{ij} = \text{Re}(\braket{\partial_i\psi(\boldsymbol{\theta}) \vert\partial_j \psi(\boldsymbol{\theta})}- \braket{\partial_i\psi(\boldsymbol{\theta} )\vert\psi(\boldsymbol{\theta})} \braket{\psi(\boldsymbol{\theta})\vert\partial_j \psi(\boldsymbol{\theta})})\,.
    \label{eq:Fubiny_Study_sup}
\end{equation}
Superior performance of the QNG compared to other gradient methods has been reported~\cite{yamamoto2019natural,stokes2020quantum} and it has been shown that it can avoid becoming stuck in local minima~\cite{wierichs2020avoiding}. It can be generalized to noisy quantum circuits~\cite{koczor2019quantum}. 
The QNG can be combined with adaptive learning rates $\eta(\theta_i^{t}$ that change for every step of gradient descent to speed up training. For hardware efficient PQCs, one can calculate adaptive learning rates using the quantum Fisher information metric~\cite{haug2021optimal}.
While the full Fubini-Study metric tensor is difficult to estimate on quantum hardware, diagonal and block-diagonal approximations can be efficiently evaluated~\cite{stokes2020quantum} and improved classical techniques to calculate the full tensor exist~\cite{jones2020efficient}. A special type of PQC, the natural PQC, has a euclidean quantum geometry such that the gradient is equivalent to the QNG close to a particular set of parameters~\cite{haug2021natural}.
    
\paragraph{Quantum imaginary time evolution.}
    Instead of using the standard gradient descent for optimization, a variational imaginary time evolution method was proposed in \cite{mcardle2019variational} to govern the evolution of parameters. They focused on many-body systems described by a $k$-local Hamiltonian 
    and considered a PQC that encodes the state $\ket{\psi(\tau)}$ as a parameterized trial state
    $\ket{\psi(\boldsymbol{\theta}(\tau))}$.
    The evolution of $\boldsymbol{\theta}(\tau)$ with respect to all the parameters can then be  obtained by solving the following differential equation
    \begin{equation}
        \sum_j A_{ij} \partial_{\tau}\theta_j(\tau) = C_i, 
    \end{equation}
    with 
        $A_{ij} = \text{Re}\left(\langle\partial_{i}\psi(\boldsymbol{\theta}(\tau))\vert\partial_{j}\psi(\boldsymbol{\theta}(\tau))\rangle\right)$ and 
        $C_i = \text{Re} \left(- \sum_\alpha c_\alpha \bra{\partial_{i}\psi(\boldsymbol{\theta}(\tau))} h_\alpha \ket{\psi(\boldsymbol{\theta}(\tau))}\right)$,
    where $h_\alpha$ and $c_\alpha$ are the Hamiltonian terms and coefficients. It was later shown in~\cite{stokes2020quantum} that the matrix $A_{ij}$ is related to the Fubini-Study metric tensor from \eqref{eq:Fubiny_Study_sup}, and the imaginary time evolution is analogous to the gradient descent via the QNG when considering infinitesimal small step sizes.

\paragraph{Hessian-aided gradient descent.}
    A recent work~\cite{huembeli2020characterizing} proposed computing the Hessian and its eigenvalues to help analyze the cost function landscapes of quantum machine learning algorithms.
    Tracking the numbers of positive, negative, and zero eigenvalues provides insight whether 
    the optimizer is heading towards a stationary point.
    The Hessian can be computed by doubly applying the parameter shift rule as \cite{mitarai2019methodology}
    \begin{multline}
        \partial_{i} \partial_{j} f(\boldsymbol\theta) = \\
        \frac{1}{2} \bigg( \expval{f(\boldsymbol\theta_{\lnot (i,j)}, \theta_i + \alpha, \theta_j + \alpha)}{} + \expval{f(\boldsymbol\theta_{\lnot (i,j)},\theta_i - \alpha, \theta_j - \alpha)}{} \\
        - \expval{f(\boldsymbol\theta_{\lnot (i,j)},\theta_i - \alpha, \theta_j + \alpha)}{} - \expval{f(\boldsymbol\theta_{\lnot (i,j)},\theta_i + \alpha, \theta_j - \alpha)}{} \bigg),
    \end{multline}
    where the shift parameter $\alpha=\frac{\pi}{4\lambda}$ for gates generated by operators with eigenvalues $\pm \lambda$. Other parameters, i.e. parameters not at the $i$-th and $j$-th indices, denoted $\boldsymbol\theta_{\lnot (i,j)}$, are fixed. 
    To improve optimization, they propose setting the learning rate to the inverse of the largest eigenvalue of the Hessian.
    Numerical simulations of the Hessian-based method and QNG both showed improvement over standard gradient descent in the ability to escape flat regions of the parameter landscape, with the former requiring fewer training epochs than QNG. 
    While a deeper analysis is necessary to more closely compare the performance, both QNG and Hessian-based methods accelerate optimization by leveraging local curvature information.
    
\paragraph{Quantum Analytic Descent.}
    A method consisting of using a classical model of the local energy landscape to estimate the gradients is proposed in \cite{koczor2020quantum}. 
    In this hybrid approach, a quantum device is used to construct an approximate ansatz landscape and the optimization towards the minima of the corresponding approximate surfaces can be carried out efficiently on a classical computer. The method considers the ansatz circuit as a product of $m$ unitary operations as 
    $U(\boldsymbol{\theta}) = U_m(\theta_m)\cdots U_1(\theta_1)$
    which, without loss of generality, can be approximated around a reference point $\theta_0$ as:
    \begin{multline}
        U(\boldsymbol{\theta}) = A(\boldsymbol{\theta}) U^{(A)} + \sum_{k=1}^v\left[B_k(\boldsymbol{\theta})U_k^{(B)} + C_k(\boldsymbol{\theta})U_k^{(C)} \right] \\
        + \sum_{l>k}^v[D_{kl}(\boldsymbol{\theta})U_{kl}^{(D)}] + O(\sin^3\delta),
    \end{multline}
    where $A$, $B_k$, $C_k$, $D_{kl}$ : $\mathbb{R}^v \rightarrow \mathbb{R}$ are products of simple univariate trigonometric functions, $U^{(A)}$, $U^{(B)}_k$, $U^{(C)}_k$, $U^{(D)}_{kl}$ are discrete mappings of the gates and $\delta$ is the absolute largest entry of the parameter vector. Using this approximate ansatz landscape, the full energy surface, gradient vector and metric tensor can be expressed in term of the ansatz parameters. The analytic descent has been shown to achieve faster convergence as compared to the QNG~\cite{koczor2020quantum}. 
    

\paragraph{Stochastic gradient descent.} 
    A major drawback of gradient-based methods is the high number of measurements.
    The stochastic gradient descent (SGD) algorithm 
    addresses this issue by replacing the normal parameter update rule with a modified version
    \begin{equation}
        \boldsymbol{\theta}^{(t+1)} = \boldsymbol{\theta}^{(t)} - \alpha \  \boldsymbol{g}(\boldsymbol{\theta}^{(t)}),
    \end{equation}
    where $\alpha$ is the learning rate and $\boldsymbol{g}$ is an unbiased estimator of the gradient of the cost function.
    There are many choices for this estimator, for instance a measurement of the gradient with a finite number of shots~\cite{harrow2019low}.
    It was also shown that it is not necessary to include all Pauli terms in the evaluation of the cost function; sampling from a subset still results in well-behaved gradient estimator.
    On top of that, it is possible to go even further by combining this technique with sampling of the parameter-shift rule terms~\cite{sweke2020stochastic}.
    
    In the doubly stochastic gradient, finite measurements are performed for only a subset of the expectation values of 
    the Hamiltonian terms.
    This sampling can be performed in the extreme situation where only one Pauli-term is evaluated at a single point in the quadrature.
    This is a very powerful method that reduces the number of measurements drastically~\cite{anand2020experimental}.
    
    This method can be extended beyond circuits that allow the parameter-shift rule by expressing the gradient as an integral~\cite{banchi2020measuring}.
    The integral can be seen as an infinite sum of terms that can be sampled. To accelerate the convergence of SGD for VQA, two optimization strategies are proposed \cite{lyu2020accelerated}: \textit{i)} Qubit-recursive, where the optimization is first performed for a smaller quantum system and is then used as initial parameter guess for a larger quantum system; and \textit{ii)} Layer-recursive, which is similar to a greedy approach, where the parameters are sequentially updated layer-by-layer in the quantum circuit.

\subsection{Gradient-free approaches}

In this section, we discuss optimization methods for VQA that do not rely on gradients measured on the quantum computer.

\paragraph{Evolutionary algorithms.}
    Evolutionary strategies \cite{rechenberg1978evolutionsstrategien,schwefel1977numerische} are black-box optimization tools for high dimensional problems that use a search distribution, from which they sample data, to estimate the gradient of the expected fitness to update the parameters in the direction of steepest ascent. More recently, natural evolutionary strategies (NES)~\cite{wierstra2014natural} have demonstrated considerable progress in solving these high dimensional optimization problems. They use natural gradient estimates for parameter updates instead of the standard gradients. They have been adapted for optimization of VQA~\cite{zhao2020natural, anand2020natural} and have been shown to have similar performance as the state-of-the-art gradient based method. 
    
    The search gradients used in NES can be estimated as
    \begin{equation}
     \nabla J(\boldsymbol\theta) \approx \frac{1}{k} \sum_{n=1}^k f(\boldsymbol{z}_n) \nabla \log \pi(\boldsymbol{z}_n|\boldsymbol\theta),
    \end{equation}
    where $J(\boldsymbol\theta) = E_{\boldsymbol\theta}[f(\boldsymbol{z})]$ is the expected fitness, $\pi(\boldsymbol{z}|\boldsymbol\theta)$ is the density of the search distribution with parameter $\boldsymbol\theta$, $f(\boldsymbol{z})$ is the fitness for the corresponding sample $\boldsymbol{z}$ drawn from the search distribution and $k$ is the different number of samples drawn from the distribution. The (classical) Fisher matrix $\mathcal{F}_\text{C}$ for the natural gradient can be estimated as
    \begin{equation}
        \mathcal{F}_\text{C} \approx \frac{1}{k} \sum_{n=1}^k \nabla\log\pi(\boldsymbol{z}_n|\boldsymbol\theta)  \nabla\log\pi(\boldsymbol{z}_n|\boldsymbol\theta)^T
    \end{equation}
    and the parameter update can then be carried out as $\boldsymbol{\theta} = \boldsymbol{\theta} + \eta \cdot  \mathcal{F}_\text{C}^{-1} \nabla J(\boldsymbol{\theta})$. In \cite{anand2020natural} it is shown that NES, along with techniques like Fitness shaping, local natural coordinates, adaptive sampling and batch optimization, can be used for optimization of deep quantum circuits.
    
\paragraph{Reinforcement learning.}
    Several authors have used reinforcement learning (RL) to optimize the QAOA parameters \cite{khairy2019reinforcement,wauters2020reinforcement,yao2020noise,yao2020policy}. This framework consists of a decision-making agent with policy $\pi_{\boldsymbol{\theta}}(a|s)$ parameterized by $\boldsymbol{\theta}$, which is a mapping from the state space $s \in \{S\} $ to an action space $a \in \{A\}$. In response to the action, the environment provides the agent with a reward $r$ from the set of rewards $\{R\}$. The goal of RL is to find a policy which maximizes the expected total discounted reward. For more details, refer to \secref{sec:applications_ML_RL}. In the context of QAOA, for example, $\{S\}$ can be the set of QAOA parameters ($\boldsymbol{\gamma},\boldsymbol{\beta}$) used, $a$ can be the value of $\gamma$ and $\beta$ for the next iteration, and the reward can be the finite difference in the QAOA objective function between two consecutive iterations. The policy can be parameterized by a deep neural network with the weights $\boldsymbol{\theta}$. The policy parameters $\boldsymbol{\theta}$ can be optimized using a variety of algorithms such as Monte-Carlo methods~\cite{hammersley2013monte,sutton2018reinforcement}, Q-Learning~\cite{watkins1992q} and policy gradient methods~\cite{sutton2018reinforcement}.
    
\paragraph{Sequential minimal optimization.}
    In machine learning, the sequential minimal optimization (SMO) method~\cite{platt1998sequential} has proven successful in optimizing the high-dimensional parameter landscape of support vector machines.
    The method breaks the optimization into smaller components for which the solution can be found analytically.
    This method has been applied to variational circuit optimization~\cite{nakanishi2020sequential}, circuit optimization with classical acceleration~\cite{parrish2019jacobi} and circuit optimization and learning with Rotosolve and Rotosolect~\cite{ostaszewski2019quantum}.
    Although these algorithms heavily rely on the parameter-shift rule, they can be considered gradient-free methods.
    They exploit the sinusoidal nature of the expectation value of a specific operator $\hat{O}$ when all but one parameters in the variational circuit are fixed:
    \begin{equation}
        \langle \hat{O}\rangle\left(\boldsymbol{\theta}\right) = A\sin\left(\boldsymbol{\theta} + B\right) + C,
    \end{equation}
    where $A$, $B$ and $C$ are parameters that can be found analytically.
    This means that only three well-chosen circuit evaluations are needed to exactly determine these coefficients and the optimal value $\theta^*$ of the parameter $\theta$ for this operator is given by
    \begin{multline}
        \theta^* =  - \text{arctan2}\Bigg[2\langle \hat{O}\rangle\left(\varphi\right) - \langle \hat{O}\rangle\left(\varphi + \frac{\pi}{2}\right)-\langle \hat{O}\rangle\left(\varphi - \frac{\pi}{2}\right),\\
        \langle \hat{O}\rangle\left(\varphi + \frac{\pi}{2}\right) - \langle\hat{O}\rangle \left(\varphi - \frac{\pi}{2}\right)\Bigg]        +2\pi k-\varphi - \frac{\pi}{2}\,,
    \end{multline}
    where $\text{arctan2}$ is the 2-argument arctangent and
    for any integer $k$ and angle $\varphi$.
    The most straightforward choice is to set $\varphi = 0$ and choose $k$ such that $\theta^*\in (-\pi,\pi]$.
    The algorithm proceeds by looping over all the variational parameters until convergence.
    The method can be generalized to optimize more than one parameter at a time~\cite{nakanishi2020sequential,parrish2019jacobi} but no general analytical expression can be found here.
    One has to resort to numerical methods to find the solutions for the free parameters.
    
    SMO offers a versatile starting point that can be combined with more advanced search acceleration algorithms 
    like Anderson acceleration~\cite{anderson1965iterative} or direct inversion of the iterative subspace (DIIS)~\cite{pulay1980convergence}.
    The same tools have also been used to optimize categorical variables like rotation axes in the Rotoselect algorithm~\cite{ostaszewski2019quantum}.
    While cost efficient, sequential parameter optimization only takes into account local information (albeit exactly), which often causes the optimization to get stuck in local minima~\cite{koczor2020quantum}.
    One has to balance the speed of a local method like SMO with the global approximate information of methods like the quantum analytic descent for specific problems.

\paragraph{Surrogate model-based optimization.} 
    When function evaluations are costly, it pays off to not only use the current function value to inform a next parameter value, but to use all previous evaluations to extract information about the search space.
    The function values in memory are used to build a surrogate model, an auxiliary function that represents the full expensive cost function based on the current information.
    The surrogate model of the cost function can be evaluated cheaply and many of these evaluations can be used to inform the next parameter value at which to compute the cost function.
    The new evaluation of the true objective function is added to the set of function values and the surrogate model is iteratively refined until convergence.
    All optimization happens on the surrogate cost landscape, so no explicit derivatives of the cost function are needed.
    Through the use of a fitted cost function, these methods are also expected to be more resilient to noise.
    
    Several classical surrogate models have been included in the scikit-quant package~\cite{lavrijsen2020classical,scikit-quant}.
    In the Bound optimization by quadratic approximation (BOBYQA) algorithm~\cite{powell2009bobyqa}, a local quadratic model is formulated from the previous function values.
    It is then minimized in the trust region to obtain a new parameter value.
    When the evaluation at this new parameter value does not result in a lower function value, the trust region is altered and the quadratic model is optimized in this new parameter space.
    It was shown that this method works well when the PQC is initialized close to the optimal parameters but has more problems with shallow optimization landscapes and gets stuck in local minima~\cite{lavrijsen2020classical}.
    The stable noisy optimization by branch and fit (SnobFit)~\cite{huyer2008snobfit} algorithm uses a branching algorithm to explore new areas in parameter space.
    In these areas it proposes several evaluation points and from the function values it fits a quadratic model.
    The algorithm combines this local search with an explorative generation of points in new areas of the parameter space.
    SnobFit performs well when tight bounds on the parameters are available~\cite{lavrijsen2020classical}.
    This can be achieved by combining the method with other optimizers that limit the size of the parameter space.
    
    In \cite{sung2020using}, a trust region with a least-squares fit to a quadratic function is used. They find that the minimum of the quadratic function often lies outside of the trust region, which causes the algorithms to constantly readjust it.
    Instead of using standard trust region optimization, they use either standard gradient descent or policy gradient descent optimization on the quadratic function to define a search direction in the parameter space.
    Both the model gradient descent (MGD) as the model policy gradient (MPG) have a hyperparameter that gradually shrinks the set of points used for the fit around the current minimum as the optimization progresses.
    The authors show that their algorithms are well suited for realistic conditions on near-term hardware.
    In particular, they study the ability of the algorithm to take into account a cloud access situation where circuits need to be uploaded in batches with a certain latency.
    The MPG outperforms MGD in the case of gate errors due to the ability to handle a large level of uncertaintity while learning its policy~\cite{sung2020using}.

\subsection{Resource-aware optimizers}

Optimization methods and strategies adopted for early demonstrations of VQA are largely general-purpose and black-box with minimal emphasis on reducing the quantum resources used in the optimization. Therefore, they are more costly and prone to errors than their classical counterparts.
Optimizers developed in more recent years are tailored to additionally minimize quantities associated with the quantum cost of the optimization, e.g. number of measurements or real hardware properties. Additionally, one can use circuit compilation methods as the ones described in \secref{sec:compilers}.
    
\paragraph{ROSALIN.} 
    While VQA leverage low-depth circuits to execute on near-term quantum processors, a significant challenge in implementing these algorithms is the prohibitive number of measurements, or shots, required to estimate each expectation value that is used to compute the objective.
    To address the challenge, \cite{arrasmith2020operator} developed a shot-frugal optimizer called ROSALIN (Random Operator Sampling for Adaptive Learning with Individual Number of shots) that effectively distributes fractions of a predefined number of shots to estimate each term of the Hamiltonian as well as each partial derivative.
    Given the expectation value of the Hamiltonian decomposed into the $h_{i}$ terms as in \eqref{eq:Pauli_string},
    the authors note several strategies for allocating shots for estimating each term $\expval{h_i}{}$. 
    While a naive strategy would allocate equal numbers of shots per term, the authors observed lower variance in the energies using weighted approaches in which the number of shots allocated to the $i$-th term $b_i$ is proportional to the corresponding Hamiltonian coefficient $c_i$. For instance, in the \emph{weighted deterministic sampling} method, $b_i = b_{\text{tot}}\frac{|c_i|}{\sum_i |c_i|}$ where $b_{\text{tot}}$ is the total number of shots. In the \emph{weighted random sampling} method, $b_i$ is drawn from a multinomial distribution with the probability of measuring the $i$-th term weighted by $|c_i|$, i.e. $p_i = \frac{|c_i|}{\sum_i |c_i|}$.
    
    In addition, ROSALIN employs iCANS (individual Coupled Adaptive Number of Shots), an optimizer that allocates shots for partial derivatives, as a subroutine \cite{kubler2020adaptive}.
    As a brief overview, the iCANs algorithm allocates measurements for each partial derivative such that the expected gain per shot is maximized. This gain depends on quantities such as the learning rate, the Lipschitz constant of the cost function, and estimates of gradient components and their variances. 
    Through VQE optimizations, ROSALIN was shown to outperform other optimizers such as iCANS and Adam especially in the presence of noise.  
    
\paragraph{SPSA.}
    In experimental realizations of VQA, the optimizer is often hindered by statistical noise.
    In \cite{kandala2017hardware} this issue is circumvented by applying the simultaneous perturbation stochastic approximation (SPSA) algorithm \cite{spall1992multivariate}, in which the algorithm hyperparameters are determined by experimental data on the level of statistical noise.
    Compared to the finite-difference gradient approximation, which requires $O(p)$ function evaluations for $p$ parameters, SPSA requires only two evaluations.
    That is, for a small positive $b_k$, the gradient at $k$-th iteration is approximated as
    \begin{equation}
        \boldsymbol{g}_k(\boldsymbol\theta_k) = \frac{\bra{\psi(\boldsymbol\theta_k^+)}{H} \ket{\psi(\boldsymbol\theta_k^+)} - \bra{\psi(\boldsymbol\theta_k^-)} {H} \ket{\psi(\boldsymbol\theta_k^-)}}{2 b_k \ \boldsymbol\Delta_k}, 
    \end{equation}
    where $\boldsymbol\Delta_k$ is a random perturbation vector and $\boldsymbol\theta_k^{\pm} = \boldsymbol\theta_k \pm b_k \boldsymbol\Delta_k$. 
    After computing the gradient estimate, the next parameter settings are updated with learning rate $a_k$.
    In general, $b_k$ and $a_k$ decrease over iterations, e.g. $b_k = \frac{b}{k^\gamma}$ and $a_k = \frac{a}{k^\alpha}$ for some fixed $\gamma$ and $\alpha$. 
    The values of $b$ and $a$ are carefully estimated to be robust against statistical noise based on samples of energy differences $|\bra{\psi(\boldsymbol\theta_k^+)} H \ket{\psi(\boldsymbol\theta_k^+)} - \bra{\psi(\boldsymbol\theta_k^-)} H \ket{\psi(\boldsymbol\theta_k^-)}|$.
    The convergence of SPSA with various types of PQCs has been studied~\cite{woitzik2020entanglement}.

\section{NISQ Applications for Finance}\label{sec:finance}
\twocolumngrid

The subject of finance deals with money and investments. The three
typical subdivisions of this field are; personal finance, corporate finance
and public finance. Due to its practical relevance, comprehensive
and rigorous investigations have been carried out to comprehend finance using
techniques from mathematics. Some of the typical applications from
finance which could potentially benefit from quantum technologies are
portfolio optimization, modelling financial markets via differential
equations and predicting market trends.
In the following lines, we present one possible approach to classify mathematical problems in finance \cite{egger2020quantum}.

\paragraph{Optimization based. } Many of the finance problems involve decision-making subject to certain constraints and consequently fall in the optimization framework. Some of the typical examples of optimization problems in finance are portfolio optimization, portfolio diversification and auctions. Concepts from convex optimization and combinatorial optimization turn out to be appropriate for the aforementioned class of problems. A considerable number of optimization centric problems can be converted to QUBO (see \eqref{eq:QUBO_1} in section \secref{subsec:quantum-annealing}). Quantum algorithms for linear systems, convex optimization and QUBO have been employed with the hope for a possible advantage. In particular, the problems which can be reduced to QUBO have been investigated extensively using quantum annealing, VQE and QAOA~\cite{bouland2020prospects,hodson2019portfolio}. 

\paragraph{Simulation based. } 
The simulation problems in finance deal with simulating potential outcomes, typical examples being simulating the influence of volatility on risk or estimating asset values for pricing. Monte-Carlo simulations and algorithms for stochastic processes are particularly beneficial for such tasks. Consequently, quantum Monte-Carlo algorithms and quantum algorithms for modelling stochastic processes have been investigated in the context of finance~\cite{kubo2020variational,bouland2020prospects,egger2020quantum,ramos2019quantum,blank2020quantum2}.

\paragraph{Machine learning based. }
Some of the standard machine learning based problems in finance require predicting a future event based on historical data, pattern and anomaly detection, and classification of the end result in categories. Sample problems are fraud detection, algorithmic trading, risk assessment and credit scoring. One can potentially hope to apply QML algorithms as the ones presented in \secref{sec:QML} for such tasks.

The complete coverage of quantum finance is outside the scope of this
review. We direct the reader to \cite{egger2020quantum,bouland2020prospects,orus2019quantum} for a comprehensive treatment.
We proceed to discuss portfolio optimization and fraud detection; two canonical examples from finance. The algorithms that appear in this subsection are listed in \tabref{tab:finance}.

\subsection{Portfolio optimization}

The mathematical notion of portfolio optimization, was first proposed in~\cite{markowitz1952portfolio}. Given some capital and set of
$m$ assets, the goal is to find the best
investment strategies under a set of constraints. Some of the typical
constraints could be non-negativity of the investment or limit on
total budget $B$. The objective function could be variance of the
whole portfolio or overall risk. Let us represent the overall portfolio
allocation by $x\in\mathbb{R}^{m}$, where $x_{k}$ denotes the investment
in the $k$-th asset. Suppose the return at time step $t$ is $c(t)\in\mathbb{R}^{m}$
with expected return $\mu\in\mathbb{R}^{m}$and covariance matrix
$\Sigma\in\mathbb{R}^{m\times m}$ given by
\begin{align}
\mu(T)=&\frac{1}{T}\sum_{t\in\left[T\right]}c(t),\label{eq:mean}\\
\Sigma(T)=&\frac{1}{T-1}\sum_{t\in\left[T\right]}\left(c(t)-\mu\right)\left(c(t)-\mu\right)^{T}.\label{eq:Covariance}
\end{align}
With some additional equality constraints captured by $A\in\mathbb{R}^{m\times m}$
and $b\in\mathbb{R}^{m},$ such as the total budget constraint $\sum x_{i}=B,$
the portfolio optimization task corresponds to the following program,
\begin{equation} \min_{x\in\mathbb{R}^{m}}x^{T}\Sigma x\numberthis\label{eq:Opt_1}\,,\hspace{0.2cm}
\text{s.t }\,c^{T}x=\mu\,,\hspace{0.3cm}
A^{T}x=b.
\end{equation}

We can introduce Lagrange multipliers $\eta$ and $\theta$ for the
equality constraints, and get the following Lagrangian corresponding
to the aforementioned program,
\begin{equation}
\mathcal{L}\left(x,\eta,\theta\right)=\frac{1}{2}x^{T}\Sigma x+\eta\left(c^{T}x-\mu\right)+\theta\left(A^{T}x-b\right).\label{eq:Lagrangian_portfolio_1}
\end{equation}
The solution to the portfolio optimization \eqref{eq:Opt_1} reduces
to solving the following linear system,
\begin{equation}
\left[\begin{array}{ccc}
0 & 0 & c^{T}\\
0 & 0 & A^{T}\\
c & A & \Sigma
\end{array}\right]\left[\begin{array}{c}
\eta\\
\theta\\
x
\end{array}\right]=\left[\begin{array}{c}
\mu\\
b\\
0
\end{array}\right]\label{eq:Linear_system_portfolio_1}
\end{equation}

In ~\cite{rebentrost2018quantum} the HHL algorithm was proposed to implement the above approach with a quantum computer. One can use near-term linear
system solvers for portfolio optimization tasks given assumptions
about hardware capabilities and input model are satisfied~\cite{huang2019nearterm}.  A modified
version of the program from \eqref{eq:Opt_1} was recently tackled using
VQE and QAOA~\cite{egger2020quantum}. Specifically, the following optimization program was
attempted,
\begin{equation}
\min_{x\in\left\{ 0,1\right\} ^{m}}qx^{T}\Sigma x-\mu^{T}x\numberthis\label{eq:obj_2_finance}\hspace{0.33cm},\text{s.t.}\,\,1^{T}x=B.
\end{equation}
Here, the portfolio vector $x\in\left\{ 0,1\right\} ^{m}$ is a vector
of binary variables where $x_{i}=1$ means the $i$th asset was selected.
$x_{i}=0$ means otherwise. Similar to the previous case, $\mu$ and
$\Sigma$ denote expected return and covariance matrix respectively.
The total budget has been denoted by $B$ and $q>0$ captures the
risk appetite of the decision maker. It is assumed that all assets
have same price and total budget has to be spent. By adding the constraint
as a penalty term $\left(1^{T}x-B\right)^{2},$ the authors convert
optimization program \eqref{eq:obj_2_finance} to a QUBO problem, which further reduces to an appropriate
Hamiltonian ground state problem. They employ both VQE and QAOA hereafter.
They also study portfolio diversification, by again converting the
same to QUBO~\cite{egger2020quantum}. 
Since problems which can be converted in QUBO can be a good fit for annealing based optimization, 
quantum annealing has been extensively applied for portfolio optimization~\cite{marzec2016portfolio,rosenberg2016solving,venturelli2019reverse,cohen2020portfolio}. For details, refer to~\cite{orus2019quantum,bouland2020prospects}. In \cite{alcazar2021enhancing}, a quantum
optimization strategy exploiting generative machine learning models to bypass the need for constructing QUBO or PUBO cost functions has been proposed. Besides acting as a black-box solver (i.e., an optimizer which is agnostic to the details of the cost function), it was shown that this quantum portfolio optimization strategy scales to industrial data sets such as the entire S\&P 500 by leveraging quantum-inspired models based on tensor networks, instead of using hardware models.

\subsection{Fraud detection}

Machine learning can be utilised to detect financial frauds. A typical
example could be somebody's credit card getting blocked because higher
than usual transaction was under process. Based on the historical
data, often a fraudulent transaction can be caught. There are, however,
false positives also. A typical example of false positive is somebody's
credit card transaction getting declined when trying it at a new
store. Based on synthetic credit card transaction data in Ref.~\cite{altman2019synthesizing}, variational quantum classification was performed in Ref.~\cite{egger2020quantum}. The example problem in Ref.~\cite{egger2020quantum} contains dataset with $100$
purchase transaction records with each transaction $k$ marked fraudulent
$\left(y_{k}=-1\right)$ or not fraudulent $\left(y_{k}=1\right)$.
The $k$-th transaction input vector $x_{k}$ contains information
about transaction amount, time, method and location. Variational quantum classification was used to predict the right label. A similar analysis
using variational QBM was carried on in Ref.~\cite{zoufal2020variational}. NISQ algorithms for anomaly detection, such as ~\cite{herr2020anomaly}, could also be used for detecting financial frauds.
\section{Unitary t-design} \label{sec:t-design}

The primary objective of unitary $t$-designs is to substitute with a finite sum the integration
over the space of unitaries. The aforementioned
approach provides an accessible way to find the average of functions
over unitaries and then prove intriguing theorems about them. The
set of unitaries $\mathbb{U}(d)$ forms a topologically compact and connected
group. Furthermore, they correspond to the set of norm-preserving
matrices in $\mathbb{C}^{d}.$ We can define a unique translation-invariant
measure, the \textit{Haar measure}, on $\mathbb{U}(d)$ which can be employed to calculate
expectation values of functions of unitaries,
\begin{equation} 
\langle f\rangle=\int_{\mathbb{U}(d)}f(U) \ dU.
\label{eq:Haar_expectation}
\end{equation}
We present the aforementioned statement in a relatively formal manner.
\begin{defn} \label{def:Haar}
Let $\mathbb{U}$ be a group of $n\times n$ unitaries. A probability measure $\mu$
on $\mathbb{U}$ is called Haar measure on $\mathbb{U}$ if for any
subset $\mathbb{S}\subseteq\mathbb{U}$ and for any fixed $K\in\mathbb{U}$,
we have 
\begin{equation}
\mu\left(K\mathbb{S}\right)=\mu\left(\mathbb{S}K\right)=\mu\left(\mathbb{S}\right)\label{eq:translation_invariance}
\end{equation}
where $K\mathbb{S}\equiv\left\{ KU:U\in\mathbb{S}\right\} $and $\mathbb{S}K\equiv\left\{ UK:U\in\mathbb{S}\right\}.$
The property in \eqref{eq:translation_invariance} is called
translation invariance. 
\end{defn}

Because of translation invariance, if $U_{1}$ is a Haar-distributed
random unitary matrix, then so are $UU_{1}$ and $U_{1}U$ for a fixed
unitary matrix $U.$ To present the concept of unitary designs, we need to first discuss
the notions of \textit{homogeneous polynomials} $\left(Hom(a,b)\right)$ and
\textit{weight functions}, defined in the context of unitary matrices. Any
polynomial of degree $a$ in the entries of $U\in \mathbb{U}(d)$ and $b$ in
$U^{\dagger}$ is called a homogeneous polynomial $Hom(a,b).$ Some
of the examples of homogeneous polynomials are $U^{\dagger}V^{\dagger}UV\in Hom(2,2)$
and $VU^{\dagger}VU\in Hom(3,1).$ A weight function on a set $S,$
$w:S\rightarrow(0,1]$, for all $U\in S$ satsfies the following two
properties: $(i)\,w(U)\ge0\text{ and  }(ii)\,\sum_{U\in S}w(U)=1.$ Having discussed the concepts of homogeneous polynomials and weight function, we proceed to define the notion of unitary t-designs. 
\begin{defn}
Unitary $t$-design: Given a finite set $S\subset \mathbb{U}(d)$ and a weight
function $w:S\rightarrow(0,1]$, the tuple $(S,w)$ is called a unitary
t-design if for all $f\in Hom(t,t)$
\begin{equation}
\sum_{U\in S}w(U)f(U)=\int_{\mathbb{U}(d)}f(U) \ dU.\label{eq:unitary_t_1}
\end{equation}
\end{defn}
We will refer to unitary $t$-design as $t$-design henceforth. Note that
to verify if a tuple $(S,w)$ forms a $t$-design, we need to check the condition from
\eqref{eq:unitary_t_1} for every function $f$ in $Hom(t,t).$ For
many cases, such an approach might be impractical and hence it requires
more tractable condition, such as the one from the following Lemma.
\begin{lem}
Given a finite set $S\subset \mathbb{U}(d)$ and a weight function $w:S\rightarrow(0,1]$,
the tuple$(S,w)$ forms a $t$-design if and only if $\sum_{U\in S}w(U)U^{\otimes t}\otimes\left(U^{\dagger}\right)^{\otimes t}=\int_{U(d)}U^{\otimes t}\otimes\left(U^{\dagger}\right)^{\otimes t}dU.$
\end{lem}


\end{appendices}

\end{document}